\documentclass[12pt]{JHEP3}
\input epsf.tex
\epsfclipon

\usepackage{epsfig}
\usepackage{epstopdf}
\usepackage{epsf}
\input{epsf.sty}
\usepackage{graphicx,amsmath,amssymb}
\usepackage{epsfig,multicol}

\usepackage{bbm,bm,amsmath,amssymb}
\def\blfootnote{\xdef\@thefnmark{}\@footnotetext}

\long\def\symbolfootnote[#1]#2{\begingroup%
\def\thefootnote{\fnsymbol{footnote}}\footnote[#1]{#2}\endgroup}

\newcommand{\be}{\begin{eqnarray}}
\newcommand{\ee}{\end{eqnarray}}
\newcommand{\ben}{\begin{eqnarray*}}
\newcommand{\een}{\end{eqnarray*}}

\newcommand{\bcent}{\begin{center}}
\newcommand{\ecent}{\end{center}}
\newcommand{\benum}{\begin{enumerate}}
\newcommand{\eenum}{\end{enumerate}}
\newcommand{\bdesc}{\begin{description}}
\newcommand{\edesc}{\end{description}}
\newcommand{\bitem}{\begin{itemize}}
\newcommand{\eitem}{\end{itemize}}
\newcommand{\bquote}{\begin{quote}}
\newcommand{\equote}{\end{quote}}
\newcommand{\bhalfp}{\begin{minipage}{0.45\textwidth}}
\newcommand{\ehalfp}{\end{minipage}}
\newcommand{\bhead}{\begin{center}\bf \Large}
\newcommand{\ehead}{\end{center}\bigskip}

 \newcommand{\bfk}{{\bf k}}

 \newcommand{\bfx}{{\bf x}}

 \newcommand{\hatpi}{{\hat{\pi}}}

 \newcommand{\ave}[1]{\langle {#1} \rangle}

\def\be{\begin{equation}}
\def\ee{\end{equation}}
\def\ba{\begin{eqnarray}}
\def\ea{\end{eqnarray}}

\newcommand{\roughly}[1]{\mathrel{\raise.3ex\hbox{$#1$\kern-0.85em
\lower1ex\hbox{$\sim$}}}}

\def\2pi{\left(2\pi\right)}
\def\ve{\overrightarrow}

\def\beq{\begin{equation}}
\def\eeq{\end{equation}}
\def\bg{\begin{eqnarray}}
\def\nd{\end{eqnarray}}
\def\bea{\begin{eqnarray}}
\def\eea{\end{eqnarray}}

\def\D3{\overline{\mbox{D3}}}

\title{Five Easy Pieces: The Dynamics of Quarks in Strongly Coupled Plasmas}
\author{Mohammed Mia, Keshav Dasgupta, Charles Gale, Sangyong Jeon\\
Ernest Rutherford Physics Building, McGill University,\\ 3600 University
Street, Montr{\'e}al QC, Canada H3A 2T8\\
\vskip.07in
{\tt miam, keshav, gale, jeon@hep.physics.mcgill.ca}}
\date{.. Jan 2009}

\abstract{We revisit the analysis of the drag a massive quark experiences and the wake it creates
at a temperature T 
while moving through a plasma using a gravity dual that captures the renormalisation group
 runnings in the dual gauge theory. 
Our gravity dual has a black hole and seven branes embedded via Ouyang embedding, but the geometry is 
a deformation of the usual conifold metric. In particular the gravity dual has squashed two spheres, and a 
small resolution at the IR. Using this background we show that the drag of a massive quark receives 
corrections that are proportional to powers of log T when compared with the drag computed using AdS/QCD 
correspondence. The massive quarks map to fundamental strings in the dual gravity theory. We use the perturbation 
produced by these strings to compute the wake and compare with the results obtained using AdS/QCD correspondence.
We also study the shear viscosity in the theory with running couplings, analyze the 
viscosity to entropy ratio and compare the result with the bound derived from AdS backgrounds. In the  
presence of higher order curvature square
corrections from the back-reactions of the embedded D7 branes, we argue the possibility of the entropy to 
viscosity bound being violated. Finally, we show that our set-up could in-principle allow us to study a family of
gauge theories at the boundary by cutting off the dual geometry respectively
at various points in the radial direction. All these 
gauge theories can have well defined UV completions, and more interestingly, we demonstrate that any thermodynamical
quantities derived from these theories would be completely independent of the cut-off scale and only depend on the 
temperature at which we define these theories. Such a result would justify the holographic renormalisabilities of these 
theories which we, in turn, also demonstrate.  
We give physical interpretations of these 
results and compare them with more realistic scenarios.}

\maketitle

\begin{document}

\section{Introduction}

There is no question that understanding the behavior of many-body QCD in the strong coupling regime
is a hard problem to solve, but this needs to be done. The wealth of intriguing experimental data obtained at the Relativistic Heavy Ion
Collider (RHIC) has made this situation abundantly clear. One of the main goals of the RHIC program is the creation and the analysis of the quark gluon plasma, a new phase of matter predicted by lattice QCD. The goal of this paper is to discuss several quantities that can be related to observables measured at RHIC, or to be measured at the LHC. Calculations are done in a regime where the gauge sector is nonperturbative, using techniques borrowed from string theory.

From the early days of the RHIC experiments, the appearance of 
a strong elliptic hydrodynamic flow was taken as a consequence of early thermalization (before 1 fm/c), and indicative of QGP formation \cite{Kolb:2003dz}.
The elliptic flow, defined as the second harmonic component of the 
momentum distribution, develops when the system undergoing the hydrodynamic
expansion has an elliptical shape with different short and long axes.  This
difference in the spatial shape causes difference in the pressure-gradient,
which in turn causes the particles to accelerate more in the short axis
direction.
The anisotropy in the acceleration then causes the final momentum
distribution to be anisotropic. The efficiency of this process of generating the momentum space
anisotropy from the spatial anisotropy, however,
depends on the size of the shear viscosity $\eta$: a quantity that  will be discussed here in some detail.  
%In the process we will be able to provide a physical reason for the 
%viscosity to the entropy bound.
Another experimental observable that has been linked to the formation of a plasma of quarks and gluons is the amount of energy lost by a fast parton travelling through this hot and dense medium. This phenomenon has also been dubbed {\it jet quenching}. Interestingly, there is a theoretical link between the concept of a small shear viscosity and that of a large jet quenching \cite{Majumder:2007zh}. The hard partons that travel through the strongly interacting plasma may also leave a wake behind, owing to the medium's response to the source which is the hard jet. We shall also discuss this in this paper. Somewhat related, the amount of energy lost by a heavy quark has been of great interest as well: the drag force can be related to the properties of the strongly interacting medium. Interestingly, the kinematics of  bound states with a heavy  quark can be observed through semileptonic decay channels. We compute the drag force experienced by a quark, as it looses energy to the surrounding medium. A brief introduction to some of these topics follows, before the general organization of the paper is outlined. 

\subsection{Shear viscosity and the viscosity to entropy bound}

The shear viscosity represents the strength of the {\em collective}
interaction between 
the two laminally flowing layers. Roughly speaking, large shear viscosity
means faster mixing of the particles in two neighboring laminas. 
Somewhat counter-intuitively, the strength of this collective interaction
is actually smaller when the microscopic interaction is stronger.
This is because the rate of mixing is controlled by the mean free path.
When the mean free path is small compared to the
typical size of the flow velocity variation, 
two laminas with different flow
velocities cannot easily mix since the exchange of particles
is limited to the small volume near the interface of the two laminas: Most 
particles in the fluid just flows
along as if there is no other layers nearby.
On the other hand, if the mean free path is comparable to the typical size
of the flow velocity variation, then mixing between different layers
can proceed relatively quickly.

When the elliptic flow develops, the fluid has anisotropic fluid
velocity distribution. 
Since the shear viscosity controls the mixing, a large shear viscosity
can quickly wash out these difference in the local fluid variables.
Therefore, one can
say that the smaller the shear viscosity, the stronger the elliptic
flow. 

In the weakly coupled QCD, the ratio $\eta/s$ is parametrically large since
it is of the order $1/\alpha_s^2\ln(\alpha_s)$ \cite{amy}. 
If one is to believe that the QGP created in relativistic heavy ion
collisions is in the weak coupling regime, one would then expect the
elliptic flow to be small.
One of the big surprises from the RHIC experiments is that this 
expectation is almost maximally violated. It turned out that
the ideal hydrodynamics where the value of $\eta$
is just set to $0$ consistently describes the elliptic flow at RHIC
pointing to a {\em strongly interacting} plasma well above the phase transition
temperature.
In fact, the QGP created at RHIC behaves like the most
perfect fluid ever observed. 

The analytic tools available to a theorist to tackle this issue,
however, have been rather limited.
Perturbation theory is obviously not valid
in the strong coupling regime and the lattice QCD study has been so far
limited to the Euclidean space where extraction of the dynamic quantities
such as the perfect-fluidity can be difficult. 

This situation changed dramatically when
Policastro, Son and Starinets \cite{pss} discovered that Maldacena conjecture
\cite{adscft} can be used to calculate the shear viscosity of a certain
strongly coupled quantum field exactly. 
More intriguingly, it turned out that in their solution the shear viscosity
and the entropy density ratio, $\eta/s$, had a minimum value at ${\hbar\over 4\pi k_B} \equiv 
{1\over 4\pi}$ \cite{pss, Kovtun1} where $k_B$ is the Boltzmann constant.
The authors \cite{Kovtun1} then 
made a conjecture that this is indeed the lower bound for any quantum field
theory which has a gravity dual.
Whether $1/4\pi$ is a true bound or not have been debated many times in the
literature \cite{Kats, violation, anindapaper}. But in all similar calculations, $\eta/s$ remains to be 
${\cal O}(1/4\pi)$ \cite{violation}. The question is thus: What can we learn
from this about strongly coupled QCD?

In this paper, we use the {\it non}-AdS/QCD theory to investigate this question
using a modified Klebanov-Strassler construction \cite{1, 4, sully, cotrone}.
In the IR limit, this theory eventually leads to a confining $SU(M)$
theory with a large $M$ and $N_f$ flavors\footnote{An alternative possibility is to get an 
approximately conformal theory with $N_f$ 
flavors under cascade. This can also be realized in our set-up but we will ignore this possibility.}. 
This is, of course not a full QCD, but it shares
many features with the real world QCD including the appearance of the
renormalisation scale\footnote{Note however that by QCD we will always mean a theory 
that resembles at IR large $N$ QCD (but UV strongly coupled and {\it almost} conformal) throughout the text because
it has a large number of colors. 
For finite $N$ there is no known gravity dual. We will however need to keep finite number of 
fundamental flavors because of certain constraints that will become clearer as we go along.}.  
There is no gravity-dual construction of QCD yet but our study may shed some
light on how the strongly coupled QCD should behave.

Before we go into the details of our calculations, it is instructional  
to consider the physical meaning of the shear viscosity and its behavior
and also why it makes sense to talk about the ratio $\eta/s$  as a measure
of its strength.

The definitions of the shear viscosity $\eta$ and the bulk viscosity $\zeta$
are given by the following constitutive equation
\be
\ave{\delta T_{ij}} 
=
-{\eta\over \varepsilon + P}
\left(\nabla_i\ave{T_j^0} + \nabla_j\ave{T_i^0} - {2\over
3}\delta_{ij}\nabla_l \ave{T^{l0}}\right)
-{\zeta\over \varepsilon + P}
\delta_{ij} \nabla_l \ave{T^{l0}}
\ee
where $\delta T^{ij}$ is the deviation from the ideal fluid stress
tensor in the fluid rest frame and $\varepsilon$ and $P$ are the local
energy density and the pressure, respectively.
Upon using the thermodynamic identity
$Ts = \varepsilon + P$ where $s$ is the entropy density,
the two coefficients can be also written as 
$\eta/Ts$ and $\zeta/Ts$.
Since the temperature is the only relevant energy scale in the highly
relativistic fluid, one can easily see that the importance of the viscous
terms depends on the size of the dimensionless ratios $\eta/s$ and $\zeta/s$.

To see what the shear viscosity signifies, consider situation where
the fluid is flowing in the $z$ direction but the speed of the flow 
varies in the $x$ direction (c.f. {\bf figure 1}).
That is, we have $\ave{T^{0x}} = \ave{T^{0y}} = 0$ everywhere, 
but $\ave{T^{0z}(x)}$ does not have to vanish everywhere.
Recall that the stress part, $T^{ij}$,
of the stress-energy tensor has the interpretation
of the $i$-th component of the current for
the conserved momentum density $T^{0j}$.   

If the value of the conserved density $\ave{T^{0z}(x)}$ at two different
points are different, then there must be a net current
between these two points.
Now consider a $z$-$y$ plane at a fixed $x$ and think about the amount of
microscopic current across this plane (c.f. {\bf figure 1}).
A particle crossing this plane from above in {\bf figure 1} has the
average $v_x$ according to the thermal distribution and average $p_z$
according to the flow velocity above the plane.
A particle crossing the plane from the below share the same (but opposite
sign) $v_x$ but the average $p_z$ in this case is proportional to 
the flow velocity below the plane.
\begin{figure}[htb]\label{layers}
		\begin{center}
\includegraphics[height=6cm]{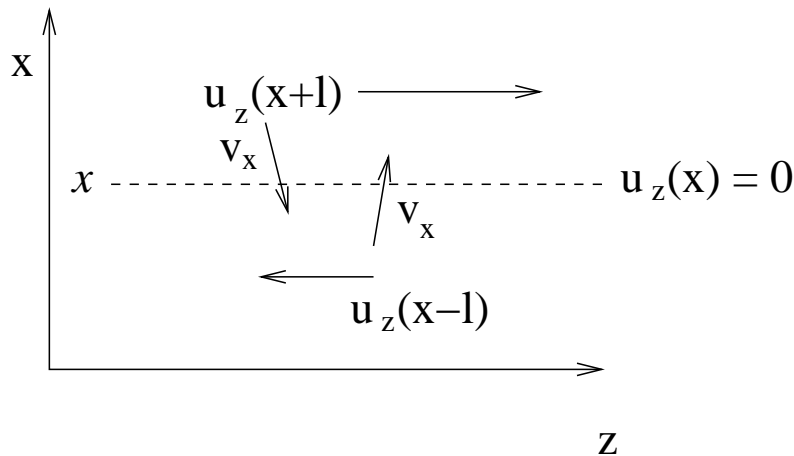}
		\caption{{ Difference of flows for a particle moving above and below the x-plane.}}
		\end{center}
		\end{figure} 
The net flow of $T^{0z}$ through this $z$-$y$ plane is then
\be
T_{xz}(x) \approx T^{0z}(x-l) \ave{|v_x|}_{\rm th}
- T_{0z}(x+l) \ave{|v_x|}_{\rm th}
\ee
Now for this expression to hold, the particles must not suffer a collision
while attempting to cross the plane. Hence the point of origin for the
border-crossing particles must not exceed the size of the mean free path.
Hence, Taylor expanding the above expression to the first order
and setting $l$ to be the mean free path $\lambda$ yields
\be
T_{xz}(x)
\sim -\lambda (\varepsilon+P) \ave{|v_x|}_{\rm th} \partial_x u_{z}(x) 
\ee
Here we used the fact that in
case of local equilibrium $T^{0i}_{\rm eq} = (\varepsilon + P)u^0 u^i$.
We also chose the frame where $u_z(x)=0$.
Comparing with the constitutive equation, one sees that
\be
\eta \sim \lambda (\varepsilon + P) \ave{v}_{\rm th}
\ee
where we have substituted $\ave{|v_x|}$ with the average speed
$\ave{v}$ for this rough estimate and discarded all $O(1)$ constants.
Upon using the thermodynamic identity
$Ts = \varepsilon + P$, this becomes 
\be
{\eta\over s}
\sim \lambda T \ave{v}_{\rm th} 
\label{eq:etaovers_estim}
\ee
{}From the discussion above, it is clear that the shear viscosity controls the
rate of the {\em momentum diffusion} in the transverse direction to the
flow. In fact, using the constitutive equation, it can be easily shown that
the momentum densities satisfy a diffusion equation with the diffusion
constant give by
\bg
D = {\eta\over \varepsilon + P} = {\eta \over Ts} \sim
\lambda\ave{v}_{\rm th}
\nd
{}From Eq.(\ref{eq:etaovers_estim}), we can also argue the existence of the
lower bound following \cite{DGu}.
The mean free path $\lambda$ is defined as the average distance between two
collisions. This length scale cannot become arbitrarily small due to the
uncertainty principle $\Delta x \Delta p \ge 1/2$.
The distance between two collisions
must be at least longer than the Compton wavelength $1/m$ and
the de Broglie wavelength $1/p$.
Since the factor $T\ave{v}_{\rm th}$ in Eq.(\ref{eq:etaovers_estim}) 
can be though of as the typical momentum scale in the 
medium, the $\eta/s$ ratio must satisfy 
\be
{\eta\over s} \ge B_{\rm low}
\ee
where $B_{\rm low}$ is a non-zero ${\cal O}(1)$ constant.
Therefore, the fact that $\eta/s$ is bounded from below by an ${\cal O}(1)$
constant is a rather robust
consequence of quantum mechanics as long as the entropy is not
completely dominated
by large chemical potential; 
cf. Ref. \cite{Cherman:2007fj}.
What the value of $B_{\rm low}$ is and whether the strongly coupled QCD has
the $\eta/s$ ratio close to this bound, of course, is the issue that we 
are discussing here.

In the theory we are using in this paper it will turn out that $\eta/s$ is smaller than the 
$1/4\pi$ bound of \cite{Kovtun1}. Of course it was argued earlier in \cite{Kats} that if one 
incorporates ${\cal O}(R^2)$ corrections (where $R$ is the curvature tensor) then the bound is 
automatically violated in the AdS space. What we find here is that if one goes beyond the AdS case by 
incorporating RG flows in the gauge theory the bound is violated from the 
curvature square corrections but the non-trivial RG flow do contribute to this also. 
We will discuss this in full details in {\bf sections 3.4} and 
{\bf 3.5} where we will argue that the required curvature squared corrections can be achieved by adding appropriate 
number of D7 branes to the background.

There are two main ingredients in our calculation 
of the $\eta/s$ ratio.
The calculation of the shear viscosity 
in a quantum field theory relies on the Kubo formula
\be
\eta = {1\over 20}\lim_{\omega\to 0}\lim_{\bfk\to 0}
{1\over \omega}
\int dt d^3 x\, e^{i\omega t - i\bfk{\cdot}\bfx}\,
\ave{[\hatpi_{ij}(t,\bfx), \hatpi^{ij}(0)]}_{\rm eq}
\ee
where $\hatpi_{ij}$ is the spin-2 part (traceless part) of the spatial
stress tensor and the average is taken in the fluid rest frame.
This Kubo formula relies only on the infra-red behavior in the linear
response theory. Hence, it is valid for an arbitrarily strong coupling
limit. 

For the entropy, we may use the Beckenstein-Hawking entropy as originally
done by 
Policastro, Son and Starinets \cite{pss}.
Alternatively, we can directly calculate the entropy by
using the thermodynamic identity $Ts = \varepsilon + P$.
When local equilibrium is reached,
the energy density and the pressure can easily be obtained
once $T^{\mu\nu}$ is known.
In the fluid rest frame, we have
\be
\ave{T^{00}(x)}_{\rm eq} = \varepsilon(x)
\ \ \ \hbox{and}
\ \ \ {1\over 3}\ave{T^i_i(x)}_{\rm eq} = P(x)
\ee
so that the product of entropy density and the equilibrium temperature is
\be
Ts = \ave{T^{00}}_{\rm eq} + {1\over 3}\ave{T^i_i}_{\rm eq}
\ee
We expect that the above result matches order by order in $g_sN_f$ with the 
Bekenstein-Hawking result where $g_s$ is the string coupling and $N_f$ is the number of flavors (we will discuss this 
in some details in {\bf section 3.5}). 
Once we combine these
ingredients with the RG flow and curvature squared corrections the $\eta/s$ ratio becomes smaller than the known 
$1/4\pi$ bound. How small the ratio becomes seems to depend on the details of the model as well as on the UV 
degrees of freedom in the corresponding dual gauge theory, although the latter dependence is exponentially suppressed. 

\subsection{Background geometry and holographic renormalisation}

Our aim in this paper is to study aspects of large $N$
thermal QCD using the dual gravity picture. Of course, as we mentioned 
earlier, there is no known dual gravitational background to thermal QCD. What we know is a gravitational background
dual to a theory with a RG flow that confines in the far IR and has {\it infinite} interacting
degrees of freedom at the far UV. At zero temperature and in the 
absence of fundamental flavors, this is the Klebanov-Tseytlin background \cite{klebtsey} with zero deformation 
parameter. 
At high temperature and in the 
presence of $N_f$
fundamental (and $M$ bi-fundamental) flavors the background is much more complicated that, as far as we know, 
has not been studied before (see \cite{cotrone1} for a different model that incorporates flavors but doesn't have a 
good UV behavior). 
In {\bf sections 3.1} and {\bf 3.3}
we show that, to lowest order in $g_sN_f$ and $g_sM^2/N$, there is some 
analytic control on the background, meaning that we can derive analytic expressions for the metric and fluxes to the
lowest order in $g_sN_f$ and $g_sM^2/N$. To higher orders in $g_sN_f$ and $g_sM^2/N$ one can only derive the expressions
for the metric and fluxes 
numerically. We clarify many previously ignored subtleties in the literature, namely the existence of small resolution
and the effect of this on the fluxes. 

Once we have infinite interacting degrees of freedom in the far UV we naturally face the question of 
holographic renormalisation. For standard AdS background this has been demonstrated beautifully in 
the series of papers \cite{Kostas-1, Kostas-2, Kostas-3, 2, 3}. For
the Klebanov-Strassler background without fundamental flavor, an equivalent treatment has been shown to apply in
\cite{ofer}. In {\bf section 3.3} we show that, modulo some subtleties, such a treatment of holographic 
renormalisation can {\it almost}
be extended to theories with fundamental flavors also. The subtleties have to do with the 
existence of non-trivial powers of ${\rm log}~r$ over and above the $1/r$ suppressions in various expressions (here 
$r$ is the radial coordinate that determines the energy scale of the gauge theory). 
In the 
limit of small $g_sN_f$ and $g_sM^2/N$ we argue that once the highest integer power of $r$ has been regularised, the 
theory is naturally holographically renormalised. On the other hand once $g_sN_f$ and $g_sM^2/N$ are large, we loose 
all control on our order-by-order expansions and 
new methods need to be devised to holographically renormalise the 
theory. 

\subsection{Organization of the paper}

The paper is organized as follows. In {\bf section 2} we summarize our main results. This section is meant for readers
who would like to know our results without going through the details of the derivations. The summary section 
also involves a discussion of the subtleties of background associated with RG flows etc. In the presence 
of non-trivial RG flows and corresponding Seiberg dualities the interpretation of gauge/gravity duality here is 
not so straightforward as in the AdS/CFT case. We point this out in some details. 

{\bf Section 3} is the main section of our paper. As promised in the title, we perform five {\it easy} computations 
in thermal QCD. In {\bf sections 3.1} and {\bf 3.3} we give a detailed derivations of our background. At 
zero temperature and in the presence of fundamental and bi-fundamental flavors the gravity dual is given by 
Ouyang \cite{4}. However
inserting a black hole 
in the Ouyang background to generate a non-zero temperature in the gauge theory changes everything. 
We can no longer argue that the fluxes, 
warp factor etc would remain unchanged. Even the internal manifold cannot remain a simple conifold any more. 
All the internal spheres would get squashed, and at $r = 0$ there could be both resolution as well as deformation
of the two and three cycles respectively. In {\bf section 3.1} we present our results to ${\cal O}(g_sN_f, g_sM^2/N)$, 
and in 
{\bf section 3.3} we give more detailed derivations and extend this to higher orders in $g_sN_f$ and  $g_sM^2/N$.
In the limit where the deformation parameter is small, we show that to ${\cal O}(g_sN_f, g_sM^2/N)$ we can 
analytically derive the background taking a resolved conifold background. The resolution parameter
depends on $g_sN_f, g_sM^2/N$ as well as on the horizon radius $r_h$. 

Our background also has D7 branes that give rise to fundamental flavors in the gauge theory. These D7 branes are embedded 
via Ouyang embedding \eqref{seven} and therefore the strings with one ends on the D7 branes and the other ends falling
through the horizon would be associated with $N_f$ thermal quarks. In {\bf section 3.2} we evaluate the mass and drag of
the quarks in this background. We point out there how our results differ slightly from the analyses done using 
AdS/QCD techniques.

In {\bf section 3.3} we give a more complete picture of the system. Although the section is dedicated to 
evaluating the wake of the quarks in a thermal medium, we study three related topics here. The first one 
is already mentioned above: we give a detailed derivation of the background geometry that fills up the gaps 
left in the discussions of {\bf section 3.1}. We then give a detailed derivation of the holographic renormalisability 
of our theory in the limit of small $g_sN_f$ and $g_sM^2/N$. Finally all these analyses are combined to evaluate the
finite energy-momentum-tensor of the background plus the quark strings. From there we evaluate the wake of the 
quarks by removing the energy-momentum-tensor contribution of the quark strings. Due to the complicated nature of the 
background we could evaluate certain formal quantities in this section without going into numerical details. In the 
appendices we provide a toy example with only diagonal metric perturbations, where a more direct analysis could 
be performed\footnote{See for example {\bf Appendices A, B} and {\bf C}. {\bf Appendix A} is not directly related 
to the main calculations of our paper, but gives the corresponding example for the AdS case.}.  
  
Another upshot of this section is the realization that we could study infinite number of gauge theories by 
cutting off the geometry at various $r = r_c$ and UV completing them by inserting ``UV caps'' at various 
$r_c$. From gauge theory side this is like inserting correct relevant, marginal or irrelevant operators; and from the 
gravity side this is like inserting non-trivial geometries from $r = r_c$ to $r = \infty$. 
We show that any thermal quantities evaluated in these gauge theories are completely {\it independent} of the 
cut-off scale $r= r_c$ and only depend on the temperature 
(the dependences on the far UV degrees of freedom are exponentially suppressed),
justifying the holographic renormalisabilities of these 
theories.

The cut-off independence is also apparent in the last two sections, {\bf sections 3.4} and {\bf 3.5} where we 
study viscosities and the ratios of the viscosities by their corresponding entropy densities. For both these cases there 
are no cut-off dependences but the ratios of the viscosities by their corresponding entropy densities are always 
smaller than the celebrated $1/4\pi$ bound \cite{Kovtun1}.  

Finally in {\bf section 4} we present our conclusions. Throughout the paper, we allude to various future directions 
that will be 
dealt in the sequel to this paper \cite{sequel}.

%\newpage

\section{Summary of the results}

Before we go into the full analysis, let us summarize the main results of our paper. This summary is meant for 
readers who would want to see the main conclusions without going through our detailed calculations. 

In this paper, as the title suggests, we have done five concrete calculations. They are, in order of 
appearance: 

\vskip.1in

\noindent $\bullet$ Construction of the gravity dual of a thermal gauge theory with running coupling 
constant and fundamental flavors. 

\noindent $\bullet$ Mass and
drag of the quark in the gauge theory from the gravity dual. 

\noindent $\bullet$ Wake left by the quark when it moves in the 
QGP medium, again from the gravity 
~~~~ dual.

\noindent $\bullet$ Shear viscosity of the QGP medium; and finally

\noindent $\bullet$ Viscosity by entropy bound 
for the quark from our dual picture.   

\vskip.1in

Let us now give some brief descriptions of each of the five calculations. {\bf First} is the gravity dual. Recall that 
most of the recent analysis have relied heavily on using AdS/CFT correspondence to study the above phenomena. However
in this paper we would like to study the gauge theory with non-trivial RG
flow, in the hope that we can capture some aspects of the strongly coupled QCD. Although our gravity dual 
will be far from realistic QCD, we will try to address certain generic issues that could lie in a class of gauge
theories that are at least IR confining.

At high temperature the situation is a little subtle, but still IR dynamics of QCD could not be extracted from the 
AdS-Black Hole (AdS-BH) picture. Here, as one may recall, a black hole is inserted in the AdS picture to account 
for a non-zero temperature in the gauge theory \cite{wittenBH}. To do a better job we need another model that 
can capture the RG flow in the gauge theory. 

The model that comes to mind immediately is the so called Klebanov-Strassler warped conifold construction 
\cite{1}\footnote{A somewhat equivalent constructions given around the same time are \cite{vafa}, \cite{mn}.}.
 The advantage of this construction is that the gravity dual $-$ which is a warped deformed conifold with 
three form type IIB fluxes $-$ captures the RG flow of the gauge theory. The KS gauge theory 
is confining in the far IR but is not asymptotically free. 
The UV of the theory is a more complicated cascading gauge theory. The other two 
duals \cite{vafa} and \cite{mn} do not have a good four-dimensional UV description. Additionally, all these 
constructions are for zero temperature gauge theories. 

The other issue with the KS picture is that the quarks therein are all in the bi-fundamental representations of 
the two possible UV gauge groups; and they eventually cascade away in the far IR. So what we need is a 
dual gravity theory that allows fundamental quarks at high temperature. 

Before inserting fundamental quark at high temperature let us point out that the situation now, even at zero 
temperature (and without fundamental flavors),  
is much more subtle than the AdS case studied earlier. Due to renormalisation group flow in the gauge theory 
side, we need to ask precisely {\it what} aspects of the dual gravity picture captures the dynamics of the 
strongly coupled gauge theory. The key question to ask here is whether the weakly coupled gravity dual sees the 
cascading Seiberg dualities or it only sees a smooth RG flow in the gauge theory side. 
Since the answer to this question lies 
in the heart of the matter, we will spend some time elaborating this. We would like to caution the readers that this 
point has been misunderstood in most of the literature and the only article, in our opinion, 
that has been able to fully explain the 
subtleties is the one by Strassler \cite{Strassler:2005qs}. What follows below is an elaboration of these subtleties. 

To start off let us ask what ${\cal N} = 4$ AdS/CFT duality tells us? Recall that due to the tight constraint from
supersymmetry all quantum corrections in ${\cal N} = 4$ theory is cancelled out, leaving us with only classical 
amplitudes. This in particular means that if we choose any gauge coupling in the theory, it stays there, with possible
small finite shifts,  
under any 
RG flow (which here means going from UV to IR). For example if we choose the gauge coupling to be very strong, the 
theory will remain at strong coupling from UV to IR. On the other hand once the gauge theory is at strong coupling 
we could analyze the theory from weakly coupled {\it supergravity} description on AdS space. 
However when the gauge theory is 
at very weak coupling there exists no weakly coupled gravity description, and the story there is captured
by the full string theory on AdS space\footnote{Alternative one could think that the {\it weakly} coupled 
${\cal N} = 4$ gauge theory is captured by string theory in {\it twistor} space. This is basically the key essence of 
\cite{Witten:2003nn}.}. 

The above conclusions also mean that every value of the coupling is a RG fixed point in the gauge theory. There is 
no coupling flow and therefore the system is simple without any inherent subtleties. What happens now if we introduce
a non-trivial RG flow in the theory? There are three cases to consider:

\noindent $\bullet$ The RG flows lead to a non-trivial fixed point (or isolated fixed points) in the theory.

\noindent $\bullet$ The RG flows lead to a non-trivial surface of fixed points in the theory; and 

\noindent $\bullet$ The RG flows lead to no fixed point(s) or fixed surface(s) in the theory. 

The first case, as far as we know, leads to no known gravity dual so will ignore this case. The second case is more 
interesting. We know of one example where non-trivial RG flows in the theory lead to a fixed RG surface. This is the 
so-called Klebanov-Witten model \cite{klebwit}. There are three couplings in the theory ($g_1, g_2, h$) 
corresponding to the two gauge couplings and the coupling associated to the quartic superpotential of the theory 
respectively. The gauge group is $SU(N) \times SU(N)$ and the three beta functions are:
\bg\label{kwbeta}
\beta_{g_1} = - {g_1^3 N\over 16\pi^2} \left({1+2\gamma_0\over 1-{g_1^2N\over 8\pi^2}}\right), ~~~
\beta_{g_2} = - {g_2^3 N\over 16\pi^2} 
\left({1+2\gamma_0\over 1-{g_2^2N\over 8\pi^2}}\right), ~~~ \beta_\eta = \eta(1+2\gamma_0)
\nd
where $\eta = h\mu$ is the dimensionless coupling of the theory for any energy scale $\mu$. As discussed in details 
in \cite{klebwit, Strassler:2005qs} all the fields in the theory have the same anomalous dimension 
$\gamma_0(g_1, g_2, h)$, and therefore the three beta functions vanish exactly when:
\bg\label{betazero1}
\gamma_0(g_1, g_2, h) ~ = ~ -{1\over 2}
\nd   
which is one equation for the three couplings. Therefore the fixed points in this theory form a {\it two-dimensional}
surface in the three-dimensional space of couplings (see {\bf figure 2} for details). 
\begin{figure}[htb]\label{rgsurface1}
		\begin{center}
\includegraphics[height=6cm]{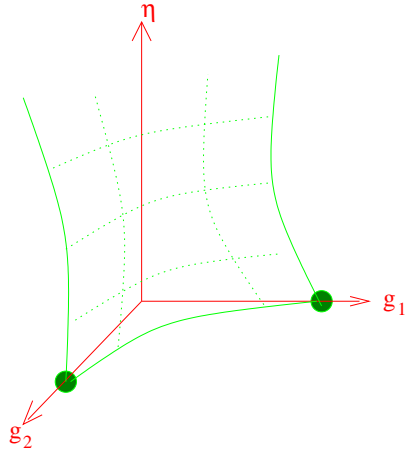}
		\caption{{The two-dimensional RG surface in the Klebanov-Witten theory.}}
		\end{center}
		\end{figure} 
\noindent This also means that the typical RG flow in the theory will take the following simple form as illustrated in 
{\bf figure 3} below.
\begin{figure}[htb]\label{rgflow1}
		\begin{center}
	\includegraphics[height=6cm]{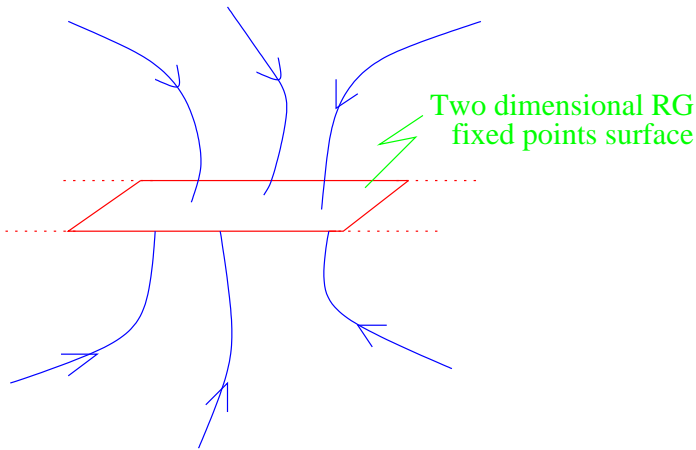}
				\caption{{ The typical RG flows in the Klebanov-Witten theory.}}
		\end{center}
		\end{figure}
\noindent Since the sign of the two beta functions are negative any arbitrary flow in the coupling constant space 
will bring us to the fixed point surface. This surface is IR stable. Notice also that at the boundary of the 
Klebanov-Witten fixed point surface
 the gauge theory is weakly coupled and so the gravity description is very strongly coupled. To 
get the weakly coupled supergravity description, we need to go towards the centre of the fixed point surface. 
The $AdS \times T^{1,1}$ description is valid here. Note also that on the surface the bi-fundamentals scalars all 
attain anomalous dimensions of $-{1\over 2}$, as obvious from the beta functions given earlier. 

The story changes quite a bit once we deform away from the AdS background. At zero temperature and in the absence of
fundamental flavors, this is of course the Klebanov-Strassler model \cite{1}. The gauge theory dual of the 
Klebanov-Strassler model is more complicated now because of the non-trivial RG flows of the two couplings lead to 
cascades of Seiberg dualities in the theory. Although the UV of the gauge theory has {\it infinite} 
degrees of freedom, the 
theory {\it is} holographically renormalisable\footnote{We will discuss later that the theory with fundamental flavors is 
also holographically renormalisable.} \cite{ofer} (see also \cite{haack}). 

To see how the story differs from the AdS case discussed above, let us take the far UV theory to be $SU(kM) \times 
SU(kM-M)$ with gauge couplings $g_k, g_{k-1}$ for the two gauge groups respectively and $\eta$ to be the other 
dimensionless coupling defined above. The three beta function now are \cite{1, Strassler:2005qs}:
\bg\label{betaksnow}
&&\beta_k = -{g_k^3 kM\over 16\pi^2}\left[ {(1+2\gamma_0) + {2\over k}(1-\gamma_0)\over 1-{g_k^2 kM\over 8\pi^2}}\right],
~~~~~ \beta_\eta = \eta(1+2\gamma_0)\\
&&\beta_{k-1} = -{g_{k-1}^3 (k-1)M\over 16\pi^2}
\left[ {(1+2\gamma_0) - {2\over k-1}(1-\gamma_0)\over 1-{g_{k-1}^2 (k-1)M\over 8\pi^2}}\right]
\nd
where we see now that they differ from \eqref{kwbeta} by ${\cal O}(1/k)$ factors. This also means that there will be 
no point in the coupling constant space where all the three beta functions could vanish exactly although there 
would be numerous points where all the beta functions could be very small. Two questions arise immediately:

\noindent $\bullet$ Since there are infinite number of gauge theories involved here, what is the gravity description 
of the theory? 

\noindent $\bullet$ At the point where the theory has a 
{\it weakly} coupled gravity description, is the gauge theory described by a smooth RG flow or the description 
involves ``choppy'' Seiberg dualities? 

Since the answer to these two questions are rather involved, we will go in small steps. It is 
clear that
 the geometrical description that involves cascade of Seiberg dualities {\it cannot} have a weakly coupled gravity
description. The RG flows in the theory have been succinctly presented in \cite{Strassler:2005qs} so we will be brief
here. The RG flows at the boundary of the Klebanov-Witten fixed point surface can be illustrated by {\bf figure 4}. 
Observe that the flow takes us from one surface to another because we are moving from one set of gauge theory 
descriptions convenient at a certain scale to another set of descriptions convenient at a different scale.
\begin{figure}[htb]\label{cascade1}
		\begin{center}
		\includegraphics[height=8cm]{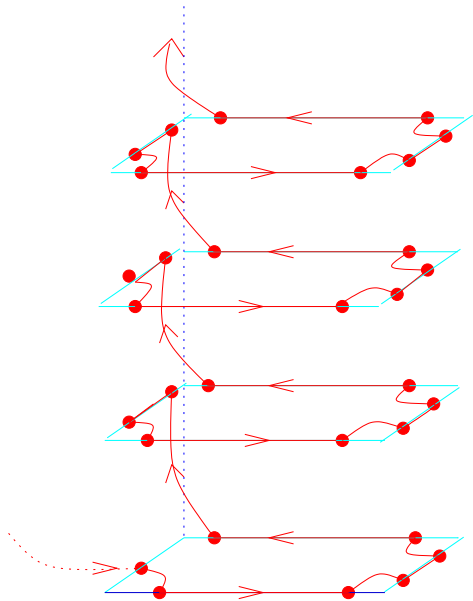}
		\caption{{ The RG flow at the boundary of the Klebanov-Witten wall for the Klebanov-Strassler model. 
The big dots are the Seiberg fixed points of the theory. The vertical distances between the planes don't signify 
anything here.}}
		\end{center}
		\end{figure}
For the RG flows illustrated in {\bf figure 4} there is no weakly coupled gravity description available. 
Therefore, contrary to popular belief, the 
usual cascade of Seiberg dualities {\it do not} have a supergravity description on a deformed conifold. In fact there 
is no simple gravity description that could capture the choppy RG flows in the dual gauge theory! So then where 
would our gravity calculations fit in? In other words, the analysis that we do in the gravity side captures what 
aspects of the gauge theory? Clearly since we fail to capture the cascades of Seiberg dualities using our weakly 
coupled gravity picture, do we then see any interesting aspects of the gauge theory now?  

Things are not really that bad once we realise that the strongly coupled gauge theory description or equivalently the 
weakly coupled gravity description becomes better as we go away from the boundary of the Klebanov-Witten fixed 
points surface. As we go towards the centre of the surface, the RG flows lose their choppy nature and tend to become
smooth. This is illustrated in {\bf figure 5}.  
\begin{figure}[htb]\label{cascade2}
		\begin{center}
		\vskip.1in
		\includegraphics[height=8cm]{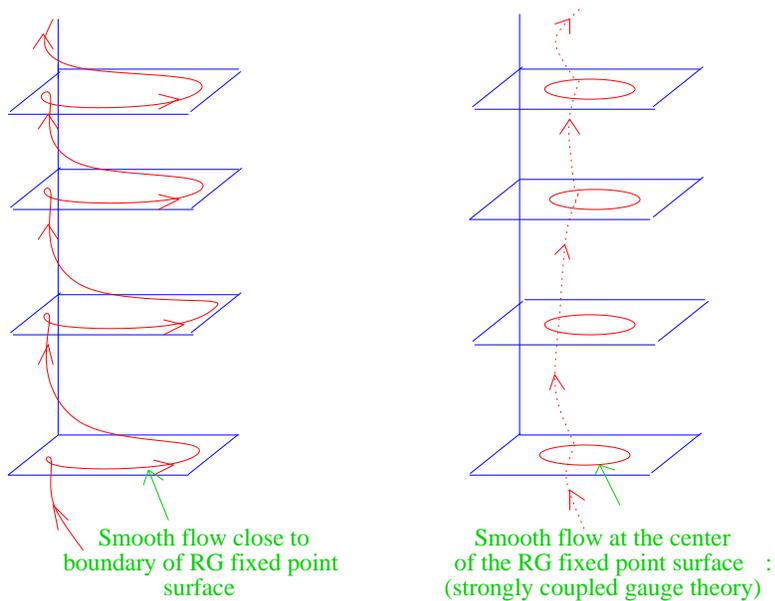}
\caption{{ The RG flow as we go inside the boundary of the RG fixed point surface. Observe that the 
gauge theory becomes strongly coupled at the centre of the surface and the RG flow loses its sharp edges and 
subsequently becomes very smooth.}}
		\end{center}
		\end{figure}
\noindent Thus in the gauge theory side we can summarize our situation by three points:

\noindent $\bullet$ The theory lies at the boundary of the RG surface and jumps from
one Seiberg fixed point to another\footnote{As we observed above, all the three beta functions do not vanish 
simultaneously. Therefore the fixed points are not the absolute Klebanov-Witten type fixed points, rather they have 
s small tilt given by ${\cal O}(1/k)$ corrections \eqref{betaksnow}.}.
This is the ``usual'' cascading and the
surface corresponds to the two-dimensional surface of fixed points (for the
corresponding Klebanov-Witten theory) in a three-dimensional space of coupling constants.

\noindent $\bullet$ The theory lies close to the boundary of the surface but never quite touches the fixed
points. So this flow is parallel to the boundary.

\noindent $\bullet$ The theory hovers at the centre of the two-dimensional surface and has a smooth RG
flow from UV to IR. The degrees of freedom of this theory changes
continuously from UV to IR (and is a well defined QFT).

\noindent Similarly in the gravity side we can also summarise the situation by three ``dual'' points:

\noindent $\bullet$ For the first case the gauge theory is weakly coupled and therefore
there is only a strongly coupled gravity description. This amounts to
saying that the tree-level dual is full string theory on a warped deformed conifold
(which also means that we cannot say anything from the gravity side other
than some protected quantities!).

\noindent $\bullet$ For the second case a somewhat similar statement can be made. There is
no weakly coupled gravity description available.

\noindent $\bullet$ For the third case there is a well defined weakly coupled gravity
description. This is supergravity defined on warped deformed conifold and
tells us clearly that in the dual QFT the colors should decrease
logarithmically (although its a little tricky to say {\it how many} colors we have).
\begin{figure}[htb]\label{correctRG}
		\begin{center}
	\includegraphics[height=12cm]{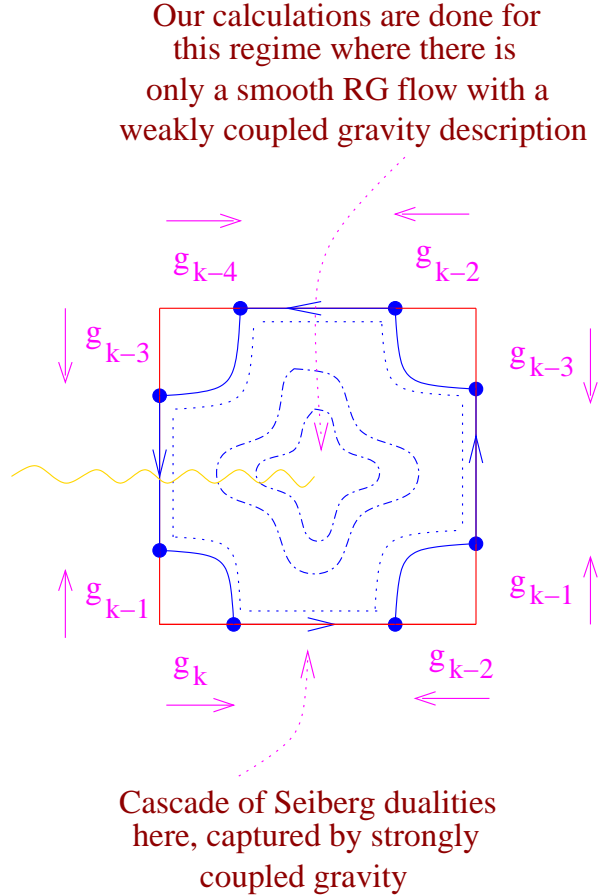}
		\caption{{The complete RG flows depicting the regime of validity of our calculations. This figure 
is taken from {\rm \cite{Strassler:2005qs}}. 
The yellow wavy line is the branch cut, and corresponds to the various planes 
in the earlier figures.}}
		\end{center}
		\end{figure}

\noindent All of our above discussions imply that the regime of smooth RG flow in the gauge theory side can be captured 
by weakly coupled supergravity description. But what supergravity description are we looking for here? We want our 
supergravity description to capture the RG flow in a gauge theory with fundamental flavor and at a 
non-zero temperature. In 
other words we want the sugra description for a thermal gauge theory with fundamental flavors.   

The first problem of having fundamental quark may be solved by inserting $N_f$ 
D7 branes in the KS geometry. However this is 
subtle as we discuss in sec (3.1). What we know so far is how to insert 
coincident D7 branes in the Klebanov-Tseytlin background  
\cite{klebtsey}. This is the Ouyang background \cite{4} that has all the type IIB fluxes switched on, including the 
axio-dilaton. Our aim here is to insert a black hole in this geometry and study the resulting gravity dual. 

Inserting a black hole in the Ouyang background changes everything. We can no longer argue that the fluxes, 
warp factor etc would remain unchanged. Even the internal manifold cannot remain a simple conifold any more. 
All the internal spheres would get squashed, and at $r = 0$ there could be both resolution as well as deformation
of the two and three cycles respectively. Our metric then should look like (see the definitions of the coordinates 
in sec (3.1)):
\bg\label{bhmet1}
ds^2 = {1\over \sqrt{h}}
\Big[-g_1(r)dt^2+dx^2+dy^2+dz^2\Big] +\sqrt{h}\Big[g_2(r)^{-1}dr^2+ d{\cal M}_5^2\Big]
\nd  
where we have been able to work out $h(r), g_i(r)$ and $d{\cal M}_5^2$ only in the limit where the resolution factor 
of the internal conifold is a constant but the deformation factor is zero\footnote{Such details won't matter when 
we study only the metric. But fluxes {\it will} carry the information of background resolution.}. 
Our results are:

\noindent $\bullet$ The warp factor $h(r)$ is given by \eqref{logr}.

\noindent $\bullet$ The black hole factors $g_i(r)$ are given by  \eqref{grdef}

\noindent $\bullet$ The internal manifold is given by a warped resolved conifold. 

\noindent In certain limits along with $g_1 = g_2 = g$, that we discuss in the text, 
we can show precisely how our metric differs from the 
AdS case (see footnote 37 for definitions):
\bg\label{leahl1}
ds^2 && = {r^2\over L^2}\left(-g dt^2 + dx^i dx_i\right) + {L^2\over g r^2}~ dr^2 + 
L^2 ~ d{\cal M}^2_5 \\
&& - \left(A ~{\rm log}~r + B~{\rm log}^2 r\right)\Bigg[{r^2\over L^2} \left(-g dt^2 + dx^i dx_i\right) -
{L^2\over g r^2}~ dr^2 + L^2 ~ d{\cal M}^2_5\Bigg]\nonumber
\nd
where the first line is the standard AdS$_5 \times T^{1,1}$ space with radius $L \equiv (4\pi g_s N)^{1/4}$, 
and the second line is the deformation of the AdS as well as the internal spaces. 
Precisely because of this deformation, 
as we mentioned above, even the fluxes change drastically from the 
ones given by Ouyang. Our fluxes capture the 
fact that the two-cycles in the internal warped resolved conifold are slightly squashed. The three-form RR flux 
is (the coordinates ($\theta_i, \phi_i, \psi$) parametrise ${\cal M}_5$ in \eqref{bhmet1}):
\bg\label{brend1}
{\widetilde F}_3 & = & 2M A_1 \left(1 + {3g_sN_f\over 2\pi}~{\rm log}~r\right) ~e_\psi \wedge 
\frac{1}{2}\left({\rm sin}~\theta_1~ d\theta_1 \wedge d\phi_1-B_1{\rm sin}~\theta_2~ d\theta_2 \wedge
d\phi_2\right)\nonumber\\
&& -{3g_s MN_f\over 4\pi}A_2~{dr\over r}\wedge e_\psi \wedge \left({\rm cot}~
{\theta_2 \over 2}~{\rm sin}~\theta_2 ~d\phi_2 
- B_2 {\rm cot}~{\theta_1 \over 2}~{\rm sin}~\theta_1 ~d\phi_1\right)\\
&& -{3g_s MN_f\over 8\pi}A_3 ~{\rm sin}~\theta_1 ~{\rm sin}~\theta_2 \left({\rm cot}~{\theta_2 \over 2}~d\theta_1 +
B_3 {\rm cot}~{\theta_1 \over 2}~d\theta_2\right)\wedge d\phi_1 \wedge d\phi_2\nonumber
\nd
and the three form NS-NS flux is:
\bg\label{brend2}
H_3 &=&  {6g_s A_4 M}\Bigg(1+\frac{9g_s N_f}{4\pi}~{\rm log}~r+\frac{g_s N_f}{2\pi} 
~{\rm log}~{\rm sin}\frac{\theta_1}{2}~
{\rm sin}\frac{\theta_2}{2}\Bigg)\frac{dr}{r}\wedge \frac{1}{2}\Big({\rm sin}~\theta_1~ 
d\theta_1 \wedge d\phi_1\nonumber\\
&& ~~~~~~ -B_4{\rm sin}~\theta_2~ d\theta_2 \wedge d\phi_2\Big)
+ \frac{3g^2_s M N_f}{8\pi} A_5 \Bigg(\frac{dr}{r}\wedge e_\psi -\frac{1}{2}de_\psi \Bigg)\nonumber\\
&& ~~~~~~~~~~~~~~~ \wedge \Bigg({\rm cot}~\frac{\theta_2}{2}~d\theta_2 
-B_5{\rm cot}~\frac{\theta_1}{2} ~d\theta_1\Bigg)
\nd
where we differ from Ouyang analysis precisely by the asymmetry factors $A_i, B_i$. These asymmetry factors incorporate 
all order corrections and are given in \eqref{asymmetry} (where we present only the first order terms). 
Finally, the axio-dilaton and the five form are given by \eqref{dilato} and \eqref{axfive} respectively. These are 
thus our first set of results related to the gravity dual on which we base all our subsequent calculations. Due to the
complicated nature of our background, we have only worked things out to  
order ${\cal O}(g_sN_f, g^2_sMN_f)$ at small $r$, and provide some conjectural solution for large $r$ from the full 
F-theory embedding of our solution.
In fact for most part of this paper our calculations are valid in the following limits of $g_s, M, N_f$ and $N$: 
\bg\label{prelim} 
\left (g_s, ~g_s N_f, ~g^2_s M N_f, ~{g_s M^2\over N}\right) ~ \to ~ 0, ~~~~~
(g_s N, ~g_s M)~ \to ~ \infty
\nd 
The precise way some of these go to zero or infinity  
is discussed towards the end of section 3.1. Note that large $N_f$ limit is probably
not realisable here because of the underlying
F-theory constraints \cite{vafaF}. 

Our {\bf second} set of calculations are related 
to the mass $M$ and drag $\nu$ of a quark from the gravity dual \eqref{leahl1}. The quark is 
identified in the dual theory
with the string whose one end is connected to the D7 branes and the other hand falls in the horizon
of the black hole. The drag of the quark is precisely the rate at which the string loses its momentum 
and energy to the black hole. This is given by\footnote{Note that we will be working in the limit where 
$\hbar = c = k_B = M_p = \alpha' \equiv 1$. Thus all the relevant QCD 
observables would appear dimensionless throughout. Besides 
these choices of scales we will also impose $r_0 \equiv 1$, 
where $r_0$ is the distance between the tip of the seven brane and the 
black hole horizon. This will be discussed more in section 3.2.}:
\bg \label{KS16b} 
\nu = \frac{T_0}{mL^2}\frac{{\cal T}^2} 
{\sqrt{1+\frac{3g_s\bar{M}^2}{2\pi N}~{\rm log}\Big[\frac{\cal T}{(1-v^2)^{{1}/{4}}}\Big]
\Big(1+\frac{3 g_s\bar{N}_f}{2\pi}~\Big\{{\rm
log}~\Big[\frac{\cal T}{(1-v^2)^{{1}/{4}}}\Big]+\frac{1}{2}\Big\}\Big)}}
\nd 
where we see that we differ from the AdS result by the ${\rm log}~{\cal T}$ and ${\rm log}~(1-v^2)$ corrections where 
${\cal T} = {r_h}$, the horizon radius, is the characteristic 
temperature and $v$ is the velocity of the string (also the quark). 
These corrections aren't very big 
and occurs precisely because of the logarithmic running in our theory. The other coefficients appearing in 
\eqref{KS16b} are defined in section (3.2).   

In the above calculations, we had taken the string stretching between D7 branes and the black hole to be a probe in the 
deformed AdS background. This means that the back reactions are completely neglected. This cannot be the full picture,
because a moving string should create some {\it disturbance} in the surrounding media. This is the premise of 
our {\bf third} set of calculations. Under some reasonable assumptions (that we elaborate in sec. (3.3) of the text)
we can use this back reaction to calculate the {\it wake} of the quark in the gauge theory from our dual gravity
background. This analysis is simple to state but involves not only a
detailed series of manipulations that include, among other 
things, writing an action and then regularising and renormalising it to extract finite gauge theory variables, but also
something much more elaborate and interesting. The interesting part is that not only we can do our wake analysis (or in 
principle all other thermodynamical quantities) for the dual geometry that we elaborated above, but also 
on the geometry that is {\it cut-off}
at $r = r_c$ and a non-trivial UV cap (or UV geometry) attached from $r = r_c$ to $r = \infty$.
Such a procedure actually creates a new UV completed gauge theory at the boundary 
that is {\it different} from the original theory (that we called the {\it parent} cascading theory). This new theory
and the original ``parent''
one differs by certain operators that we can define carefully at the cut-off. Thus the properties 
of the new theory are not the universal properties of the parent cascading theory. The parent cascading theory has 
infinite number of degrees of freedom at the far UV. Our new theory should also have very large degrees
of freedom at the UV. We can call these degrees of freedom at the boundary (i.e at $r = \infty$) as ${\cal N}_{uv}$ 
such that when ${\cal N}_{uv} = \epsilon^{-n}, \epsilon \to 0, n >> 1$ 
we are studying the parent cascading theory, and the other boundary theories have 
$n \ge 1$ (see {\bf figure 10} for 
details). An elaborate discussion of this is given in section 3.3 which we would refer the readers for details. 
In fact the 
result of studying these gauge theories are both remarkable and instructive. The remarkable thing is that no matter 
what thermodynamical quantities we want to extract from these theories, the final results are 
{\it completely independent} of the cut-off $r = r_c$ that we imposed to derive these theories. The results only depend
on the characteristic temperature ${\cal T}$ (that we fix once and for all) and on the UV degrees of freedom via 
$e^{-{\cal N}_{uv}}$. Since ${\cal N}_{uv}$ is always infinite for us 
the results are only sensitive to 
${\cal T}$. Our third set of calculations related to the wake demonstrates this in full details. No matter how involved 
are the UV descriptions, or the procedure to holographically renormalise these theories, 
the 
final answers for the energy-momentum tensors for the gauge theories are remarkably clean and can be stated 
as\footnote{Repeated indices of the form $a^n b_n, a^n b_{nn}$ are summed over $n$, but $a_nb_n, (ab)^{nn}$ etc
may not be.}:
\bg\label{wakeup1}
&& T^{mm}_{{\rm medium} + {\rm quark}}    
 = \int \frac{d^4q}{(2\pi)^4}\sum_{\alpha, \beta}
\Bigg\{({H}_{\vert\alpha\vert}^{mn}+ {H}_{\vert\alpha\vert}^{nm})s_{nn}^{(4)[\beta]} 
-4({K}_{\vert\alpha\vert}^{mn}+ {K}_{\vert\alpha\vert}^{nm})s_{nn}^{(4)[\beta]}\nonumber
\nd
\bg
&& ~~~~~~ +({K}_{\vert\alpha\vert}^{mn}+ {K}_{\vert\alpha\vert}^{nm})s_{nn}^{(5)[\beta]}
+\sum_{j=0}^{\infty}~\hat{b}^{(\alpha)}_{n(j)} \widetilde{J}^n  
\delta_{nm}  e^{-j{\cal N}_{\rm uv}} + {\cal O}({\cal T} e^{-{\cal N}_{uv}})\Bigg\}
\nd 
{}from where the wake of the quark can be computed using the relation \eqref{goal} which equivalently means
 we subtract the 
energy momentum tensors of the quark from \eqref{wakeup1} above. All the functions appearing above are described 
in section (3.3), with $\alpha$ denoting the effects of the flavors. 
Here ($\hat{b}^{(\alpha)}_{n(j)}, {\cal N}_{\rm uv}$) together specify the full boundary theory for a specific
UV complete theory (for a discussion on the back reactions, see after \eqref{KS7a}). 
Note that the result has no dependence on $r = r_c$. In the limit where 
${\cal N}_{uv} = \epsilon^{-n}, n >> 1$
we reproduce the result for the parent cascading theory, and for ${\cal N}_{uv} \to \epsilon^{-n}, n \ge 1$ 
we have the UV 
dependences specified above. 

%jaslea

The instructive thing about our result is the way we differ from the AdS case. The energy momentum tensor of the 
system from AdS/CFT case is the first part of the above formula given by $s_{mm}^{(4)[0]}$ 
(see \eqref{StressTensor_gs=0} for details). The rest is the deformation from the AdS result (we have to 
remove some constant pieces from \eqref{wakeup1} to account for the stress tensor properly). This way we can again
see how different UV completed gauge theories can change some of these calculations. Of course all these theories do 
inherit some of the universal properties of the parent cascading theory. 

Our {\bf fourth} set of calculations is related to studying the shear viscosity with and without curvature squared 
corrections. Without curvature squared corrections, but involving the RG flow, the result 
for shear viscosity $\eta$ is easy to state:
\bg \label{SV-15}
\eta ~ = ~ {{\cal T}^3 L^2\over 2 g_s^2 G_N} \Bigg[{1 + \sum_{k = 1}^\infty
\alpha_k e^{-4k {\cal N}_{\rm uv}} \over 4\pi + {1\over \pi}~{\rm log}^2 \left(1 - 
{\cal T}^4 e^{-4{\cal N}_{uv}}\right)}\Bigg]
\nd
where $\alpha_k$ are functions of ${\cal T}$ that are given in section 3.4. 
Note again that the result is independent of the cut-off at $r = r_c$. In the limit where ${\cal N}_{uv} = \infty$
the shear viscosity has a simple form. This is related to the AdS result also. On the other hand the viscosity 
to the entropy ratio, which is our {\bf final} set of calculations taken in the limit where we switch on all the 
ingredients i.e the  
RG flows, curvature squared corrections as well as the contributions from the 
UV caps, has the following form:
\bg \label{final1}
\frac{\eta}{s} &=&~ {\left[{1 + \sum_{k = 1}^\infty
\alpha_k e^{-4k {\cal N}_{\rm uv}} \over 4\pi + {1\over \pi}~{\rm log}^2 \left(1 - 
{\cal T}^4 e^{-4{\cal N}_{uv}}\right)}\right]}\nonumber\\
&-&\frac{c_3\kappa}{3 L^2 \left(1- {\cal T}^4 e^{-4{\cal N}_{uv}}\right)^{3/2}}
 \left[\frac{{B_o}(4\pi^2-{\rm log}^2 ~C_o)+4\pi{A_o}~{\rm log}~C_o}{\Big(4\pi^2-{\rm
log}^2~C_o\Big)^2+16\pi^2~{\rm
log}^2~C_o}\right]
\nd
where ($A_o, B_o, C_o$) are given in \eqref{constants} and $c_3$ is given in 
\eqref{coeffeven}. We see that the ratio is again independent of any cut-off; and    
in the limit ${\cal N}_{uv} \to \epsilon^{-n}$ and $c_3 \to 0$ \eqref{final1}
the bound is exactly saturated. However in the limit 
$c_3 \ne 0$ we have a violation of the bound. The plot of the behavior of $\eta/s$ is given in sec (3.5).

%\newpage

\section{Dynamics of quarks in strongly coupled plasmas}

This section contains the five elements alluded to in the title of this paper. 
It is now time to construct a more detailed scenario wherein certain aspects of QCD calculations
could be performed. As we 
discussed in the introduction and in the summary above, 
most of the previous analysis relied on Anti-deSitter (AdS) spaces whose dual
is a conformal field theory (CFT) with no running couplings. Many of the recent works in this field have focussed on 
AdS/CFT correspondence to study certain behaviors of QCD. Our aim in this section would be to use 
non-AdS backgrounds to study properties of QCD.  

\subsection{Construction of the gravity dual}

It turns out if we embed D branes in certain geometric background, the gauge theory that lives on the D branes may become
confining and exhibit logarithmic running coupling. One of the most popular background to achieve this property is the 
Klebanov-Strassler background (see also \cite{vafa,mn}). 
In Klebanov-Strassler (KS) model \cite{1} $N$  D3 branes are
placed at the tip of a six dimensional conifold and  M D5 branes are placed in such a way that they wrap a 
2-cycle on the conifold base
$T^{1,1}$. The D3 and unwrapped part of the D5 branes extend  in four spacetime directions orthogonal to the conifold. 
The $SU(N) \times SU(N+M)$ 
gauge theory living on these branes contain matter fields that transform as bifundamental color
representation $(N, \bar N+ \bar M)_c$ and 
$(\bar N,N+M)_c$ of the group $SU(N)\times SU(N+M)$. Under a Renormalisation 
Group flow the gauge group cascades down to $SU(M)$ group in IR. 
The RG flow and fixed points surface for this model have been shown earlier. In the regime where we have a smooth RG flow,
the theory confines in the far IR. Thus there is a 
small regime of the theory that gives us a weakly coupled
gravity dual to a confining $SU(M)$ theory. Of course since the UV 
behavior is not asymptotically free\footnote{Plus it has infinite number of degrees of freedom.}, 
KS model doesn't give us the full gravity dual for QCD. However it does 
come close in giving us at least a dual model that has a running coupling constant. Similar behavior can 
also be argued for \cite{vafa} and \cite{mn}, although the UV behaviors of \cite{vafa} and \cite{mn} 
are six-dimensional theories and may develop baryonic branches \cite{minasian}. 
On the other hand, in far IR, one can show that many BPS quantities have one-to-one correspondences \cite{ohpapers}. 

The original KS model (as well as \cite{vafa,mn}) do not have quarks in the fundamental representation of the gauge
group. To introduce fundamental quarks, we need D7 branes in the gravity dual. However this is subtle because 
the full global solution that incorporate 
back reactions of the D7 branes on the KS background has not yet been computed. What we have computed is the 
local metric that incorporates the deformations of the seven branes when these branes are moved far away from the 
regime of interest (see \cite{gtpapers, KG} for more details). This is given by 
\begin{eqnarray}\label{locmet} \nonumber
&&ds^2 ~ =  ~~
h_1~[d{\bf z} + a_1~ d{\bfx} + a_2~ d{\bf y}]^2 + h_2~[d{\bf y}^2 + d\widetilde\theta_2^2] +
h_4~[d{{\bf x}}^2 + h_3~d\widetilde\theta_1^2]~ + \\ \nonumber
&& ~~~~~~~~~~~~~~~ + h_5~{\rm
sin}~\widetilde\psi~[d{\bf x} ~d\widetilde\theta_2 + d{\bf y} ~d\widetilde\theta_1] + h_5~{\rm
cos}~\widetilde\psi~[d\widetilde\theta_1 d\widetilde\theta_2 - d{\bf x}~ d{\bf y}] \\ 
&& B_{NS} =
b_{{\bf x}\widetilde\theta_1}~d{\bf x} \wedge d\widetilde\theta_1 + b_{{\bf y}\widetilde\theta_2}~d{\bf y} \wedge
d\widetilde\theta_2, ~ B_{RR} = -2A~d{\bf y} \wedge d{\bf z}, ~~ \phi = \widetilde\phi = 0
\end{eqnarray}
where (${\bf x}, {\bf y}, {\bf z}, \widetilde\theta_i, \widetilde\psi, \widetilde r$) 
are the small local regime around the point ($\langle\phi_i\rangle, 
\langle\psi\rangle, \langle\theta_i\rangle, r_0$) given in the following way:
\begin{eqnarray}\label{coordef} \nonumber
&&\psi = \langle\psi\rangle + {2 {\bf z}\over
\sqrt{\gamma'_0\sqrt{h_0}}}, ~~~~~~~~~~~~ \phi_2 =
\langle\phi_2\rangle +{2{\bf y} \over \sqrt{(\gamma_0+4a^2)\sqrt{h_0}}~{\rm
sin}~ \langle\theta_2\rangle} \\ \nonumber
&& \phi_1 =
\langle\phi_1\rangle +{2{\bf x} \over \sqrt{\gamma_0\sqrt{h_0}}~{\rm
sin}~\langle\theta_1\rangle}, ~~~~~~~~~~~ r = r_0 + {\widetilde r\over
\sqrt{\gamma'_0\sqrt{h_0}}} \\ && \theta_1 =
\langle\theta_1\rangle + {2\widetilde\theta_1\over \sqrt{\gamma_0\sqrt{h_0}}},
~~~~~~~~ ~~~~ \theta_2 = \langle\theta_2\rangle + {2\widetilde\theta_2
\over \sqrt{(\gamma_0+4a^2)\sqrt{h_0}}}
\end{eqnarray}
where $\gamma_0(r_0)$ and $h_0(r_0)$ are some constant functions of $r_0$ (see \cite{gtpapers, KG} for details); and the
un-tilded coordinates are used to write the standard metric of the conifold in the following way:
\bg \label{conifold}
ds^2 = {dr^2} + r^2\left({1\over 6} \sum_{i=1}^2 (d\theta_i^2 + {\rm sin}^2 \theta_i ~d\phi_i^2) + {1\over 9} 
(d\psi + {\rm cos}~\theta_1~d\phi_1 + {\rm cos}~\theta_2~d\phi_2)^2\right)
\nd   
There are a few key issues that we want to point out regarding the metric \eqref{locmet}:

\noindent $\bullet$ First, since the seven branes
are kept far away the axion-dilaton vanish for the background {\it locally}. Globally there will be non-zero 
axion-dilaton. 

\noindent $\bullet$ Secondly, observe that the two spheres (parametrised 
originally by ($\widetilde\phi_i, \widetilde\theta_i$)) are 
replaced by two-tori locally. This issue has already been explained in details in \cite{gtpapers}. Furthermore the 
two tori do not appear with the same coefficients. In fact there is a squashing factor associated with the two tori. 
We believe that globally such squashing factor should also show up for the two spheres. 

\noindent $\bullet$ Thirdly, the local 
back reactions on the metric due to fluxes and seven branes would modify the warp factors once we go to the full 
global scenario. 

\noindent $\bullet$ Finally, the above local (and the subsequent global) picture is dual only to zero temperature
gauge theory. What we need for our purpose is high temperature gauge theory. This would mean that the global 
extension of the above background \eqref{locmet} should contain a black hole whose horizon size should correspond 
to the temperature in the dual gauge theory at far UV. Since the number of effective degrees of freedom are changing as 
we go from UV to IR, the entropy and temperature will depend crucially on what UV 
degrees of freedom we are considering at 
a given cutoff. This is a subtle issue and we will discuss this carefully a little later. 

Thus putting a black hole in the global metric is non-trivial
because of all the above considerations. However we can formally 
write the metric in the following way (although we will use Minkowski coordinates throughout, unless mentioned otherwise, 
one may write everything in Euclidean coordinates also):
\bg\label{bhmet}
ds^2 = {1\over \sqrt{h}}
\left(-g_1 dt^2+dx^2+dy^2+dz^2\right)+\sqrt{h}\Big[g_2^{-1}dr^2+r^2 d{\cal M}_5^2\Big]
\nd
where ($x, y, z$) not to be confused with (${\bf x}, {\bf y}, {\bf z}$) discussed above,
 $g_i$ are functions\footnote{They would in general be functions of ($r, \theta_i$). We will discuss this later.} 
that determine the presence of the black hole, $h$ is the $10d$ warp factor that could be a 
function of all the internal coordinates and $d{\cal M}_5^2$ is given by:
\begin{eqnarray}\label{bhmet2}\nonumber
d{\cal M}_5^2 = && h_1 (d\psi + {\rm cos}~\theta_1~d\phi_1 + {\rm cos}~\theta_2~d\phi_2)^2 + 
h_2 (d\theta_1^2 + {\rm sin}^2 \theta_1 ~d\phi_1^2) + \\ \nonumber 
&& + h_4 (h_3 d\theta_2^2 + {\rm sin}^2 \theta_2 ~d\phi_2^2) + h_5~{\rm cos}~\psi \left(d\theta_1 d\theta_2 - 
{\rm sin}~\theta_1 {\rm sin}~\theta_2 d\phi_1 d\phi_2\right) + \\ 
&& ~~~~~~~~~~~~~~~~~~~~~~~~ + h_5 ~{\rm sin}~\psi \left({\rm sin}~\theta_1~d\theta_2 d\phi_1 - 
{\rm sin}~\theta_2~d\theta_1 d\phi_2\right)
\end{eqnarray}
with $h_i$ being the six-dimensional warp factors. One advantage of writing the background in the above form 
is that it includes all possible deformations in the presence of seven branes and fluxes. The difficulty however 
is that the equations for the warp factors $h_i$ are coupled higher order differential equations that do not 
have simple analytical solutions. The original KS solution is in the limit 
\bg \label{ksagain} h_3 = g_i = 1, ~~ h_i = {\rm fixed} \nd
but has no seven branes. In the presence of seven branes
we can do slightly better in the limit:
\bg \label{resconi} 
h_5 = 0, ~~~ h_3 = 1, ~~~ h_4 - h_2 = a, ~~~ g_i = 1 \nd
which puts a seven brane in a resolved conifold background with $a$ = constant \cite{sully} 
using the so-called Ouyang embedding \cite{4}. A black hole could be inserted in this background by switching on 
non-trivial $g_i$. However a naive choice of fluxes in 
\eqref{resconi} is known to break supersymmetry \cite{sully}.

Our next choice would then be to go to the limit where the resolution parameter $a$ in \eqref{resconi} is vanishing. 
This is the Ouyang background \cite{4} with $g_i(r) = 1$ i.e with seven branes but no black holes. Even in the absence 
of black holes, supersymmetry is an issue here as was pointed out in \cite{sully}. Supersymmetry is
spontaneously broken {\it but could be restored} by switching on appropriate gauge fluxes on the seven branes 
\cite{sully, ouyangshiu}. The seven branes are embedded via the following equation (see also \cite{4, sully}):
\bg \label{seven} 
r^{3\over 2} {\rm exp}\left[i(\psi-\phi_1 - \phi_2)\over 2\right] {\rm sin}~{\theta_1 \over 2} 
~{\rm sin}~{\theta_2 \over 2} = \mu
\nd
where $\mu$ is a complex quantity. In the limit where $\mu \to 0$, the seven branes 
are oriented along the two branches:
\begin{eqnarray}\label{branches} \nonumber 
&&{\rm Branch ~1}: ~~\theta_1 = 0, ~~\phi_1 = 0 \\
&&{\rm Branch ~2}: ~~ \theta_2 = 0, ~~\phi_2 = 0
\end{eqnarray}
{}From above it is easy to see that the seven branes in Branch 1 wrap a four cycle ($\theta_2, \phi_2$) and 
($\psi, r$) in the internal space and is stretched along the spacetime directions ($t, x, y, z$). Similarly in 
Branch 2 the seven branes would wrap a four-cycle ($\theta_1, \phi_1, r, \psi$). 

{}From the above discussion,
one might get a little concerned by the fact that the seven branes wrap a non-compact four cycle in the internal
space. This would suggest a violation of the Gauss' law as the axion charges of the seven branes have no place
to escape. For the time being 
this apparent paradox can  be resolved by allowing the seven brane to wrap a topologically trivial 
cycle so that it would end {\it abruptly} at some $r = r_0$ when the embedding is \eqref{seven}. 
This is similar to the seven brane configuration
of \cite{karch}. We will discuss a better embedding later.  

Once the above issues are resolved, we can insert a black hole in the modified 
Ouyang background \eqref{resconi} by switching on appropriate 
$g_i(r)$. Clearly we do not expect $h_i$ to remain constant anymore. 
We also expect $M$ and $N_f$ to be given by some $M_{\rm eff}$ and $N_{f}^{\rm eff}$ respectively.
Our first approximation would then be to take 
the following ansatze for the $h_i, M_{\rm eff}$ and $N_{f}^{\rm eff}$:
\bg \label{hi}
&&h_1 = {1\over 9} + {\cal O}(g_s), ~~~~~ h_2 = h_4 = {1\over 6} + {\cal O}(g_s), ~~~~~ h_3 = 1 + {\cal O}(g_s)\\
&&M_{\rm eff} = M + \sum_{m\ge n} a_{mn} (g_sN_f)^m (g_sM)^n, ~~~~~ 
N_{f}^{\rm eff} = N_f + \sum_{m \ge n} b_{mn} (g_sN_f)^m (g_sM)^n\nonumber 
\nd
with $a_{mn}, b_{mn}$ could in principle be functions of the CY coordinates. 
Note that we have made $m \ge n$ in the above expansions because the precise limits for which our series would be  
valid are: 
\bg\label{prelim2} 
\left (g_s, ~g_s N_f, ~g^2_s M N_f, ~{g_s M^2\over N}\right) ~ \to ~ 0, ~~~~~
(g_s N, ~g_s M)~ \to ~ \infty
\nd  
These limits of the variables (which we will concentrate on from now on), 
bring us closer to the Ouyang solution with very little squashing of the two spheres. This also means 
that the warp factor $h$ in \eqref{bhmet} can be written as:
\bg \label{hvalue}
h =\frac{L^4}{r^4}\Bigg[1+\frac{3g_sM_{\rm eff}^2}{2\pi N}{\rm log}r\left\{1+\frac{3g_sN^{\rm eff}_f}{2\pi}\left({\rm
log}r+\frac{1}{2}\right)+\frac{g_sN^{\rm eff}_f}{4\pi}{\rm log}\left({\rm sin}\frac{\theta_1}{2}
{\rm sin}\frac{\theta_2}{2}\right)\right\}\Bigg]\nonumber\\
\nd
with $d{\cal M}_5$ in \eqref{bhmet2} can be approximated by the angular part of the conifold metric \eqref{conifold}. 
We will however modify this further soon to get the result for large $r$.

Question now is what choices of $g_i$ are we allowed to take to insert the black hole in this geometry?  
It turns out that to solve EOMs the $g_i$'s have to be functions of ($r,\theta_1,\theta_2$). Our ansatze therefore 
would be:
\bg \label{grdef} g_1(r,\theta_1,\theta_2)= 1-\frac{r_h^4}{r^4} + {\cal O}(g^2_sM N_f), 
~~~~ g_2(r,\theta_1,\theta_2) = 1-\frac{r_h^4}{r^4} + {\cal O}(g^2_sM N_f)\nd
where $r_h$ is the horizon, and the ($\theta_1, \theta_2$) dependences come from the ${\cal O}(g^2_sM N_f)$ 
corrections. We also expect $a = a(r_h) + {\cal O}(g_s^2 M N_f)$ as the full resolution parameter. Therefore
in order to extract temperature from the geometry, we look at the metric in (\ref{bhmet}) in the near horizon limit
$r\rightarrow r_h$ and take a five dimensional slice obtained by setting $\theta_i=\pi,\phi_i=0,\psi=0$. To be exact, we
really need to start from the ten dimensional supergravity action and then integrate out the 
internal directions to obtain a
five dimensional effective action. Minimization of that five dimensional effective action will give the five dimensional
effective metric. Here we approximate this effective metric by taking the slice 
\bg\label{sol}\theta_1~ = ~ \theta_2 ~ = ~ \pi,~~~~~~\phi_i~ = ~ 0,~~~~~~~\psi~ = ~0 \nd
Such a choice can be justified by observing that the flavor D7 branes are all along this slice 
(see details below)\footnote{Note also that all the analysis presented in the appendices are done {\it without}
taking such a slice. What we find that taking a slice doesn't give us results very different from the full analysis. 
It only makes our calculations a little easier to handle. 
As an example we can quote \eqref{SV-11b} or \eqref{e2a} where the entropy is 
computed without taking the slice. We can see that the result differs from the one with slice by 
an ${\cal O}(g_sN_f)$ term, so is very small. Similar statements can be made for the case with dilaton also. The 
dilaton profile doesn't vary too much throughout our regime of interest (i.e small $r$). So taking a constant 
dilaton is meaningful here. The full analysis is underway where we plan to take the effects of dilaton and 
curvatures carefully.}.
   
Now, looking at the $r,t$ direction of the metric and by change of variable, under the assumption that 
$g_1(r, \pi, \pi) \approx g_2(r, \pi, \pi) \equiv g(r)$,
we can define $\rho^2$ as: 
\bg \label{Temp1}
\rho^2=\frac{4\sqrt{h(r_h)}g(r)}{[g'(r_h)]^2}
\nd
so that the near horizon limit of five dimensional effective metric takes the following Rindler form: 
\bg \label{Temp2}
ds^2=-\rho^2 \frac{g'(r_h)^2}{4h(r_h)g(r_c)}dt_c^2+d\rho^2
\nd
where prime denotes differentiation with respect to $r$ and we only wrote the 
$r,t$ part of the metric in terms of new variable $\rho$ and $t_c \equiv \sqrt{g(r_c)}t$.  
The
reason behind rescaling time at fixed $r_c$ is that with this time coordinate $t_c$, the five dimensional metric induces a
four dimensional Minkowski metric at every $r_c$. 

Now the temperature observed by the field theory with time coordinate
$t_c$ can be extracted by writing the metric in \eqref{Temp2} in the following form 
\bg \label{Temp3}
ds^2=-4\pi^2 T_c^2\rho^2 ~dt_c^2~ + ~ d\rho^2
\nd   
Thus comparing \eqref{Temp2} and \eqref{Temp3}, we obtain the temperature $T_c$ as:
\bg \label{Temp4}
T_c~= ~ \frac{g'(r_h)}{4\pi\sqrt{h(r_h)g(r_c)}}
\nd   
In the limit where we have $g_1 = g_2 = g = 1-{r_h^4\over r^4}$, we can easily compute the corresponding temperature using
the above formula \eqref{Temp4}. This is given by:
\bg\label{Temp5}
T_c ~ = ~ {r_h \over \pi L^2} ~ + ~ {r_h^5\over 2\pi L^2 r_c^4} + \sum_{m,n,p} c_{mnp} {r_h^{m} {\rm log}^n r_h\over 
r_c^{p}}
~ \equiv  ~ {T}_b ~+~ {\cal O}(1/r_c)
\nd
where $L$ is defined earlier, $c_{mnp}$ is in general functions of ($g_s, M, N, N_f$), 
and ${T}_b > T_{\rm deconf}$ (where $T_{\rm deconf}$ is the deconfinement 
temperature) is the temperature at $r_c \to \infty$ i.e
\bg\label{bndtemp}
{T}_b ~\equiv~ T_{\rm boundary} ~=~ {g'(r_h)\over 4\pi \sqrt{h(r_h)}} ~~~~~ \implies ~~~ r_h ~\equiv~ F({T_b}) 
~\equiv ~{\cal T}
\nd
where $F(T_b)$ is the inverse transform of the above expression once we know the black hole 
factor $g(r_h)$ as well as the warp factor $h(r_h)$ to all orders in $g_sN_f, g_s M$. The corresponding 
temperature function ${\cal T}$ will then be the characteristic temperature of the boundary theory which is fixed 
once and for all. 
This temperature function is a scale unique to our cascading theory and will be assumed to be greater than the 
deconfinement temperature henceforth. 
 
For the attentive readers, something about equations \eqref{Temp4} and \eqref{Temp5} may strike as odd. There seems to be
an ${\cal O}(1/r_c)$ dependences. What does it mean for a physical variable in our theory to have an ${\cal O}(1/r_c)$
i.e the cutoff dependence? Shouldn't we make $r_c \to \infty$ to get a result that is independent of the cutoff? In other
words does it make sense to have an explicit cutoff dependence in the physical variables? 

Of course in the standard Wilsonian renormalisation we do get results that depend on the explicit UV cutoff. In the limit 
where we make $r_c \to \infty$ we reproduce exactly the right temperature for the {\it boundary} theory. 
The fact that the temperature
{\it does not} have the cutoff dependence going as positive powers of $r_c$ should be pleasing: this has to do with 
the holographic renormalisability of the theory (that we elaborate in details later). However the fact that there is 
{\it some} cutoff dependence should signal something quite different from what we have in the AdS/QCD case. 

To see what is different here, let us pause for a while and ask what would the presence of a black hole signify 
in the dual gauge theory. In the standard AdS/QCD case we only needed to look at the boundary of the AdS Black-Hole 
(AdS-BH) geometry to study the properties of the dual thermal QCD. Here the situation differs crucially because we can 
study the weakly coupled gravity description not only at $r \to \infty$ but also at any arbitrary $r = r_c$. What does 
it mean to study the theory at $r = r_c$ and not $r \to \infty$? Of course, as we commented before, 
we can always make $r_c \to \infty$ to get the far UV results, but cutting off the theory at $r = r_c$ 
means we are putting a Wilsonian 
cutoff in the theory. {\it Such a procedure would make sense if and only if
we can carefully describe the degrees of 
freedom at} $r = r_c$. 
Furthermore $-$ and this is one of the most crucial point $-$ since the theory with flavor is holographically
renormalisable (as we will show soon) all the cutoff dependences will come as ${\cal O}(1/r_c)$. Such a state of affairs
lead to two interesting conclusions, one of which is obvious and the other not so obvious. The obvious point is that
as long as we put our cutoff at high energy, the low energy dynamics remain completely unaffected by 
our choice of the 
cutoff. The not-so-obvious point is that  
the shear-viscosity etc that we will discuss in detail soon for geometries that are cut-off at $r = r_c$ will in fact
be {independent} of the cut-off once we study them from boundary point of view! This conclusion is surprising 
because we will {\it not} be making $r_c \to \infty$ to study the boundary theory, rather we will add non-trivial 
UV ``cap'' to the geometry from $r = r_c$ to $r = \infty$. 

So the point that we want to emphasise here is that once we introduce a cutoff at $r = r_c$ we are in principle 
introducing non-trivial high energy degrees of freedom that will in general take us away from the usual cascading 
dynamics of the parent theory! {\it Adding such non-trivial degrees of freedom at $r = r_c$ is {equivalent} to adding
a UV cap to the geometry}. 
Thus it all depends 
on what we do at $r_c$ i.e which boundary conditions we choose
there. Therefore they are {\it not} universal properties of the cascading 
theory\footnote{We thank Ofer Aharony for emphasising 
this point, and clarifying other details about the cutoff dependences.}.
Universal properties arise only as $r_c \to \infty$ i.e at the boundary\footnote{Alternatively this means that we are  
not adding any UV cap because the geometry can be thought of as being ``cut-off'' at $r_c = \infty$. One issue here is 
the connection to the work of \cite{balkrauss}. The question is 
can we write a boundary theory at $r = r_c$ itself instead of going to the 
actual $r = \infty$ boundary? As shown in \cite{balkrauss} this is possible in AdS case because the 
theory on the surface of a ball in AdS space doesn't have to be a local quantum field theory. The non-local behavior
in such a ``boundary'' theory is completely captured by the Wilsonian effective action at the so called boundary. 
As discussed in the text earlier,
this is tricky in the Klebanov-Strassler model precisely because there is no unique strongly 
coupled gauge theory here. Again, as we saw in our RG flow pictures, when the dual gravity theory is weakly coupled 
we have a smooth RG flow in the gauge theory side, but the theory at any given scale can be given by 
infinite number of representative theories none of which completely capture the full dynamics. Thus it makes more 
sense to 
define the theories at $r = \infty$ boundary only and not at any generic $r = r_c$.}. 
For example we may think of attaching 
${\cal N} = 2, 4$ degrees of freedom at $r = r_c$ (somewhat along the lines of \cite{Hollowood:2004ek}) alongwith the 
remnants of the D7 brane degrees of freedom to have the full F-theory picture (in fact our UV completions should 
always have $24 - N_f$ seven branes attached to the UV caps)\footnote{The fact that there are infinite F-theory 
backgrounds gluing to our IR solutions will be discussed a little later.}. 
Clearly 
adding such degrees of freedom will give us a new theory that differs from the UV degrees of freedom of the parent 
cascading theory. In particular if $r_c$ is finite then this procedure can give us an almost free UV theory with 
confining IR dynamics. Also since at any given scale there are an infinite number of possible gauge theory descriptions 
available with different UV degrees of freedom, we can specify a particular set of degrees of freedom at 
$r= r_c$ to define the 
UV of this theory. Once this is specified the RG flow will take us to the IR. A sketch of the situation is depicted in
{\bf figure 7} below where the gravity description 
captures the smooth RG flow. 
\begin{figure}[htb]\label{duality_cascade1}
		\begin{center}		
\includegraphics[height=8cm]{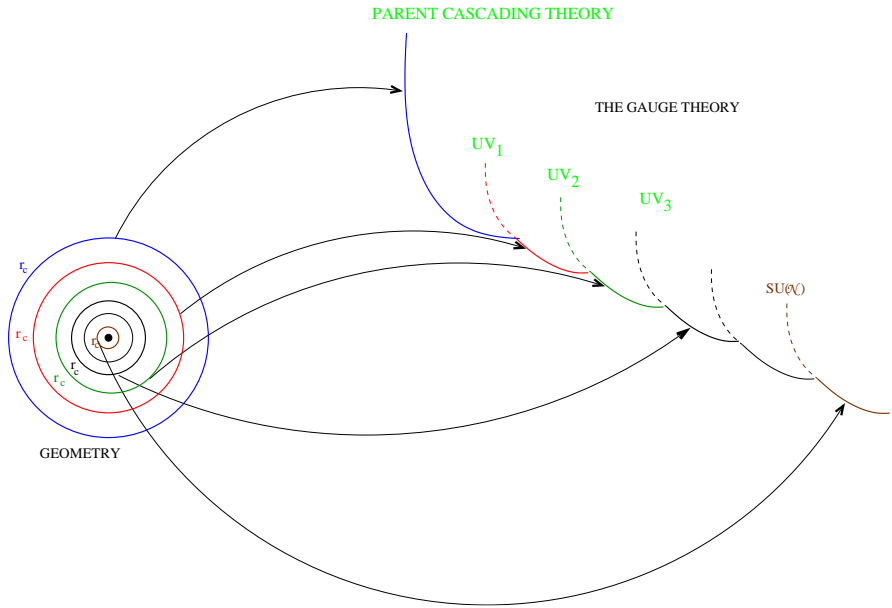}
		\caption{{ An oversimplified picture of the gravity 
description of the various gauge theories involved. 
The onion-ring model is only a caricature of a much more involved picture that we gave earlier. The dotted lines 
emanating from various points (i.e various gauge theories)
specify the distinct UV completions of these gauge theories. Once we specify the UV 
degrees of freedom at various $r = r_c$ the RG flows of each theories should be assumed to bring us to $r = r_c$ 
 points on the 
smooth RG flow of the original cascading theory. Note that at any 
point on the gauge theory side (i.e the vertical axis)
there are {\it infinite} number of possible gauge-theory descriptions available (the horizontal axis).}}  
		\end{center}
		\end{figure}  
We can view the dotted lines coming out from any 
$r= r_c$ and going to infinity (in the gauge theory side)
as describing the UV degrees of freedom for a theory at $r = r_c$. Thus in principle there is a possibility of 
defining infinite number of theories here by cutting off the geometry at various $r = r_c$ and then ``filling'' the 
$r= r_c$ to $r = \infty$ spaces
by UV caps that specify the UV degrees of freedom of these gauge theories (of course we expect small changes in the 
IR geometries every time we attach UV caps, although the far IR geometries wouldn't change). 
Under RG flows these 
UV theories (that are obviously quite different from the parent
cascading theory) meet the RG flow of the cascading theory 
at various $r = r_c$ (where of course the degrees of freedom match but not the whole 
set of marginal, relevant and irrelevant operators). 
The effective degrees of freedom at $r = r_c$ are:
\bg\label{edofk}
N_{\rm eff} = N ~ + ~ {3g_s M^2\over 2\pi} ~{\rm log}~r_c + {9g^2_s M^2 N_f\over 4\pi^2}~{\rm log}^2 r_c
\nd  
Therefore $r_c$ dependences in the physical result for our case should be replaced by the effective number 
of degrees of freedom at that scale, i.e by:
\bg\label{rcinversion}
r_c &~ = ~& {\rm exp}\left[{\pi\over 3g_s}\sqrt{{1\over N_f^2} 
- {4(N-N_{\rm eff})\over M^2 N_f}} - {\pi\over 3g_s N_f}\right]\nonumber\\
&~\approx ~& {\rm exp}\left[\alpha + \beta N_{\rm eff}\right]
\nd
which in turn
means that the right UV degrees of freedom should flow to this value at $r = r_c$\footnote{Even more interestingly, 
the final results for any physical quantities from these theories will only depend on the boundary degrees of freedom
and not on ${\cal N}_{\rm eff}$. The ${\cal N}_{\rm eff}$ degrees of freedom are only intermediate. However there are 
special cases where the boundary degrees of freedom could be close to ${\cal N}_{\rm eff}$. We will discuss them later.}. 
We have also defined
 $\alpha = {\pi\over 3g_s M}\left({\sqrt{M^2 - 4NN_f^2}\over N_f} -1\right)$ and $ \beta = 
{2\pi\over 3g_s M\sqrt{M^2 - 4N N_f}}$ 
in the limit where $N_{\rm eff}$ is small compared to the original 
colors and bi-fundamental flavors in the theory i.e $N > M > N_{\rm eff} > N_f$ precisely. 
In general then\footnote{There is another flavor-dependent degrees of freedom that one could define in this 
background. We will define this via $r_{c(\alpha)} = e^{{\cal N}_{\rm eff}[1-\epsilon_{(\alpha)}]}$, where the variables
are defined in \eqref{logr} and \eqref{epde}.} 
\bg\label{doa}
r_c ~\equiv ~ e^{{\cal N}_{\rm eff}}
\nd 
with ${\cal N}_{\rm eff}(\Lambda_c)$ 
being the effective degrees of freedom at a given scale $\Lambda_c$. 
Thus to summarise:

\noindent $\bullet$ Making the cutoff to infinity i.e $r_c \to \infty$ gives us the property of the original cascading 
theory with a particular UV behavior and a smooth RG flow to IR.

\noindent $\bullet$ Defining the theory with a cutoff at $r = r_c$ means that we specify the UV degrees of freedom 
of a particular gauge theory at that scale or equivalently introduce a UV cap in the geometry so that we are always 
defining the theory at the boundary\footnote{It should be clear that unless we specify clearly the UV degrees 
of freedom, we cannot pin-point the gauge theory there. The parent cascading theory allows an {\it infinite} number of 
gauge theory description at any given scale. At the point where we have smooth RG flow, none of these descriptions 
capture the full picture there. Thus to emphasise again, the gauge theory or the UV behavior
that we want to specify at that scale is 
distinct from the UV of the cascading theory as it could have finite (but large) number of degrees of freedom.}.  
This theory also has a smooth RG flow but the 
UV behavior (shown by dotted lines) is quite different
from the parent cascading theory.
Only at certain IR scale (i.e $r = r_c$ from the gravity side) do these theories match (wrt the 
degrees of freedom and certain set of marginal and relevant 
operators). 

%popo2

Once this is settled, let us now introduce fundamental matters in our geometry. 
As we discussed above, the seven branes introduce fundamental
matter with $N_f$ flavors. Since the above background is a deformation of both KS and Ouyang backgrounds, we will 
refer the resulting geometry as Ouyang-Klebanov-Strassler (OKS) background. With the black hole, this will 
henceforth be called as OKS-BH background. 

In the OKS-BH background, to order ${\cal O}(g_S N_f)$ and small $r$,  
let us start by considering the warp factor choice \eqref{resconi} but keeping 
$g_i(r)$ as \eqref{grdef} instead of 1. The background RR three and five-form fluxes can be 
succinctly written as\footnote{As we mentioned earlier, at large $r$ we expect all $c_i$ to be finite. This means that 
$c_i$ should become functions of inverse powers of $r$ at large $r$. This is expected from F-theory considerations and 
would help us remove the Landau pole.}: 
\bg\label{h3f3}\nonumber
  H_3 &=& dr\wedge e_\psi\wedge(c_1\,d\theta_1+c_2\,d\theta_2) + dr\wedge(c_3\sin\theta_1\,d\theta_1\wedge d\phi_1-c_4\sin\theta_2\,d\theta_2\wedge d\phi_2)\\
  & & +\left(\frac{r^2+6a^2}{2r}\,c_1\sin\theta_2\,d\phi_2 
    -\frac{r}{2}\,c_2\sin\theta_1\,d\phi_1\right)\wedge d\theta_1\wedge d\theta_2\,,\nonumber\\
  \widetilde{F}_3 &=& -\frac{1}{g_s}\,dr\wedge e_\psi\wedge(c_1\sin\theta_1\,d\phi_1+c_2\sin\theta_2\,d\phi_2)\nonumber\\
  & &  +\frac{1}{g_s}\,e_\psi\wedge(c_5\sin\theta_1\,d\theta_1\wedge d\phi_1-c_6\sin\theta_2\,d\theta_2\wedge 
d\phi_2)\nonumber\\ 
  & & -\frac{1}{g_s}\,\sin\theta_1\sin\theta_2\left(\frac{r}{2}\,c_2 \,d\theta_1-\frac{r^2+6a^2}{2r}\,c_1\,d\theta_2 \right)
    \wedge d\phi_1\wedge d\phi_2\,.
\nd
where $H_3$ is closed and $\widetilde F_3 \equiv F_3 - C_0 H_3$, $C_0$ being the ten dimensional axion. 
The derivations of the 
coefficients appearing 
in \eqref{h3f3} are rather involved, and their dependences on the resolution factor etc. 
will be described in section 3.3\footnote{It is instructive to note that the background 
EOMs cannot be trivially worked out by solving SUGRA EOMs with fluxes and seven branes sources. This is because, even if 
we know the energy momentum tensors for the fluxes, the energy momentum tensors for $N_f$ coincident 
seven branes are not 
known in the literature! The difficulty lies in finding a {\it non-abelian} Born-Infeld action for $N_f$ 
seven branes on a {\it curved} background. 
As far as we are aware of, this problem has remained unsolved till now. So in the absence of such direct approach, 
we use an alternative method to derive the EOMs. This method uses the ISD (imaginary self-duality) properties of the 
background fluxes and fields. Details on this have already appeared in \cite{4}\cite{sully}, 
so we would refer the readers 
there for a complete analysis. 
Our present analysis is however more involved than \cite{4}\cite{sully} 
because we have a black hole and no supersymmetry. 
We do however find that even for this scenario, one could find consistent solutions to EOMs using similar arguments. 
For example see equations 
\eqref{ouysol} to equation \eqref{axfive} for more detailed derivations.}. 
For the present purpose, let us just quote the 
results: 
\bg\label{defc}\nonumber
  c_1 &=& \frac{g_s^2MN_f}{4\pi r(r^2+6a^2)^2}\,\big(72 a^4-3r^4-56a^2r^2\log r+a^2 r^2\log(r^2+9a^2)\big)\,
    \cot\frac{\theta_1}{2}\nonumber
\nd
\bg 
  c_2 &=& \frac{3g_s^2MN_f}{4\pi r^3}\,\big(r^2-9a^2\log(r^2+9a^2)\big)\, \cot\frac{\theta_2}{2}\\ \nonumber
%\bg\label{defc2}\nonumber
  c_3 &=& \frac{3g_sM r}{r^2+9a^2}+\frac{g_s^2MN_f}{8\pi r(r^2+9a^2)}\,\Big[-36a^2-36 r^2\log a+34 r^2\log r\\ \nonumber
    & & \qquad\qquad\qquad\qquad +(10r^2+81a^2)\log(r^2+9a^2)+12r^2\log\left(\sin\frac{\theta_1}{2}\sin\frac{\theta_2}{2}\right)\Big]\\ \nonumber
 % \nd
%\bg\label{defc3}\nonumber
c_4 &=& \frac{3g_sM(r^2+6a^2)}{\kappa r^3}+\frac{g_s^2MN_f}{8\pi\kappa r^3}\,\Big[18a^2-36(r^2+6a^2)\log a+(34 r^2+36a^2) 
    \log r\\ \nonumber 
  & & \qquad\qquad + (10 r^2+63a^2)\log(r^2+9a^2)+(12 r^2+72a^2)\log\left(\sin\frac{\theta_1}{2}\sin\frac{\theta_2}{2}\right)\Big] \\ \nonumber
%\nd
%\bg\label{defc4}\nonumber
c_5 &=& g_sM+\frac{g_s^2MN_f}{24\pi(r^2+6a^2)}\,\Big[18a^2-36(r^2+6a^2)\log a+8(2r^2-9a^2)\log r\\ \nonumber
  & & \qquad\qquad\qquad\qquad\qquad\qquad\qquad\qquad\qquad\qquad\;\;+(10r^2+63a^2)\log(r^2+9a^2)
    \Big]\\ \nonumber
  c_6 &=& g_sM+\frac{g_s^2MN_f}{24\pi r^2}\,\Big[-36a^2-36r^2\log a+16r^2\log r+(10r^2+81a^2)\log(r^2+9a^2)\Big]
\end{eqnarray}
with $\kappa = \frac{r^2 + 9a^2}{r^2 + 6a^2}$. All the above coefficients have further corrections that we will discuss
later. Finally,
this allows us to write the NS 2--form potential:
\begin{eqnarray}\label{btwo}
  B_2 &=& \left(b_1(r)\cot\frac{\theta_1}{2}\,d\theta_1+b_2(r)\cot\frac{\theta_2}{2}\,d\theta_2\right)\wedge e_\psi\\ \nonumber
  & + &\left[\frac{3g_s^2MN_f}{4\pi}\,\left(1+\log(r^2+9a^2)\right)\log\left(\sin\frac{\theta_1}{2}\sin\frac{\theta_2}{2}\right)
    +b_3(r)\right]\sin\theta_1\,d\theta_1\wedge d\phi_1\\ \nonumber 
  & - & \left[\frac{g_s^2MN_f}{12\pi r^2}\left(-36a^2+9r^2+16r^2\log r+r^2\log(r^2+9a^2)\right)
    \log\left(\sin\frac{\theta_1}{2}\sin\frac{\theta_2}{2}\right)+b_4(r)\right]\\ \nonumber
  & & \qquad\qquad \times \sin\theta_2\,d\theta_2\wedge d\phi_2
\end{eqnarray}
with the $r$-dependent functions 
\begin{eqnarray}\label{defb}\nonumber
  b_1(r) &=& \frac{g_S^2MN_f}{24\pi(r^2+6a^2)}\big(18a^2+(16r^2-72a^2)\log r+(r^2+9a^2)\log(r^2+9a^2)\big)\\
  b_2(r) &=& -\frac{3g_s^2MN_f}{8\pi r^2}\big(r^2+9a^2\big)\log(r^2+9a^2)
\end{eqnarray}
and $b_3(r)$ and $b_4(r)$ are given by the first order differential equations
\bg\label{gdas}
  b_3'(r) &=& \frac{3g_sMr}{r^2+9a^2} + \frac{g_s^2MN_f}{8\pi r(r^2+9a^2)}\Big[-36a^2-36a^2\log a
+34 r^2\log r\\ \nonumber
  & & \qquad\qquad\qquad\qquad\qquad\qquad+(10 r^2+81a^2)
    \log(r^2+9a^2)\Big]\nonumber\\
  b_4'(r) &=& -\frac{3g_sM(r^2+6a^2)}{\kappa r^3} - \frac{g_s^2MN_f}{8\pi\kappa r^3}\Big[18a^2-36(r^2+6a^2)\log a\\ \nonumber 
  & & \qquad\qquad\qquad+(34 r^2+36a^2)\log r +(10r^2+63a^2)\log(r^2+9a^2)\Big]
\nd
where $M$ denote the remnant of the number of fractional three branes (in the gauge theory side)
and $N_f$ denote the number of flavors or seven branes in the dual gravity side. Therefore
once we know $B_2$ and the string coupling $e^{-\Phi}$ then it is easy to determine the two couplings at the 
UV of our dual gauge theory (see also \cite{klebwit}):
\begin{eqnarray} \label{twocoup} \nonumber
&&\frac{8\pi^2}{g_1^2} = e^{-\Phi}\left[\pi - {1\over 2} + \frac{1}{2\pi} \left(\int_{S^2} 
B_2\right)\right] \\  
&& \frac{8\pi^2}{g_2^2} = e^{-\Phi}\left[\pi + {1\over 2} -
 \frac{1}{2\pi} \left(\int_{S^2} B_2\right)\right]  
\end{eqnarray}
The string coupling can be determined easily from the monodromy around the seven brane to be:
\bg\label{dilato}
e^{-\Phi} = {1\over g_s} -\frac{N_f}{8\pi} ~{\rm log} \left(r^6 + 9a^2 r^4\right) - 
\frac{N_f}{2\pi} {\rm log} \left({\rm sin}~{\theta_1\over 2} ~ {\rm sin}~{\theta_2\over 2}\right)
\nd
which immediately gives us:
\begin{eqnarray} \label{SV} \nonumber
&&\frac{\partial}{\partial{\rm log}~\Lambda}\left[\frac{4\pi^2}{g_1^2} + \frac{4\pi^2}{g_2^2}\right] = 
- \frac{N_f}{8}\left(\frac{6r^6 + 36 a^2 r^4}{r^6 + 9a^2 r^4}\right)\\
&& \frac{\partial}{\partial{\rm log}~\Lambda}\left[\frac{4\pi^2}{g_1^2} - \frac{4\pi^2}{g_2^2}\right] = 
3M\left(1 + \frac{3g_s N_f}{4\pi} ~{\rm log}(r^2 + 9a^2) + ....\right)
\end{eqnarray}
which to the leading order is consistent with the Shifman-Vainshtein $\beta$-function 
\cite{SV}\footnote{Note that the LHS of the equations 
\eqref{SV} involve only gauge theory variables whereas the RHS of the equations involve gravity variables. In particular
$\Lambda$ should be identified with the radial coordinate $r$ of the geometry. Thus the equations \eqref{SV} capture
the essence of gauge/gravity duality here.}.    
The behavior at 
subleading order will tell us how the color changes as we cascade down from UV to IR. For example, in the presence 
of $N_f$ flavors the $SU(N+M)$ gauge group has $2N + N_f$ effective flavors. Under RG flow, Seiberg duality will 
tell us that the weakly coupled gauge group will become $SU(N - M + N_f) \times SU(N)$. This also means that the 
cascade will slow down quite a bit as we approach the IR and therefore the end point of the cascade could either be 
a conformal theory or a confining theory. 
As pointed out also in 
\cite{4} if in the end of the cascade $N$ decreases to zero with finite $M$ left over, we would have $SU(M)$ SYM 
with $N_f$ flavors in the IR. Extending this to the centre of the RG fixed points surface will allow us to 
analyse this using weakly coupled supergravity. 
This is of course the theory we are aiming for.  
Finally, 
the five--form flux is as usual given by 
\begin{equation}
  \widehat{F}_5 \,=\, (1+ {\ast}_{10})(dh^{-1}\wedge d^4x)\,.
\end{equation}
with $h$ being the warp factor described above. 

Before we end this section we want to point out few subtleties about the background. First as we mentioned before, 
to maintain Gauss law constraint, we need to embed our model in the full F-theory \cite{vafaF} setup. This means that the 
background configuration that we presented above should be understood as an F-theory on a four-fold where all but 
$N_f$ of the seven branes have been moved to infinity. This means that $N_f < 24$, and the four-fold is a non-trivial
torus fibration over a resolved conifold base. Such a four-fold has already been constructed in \cite{gtpapers, sully}
and so we can direct the readers to those papers for details. What is interesting here is that due to F-theory 
embedding, and with modified three-form fluxes, we expect the background to 
{\it not} have any naked singularities or Landau poles that are often associated with these backgrounds
(see for example \cite{vaman}, \cite{ingo} for some details). 
To ${\cal O}(g_sN_f)$ our result that we presented may indicate 
the presence of ${\rm log}~r$ type singularities. Additionally at:
\bg\label{dilsing}
r~\approx~{\rm exp} \left({4\pi \over 3 g_sN_f}\right), ~~~~~ {\rm for}~~~a << g_sN_f
\nd 
we might think that the dilaton is blowing up on the given slice \eqref{sol} leading to some kind of naked 
singularity. However note that
we are seeing such behavior because we have evaluated the background locally near the seven branes, and 
upto ${\cal O}(g_sN_f)$. Clearly the metric, dilaton and the fluxes have to have a good behavior at 
infinity to be a F-theory solution. 
One way to show this would be to rewrite the 
warp factor \eqref{hvalue}
completely in terms of power series in $r$ in the following way:
\bg\label{logr}
h~=~ {L^4\over r^{4- \epsilon_1}} ~+~ {L^4\over r^{4- 2\epsilon_2}} ~-~ {2L^4\over r^{4- \epsilon_2}} ~+~ {L^4 \over 
r^{4-r^{\epsilon^2_2/2}}} 
~\equiv~ \sum_{\alpha=1}^4 {L^4_{(\alpha)}\over r^4_{(\alpha)}}
\nd
where $\epsilon_i, r_{(\alpha)}$ etc are defined as:
\bg\label{epde}
&&\epsilon_1 ~ = ~ {3g_s M^2\over 2\pi N} + {g_s^2 M^2 N_f\over 8\pi^2 N} + {3g_s^2 M^2 N_f \over 8\pi N} ~
{\rm log}\left({\rm sin}~{\theta_1\over 2} {\rm sin}~{\theta_2\over 2}\right),  
~~\epsilon_2 ~ = ~ {g_s M \over \pi}\sqrt{2N_f\over N}\nonumber\\
&& ~~~~~~~~~~~ r_{(\alpha)} = r^{1-\epsilon_{(\alpha)}}, ~~ \epsilon_{(1)} = {\epsilon_1\over 4}, ~~ 
\epsilon_{(2)} = {\epsilon_2\over 2}, ~~ \epsilon_{(3)} = {\epsilon_2\over 4}, ~~ \epsilon_{(4)} = 
{\epsilon^2_2\over 8}\nonumber\\
&& ~~~~~~~~~~~ r_{(\pm\alpha)} = r^{1\mp\epsilon_{(\alpha)}}, ~~~L_{(1)} = L_{(2)} = L_{(4)} = L^4, ~~~ L_{(3)} = -2L^4 
\nd
which makes sense because we can make $\epsilon_i$ to be very small. The angles $\theta_i$ take fixed values on the
given slice \eqref{sol}. 
Note that the choice of $\epsilon_i$ doesn't require us to have 
$g_s N_f$ small (although we consider it here). 
In fact we {\it can} have all ($N, M, N_f$) large but $\epsilon_i$ 
small. A simple way to achieve this would be to have the following scaling behaviors of 
($g_s, N, M, N_f$):
\bg\label{scalbet}
g_s ~\to ~\epsilon^\alpha, ~~~~~M ~\to ~\epsilon^{-\beta}, ~~~~~ 
N_f ~\to ~\epsilon^{-\kappa}, ~~~~~ N ~\to ~\epsilon^{-\gamma} 
\nd
where $\epsilon \to 0$ is  the tunable parameter. Therefore all we require to achieve that is to allow:
\bg\label{allow}
\alpha ~+~ \gamma ~>~ 2\beta ~+~ \kappa, ~~~~~~~ \alpha ~>~ \kappa, ~~~~~~~~ \gamma ~>~ \alpha
\nd 
where the last inequality can keep $g_s N_f$ small. Thus $g_s N, g_s M$ are very large, but $g_s, 
{g_s M^2 \over N}, g_s N_f^{}$ are all very small to justify our expansions (and the choice of 
supergravity background)\footnote{For example we can have $g_s$ going to zero as  $g_s \to \epsilon^{5/2}$ and 
($N, M_{}, N_f$) going to infinities as ($\epsilon^{-8}, \epsilon^{-3}, \epsilon^{-1}$) respectively. 
This 
means ($g_s N, g_s M$) go to infinities as ($ \epsilon^{-11/2}, \epsilon^{-1/2}$) respectively, and 
($g_sN_f, g^2_sM N_f, g_sM^2/N$) 
go to zero as ($\epsilon^{3/2}, \epsilon, \epsilon^{9/2}$) respectively. 
This is one limit where we can have well defined UV completed 
gauge theories. Note however that for the kind of background that we have been studying one cannot make $N_f$ large 
because of the underlying F-theory constraints \cite{vafaF}. Since we only require $g_sN_f$ small, large or small 
$N_f$ choices do not change any of our results.}. 

The warp factor \eqref{logr} has a good behavior at infinity and reproduces the ${\cal O}(g_sN_f)$ result locally.
Our conjecture then would be the complete form of the warp factor at large $r$
will be given by sum over $\alpha$ as in 
\eqref{epde} but now $\alpha$ can take values $1\le \alpha \le \infty$. We will use this conjecture to justify the 
holographic renormalisability of our boundary theory. We will also discuss the case when $\epsilon_{(\alpha)}$ 
becomes an integer at large $r$. 

Similarly one would also expect the dilaton to behave in an identical way. Asymptotically the dilaton should go to 
a constant. So on the given slice \eqref{sol}, and near one of the seven brane, we expect:
\bg\label{dilbar}
e^{-\Phi} ~ =~ {1\over 2g_s}\left[{1\over r^{\epsilon_a}} - {3\epsilon_a a^2\over 2 r^2} + {\rm constant}\right], 
~~~~ \epsilon_a = {3g_sN_f \over 4\pi}
\nd
where the constant could be determined from the full F-theory picture. 
Somewhat similar discussion has been given in \cite{ingo}. Our conjecture for the large $r$ behavior stems from the 
finiteness of F-theory. In fact this also gives us an argument to realise
the possibility of infinite F-theory backgrounds gluing 
to our IR solutions. In the presence of $24-N_f$ seven branes there are infinite possible ways by which we 
can adjust the positions of the seven branes. For every possible configurations of seven branes there would be 
non-trivial background axio-dilaton $\tau$,
realised from the degree eight and degree twelve polynomials $f$ and $g$ respectively 
that appear in the Weierstrass equation (see \cite{vafaF}), via:
\bg\label{jtau}
j(\tau) ~ = ~ {55926 f^3\over 4f^3 + 27 g^2} ~ \propto ~ {\prod_{i = 1}^8~(z - a_i)^3\over \prod_{j = 1}^{24} ~
(z-b_j)}
\nd
where $z$ is a complex coordinate orthogonal to the seven branes, and 
$a _i \ne b_j$ in general. Since $a_i, b_j$ take continuous values, there are infinite possible configurations 
of $\tau$ here (modulo $SL(2, {\bf Z})$ transformations). For certain choices of $a_i, b_j$, $j(\tau)$ is a constant
implying that 
there are possible configurations of seven branes that give rise 
to zero axio-dilaton (see the second and the third references of \cite{vafaF}). Thus for a generic configuration 
of seven branes we expect axio-dilaton $\tau$ to behave as $\tau = \sum_i {C_i \over r^{\epsilon_{(i)}}}$, where 
$\{C_i\}$ take a particular set of values for a given configuration of seven branes. Plugging these values of 
axio-dilaton in sugra equations of motion alongwith a similar configuration of fluxes, 
we can easily argue the existence of infinite configurations of warp factors 
of the form \eqref{logr}, justifying the IR changes from UV caps. 
A particular configuration of F-theory background will glue to 
our IR solution to give us the required UV completion\footnote{More precisely, the non-trivial geometry of the UV
will capture the effects of the operators defined at the cut-off; and the F-theory seven-branes will capture the 
effects of the ultra massive fundamental quarks that help us remove the Landau poles.}. 
In fact the configuration with constant coupling 
(like the last two references of \cite{vafaF}) will give rise to AdS completions of our IR backgrounds! More details 
will be presented elsewhere.

One caveat is that 
a full analysis incorporating non-perturbative effects still needs to be performed to justify 
the whole scenario. However we expect this to be very involved because at large $r$ we have to consider not only 
the effects of all the ($p, q$) seven branes (as discussed above)
but also the back-reactions from fluxes etc.\footnote{We thank Ingo Kirsch
and Diana Vaman for discussion on this issue, and for pointing out the reference \cite{ingo}.}. 
We will address this in the 
sequel \cite{sequel}. Therefore with this assumption, all the ${\rm log}~r$ dependences 
of the fluxes should also be replaced by inverse powers of $r$ at large $r$. 
We will however, for the purpose of concrete calculations,
only work to ${\cal O}(g_sN_f)$ locally
for many representative examples unless mentioned otherwise. 
The singularities appearing in these examples at large $r$ should
then be considered as an artifact of the order at which we do the analysis. This will also be clear from the holographic 
renormalisabilities of these theories\footnote{One might wonder whether the procedure of holographic renormalisability 
should work for the full F-theory picture. It is easy to see why this should proceed without any complication: F-theory 
background simply gives us the non-perturbative completion of a given IIB background. These non-perturbative objects are 
the seven branes and they contribute to the bulk lagrangian as some matter multiplets. As long as the effects of the 
matter multiplets are not too large (as discussed after \eqref{listman} and before \eqref{KS7bi}) 
the procedure of holographic renormalisability 
should proceed as in the usual supergravity case. However if we incorporate these multiplets they still don't change 
anything because the procedure of holographic renormalisability requires derivative interactions that come exclusively 
from $\sqrt{-G} R$ term of the lagrangian. All the fluxes etc contribute to polynomial interactions, as discussed 
around \eqref{phim}. Of course the effective values of the metric fluctuations {\it are} affected by the background
seven branes.}.  
    
%\newpage

\subsection{Quark mass and drag coefficient}

We now embed a string in OKS-BH background with one end attached to one of the D7 branes and other end 
going into the black hole (see {\bf figures 8} and {\bf 9} below).
\begin{figure}[htb]\label{fepA}
		\begin{center}
\includegraphics[height= 3.5cm]{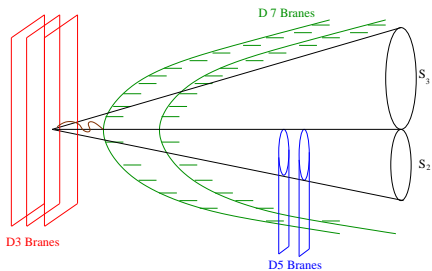}
		\caption{{Brane construction that we used to describe the zero temperature gauge theory. The 
fundamental flavors come from the wrapped D7 branes whereas the bi-fundamental flavors come from the wrapped D5 branes.
To introduce temperature all we need to do is to Euclideanise and identify the time coordinate in the gauge theory.}}
		\end{center}
		\end{figure}
\begin{figure}[htb]\label{fepB}
		\begin{center}
\vskip.15in	
\includegraphics[height= 3.5cm]{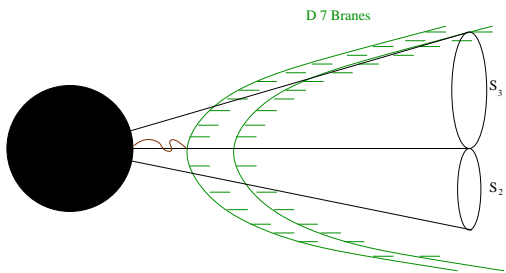}
		\caption{{The dual gravity picture for the high temperature gauge theory. The branes (except the 
D7 branes) are replaced by fluxes and the quark string has one end on the D7 branes and the other end going into  
the black hole. The conifold is replaced by a non-trivial geometry that we discussed in the previous section.}}
		\end{center}
		\end{figure}
If $X^{i}(\sigma,\tau)$ is a map from world sheet coordinates $\sigma,\tau$ to 10 dimensional space time, 
then string action or fundamental string Born Infeld action is (see for example \cite{leigh}):
\bg \label{KS3}
&& S_{\rm string}= T_0\int d\sigma d\tau \Big[\sqrt{-{\rm det}(f_{\alpha\beta} + \partial_\alpha\phi \partial_\beta\phi)} 
+ {1\over 2} \epsilon^{ab} B_{ab}
+ J(\phi)\nonumber\\
 && ~~~~~~~~+ \partial X^m \partial X^n ~\bar\Theta~ \Gamma_m \Gamma^{abc....} \Gamma_n ~\Theta
~F_{abc....} + {\cal O}(\Theta^4)\Big] =\int d^{10}x ~ {\cal L}_{\rm string}(x)\nonumber
\nd
\bg
 {\cal L}_{\rm string}(x) && = T_0\int d\sigma d\tau \Big[\sqrt{-{\rm det}(f_{\alpha\beta} + 
\partial_\alpha\phi \partial_\beta\phi)} + 
{1\over 2} \epsilon^{ab} B_{ab} + J(\phi)\nonumber\\
 && ~~~~+ \partial X^m \partial X^n ~\bar\Theta~ \Gamma_m \Gamma^{abc....} \Gamma_n ~\Theta
~F_{abc....} + {\cal O}(\Theta^4)\Big]
~\delta^{10}(X-x)
\nd
where $J(\phi)$ is the additional coupling of the dilaton $\phi$ to the string world-sheet, $T_0$ is the string tension, 
$X^n$ are the ten bosonic coordinates, $\Theta$ is a 32 component spinor, $F_{abc...} = [dC]_{abc..}$ with 
$C_{abc..}$ being the background RR form potentials,
and
$f_{\gamma\delta}$ is world sheet metric, given by the standard pull-back of the spacetime metric on the 
world-sheet:
\begin{equation} \label{fab}
f = \begin{pmatrix} \dot X\cdot \dot X & \dot X \cdot X' \\ \dot X \cdot X' & X'\cdot X'\end{pmatrix}
= \begin{pmatrix} {\dot X^2 \over \sqrt{h}} - {g_1\over \sqrt{h}} & {\dot X X'\over \sqrt{h}} \\ 
{\dot X X'\over \sqrt{h}} & {\sqrt{h} \over g_2} + {X'^2\over \sqrt{h}} \end{pmatrix}
\end{equation}
with 
$\gamma  \;{\rm or}\;\delta =0,1$, $\eta^0=\tau,\eta^1=\sigma$ and $B_{ab}$ is the pull-back of the 
NS two form field. In the ensuing analysis we will keep both $B_{ab}$ as well as $\partial_a\phi$
zero. The former will be {\it imposed} (see discussion after \eqref{otwo}). 
The latter case will be addressed soon. The interesting thing however 
is to do with the background RR forms. Note that the RR forms {\it always} couple to the 32 component spinor. Therefore
once we switch-off the fermionic parts in \eqref{KS3}, the fundamental string is completely unaffected by the background 
RR forms\footnote{This is of course the familiar statement that the RR fields do not couple 
in a simple way to the fundamental string.}. Thus in the ensuing analysis, for the mass and drag of the quark, we can 
safely ignore the RR fields.    
We have also defined:
\bg \label{KS4}
&&{\rm det}~f=-G_{ij}\dot{X}^{i}X'^{j}+(G_{ij}X'^{i}X'^{j})(G_{kl}\dot{X}^{k}\dot{X}^{l})\nonumber\\  
&& X'^{i}=\frac{\partial X^i}{\partial \sigma}, ~~~~~~
\dot{X}^{i}=\frac{\partial X^i}{\partial \tau}  \nd
where $G_{ij}$ is more generic than the background metric, and could involve the back reaction of the 
fundamental string on the geometry. The analysis is very similar to the AdS case discussed in \cite{HK} so we 
will be brief in the following. However, since our background involves running couplings, the results will 
differ from the ones of \cite{HK}. 

At this point it might be interesting to point out an equivalent classical calculation to evaluate the 
effect of a drag force on 
a point charge particle moving through a media. One can consider the point charge to be the endpoint of the 
fundamental string on the D7 brane. The mass of the point charge is given by the length of the string ${l}$ as 
$m = T_0{l}$ where $T_0$ is the tension of the string. In the presence of a constant electric flux ${\cal F}_{0i}$ 
one can show that the total drag force $F_i$ on the point charge in the dual gauge theory is given by:
\bg \label{drag}
F_i = - q {\cal F}_{0i} \left(e^{-t/\tau} - 1\right)
\nd
where $\tau$ is the so-called collision time. This can be easily determined from the mean free path of the 
collision. Viewing the end points of the open strings on the D7 brane as a gas of charged particles,
we can study the number of collisions at any given point in the gas of particles at a particular temperature, and
from there evaluate the collision length between them. The number of collisions ${\cal Z}$ 
happening per second 
per unit volume at a point $\widehat{\bf r}$ (with $\widehat r$ not to be confused with the radial coordinate $r$ in the 
gravity picture)
in the gas can be written as:
\bg \label{collision}
{\cal Z} ~ = ~ \int d^3p_1 \int d^3p_2 ~\sigma~ \vert {\bf v}_1 ~ - ~ {\bf v}_2 \vert 
~g(\widehat{\bf r}, {\bf v}_1, t)~ g(\widehat{\bf r}, {\bf v}_2, t)
\nd
where $\widehat{\bf r}, {\bf v}$ are vectors and $g(\widehat{\bf r}, {\bf v}, t)$ is the distribution function. 
The quantity 
$\sigma$ is the collision cross-section, and is a constant for our purpose. 
The distribution function $g(\widehat{\bf r}, {\bf v}, t)$ can be anything generic, and only 
becomes Maxwell-Boltzmann when the 
gas attains equilibrium. If there are $n$ strings per unit volume of the gas, one can show that the mean 
free path ${l_m}$ in the dual theory is given by the following expression:
\bg \label{mfp}
l_m ~ = ~ {n\over 2{\cal Z}} {v}_p
\nd
where ${v}_p$ is the most probable velocity which differs from the average velocity $\langle v\rangle$ by a 
numerical constant. 

Finally when the system attains equilibrium the distribution function, as discussed above, becomes MB. In
that case there are no $\widehat{\bf r}, t$ dependences in $g(\widehat{\bf r}, {\bf v}, t)$, and we have 
\bg \label{mbdi}
g(\widehat{\bf r}, {\bf v}, t) ~\equiv ~ g({\bf p}) ~ = ~ {n\over (2\pi m kT)^{3/2}} ~e^{-{\bf p}^2/2mkT}
\nd
where $m$ is the mass of a single particle in the gas, and $k$ the Boltzmann constant. Using this distribution 
it is straightforward to show that the mean free path $l_m$ is independent of temperature and is given by:
\bg \label{mfmp}
l_m ~ = ~ {\sqrt{\pi \over 8}}\cdot {1\over n \sigma}
\nd
{}from which the collision time $\tau$ can be determined to be $\tau = \frac{l_m}{\langle v\rangle}$. Plugging 
this in \eqref{drag} gives us the drag force on a particle in the dual theory.  

The above analysis is clean and simple, but doesn't completely give us the full answer. This is because, the 
above analysis ignores higher order quantum corrections
on the collision process. Such corrections are in fact captured by the classical back reactions 
of the underlying geometry in the gravity side\footnote{This is of course pure supergravity analysis valid for 
gauge theory at very strong coupling. 
There would be 
stringy corrections over and above this result once we go closer to the boundary of the RG surface. 
This will additionally introduce ${\cal O}(g_s)$ corrections to the 
existing result. For the time being we will ignore these effects as we will always be in the regime shown in 
{\bf figure 6}.}. 
The important fact that 
the D7 brane is wrapped on a curved manifold, changes much of the above analysis. To see the effect of the background 
geometry on the process, let us evaluate using the technique of \cite{HK}. 

To begin, we need the embedding of the D7 brane in our set-up. This has already appeared above as \eqref{seven}. We can
use the embedding equation to determine $r_0$, the distance upto which the D7 ends in the throat. 
The embedding equation \eqref{seven} implies:
\bg\label{embedim}
 r = \Bigg(\frac{\vert\mu\vert^2}{{\rm sin}^2~{\theta_1 \over 2}~{\rm sin}^2~{\theta_2 \over 2}}\Bigg)^{1\over 3}
\nd
 {}from where, by minimising $r$ with respect to the 
angular coordinates $\theta_i$, we obtain $\theta_1 = \theta_2 = \pi$ and from here
one can see that $r_0 = \vert\mu\vert^{2/3} \equiv 1$.\footnote{This is done to make ${r\over r_0} = r$ dimensionless. 
Note that this is another scale in our problem, which we identify to 1. As alluded to before, this will help us 
to make most of our QCD variables dimensionless.}  

As we said before, a fundamental quark will be a string starting from $r_0$ on the D7 brane to the horizon $r_h$ 
of the black hole.   
For simplicity of the calculation, we will then restrict to the case when 
\bg\label{lcone} \nonumber
&&X^0=t,~~X^4=r,~~X^1=x(\sigma,\tau),~~X^k=0~(k=2,3,5,6,7)\nonumber\\
&& (X^8, X^9) = (\theta_1, \theta_2) =\pi, ~~~~ \Theta = \bar\Theta = 0 
\nd
and we choose parametization $\tau=t,\sigma=r$ also known as the static gauge. Thus
we are only considering the case when the string extends in the $r$ direction, does not interact with the RR fields, 
and moves in the $x$ direction of 
our
manifold. More general string profile, while being 
computationally challenging, does not introduce any new physics and hence
our simplification is a reasonable one.

Before moving further, let us consider two points. First is the effect of the black hole on the {\it shape} 
of the D7 brane. We expect due to gravitational effects the D7 brane will sag towards the black hole and eventually the 
string would come very close to the horizon. In fact, as is well known, putting a point charge on the D-brane tends
to create a long thin tube on the D-brane that in general extends to infinity. The end point of the string being a 
source of point charge should show similar effects (see \cite{anirban1} for a discussion of a somewhat similar 
scenario)\footnote{Even in the supersymmetric case, putting a point charge on a D-brane tends to create a long 
tube that extends to infinity. One can then view an open string to lie at the end of the thin tube. This effect 
is somewhat similar to the one discussed in \cite{calmal}.}. 
In this paper we will ignore this effect altogether and discuss it in details in the sequel to this 
paper \cite{sequel}. 

The second point is to see how the background varying dilaton effects the string. Near the 
local region around the string, we 
expect the following behavior of the dilaton:
\bg\label{dilbe}
\phi ~\approx~ {\rm log}~g_s + \frac{3g_s N_f}{4\pi} ~{\rm log}~{r} + \frac{9g_s N_f}{8\pi} ~{a^2\over r^2}
\nd
where we could insert $\mu$ to make $r$ dimensionless. In the limit where the resolution parameter $a$ goes to zero,
it is easy to see that near $r_h$ and $r_0 \equiv 1$, the dilaton behaves respectively as (inserting the scale $\mu$):
\bg\label{dilsev}
&&\phi ~\approx ~ {\rm log}~g_s + \frac{3g_s N_f}{4\pi} ~{\rm log}~{{\cal T} \over \vert\mu\vert^{2/3}} \nonumber\\
&& \phi ~\approx ~ {\rm log}~g_s - \frac{g_s N_f}{4\pi} ~{\rm log}~{\vert\mu\vert}
\nd 
which means that near the seven brane the dilaton behavior is almost a constant although there is a log ${\cal T}$ 
dependence of the dilaton near the horizon, where ${\cal T}$ is defined in \eqref{bndtemp}. 
This behavior is quite different from the AdS case where there 
is no profile of the dilaton. As we will show later, there will be additional log ${\cal T}$ 
behavior coming from the warp 
factor also, which for small $g_s N_f$ will have dependence on ${g_s M^2 \over N}$. Thus the log ${\cal T}$ 
dilaton dependence
can only contribute to order ${\cal O}(g_s N_f)$ if we are close to the horizon. Away from the horizon and near the
seven brane, the dilaton is approximately constant. Therefore to simplify our ensuing analysis, we will take the 
dilaton to be a constant and only consider the detail implication in the sequel. 
In the following analysis we want to convince the reader that even with these simplifications: $B_{ab} = J(\phi) = 0$
in \eqref{KS3},
our system will have interesting new physics. For more details on the dilaton behavior see \cite{mcnees}.  

Now introducing a fundamental string in the geometry will change the IR geometry. We expect the back reaction to be 
small, i.e we can specify the back reaction via the perturbed metric $G_{ij} = g_{ij} + \kappa l_{ij}$ where 
$g_{ij}$ is the original metric and $\kappa l_{ij}$ is the back reaction.  
With our choice of parametization and string profile, it is easy to verify: 
\bg \label{KS5}
-{\rm det}~f=\frac{g_1(r)}{g_2(r)}+\frac{g_1(r)}{h(r,\pi,\pi)}x'^2-g_1(r)^{-1}\dot{x}^2+{\cal O}(\kappa)
+{\cal O}(\kappa^2)
\nd
where the warp factor $h(r, \theta_1, \theta_2) = h(r, \pi, \pi)$ with ($a_{mn}, b_{mn}$) $\approx 0$ 
in \eqref{hvalue} henceforth (unless
mentioned otherwise); 
and the back reaction of the string appear 
as ${\cal O}(\kappa)$ effect. 

With this, the rest of the analysis is a straightforward extension of \cite{HK}. 
The Euler-Lagrangian equation for $X^1=x(t,r)$ derived from the action (\ref{KS3}) and the associated canonical 
momenta are:
\begin{eqnarray} \label{KS6} \nonumber 
&& \frac{1}{g_2}\frac{d}{dt}\Big(\frac{\dot{x}}{\sqrt{-{\rm det}~f}}\Big)+\frac{d}{dr}\Big(\frac{g_1x'}
{h\sqrt{-{\rm det}~f}}\Big)=0 
\\\nonumber
&& \Pi_i^0= -T_0 G_{ij}\frac{(\dot{X}\cdot X')(X^{j})'-(X')^2(\dot{X}^j)}{\sqrt{-{\rm det}~f}}\nonumber\\
&& \Pi_i^1 = -T_0 G_{ij}\frac{(\dot{X}\cdot X')(\dot{X}^{j})-(\dot{X})^2(X^j)'}{\sqrt{-{\rm det}~f}}
\end{eqnarray}
If we consider a static string configuration, i.e. $x(\sigma,\tau)=b={\rm constant}$, then energy can be interpreted as
the thermal mass of the quark in the dual gauge theory. Using the static solution in (\ref{KS6}), we obtain the mass
using $E = -\int d\sigma ~\Pi^0_t$, as
\bg \label{KS15b}
m({\cal T})=T_0(r_0-r_h)=T_0\left(\vert\mu\vert^{2/3}- {\cal T}\right) \equiv T_0\left(1-{\cal T}\right)
\nd 
Now recounting our earlier analysis of the drag force \eqref{drag} we see that the velocity of a particle in the gas 
(given by the end point of the 
string) will eventually approach a constant velocity $v$ as 
\bg\label{drift} 
{\bf V}(t) = {\bf v}\left(1 - e^{-t/\tau}\right)
\nd
where $v_i = \frac{q{\cal F}_{0i} \tau}{m}$ with $m$ derived at a constant temperature. Thus once the string (or the 
fundamental quark in the dual gauge theory) moves with a constant velocity, we need to apply force constantly to 
keep it in that state. This gives us the drag force. 
To obtain the drag coefficient $\nu$, 
we consider strings moving with constant velocity:
\bg \label{KS10}
x(t,r)=\bar{x}(r)+vt
\nd
where $v \equiv v_1$ from \eqref{drift}, as after large enough time $V(t) \to v$ and the string moves at a uniform 
speed.  
Then from (\ref{KS6}), noting that $f$ is independent of time, we can solve the equation of motion to get: 
\bg \label{KS12}
\bar{x}'^2= \frac{h^2C^2v^2}{g_1g_2}\cdot \frac{g_1-{v^2}}{g_1-{h C^2v^2}}
\nd
where $C$ is a constant of integration that can be determined by demanding that 
$-{\rm det}~f$ is always positive. Using the value of $\bar{x}'^2$ from \eqref{KS12} we can give an explicit expression
for the determinant of $f$ as:
\bg \label{KS13}
-{\rm det}~f=\frac{g_1}{g_2}\cdot \frac{g_1-v^2}{g_1-h C^2v^2}
\nd
For $-{\rm det}~f$ to remain positive for all $r$, we need both 
numerator and denominator to change sign at same value of $r$. This is the same argument as in \cite{HK}. 
The
numerator changes sign at\footnote{Note that by ${\cal O}(g_sN_f,g_sM)$ we will always mean 
${\cal O}(g_sN_f,g^2_sMN_f, g_sM^2/N)$ unless mentioned otherwise.}  
\bg \label{cvaluea}
r^2=\frac{r_h^2}{\sqrt{1-v^2}} + {\cal O}(g_sN_f,g_sM)
\nd  
Requiring that denominator also change sign at that value fixes $C$ to be:
\bg \label{cvalue}
C =\frac{r_h^2 L^{-2}}{\sqrt{1-v^2}}\cdot
\frac{1}{\sqrt{1+\frac{3 g_s\bar{M}^2}{2\pi N}~{\rm log}\Big[\frac{r_h}{(1-v^2)^{{1}/{4}}}\Big]
\Big(1+\frac{3 g_s\bar{N}_f}{2\pi}~\Big\{{\rm
log}~\Big[\frac{r_h}{(1-v^2)^{{1}/{4}}}\Big]+\frac{1}{2}\Big\}\Big)}}
\nd
where $\bar{M}$ and $\bar{N}_f$ differs from $M,N_f$ due to the ${\cal O}(g_sN_f,g_sM)$ terms in (\ref{cvaluea}). 
The first part of $C$ is the one derived in \cite{HK}. The next part is new. 
Now the rate at which momentum is lost to the black hole is given by the momentum density at horizon
\bg \label{KS15a}
\Pi_1^x(r=r_h)= - T_0 Cv
\nd
while the force  quark experiences due to friction with the  plasma is $\frac{dp}{dt}=-\nu p$ with 
$p=mv/\sqrt{1-v^2}$. To keep the quark moving at
constant velocity, an external field ${\cal E}_i$ does work and the equivalent energy is dumped into the 
medium \cite{HK}. This external field ${\cal E}_i$ is exactly the flux ${\cal F}_{0i}$ discussed above. Thus 
the rate at which a quark dumps energy and momentum into the thermal medium is 
precisely the rate at which the string loses energy and momentum to the black hole. Thus upto 
${\cal O}(g_sN_f, g_s M)$ we have ${\nu m
v \over \sqrt{1-v^2}} = - \Pi_1^x(r=r_h)$ and 
\bg \label{KS15b} 
\nu &= & \frac{T_0 C \sqrt{1-v^2}}{m}\\
&=&\frac{T_0}{mL^2}\frac{{\cal T}^2} 
{\sqrt{1+\frac{3 g_s\bar{M}^2}{2\pi N}~{\rm log}\Big[\frac{{\cal T}}{(1-v^2)^{{1}/{4}}}\Big]
\Big(1+\frac{3 g_s\bar{N}_f}{2\pi}~\Big\{{\rm
log}~\Big[\frac{{\cal T}}{(1-v^2)^{{1}/{4}}}\Big]+\frac{1}{2}\Big\}\Big)}}\nonumber
\nd
which should now be compared with the AdS result \cite{HK}.
%\footnote{To make a precise comparison, one has to 
%first insert back $\mu, \hbar, c, r_0$ etc. As it stands, all the variables appearing in \eqref{KS15b} are 
%dimensionless now.}. 
In the AdS case the drag coefficient $\nu$ is 
proportional to ${\cal T}^2$. For our case, when we incorporate RG flow in the gravity dual, 
we obtain ${\cal O}\left(1/\sqrt{A ~{\rm log}~{\cal T}+B ~{\rm log}^2 {\cal T}}\right)$ 
correction to the drag coefficient computed
 using AdS/CFT correspondence \cite{HK} \cite{Gubser}.  

%\newpage

\subsection{Wake created by the moving quark}

In the previous section we computed the drag force on the quark. Clearly a moving quark should leave some disturbance 
in the surrounding media. This disturbance is called the {\it wake} of the 
quark.
Thus, 
in order to compute the wake left behind by a fast moving quark in the Quark Gluon Plasma, we need to compute 
the stress tensor $T^{pq}$, $p,q=0,1,2,3$ of the entire system. Our goal therefore would be to 
compute\footnote{In the published version this section is slightly abridged by 
emphasising mostly on the holographic renormalisability of the theory with fundamental flavors
and very little on the actual wake.}:
\bg\label{goal}
T^{pq}_{{\rm medium} + {\rm quark}} ~- ~ T^{pq}_{\rm quark}
\nd
where the first term is basically the energy-momentum tensor of OKS-BH background plus string i.e
$T^{pq}_{{\rm background} + {\rm string}}$
restricted to 
four-dimensional space-time. Similarly the second term is the energy momentum tensor of the string 
i.e $T^{pq}_{\rm string}$ restricted to 
four-dimensional space-time. This is similar to the 
analysis done in \cite{Yaffe-1} for the AdS case. For our case
the above idea, although very simple to state, will be rather technical because of the underlying RG 
flow in the dual gauge theory side. Our second goal would then be to see how much we differ from the AdS results 
once we go from CFT to theories with running coupling constants.   

For a strongly coupled QGP, we will apply the gauge/gravity duality
to compute $T^{pq}$ of QGP using the supergravity action $S_{\rm total}$. 
The supergravity action will be 
defined as a functional 
of the perturbation ${l}_{pq}$ from the string on the background metric
$g_{pq}$. Making use of the duality, we expect that the Hilbert space of strongly coupled
QCD to be mostly contained in the
Hilbert space of low energy weakly coupled
{\it classical} Supergravity i.e. the OKS-BH geometry of (\ref{bhmet}), or alternatively,
the full Hilbert space of QCD should be contained in the Hilbert space of {\it string theory}
in the OKS-BH background. However there is a subtlety here. The standard supergravity analysis 
in this theory will lead to actions that blow up at the boundary (i.e taking $r = r_c \to \infty$). The reason for this 
is rather simple to state (see also \cite{ahabu}). The UV completion of cascading type theories require {\it infinite} 
degrees of freedom $-$ much like string theories. This is of course another reason why the dual of 
cascading theories are given by string theories. Once we require infinite degrees of freedom at the UV, we no longer 
expect a finite boundary action from supergravity analysis! What we need is to regularise and renormalise the 
supergravity boundary action so that finite correlation functions could be extracted. This would also mean that 
the usual Witten type proposal \cite{Witt1} for the AdS/CFT correspondence can be re-expressed
in terms of the boundary variables to give us the complete picture.    
Therefore we can rewrite the ansatze proposed by Witten {\it et al} \cite{Witt1}
for our OKS-BH geometry to take the following Wilsonian form (also as mentioned in footnote 2, 
by QCD we will only mean a theory
that approximates large $N$ IR QCD, but with a strongly coupled and {almost} conformal UV):
\bg \label{KS16}
{\cal Z}_{\rm QCD}[\phi_0] &~\equiv &~ \langle{\rm exp}\int_{M^4} \phi_0 {\cal O}\rangle
~=~ {\cal Z}_{\rm total}[\phi_0] \nonumber\\
&~ \equiv &  ~{\rm exp}(S_{\rm total}[\phi_0] + S_{\rm GH} + S_{\rm counterterm})
\nd
where $M^4$ is Minkowski manifold, $S_{\rm GH}$ is the Gibbons-Hawking boundary term \cite{Gibbons-Hawking}, 
$\phi_0$ should be understood as a fluctuation over a given configuration of field,
and $S_{\rm counterterm}$ is the 
counter-term action added to renormalise the action. 
Observe that in the usual AdS/CFT case we consider the action at the boundary to 
map it directly to the dual gauge theory side. For the OKS-BH background, as we discussed above, there are many 
possibilities of defining 
different gauge theories at the boundary depending on how we cut-off the geometry and add UV
caps. 
Taking the radial coordinate as 
setting the energy scale $\Lambda$ 
the gauge theory side would make sense at that scale once we 
define the UV degrees of freedom there. This is what we called ${\cal N}_{\rm eff}(\Lambda)$ in \eqref{doa}.  
Therefore in general the action at any point $r = r_c$ in the OKS-BH geometry 
will map to the dual gauge theory with ${\cal N}_{\rm eff}(\Lambda)$ degrees of freedom at that
energy scale. The properties of this dual gauge theory may not necessarily coincide with the universal properties of the 
parent cascading theory when both are studied from the boundary. 
However 
the RG flow of this theory will eventually catch up with the RG flow of the smooth cascading theory at 
that scale fixed by our choice $r = r_c$ (see {\bf figure 7}). 
For large enough $r_c$ the above correspondence \eqref{KS16} should give us finite boundary 
action. Our procedure then would be to fix the boundary action for large $r_c$ (typically $r_c \to \infty$) by 
adding the corresponding $S_{\rm GH}$ and $S_{\rm counterterm}$, and then extrapolate this to smaller 
radii\footnote{Once the boundary theory is holographically renormalised, all $r_c$ dependences would go as 
$r_c^{-1}$. In the dual gauge theories this would mean that in addition to the universal properties inherited by 
each of these gauge theories, there would be corrections associated with the 
UV degrees of freedom that we needed to provide to define these theories at $r = r_c$. These are precisely the corrections
that take us away from the parent cascading theory and give us results associated with new gauge theories. Needless to 
say, such extravagant richness of physical theories are not available to us in the AdS/QCD picture. Note however that 
even for the AdS/QCD case there might arise situations where we would require ${\cal N} = 2$ degrees of freedom to 
UV complete a ${\cal N} =1$ configuration at IR. The issue of holographic renormalisation is not much affected 
by this because such a UV completion only affects the internal space (here it may change $T^{1,1}$ geometry to 
$S^5/{\bf Z}_2$). For our case the full F-theory completion of the parent cascading 
theory would also be UV complete.}.
Thus,
for computing stress tensor of QCD, we have ${\cal O}=T^{pq}$ and
$\phi_0=\kappa {l}_{pq}$. It follows that  
\bg \label{KS17}
\langle T^{pq}\rangle ~= ~ {1\over \kappa} 
\frac{\delta S}{\delta {l}_{pq}}\Bigg{|}_{\kappa {l}_{pq}=0}
\nd 
where $S \equiv S_{\rm total} + S_{\rm GH} + S_{\rm counterterm}$ and $S_{\rm total}$  
is the low energy Type IIB supergravity action in ten dimensions defined in string frame as: 
\bg \label{KS77}
S_{\rm total}&=&\frac{1}{2\kappa_{10}^2}\Bigg[\int d^{10}x~e^{-2\Phi} \sqrt{-G}
\Bigg(R-4\partial_i \Phi \partial^j \Phi-\frac{1}{2}|H_3|^2\Bigg)\nonumber\\
&-&\frac{1}{2} \int d^{10}x \sqrt{-G}\Bigg(|F_1|^2+|\widetilde{F}_3|^2+\frac{1}{2}|\widetilde{F_5}|^2\Bigg)
- \frac{1}{2}\int C_4\wedge H_3\wedge F_3\Bigg]+S_{\rm
solitons}\nonumber\\
&\equiv&S_{\rm SUGRA}~ + ~ S_{\rm string}~ + ~ S_{\rm D7}
\nd
where $S_{\rm SUGRA}$ is the background supergravity action and 
$S_{\rm solitons}$ is the action of the solitonic objects in our theory, namely the fundamental string
and the D7 brane. Recall that the OKS-BH background is constructed by inserting the D7 brane in the supergravity 
action with an additional black hole singularity. Therefore we will define 
\bg \label{oks}
S_{\rm OKS-BH} \equiv S_{\rm SUGRA} + S_{\rm D7}
\nd
so that the background solutions that we gave in the previous subsection will correspond to $S_{\rm OKS-BH}$. Once we
introduce an additional fundamental string we expect some of the background values to change. The change in the 
metric will take the following form:
\bg\label{metchange}
G_{ij}~&= & ~g_{ij}+\kappa l_{ij}\nonumber\\
 l_{ij}~ & \equiv & ~ {l}_{ij}(r,x,y,z,t)
\nd  
where $g_{ij}$ is the OKS-BH metric and  $l_{ij}$ ($i,j=0,..,9$)
denote the perturbation from the moving string source (with $\kappa \to 0$). 
We also expect 
the NS two form $B_2$ defined in \eqref{btwo} to pick up an additional component along the ($0r$) direction. The 
three and five forms RR field strengths defined as:
\bg \label{threefive}
\widetilde{F}_3 ~ = ~ F_3-C_0\wedge H_3, ~~~~~ \widetilde{F}_5 ~ = ~ F_5+\frac{1}{2}B_2\wedge F_3
\nd 
would change from the values defined earlier because $H_3$ changes. In the absence of the string,
$F_3$ is the three form sourced by D5 branes and $F_5$ is the five form sourced
by D3 branes. 

To proceed, let us first figure out the possible changes in $H_3$, the NS three form field strength from \eqref{h3f3} 
given earlier. 
To order 
${\cal O}(g_s N_f)$ locally the background value of $H_3$ in the absence of the fundamental string 
for a Klebanov-Tseytlin type geometry \cite{klebts} can be given by 
the following values of $c_i$ \cite{4, 5, 6, 7, ofer}:
\bg\label{ouysol}
&&c_1 = -\frac{3g_s^2 M N_f}{4\pi r} ~{\rm cot}~{\theta_1 \over 2} + {\cal O}(g_s^2N_f^2), ~ 
c_2 = -\frac{3g_s^2 M N_f}{4\pi r} ~{\rm cot}~{\theta_2 \over 2} + {\cal O}(g_s^2N_f^2)\\
&& c_3 = c_4 = {3g_sM\over r}\Bigg[1 + {g_sN_f\over 4\pi}\left\{9~ {\rm log}~r + 
2~{\rm log}\left({\rm sin}~{\theta_1\over 2}
~{\rm sin}~{\theta_2\over 2}\right)\right\}\Bigg] + {\cal O}(g_s^2N_f^2)\nonumber
\nd
where the ${\cal O}(g_s^2N_f^2)$ local corrections are typically of the following form (see also \cite{5, 6, 7, ofer}):
\bg\label{corr}
\sum_{n\ge m;p} a_{mnp} (g_s M)^m (g_s N_f)^{n+1} ({\rm log}~r)^{p+1}
\nd
with $a_{mnp}$ are not in general constants. For certain examples studied in \cite{5, 6, 7, ofer} without D7 branes, $a_{mnp}$ are
functions of the radial coordinate $r$. With the D7 branes these corrections have not been computed. 

In addition to the ${\cal O}(g_s^2N_f^2)$ corrections we have another set of corrections already to order 
${\cal O}(g_s N_f)$ that appear because of our choice of background \eqref{resconi}. These ${\cal O}(g_s N_f)$ 
corrections are in general difficult to work out if in \eqref{resconi} the parameter $a$ is not a constant. When 
$a$ is a constant then this is a small resolution of the conifold and changes the coefficients $c_i$ of 
\eqref{ouysol} in the following way:
\bg\label{cichange}
&& {\Delta c_1 \over c_1} = 6a^2 \Bigg({{\rm log}~r^3 - 2 \over r^2}\Bigg) + {\cal O}(a^2 {\rm log}~a) + 
{\cal O}(g_s^2N_f^2)\nonumber\\
&& {\Delta c_2 \over c_2} = -18 a^2 \Bigg({{\rm log}~r \over r^2}\Bigg) + {\cal O}(a^3) + 
{\cal O}(g_s^2N_f^2)\\
&& {\Delta c_3 \over c_3} = - {9a^2 \over r^2} - {3g_s N_f a^2\over 4r^2} \Bigg[\frac{8+9~{\rm log}~r 
-{2r^2\over a^2}~{\rm log}~a}{1 + {g_sN_f\over 4\pi}\left(9~{\rm log}~r + 2~{\rm log} ~{\rm sin}~{\theta_1\over 2}
~{\rm sin}~{\theta_2\over 2}\right)}\Bigg] + {\cal O}(a^2 {\rm log}~a, g_s^2N_f^2)\nonumber\\
&& {\Delta c_4 \over c_4} = {3a^2 \over r^2} \cdot \frac{1 + {g_sN_f\over 2\pi}\left(3 +
{\rm log} ~{\rm sin}~{\theta_1\over 2}~{\rm sin}~{\theta_2\over 2}\right)}
{1 + {g_sN_f\over 4\pi}\left(9~{\rm log}~r + 2~{\rm log} ~{\rm sin}~{\theta_1\over 2}
~{\rm sin}~{\theta_2\over 2}\right)} + {\cal O}(a^2 {\rm log}~a) + {\cal O}(g_s^2N_f^2) \nonumber
\nd
Putting everything together we see that the NS three form changes a bit from what is given in \cite{4, 5, 6, 7, ofer}. To 
order $g_sN_f$ the local three form is given by:
\bg\label{hthree}
 H_3 &&=  {6g_s M}\Bigg(1+\frac{9g_s N_f}{4\pi}~{\rm log}~r+\frac{g_s N_f}{2\pi} 
~{\rm log}~{\rm sin}\frac{\theta_1}{2}~
{\rm sin}\frac{\theta_2}{2}\Bigg)\frac{dr}{r}\wedge \omega_2  \nonumber\\
&&+ \frac{3g^2_s M N_f}{8\pi}\Bigg(\frac{dr}{r}\wedge e_\psi -\frac{1}{2}de_\psi \Bigg)\wedge
\Bigg({\rm cot}~\frac{\theta_2}{2}~d\theta_2-{\rm cot}~\frac{\theta_1}{2} ~d\theta_1\Bigg)  \nonumber\\
&& - \frac{18 g_s M a^2}{r^2}\Bigg(1+\frac{9g_s N_f}{4\pi}~{\rm log}~r+\frac{g_s N_f}{2\pi} 
~{\rm log}~{\rm sin}\frac{\theta_1}{2}~
{\rm sin}\frac{\theta_2}{2}\Bigg)\frac{dr}{r}\wedge \Big(\omega_2 + {\rm sin}~\theta_2 ~d\theta_2 
\wedge d\phi_2\Big) \nonumber\\
&& + \frac{27 g_s^2 MN_f}{2\pi}\cdot 
\frac{a^2{\rm log}~r}{r}\Bigg(\frac{dr}{r}\wedge e_\psi -\frac{1}{2}de_\psi \Bigg)\wedge
\Bigg({\rm cot}~\frac{\theta_2}{2}~d\theta_2 +{\rm cot}~\frac{\theta_1}{2} ~d\theta_1\Bigg) \nonumber\\
&& +~ \Big[{\cal O}(a^2g_s N_f) ~+~ {\cal O}(g_s^2 N_f^2)\Big] \wedge d\omega_3
\nd
where we only pointed out those ${\cal O}(a^2g_s N_f)$ terms that have similar forms as the terms of $H_3$ with 
$a = 0$. There are a few more ${\cal O}(a^2g_s N_f)$ terms described by generic three form $d\omega_3$
in the OKS-BH geometry, that we do not write here but could be easily worked out.  
In addition to that we have ${\cal O}(g_s^2 N_f^2)$ terms of the form \eqref{corr} that we need to incorporate. The
two form $\omega_2$ is defined above as:
\bg\label{otwo}
\omega_2 =\frac{1}{2}\left({\rm sin}~\theta_1~ d\theta_1 \wedge d\phi_1-{\rm sin}~\theta_2~ d\theta_2 \wedge
d\phi_2\right)
\nd 
Notice also that the fourth term in \eqref{hthree} has a relative sign difference from the second term of the angular 
forms. Similarly the third term has an unequal distribution of the three form on the base two spheres. In additional to
that $-$
for $d\omega_3$ given 
exclusively by the internal three forms $-$ we can always define a $B_2$ field that lie only along the angular directions
($\theta_i, \phi_i, \psi$). In that case the dynamics of a fundamental string {\it will} get  
influenced by the background
NS field (so to avoid issues like \cite{cschu} we will take the slice \eqref{sol}
 where the pull-back $B_{ab}$ in \eqref{KS3} is zero). Also 
in this paper we will only consider the  
case where $B_2$ can be made to lie in the internal angular directions by appropriate gauge transformations. We 
should however remind the readers that this is {\it not} generic. In the presence of branes and $H_3$ fluxes 
 such procedure 
cannot always be done without generating noncommutative theories on the branes. 

Let us now determine the RR three form ${\widetilde F}_3$ 
which is a combination of $F_3$ and $H_3$. For a Klebanov-Tseytlin
\cite{klebts} kind of background we expect, in addition to the $c_i$ defined earlier in \eqref{ouysol}, there 
would be two more $c_i$ given by:
\bg\label{ouysol2}
c_5 = c_6 = g_s M\left(1 + \frac{3g_s N_f}{2\pi} ~{\rm log}~r\right) + {\cal O}(g_s^2 N_f^2)
\nd
As mentioned earlier, the above choices of $c_i$ are still incomplete. For constant small resolution $a$ we expect 
$\Delta c_i$ to be given by:
\bg\label{ciratio}
&&{\Delta c_5 \over c_5} = \frac{9g_s N_f}{4\pi}\cdot {a^2 \over r^2} \Bigg(\frac{2 - 3~{\rm log}~r}
{1 + {3g_sN_f \over 2\pi}~{\rm log}~r}\Bigg) + {\cal O}(a^2 {\rm log}~a) + 
{\cal O}(g_s^2N_f^2)\nonumber\\
&& {\Delta c_6 \over c_6} = \frac{9g_s N_f}{4\pi}\cdot {a^2 \over r^2} \Bigg(\frac{1 + 3~{\rm log}~r}
{1 + {3g_sN_f \over 2\pi}~{\rm log}~r}\Bigg) + {\cal O}(a^2 {\rm log}~a) + 
{\cal O}(g_s^2N_f^2) 
\nd
Combining everything together, we then get the following value of the background RR three form ${\widetilde F}_3$:
\bg \label{tilf}
{\widetilde F}_3 & = & 2M \left(1 + {3g_sN_f\over 2\pi}~{\rm log}~r\right) ~e_\psi \wedge \omega_2\nonumber\\
&& -{3g_s MN_f\over 4\pi}~{dr\over r}\wedge e_\psi \wedge \left({\rm cot}~{\theta_2 \over 2}~{\rm sin}~\theta_2 ~d\phi_2 
- {\rm cot}~{\theta_1 \over 2}~{\rm sin}~\theta_1 ~d\phi_1\right)\nonumber\\
&& -{3g_s MN_f\over 8\pi} ~{\rm sin}~\theta_1 ~{\rm sin}~\theta_2 \left({\rm cot}~{\theta_2 \over 2}~d\theta_1 +
{\rm cot}~{\theta_1 \over 2}~d\theta_2\right)\wedge d\phi_1 \wedge d\phi_2\nonumber\\
&&+{9g_s MN_f\over 2\pi} \left(1 + {3g_sN_f\over 2\pi}~{\rm log}~r\right) {a^2 \over r^2}~e_\psi \wedge
\left[\left(2 - 3~{\rm log}~r\right) \omega_2 -{9\over 2}~{\rm log}~r ~{\rm sin}~\theta_2 ~d\theta_2 \wedge 
d\phi_2\right]\nonumber\\
&& +{27 g_s MN_f\over 2\pi}\cdot {a^2 {\rm log}~r \over r^3} ~dr \wedge e_\psi \wedge 
\left({\rm cot}~{\theta_2 \over 2}~{\rm sin}~\theta_2 ~d\phi_2 
+ {\rm cot}~{\theta_1 \over 2}~{\rm sin}~\theta_1 ~d\phi_1\right)\nonumber
\nd
\bg
&&  +{27 g_s MN_f\over 4\pi}\cdot {a^2 {\rm log}~r \over r^2}
 ~{\rm sin}~\theta_1 ~{\rm sin}~\theta_2 \left({\rm cot}~{\theta_2 \over 2}~d\theta_1 -
{\rm cot}~{\theta_1 \over 2}~d\theta_2\right)\wedge d\phi_1 \wedge d\phi_2\nonumber\\
&& +~ \Big[{\cal O}(a^2g_s N_f) ~+~ {\cal O}(g_s^2 N_f^2)\Big] \wedge d\omega_3
\nd 
where as before we considered only the ${\cal O}(a^2g_s N_f)$ terms that are proportional to the existing resolution 
free terms that appeared in \cite{4, ofer}. However our form is more involved than the ones considered earlier even if
we ignore the ${\cal O}(g_s^2 N_f^2)$ corrections of the form \eqref{corr}. The fact that the background 
has a resolution changes much of the details. Notice also the fact that these corrections cannot be absorbed as a 
renormalisation of the $a = 0$ terms because the form structures have relative sign differences. 

In the presence of a non-trivial dilaton, these three forms combine together to give us $G_3 \equiv {\widetilde F}_3 - i e^{-\phi} H_3$. 
It is instructive to construct $G_3$ for our background because most of the details can be expressed directly in terms of $G_3$. The 
full expression for $G_3$ is involved, but could be combined succinctly using certain one and three forms to take the following 
background value:
\bg\label{G3baaz}
G_3 & = & 2M\left(1 + {3g_s N_f \over 2\pi}\right) \left(e_\psi - {3i dr\over r}\right) \wedge \omega_2 \nonumber\\
&& - \frac{3g_sMN_f}{4\pi} ~{\rm cot}~{\theta_2\over 2} \left({\rm sin}~\theta_2~d\phi_2 - {i\over 2} d\theta_2\right) \wedge 
\left({dr\over r} \wedge e_\psi + {1\over 2}{\rm sin}~\theta_1~d\theta_1 \wedge d\phi_1\right) \nonumber\\
&& + \frac{3g_sMN_f}{4\pi} ~{\rm cot}~{\theta_1\over 2} \left({\rm sin}~\theta_1~d\phi_1 - {i\over 2} d\theta_1\right) \wedge 
\left({dr\over r} \wedge e_\psi + {1\over 2}{\rm sin}~\theta_2~d\theta_2 \wedge d\phi_2\right) \nonumber\\
&& +{3i g_s^2 M N_f^2 \over 16 \pi^2}~{\rm log}~\left(r^{3/2}{\rm sin}~{\theta_1\over 2} ~{\rm sin}~{\theta_2 \over 2}\right) 
\left({\rm cot}~{\theta_2\over 2} ~d\theta_2 -{\rm cot}~{\theta_1\over 2} ~d\theta_1\right)\wedge \left({dr\over r}\wedge e_\psi
- {1\over 2} de_\psi\right)\nonumber\\
&& +{iMg_s^2 N_f^2 \over 2\pi^2}\left[{27 \over 8}~{\rm log}^2 r + 3~{\rm log}~r ~{\rm log}\left({\rm sin}~{\theta_1\over 2} ~{\rm sin}~{\theta_2 \over 2}\right) + {1\over 2} {\rm log}^2\left({\rm sin}~{\theta_1\over 2} ~{\rm sin}~{\theta_2 \over 2}\right)\right]\nonumber\\
&& \times \left({dr\over r} \wedge \omega_2\right)
+ 2 \widetilde M \left(1 + {3g_s N_f \over 2\pi}\right) \left(e_\psi \wedge {\widetilde \omega}_2 + {3i \widehat g_s \over g_s} ~{dr\over r}\wedge 
\widehat\omega_2\right)\nonumber\\
%\nd
%\bg
&& + {3N_f \over 4\pi}~{\rm cot}~{\theta_2\over 2}\left(\widetilde g_s \widetilde M~{\rm sin}~\theta_2~d\phi_2 - {i\over 2} ~ {\widehat M\widehat g_s^2 \over
g_s} ~d\theta_2\right)\wedge \left({dr\over r} \wedge e_\psi + {1\over 2}~{\rm log}~r~{\rm sin}~\theta_1~d\theta_1 \wedge d\phi_1\right)\nonumber\\
&& -{3N_f \over 4\pi}~{\rm cot}~{\theta_1\over 2}\left(\widetilde g_s \widetilde M~{\rm sin}~\theta_1~d\phi_1 - {i\over 2} ~ {\widehat M\widehat g_s^2 \over
g_s} ~d\theta_1\right)\wedge \left({dr\over r} \wedge e_\psi + {1\over 2}~{\rm log}~r~{\rm sin}~\theta_2~d\theta_2 \wedge d\phi_2\right)\nonumber\\
&& -{3i \widehat g_s^2 \widehat M N_f^2 \over 16 \pi^2} {\rm log}~\left(r^{3/2}{\rm sin}~{\theta_1\over 2} ~{\rm sin}~{\theta_2 \over 2}\right) 
\left({\rm cot}~{\theta_2\over 2} ~d\theta_2 +{\rm cot}~{\theta_1\over 2} ~d\theta_1\right)\wedge \left({dr\over r}\wedge e_\psi
- {1\over 2} de_\psi\right)\nonumber\\
&& -{i g_s \widehat g_s\widehat M N_f^2 \over 2\pi^2}\left[{27 \over 8}~{\rm log}^2 r + 3~{\rm log}~r ~{\rm log}\left({\rm sin}~{\theta_1\over 2} ~{\rm sin}~{\theta_2 \over 2}\right) + {1\over 2} {\rm log}^2\left({\rm sin}~{\theta_1\over 2} ~{\rm sin}~{\theta_2 \over 2}\right)\right]\nonumber\\ 
&& \times \left({dr\over r}\wedge {\widehat\omega}_2\right)
+~ \Big[{\cal O}(a^2g_s N_f) ~+~ {\cal O}(g_s^2 N_f^2)\Big] \wedge d\omega_3 ~\left(1 - {i\over g_s}\right)
\nd
where we have used two new two-forms $\widetilde\omega_2$ and $\widehat\omega_2$ defined in terms of $\omega_2$ in the following way:
\bg \label{2for}
&& \widetilde\omega_2 \equiv (2 - 3 {\rm log}~r) \omega_2 - {9\over 2} ~{\rm log}~r ~{\rm sin}~\theta_2 ~d\theta_2 \wedge d\phi_2\nonumber\\
&& \widehat\omega_2 \equiv \omega_2 + {\rm sin}~\theta_2 ~d\theta_2 \wedge d\phi_2
\nd
These forms help us to express the deformation of $G_3$ once we take the resolution etc into account. The other effective $M$ and 
$g_s$ are then defined in terms of ($M, g_s, N_f$) in the following way\footnote{Since upto ${\cal O}(g_s N_f)$ our 
metric remains similar to the Ouyang metric, we then expect the effective number of fiveform flux to change as
$N_{\rm eff} = N + {3g_s M^2\over 2\pi}\Bigg({\rm log}~r + {3g_s N_f \over 2\pi} {\rm log}^2 r\Bigg)$. This 
would imply that under radial rescaling $r \to e^{-2\pi/3g_s M} r$, $N_{\rm eff}$ decreases by $M - N_f$ units, exactly 
as we discussed earlier (see also \cite{4}).}:
\bg \label{mgs}
\widetilde M = {9g_s M N_f \over 4\pi} \cdot {a^2 \over r^2}, ~~~ \widehat M = {M a^2 \over 4 r^3}, ~~~ \widetilde g_s = {8\pi \over g_s N_f}, ~~~
\widehat g_s = 12 g_s r
\nd
Using these definitions we can see how the back reactions effect the three forms in our setup. Note that the deformations appear in $G_3$ almost exactly like the undeformed forms, but there are crucial relative signs and extra ($r, \theta_i$) dependent factors. However the way
we have written the backgrounds is not very productive because of many complicated terms. But there exist an alternative way to rewrite 
the above background which would tell us exactly how the black hole modifies the original Ouyang setup. This can be 
presented 
in the following way:
\begin{eqnarray}
{\widetilde F}_3 & = & 2M {\bf A_1} \left(1 + {3g_sN_f\over 2\pi}~{\rm log}~r\right) ~e_\psi \wedge 
\frac{1}{2}\left({\rm sin}~\theta_1~ d\theta_1 \wedge d\phi_1-{\bf B_1}~{\rm sin}~\theta_2~ d\theta_2 \wedge
d\phi_2\right)\nonumber\\
&& -{3g_s MN_f\over 4\pi} {\bf A_2}~{dr\over r}\wedge e_\psi \wedge \left({\rm cot}~{\theta_2 \over 2}~{\rm sin}~\theta_2 ~d\phi_2 
- {\bf B_2}~ {\rm cot}~{\theta_1 \over 2}~{\rm sin}~\theta_1 ~d\phi_1\right)\nonumber \\
&& -{3g_s MN_f\over 8\pi}{\bf A_3} ~{\rm sin}~\theta_1 ~{\rm sin}~\theta_2 \left({\rm cot}~{\theta_2 \over 2}~d\theta_1 +
{\bf B_3}~ {\rm cot}~{\theta_1 \over 2}~d\theta_2\right)\wedge d\phi_1 \wedge d\phi_2\label{brend} \\
H_3 &=&  {6g_s {\bf A_4} M}\Bigg(1+\frac{9g_s N_f}{4\pi}~{\rm log}~r+\frac{g_s N_f}{2\pi} 
~{\rm log}~{\rm sin}\frac{\theta_1}{2}~
{\rm sin}\frac{\theta_2}{2}\Bigg)\frac{dr}{r}\nonumber \\
&& \wedge \frac{1}{2}\Bigg({\rm sin}~\theta_1~ d\theta_1 \wedge d\phi_1
- {\bf B_4}~{\rm sin}~\theta_2~ d\theta_2 \wedge d\phi_2\Bigg)
+ \frac{3g^2_s M N_f}{8\pi} {\bf A_5} \Bigg(\frac{dr}{r}\wedge e_\psi -\frac{1}{2}de_\psi \Bigg)\nonumber  \\
&& \hspace*{1.5cm} \wedge \Bigg({\rm cot}~\frac{\theta_2}{2}~d\theta_2 
-{\bf B_5}~{\rm cot}~\frac{\theta_1}{2} ~d\theta_1\Bigg)\nonumber
\end{eqnarray}
where we see that the background is exactly of the form presented in \cite{4} except that there are asymmetry factors ${\bf A_i}, {\bf B_i}$. These
asymmetry factors contain all the informations of the black hole etc in our background\footnote{One can easily see from 
these asymmetry factors that one of the two spheres is squashed. As we mentioned before, this squashing factor is of 
order ${\cal O}(g_s N_f)$ and therefore could have a perturbative expansion. Note also that although the resolution 
factor in the metric is hidden behind the horizon of the black hole the effect of this shows up in the fluxes. As far 
as we know, these details have not been considered previously.}. 
To order ${\cal O}(g_sN_f)$ these 
asymmetry factors are given by:
\bg\label{asymmetry}
&& {\bf A_1} ~=~ 1 + {9g_s N_f \over 4\pi} \cdot {a^2\over r^2}\cdot (2 - 3~{\rm log}~r) + {\cal O}(a^2 g_s^2 N_f^2) \nonumber\\
&& {\bf B_2} ~=~ 1 + {36 a^2~{\rm log}~r \over r^3 + 18 a^2 r ~{\rm log}~r} + {\cal O}(a^2 g_s^2 N_f^2)\\
&& {\bf A_2} ~= ~1 + {18 a^2 \over r^2} \cdot {\rm log}~r + {\cal O}(a^2 g_s^2 N_f^2) \nonumber\\ 
&& {\bf B_1} ~=~ 1 + {81\over 2} \cdot 
{g_s N_f a^2 {\rm log}~r \over 4\pi r^2 + 9 g_s N_f a^2 (2 - 3~{\rm log}~r)} + {\cal O}(a^2 g_s^2 N_f^2)\nonumber\\ 
&& {\bf A_3} ~=~ 1 - {18 a^2 \over r^2}\cdot {\rm log}~r +  {\cal O}(a^2 g_s^2 N_f^2)\nonumber\\ 
&& {\bf B_3} ~ = ~ 1 + {36 a^2 {\rm log}~r \over r^2 - 18 a^2 {\rm log}~r} + {\cal O}(a^2 g_s^2 N_f^2)\nonumber\\ 
&& {\bf A_4} ~ = ~ 1 - {3a^2 \over r^2} + {\cal O}(a^2 g_s^2 N_f^2), ~~~~~ {\bf B_4} ~ = ~ 1 + {3g_s a^2 \over r^2 - 3 a^2} + {\cal O}(a^2 g_s^2 N_f^2)\nonumber\\ 
&& {\bf A_5} ~ = ~ 1 + {36 a^2 {\rm log}~r \over r} +  {\cal O}(a^2 g_s^2 N_f^2), ~~~~ 
{\bf B_5} ~ = ~ 1 + {72 a^2 {\rm log}~r \over r + 36 a^2 {\rm log}~r} + {\cal O}(a^2 g_s^2 N_f^2)\nonumber
\nd
These asymmetry factors tell us that corrections to Ouyang background \cite{4} come from ${\cal O}(a^2/r^2)$ onwards. Thus to complete the  
picture all we now need are the values for the axion $C_0$ and the five form $F_5$. They are given by:
\bg\label{axfive}
&&C_0 ~ = ~ {N_f \over 4\pi} (\psi - \phi_1 - \phi_2)\nonumber\\
&& F_5 ~ = ~ {1\over g_s} \left[ d^4 x \wedge d h^{-1} + \ast(d^4 x \wedge dh^{-1})\right]
\nd
with the dilaton to be taken as approximately a constant near the D7 brane and $h$ is the ten dimensional warp factor discussed above. 
Thus combining \eqref{brend} and \eqref{axfive} our background can be written almost like the Ouyang background 
\cite{4} with deviations 
given by \eqref{asymmetry}.    

So far we got the background without taking the back reaction of the string. The back reaction of the string can be 
computed from its energy momentum tensor. 
Using the action (\ref{KS3}) we can obtain the energy-momentum tensor of the
string as:
\bg \label{KS8}
T^{ij}_{\rm string}(x)&=& \frac{\delta {S}_{\rm string}}{\delta G_{ij}} \\
&=& \int d\sigma d\tau ~\Bigg(\frac{2 \dot{X}\cdot X' \dot{X}^iX'^j-X'^iX'^j \dot{X}^2-\dot{X}^i\dot{X}^jX'^2}{2\sqrt{-{\rm det}f}}\Bigg)\;\delta^{10}(X-x)\nonumber
\nd
where all the variables have been defined earlier.

To study the back reaction without incorporating the backreactions from other F-theory seven-branes, 
we need to analyse the geometry close to the string. This will be rather involved because our
metric \eqref{bhmet} with the choices of warp factor \eqref{hvalue} and the black hole functions \eqref{grdef} are 
complicated. However we can impose two immediate simplifications: 
\bg\label{simlificazion}
h(r, \theta_1, \theta_2) = h (r, \pi, \pi), ~~~~~ g_1(r) \approx g_2(r) \equiv g(r) = 1-\frac{r_h^4}{r^4}
\nd
which is motivated from the fact that we are close to the string and the ${\cal O}(g_s^2 M N_f)$ corrections to the 
black hole factor are subleading (as we saw in the previous section). Henceforth we will stick with these choices 
throughout our analysis.  
With this therefore we see that the local metric is given from
\eqref{bhmet} and \eqref{hvalue} (or \eqref{logr}) as:
\bg\label{allura}
ds^2 & = & {r^2\over L^2}\left(1 - A ~{\rm log}~r - B~{\rm log}^2 r\right)\left(-g dt^2 + dx^i dx_i\right) + 
{1\over g}\cdot {L^2\over r^2} \cdot dr^2 \big(1 + A ~{\rm log}~r \nonumber\\
&& + B~{\rm log}^2 r\big) + \left(L^2 + A L^2 {\rm log}~r + B L^2 {\rm log}^2 r\right) d{\cal M}^2_5
\nd  
where $A$ and $B$ are defined in terms of $g_s, N^{\rm eff}_f, N$ and $M_{\rm eff}$ as:
\bg\label{abngm}
A = {3g_s M_{\rm eff}^2\over 4\pi N} \Big(1 + {3g_s N^{\rm eff}_f \over 4\pi}\Big), 
~~~~~~ B = {9g_s^2 M_{\rm eff}^2 N^{\rm eff}_f \over 8\pi^2 N}
\nd 
with $M_{\rm eff}$ and $N_f^{\rm eff}$ are defined in \eqref{hi};
and $L^2 = \sqrt{4\pi g_s N}$ being the usual definition. For supergravity to be valid we require weak string 
coupling but large $N$ such that:
\bg\label{limit}
g_s \to 0, ~~~~ L^2 >> 1, ~~~~ {N_f\over N} << 1, ~~~~ {M\over N} < 1, ~~~~ N >> 1, ~~~~g_s M >> 1
\nd
Using these limits (see also footnote 23 for more precise parametrisation) 
one can easily show that both $A$ and $B$ can be made very small, and 
\bg\label{lim2}
\Big({A\over L^2},~ {B\over L^2}\Big) ~ << ~ 1, ~~~~~~ \left(AL^2, ~BL^2\right) ~ << ~ 1
\nd
The above equation is very useful for us because we can recast our metric using \eqref{lim2} to show how much we 
deform from the AdS black hole metric. Since $L^2 >>  (AL^2,~ BL^2)$ we can rewrite \eqref{allura} 
as\footnote{Using the parametrisation given in footnote 23, observe that $A \to \epsilon^{9/2}, B \to \epsilon^6$ and 
$L^2 \to \epsilon^{-11/4}$. Thus $AL^2 \to \epsilon^{7/4}, BL^2 \to \epsilon^{13/4}, {A\over L^2} \to \epsilon^{29/4}$ 
and ${B\over L^2} \to \epsilon^{35/4}$ for $\epsilon \to 0$. This would justify the limits considered above.} 
\bg\label{leahl}
ds^2 && = {r^2\over L^2}\left(-g dt^2 + dx^i dx_i\right) + {L^2\over g r^2}~ dr^2 + 
L^2 ~ d{\cal M}^2_5 \\
&& - \left(A ~{\rm log}~r + B~{\rm log}^2 r\right)\Bigg[{r^2\over L^2} \left(-g dt^2 + dx^i dx_i\right) -
{L^2\over g r^2}~ dr^2  + L^2 ~ d{\cal M}^2_5\Bigg]\nonumber
\nd
where the first line is the $AdS_5 \times {\cal M}_5$ black hole solution, and the second line is the 
deformation of the $AdS_5$ and the internal 
geometries. Observe that (a) the internal space and the radial direction are very
mildly deformed in our limit for regions
close to the string\footnote{This is because the coefficients of $dr^2$ and the internal space are proportional to 
$L^2$ which go to infinity as $\epsilon^{-11/4}$, whereas the other parts are proportional to $L^{-2}$ which go to
zero. Clearly then the $dr^2$ and the internal space are strongly dominated by $L^2$ whereas the other parts 
can have small ${\rm log}~r$ deviations. Additionally any UV modifications will not affect the IR geometry.}, 
and (b) the deformation of the AdS geometry is not another AdS space.
These conclusions also imply that we can integrate 
over the internal coordinates $\psi,\phi_1,\phi_2,\theta_1,\theta_2$ in (\ref{KS77}), 
to obtain the  five dimensional effective action $S^{\rm eff}_{\rm total}$ in the following way: 
\bg\label{effaction}
S^{\rm eff}_{\rm total} &= & {1\over 8\pi G_N}\int d^5 x \sqrt{-G} 
\Big[e^{-2\phi} R(G) - 2\Lambda(r) + {\cal G}_{ij} \partial \Phi^i \cdot \partial \Phi^j  - 
m_{ij} \Phi^i \Phi^j + .....\Big] \nonumber\\
&& + T_5 \int d^5 x \Big[F \wedge \ast F + {\rm scalars} + \sum_{n \ge 2} c_n R^n + 
\sum_{n \ge 1} b_n {\rm tr}~(R \wedge R)^n \Big] - S^{\rm eff}_{\rm string} \nonumber\\ 
& = & S^{\rm eff}_{\rm OKS-BH} - S^{\rm eff}_{\rm string}
\nd
where $G_N$ is the Newton's constant, $T_5$ is the effective tension, $\Lambda(r)$ is the cosmological ``constant''
 coming from the contributions of the background fields and  
$\Phi^i$ are the scalars that we get by dimensionally reducing IIB supergravity fields over the internal manifold.
Some of these scalars come from the metric fluctuations of the internal five-manifold $g_{mn}(y)$ as:
\bg\label{koroth}
G^{(10)}_{mn}(x,y) := G_{\mu\nu}(x)~ \oplus ~ \sum_i \Phi^i(x) \Omega^i_{mn}(y) ~\oplus ~ g_{mn}(y) 
\nd
where ($x, y$) are the five-dimensional spacetime and the internal indices respectively, $G_{mn}(x,y)$ would incorporate
the 
background geometry \eqref{leahl} as well as string contribution (to be discussed below),
$\Omega^j$ are the normalisable $p$-form satisfying $\int \Omega^i \wedge \Omega^j = \delta^{ij}$. 
Similarly the other scalars (excluding the ones that come directly from the axio-dilaton and the five-form) come  
from the two two-form fields $B$ in the
following way:
\bg\label{decom}
\Phi^j = \int \Big(B -\langle B \rangle\Big) \wedge \Omega^j
\nd
where
$\langle B \rangle$ are the background $B_{NS}$ and $B_{RR}$ fields that we derived earlier. 
The rest of the scalars 
and the gauge fields are from the wrapped D7 brane.
We have also included the higher derivative $R^2$ and 
${\rm tr}~(R \wedge R)$ type terms that come from the D7 brane back reaction and world volume Cherns-Simons terms
respectively. These terms will have important implications that we will discuss in the next section. For the 
time being we want to point out that ${\cal O}(R^4)$ terms have been studied recently in the context of the 
$\eta/S$ bound, namely the viscosity over entropy bound in \cite{anindya2}. It was found therein that the $\eta/S$ 
receive quantum corrections that {\it doesn't} lower the known ${1\over 4\pi}$ bound of 
Kovtun-Son-Starinets \cite{Kovtun1}. However an ${\cal O}(R^2)$ term should lower the bound 
as discussed in \cite{Kats},
although one needs to be careful about two issues: the sign of the $R^2$ term and the relative strength of the 
 $R^4$ and $R^2$ 
terms\footnote{We thank Aninda Sinha for pointing this out to us.}.  

Once we have our effective five dimensional action, 
we can derive linearized Einstein equation 
using this effective action. 
At this point we can parametrise the string contribution 
as $\kappa l_{\mu\nu}$ with $\kappa \to 0$. This means that the perturbations
are small, which is a reasonable assumption for our case. Thus the total metric is $G_{\mu\nu} = g_{\mu\nu} + 
\kappa l_{\mu\nu}$ with $g_{\mu\nu}$ given by \eqref{leahl}. The equation of motion for $G_{\mu\nu}$ would be: 
\bg \label{KS7a}
R_{\mu\nu}\left(g_{\alpha\beta} + \kappa l_{\alpha\beta}\right) - {1\over 2}
\left(g_{\mu\nu} + \kappa l_{\mu\nu}\right) R\left(g_{\alpha\beta} + \kappa l_{\alpha\beta}\right) = 
T_{\mu\nu}^{\rm string} + T_{\mu\nu}^{\rm fluxes} + T_{\mu\nu}^{(p, q)7}
\nd
where the $T_{\mu\nu}^{\rm fluxes}$ come from the five-form fluxes $F_{(5)}$ (that give rise to the $AdS_5$ part) and the 
remnant of the $H_{NS}, H_{RR}$ and the axio-dilaton along the radial $r$ direction (that give rise to the 
deformation of the $AdS_5$ part). In five-dimensional space these fluxes would appear as one-forms $F_r^{(i)}$ with 
$i = 1,..., 4$. The effect of $T_{\mu\nu}^{(p, q)7}$ will not be substantial if we take it as a probe in this 
background. 

At this point we can approach the problem in two ways. The first way is to assume that the fluxes contribute to the 
five dimensional cosmological constant $\Lambda(r)$ as given in \eqref{effaction}. In this set-up we have an effective
five-dimensional theory \eqref{effaction} and we put a string in this background to study the perturbation in the 
metric. Here the assumption is that the string do not back-react on the cosmological constant $\Lambda(r)$. This 
seems to be the general approach in the literature. 

The second way is to actually consider the back reaction of the string on the five dimensional cosmological 
constant. This can be worked out if one considers the effects of all the background fluxes in our theory. The 
result of such an analysis can be presented in powers of $\kappa$. For our case we are only interested in 
back reactions that are linear in $\kappa$. To this order 
the equation of motion satisfied by $l_{\alpha\beta}$ is determined 
by expanding \eqref{KS7a} in the following way:
\bg\label{lalbe}
\kappa\Big(\triangle_{\mu\nu}^{\alpha\beta} - {\cal B}_{\mu\nu}^{\alpha\beta} - {\cal A}_{\mu\nu}^{\alpha\beta}\Big)
l_{\alpha\beta} = T_{\mu\nu}^{\rm string}
\nd
where $\triangle_{\mu\nu}^{\alpha\beta}$ is an operator whereas ${\cal B}_{\mu\nu}^{\alpha\beta}$ and 
${\cal A}_{\mu\nu}^{\alpha\beta}$ are functions of $r$, the radial coordinate\footnote{There will be another 
contribution from the ($p, q$) seven branes in the background, although for small $g_s N_f$ these are subleading.}.  
We have been able to determine 
the form for the operator $\triangle_{\mu\nu}^{\alpha\beta}$
 for any generic perturbation $l_{\alpha\beta}$. The resulting equations are rather long and 
involved; and we give them in the {\bf Appendix B}. For the functions 
${\cal A}_{\mu\nu}^{\alpha\beta}$
and ${\cal B}_{\mu\nu}^{\alpha\beta}$
we have worked out a toy example in 
{\bf Appendix C} with only diagonal perturbations. For off diagonal perturbations we need
to take an inverse of a $5\times 5$ matrix to determine the functional form. We shall provide details of this in the 
following. The variables defined in \eqref{lalbe} are given as:
\bg\label{listman}
&& \triangle_{\mu\nu}^{\alpha\beta} = 
\left({\delta R_{\mu\nu}\over \delta g_{\alpha\beta}}\right) - {1\over 2} ~g_{\mu\nu} 
\left({\delta R\over \delta g_{\alpha\beta}}\right) - {1\over 2}~R~\delta_{\mu\alpha} \delta_{\nu \beta}\nonumber\\ 
&& {\cal A}_{\mu\nu}^{\alpha\beta} = 5\sum_{b,c,d, ..} F_{(5)\mu bcd a}F_{(5)\nu b'c'd'a'}g^{bb'}g^{cc'}g^{dd'} 
g^{a\alpha}g^{a'\beta}\nonumber\\
&& {\cal B}_{\mu\nu}^{\alpha\beta} = -{5\over 8} \sum_{a,b,c,d, ..} F_{(5)n abcd} F_{(5)n' a'b'c'd'} g^{aa'} g^{bb'}
g^{cc'}g^{dd'} g^{nn'}
(g_{\mu\nu}g^{\alpha\beta} - \delta_\mu^\alpha\delta_\nu^\beta)\nonumber\\
&&~~~~~~~~ -{1\over 4}\sum_{i=1}^4 g_{\mu\nu} F_a^{(i)} F_b^{(i)} g^{a\alpha}g^{b\beta} + 
\sum_{i=1}^4 F_r^{(i)} F_r^{(i)} g^{rr} \delta_{\mu}^\alpha \delta_{\nu}^\beta 
\nd
where we have given the most generic form in \eqref{listman} above. Furthermore, 
${\delta R_{\mu\nu}\over \delta g_{\alpha\beta}}$ and ${\delta R\over \delta g_{\alpha\beta}}$ are operators
and not functions. As usual $g_{ab}$ is the metric of the OKS-BH background. 

At this point note that the first approach of ignoring the back reaction of the string on the cosmological 
constant $\Lambda(r)$ can only be achieved in the limit where there are no additional metric-derivatives and the
corrections to  
$F_{(5)}$ and $F_r^{(i)}$ are very small. Otherwise
this assumption would clearly fail. In this paper we will work out one concrete example (given in {\bf Appendix C}) 
using the first approach only and 
give the full analysis in the sequel \cite{sequel}. Therefore in the limit where the background fluxes, including the 
effects of the D7 brane, 
are very small
the above equation \eqref{lalbe}
can be presented as an operator equation of the following form:
\bg \label{KS7bi}
\kappa\; \triangle_{\mu\nu}^{\alpha\beta} l_{\alpha\beta}(x) \approx 
T_{\mu\nu}^{\rm string}(x)
\nd
where $x$ is a generic five-dimensional coordinate. In this final form,
the operator $\triangle_{\mu\nu}^{\alpha\beta}$ is a second order differential operator derived from
(\ref{KS7a}), with $\mu,\nu,\alpha, \beta = 0,1,2,3,4$.

Before we solve the above equation, we need to determine the relevant number of free components of $\l_{\mu\nu}$.  
Notice that as $l_{\mu\nu}$ is a symmetric tensor, it has fifteen degrees of freedom.
We can use coordinate transformations to 
fix ten degrees of
freedom. As an example consider the coordinate transformation: 
$x^a\rightarrow x'^a=x^a+g_s e^a$, $l_{\mu\nu}$ will transform as:
\bg\label{lmunu}
l_{\mu\nu}~ \rightarrow ~ l_{\mu\nu}-D_\mu e_\nu-D_\nu e_\mu
\nd
where $D_{\mu}$ is the covariant derivative.
Using this we can fix
five components, say: $l_{4\mu}=0,\mu=0,1,2,3,4$. As one might have expected for a similar example in
electromagnetism, this does not completely fix the gauge. The residual gauge transformation allow us to 
eliminate another five degrees of
freedom and we end up with five physical degrees of freedom i.e. five independent metric perturbation using
which all other components could be expressed. Alternatively one can use certain combinations of the fifteen 
components to write five independent degrees of freedom for the metric fluctuations.

The above result is easy to demonstrate for the AdS space, as has already been discussed in \cite{Yaffe-1}. For non-AdS
spaces this is not so easy to construct. Therefore
in the following we will take the complete set of ten components and using them we will determine the 
triangle operator $\triangle_{\mu\nu}^{\alpha\beta}$ in \eqref{lalbe}. If the ten components that can be 
labelled as a set: 
\bg\label{tencomp}
l_n = \big\{l_{00},~~l_{01},~~l_{02},~~l_{03},~~l_{11},~~l_{12},~~l_{13},~~l_{22},~~l_{23},~~l_{33}\big\}
\nd
then the operator $\triangle_{\mu\nu}^{\alpha\beta}$ in \eqref{lalbe} gives rise to 77 equations that we present 
in {\bf Appendix B}. The warp factor $h$ appearing in these equations can be  
taken to be $h=h(r,\pi,\pi)$ i.e \eqref{hvalue} with the slice condition \eqref{sol},
because we will analyse fluctuations close to the string. 

Before analysing \eqref{lalbe}, we make a coordinate transformation $r=1/u$ which will be useful in the wake analysis at
zeroth order in $g_s$. Since the internal space is now independent of warping or more appropriately, has very mild 
warping (see \eqref{leahl}), 
the local OKS-BH metric under this transformation will become:
\bg \label{metric1}
&&ds_{5}^2 = {-g(u)dt^2+dx^2+dy^2+dz^2 \over \sqrt{h(u, \pi, \pi)}} 
+\frac{\sqrt{h(u,\pi,\pi)}} {u^4 ~g(u)}du^2\nonumber\\
&& h(u,\pi,\pi) = u^4 L^4\left[1-{A}~{\rm log}~u -{B}~{\rm log}^2 u \right], ~~~ 
g(u)= 1-{u^4 \over u_h^4} \nonumber\\
&& r_h={1 \over u_h}, ~~~
{A} \approx  \frac{3g_s M^2}{4\pi N}\left(1+\frac{3g_s N_f}{4\pi}\right), ~~~
{B} \approx  \frac{9g_s^2 N_f M^2}{16\pi^2 N}
\nd
We will now solve (\ref{lalbe}) order by order in
$g_sN_f$ and $g_sM^2/N$.
At zeroth order in $g_sN_f,g_sM^2/N$, the warp factor becomes $h(u)=L^4u^4$ and the metric (\ref{metric1}) 
reduces to that of $AdS_5$. Hence at zeroth order in
$g_sN_f,g_sM^2/N$ our analysis will be similar to the AdS/CFT calculations \cite{Yaffe-1} but with certain crucial
differences that we will elaborate below.    
As (\ref{lalbe}) is a second order non linear partial differential equation, 
we can solve it by Fourier decomposing $x,y,z,t$ dependence of $l_{\mu\nu}$ and 
writing it as a Taylor series in $u$. To start off then let us decompose $l_{\mu\nu}$ as:
\bg \label{metricSolu} 
l_{\mu\nu} ~ = ~ l_{\mu\nu}^{[0]}+l_{\mu\nu}^{[1]}
\nd
where the subscript $[0], [1]$ refer to the zeroth and the first order in ($g_sN_f, g_s M^2/N$). To express the zeroth 
order fluctuation $l_{\mu\nu}^{[0]}$ in a Fourier series one has to be careful about the fact that our underlying 
space is not flat. In a curved space the Fourier series is expressed in terms of the corresponding harmonic 
forms in the space. These forms satisfy a Klein-Gordon type equation that one could easily determine for a 
given choice of the background metrics. Although in general the spatial parts of these forms could be complicated,
the temporal part however is always of the form $e^{-i\omega \tau}$ where $\tau$ is the time in the curved space. 
On the other hand, once we take the radial coordinate very large i.e $r = r_c \to \infty$, then we can take the 
harmonic forms on the boundary to be approximately wave like with $\tau$ defined as:
\bg\label{taudef}
\tau = \sqrt{g(u_c)}~ t
\nd
Of course this is not a generic picture at all points away from the boundary, but will nevertheless
suffice for us because we 
will eventually address the theory on the boundary only. Such a procedure will help us extract the universal features of 
the cascading theory. For all other theories that we will define by specifying UV degrees of freedom will all 
inherit some aspects of the universal features (plus additional dependences on the UV degrees of freedom). 
Therefore
in this limit, the zeroth order can 
be succinctly presented as a Fourier series in the following way:
\bg\label{fseries}
l_{\mu\nu}^{[0]}(t,u,x,y,z)~= ~ \sum_{k=0}^{\infty}\int \frac{d^3q d\omega}{(2\pi)^4\sqrt{g(u_c)}} 
\Bigg[e^{-i(\omega\sqrt{g}t-q_1x-q_2y-q_3z)} {\widetilde s}^{(k)[0]}_{\mu\nu}(\omega,q_1,q_2,q_3)u^k\Bigg]\nonumber\\
\nd
where 
${\widetilde s}^{(k)[0]}_{\mu\nu}$ 
are expansion coefficients of the solution $l_{\mu\nu}^{[0]}$. The precise deviations from the wave-like 
behavior for 
 $l_{\mu\nu}^{[0]}$ will not be very relevant for the present analysis. 

Similarly, we can also write the source in Fourier space as:
\bg \label{sourceF}
T_{\mu\nu}^{\rm string}~= ~T_{\mu\nu}^{[0]{\rm string}}+T_{\mu\nu}^{[1]{\rm string}}
\nd
where as before, [0, 1] refer to the zeroth and first orders in ($g_sN_f, g_s M^2/N$) respectively. The zeroth order 
can then be written as:
\bg\label{zert}
T_{\mu\nu}^{[0]{\rm string}}(t,u,x,y,z) =
\int \frac{d^3q d\omega}{(2\pi)^4 \sqrt{g(u_c)}} \;\;e^{-i(\omega \sqrt{g}t-q_1x-q_2y-q_3z)} 
t_{\mu\nu}^{[0]}(\omega,u,q_1,q_2,q_3)\nonumber\\
\nd
where $t_{\mu\nu}^{[0]}$ are expansion coefficients of source  
$T_{\mu\nu}^{[0]{\rm string}}$ at zeroth order in $g_sN_f,g_sM^2/N$. 
These coefficients are obtained by using explicit expressions 
for $T_{\mu\nu}^{\rm string}(x^\mu)$ that can be extracted from
\eqref{KS3}. One may also note that since $g$ is defined 
at $r = r_c$, we will henceforth be analysing the theory there with appropriate degrees of freedom
to be added later so as to have the full UV description. As mentioned before, this 
theory will correspond to a certain gauge theory with a RG flow that would inherit some properties of the 
$r_c \to \infty$ cascading theory, but in general will be different from the parent cascading theory 
(see {\bf figure 10})\footnote{Note however that even if these theories differ significantly in the UV, they all
confine in the far IR at zero temperature. 
This is illustrated in {\bf figure 10} by the existence of $r_{\rm min}$ for both the 
gravity duals.}. 
\begin{figure}[htb]\label{correctboundary}
		\begin{center}
	\includegraphics[height= 6cm]{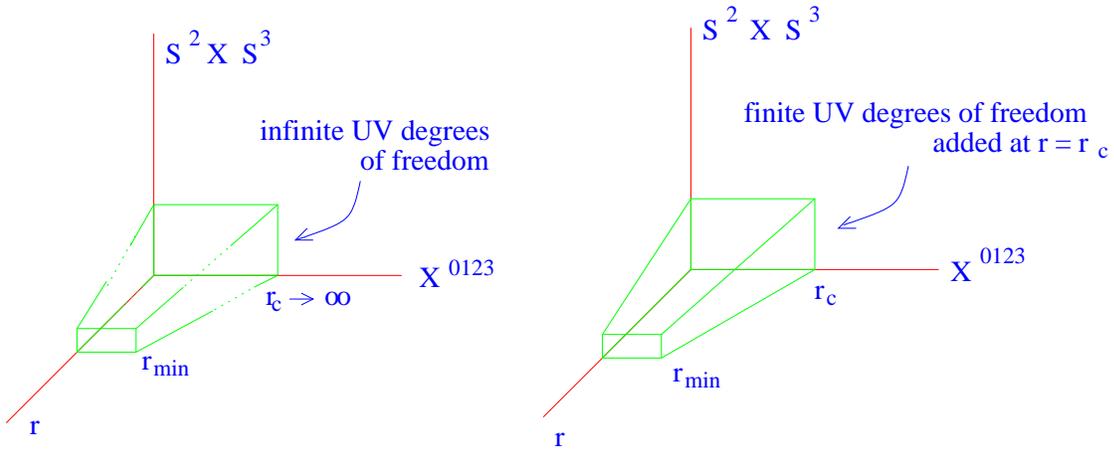}
		\caption{{ The full ten-dimensional picture of the gravity dual. The figure on the left is the 
actual dual to the cascading theory with infinite (gauge theory) degrees of freedom (dof) 
at the boundary. The figure on the 
right is the one with a cutoff at $r = r_c$ where we add a finite (but large) number of dof
to define the UV of that theory. Once the UV dof are specified we can describe this theory at the boundary. The IR 
physics changes a bit by inclusion of irrelevant operators.
Clearly the gravity 
dual on the right can capture only some universal properties of the parent theory (on the left). Nevertheless the theory 
on the right is an interesting theory (with good UV behavior) that can be studied directly from our analysis. The
RG flows of both these theories have been discussed earlier. 
The existence of $r_{\rm min}$ signify the confining 
nature of these theories in the far IR at zero temperature. At high temperature we should view the blackhole covering
the region near $r_{\rm min}$.}}  
		\end{center}
		\end{figure}

One may also analyse similarly  
the metric perturbation at linear order in $g_sN_f$ and $g_sM^2/N$ where it is easiest to switch to 
coordinate $u=\frac{1}{r_c(1-\zeta)}$, 
so that the entire manifold (from $r = 0$ to $r = r_c$)
is now described by $0 \leq \zeta \leq 1$ with $r_c$ arbitrarily large and 
we can get a meaningful 
Taylor series expansion of the logarithms and other functions appearing in the equation. 
As we did for the zeroth order cases, we can decompose the first order 
in $g_sN_f, g_sM^2/N$ fluctuations via the following Fourier series:
\bg \label{perteq11a}
l_{\mu\nu}^{[1]}(t,\zeta,x,y,z)&=&\sum_{k=0}^{\infty}\int \frac{d^3q d\omega}{(2\pi)^4 \sqrt{g(u_c)}} 
\Bigg[e^{-i(\omega \sqrt{g}t-q_1x-q_2y-q_3z)} {\widetilde s}^{(k)[1]}_{\mu\nu}(\omega,q_1,q_2,q_3)\zeta^k\Bigg]\nonumber\\
\nd
where ${\widetilde s}_{\mu\nu}^{(k)[1]}$ are the corresponding Fourier modes. 
These Fourier modes will 
eventually appear in the final equations for $l_{\mu\nu}^{[1]}$.

However, as we discussed above, these modes are not all independent. There are only five independent metric fluctuations
which could be written as some linear combinations of the ten components \eqref{tencomp}. The zeroth 
and first order independent fluctuations can then be expressed as:
\bg\label{oand1}
&&L_{mm}^{[0]} = \sum_{n = 1}^{10} a_{mn} l_n^{[0]} = \sum_{k = 0}^\infty \int {d^3q d\omega\over (2\pi)^4 \sqrt{g(u_c)}} 
\Bigg[e^{-i(\omega \sqrt{g}t-q_ix^i)} {s}^{(k)[0]}_{mm}(\omega,\ve{q})u^k\Bigg]\nonumber\\
&&L_{mm}^{[1]} = \sum_{n = 1}^{10} b_{mn} l_n^{[1]} = \sum_{k = 0}^\infty \int {d^3q d\omega\over (2\pi)^4 \sqrt{g(u_c)}} 
\Bigg[e^{-i(\omega \sqrt{g}t-q_ix^i)} {s}^{(k)[1]}_{mm}(\omega,\ve{q})\zeta^k\Bigg]
\nd
where $a_{mn}$ and $b_{mn}$ are functions of all the coordinates in general with $m$ running from 1 to 5. 
Notice that the way we are expressing
the independent modes is different from the way it is presented in \cite{Yaffe-1}. It should also be clear that 
\bg\label{siide}
&&s_{mm}^{(k)[0]} = a_{m1} {\widetilde s}_{00}^{(k)[0]} + a_{m2} {\widetilde s}_{01}^{(k)[0]} + 
a_{m3} {\widetilde s}_{02}^{(k)[0]} + a_{m4} {\widetilde s}_{11}^{(k)[0]} + .... \nonumber\\
&&s_{mm}^{(k)[1]} = b_{m1} {\widetilde s}_{00}^{(k)[1]} + b_{m2} {\widetilde s}_{01}^{(k)[1]} + 
b_{m3} {\widetilde s}_{02}^{(k)[1]} + b_{m4} {\widetilde s}_{11}^{(k)[1]} + .... 
\nd
Since there are five independent metric components, we expect that there should also be five independent sources.
These sources, coming from the string, should again be some linear combination of the ten possible components. We 
can write them as ${\widetilde T}_{mm}^{[0]}$ and ${\widetilde T}_{mm}^{[1]}$ to denote the zeroth and first 
orders in ($g_sN_f, g_sM^2/N$) respectively. The above considerations also imply that the operator equation \eqref{lalbe}
will become:
\bg\label{lalbenow}
\kappa \sum_{n=1}^5 \square^{nn}_{mm} L_{nn} = {\widetilde T}_{mm}
\nd
where the form of $\square^{nn}_{mm}$ can be easily derived from \eqref{lalbe}.

Knowing the metric perturbations $L_{mm}$, we can now compute the effective supergravity action in five dimensions 
and use this action to compute the four dimensional boundary action. Recall that our space is a deformation of the 
AdS space \eqref{leahl} and from all our previous discussions, we need to be careful when we 
want to give a {\it boundary} description. One of the simplest boundary description that appears naturally in this 
framework is the one with an infinite degrees of freedom. The other alternative descriptions are defined through the 
cutoffs imposed at various
$r_c$. All these infinite possible descriptions may have finite (but very large) 
UV degrees of freedom and flow to $N_{\rm eff}$ 
degrees of freedom at $r = r_c$ (see \eqref{edofk}). Once we specify this 
we expect that the boundary will
capture the UV of the corresponding gauge theory. In the limit where $N >> M$ one can even compute the spectrum of the 
operators from the boundary description (see for example \cite{Hollowood:2004ek}
for a zero-temperature example without flavor). To 
describe the boundary action we will in principle  
take a
 ``functional derivative'' of our action with respect to the 
perturbation ${L}_{mm}$ $-$ meaning that we will extract only the linear coefficient of the perturbation $L_{mm}$ $-$
obtained by a fixing a value for the radial coordinate ($u$=fixed or
 $\zeta$=fixed). 
Using the boundary action our aim then is to compute \eqref{goal} which is the {\it wake} left 
behind by a fast moving quark. 

For comparison with AdS/CFT result, a simple way would be to split the five dimensional effective action into
two parts: 
 \bg \label{actionAdS+KS}
 S^{\rm eff}_{\rm total} &=& S^{\rm eff({\rm AdS+string})}_{\rm total}+S^{\rm eff({\rm OKS+run})}_{\rm total}
 \nd 
where the first part is the vanilla AdS with a quark string and the second part is the deformation that 
captures the running of the gauge theory in the dual picture. The AdS-string part is measured at zeroth 
order in $g_s N_f$ and $g_s M^2/N$ whereas the OKS-run part is measured at first order in $g_s N_f$ and $g_s M^2/N$.  
It should also be clear that in the limit
($g_sN_f, g_sM^2/N$) $\rightarrow 0$, our geometry is $AdS_5\times T^{11}$ Black-Hole with
 back reaction from the string\footnote{We will discuss the back reaction from the D7 brane
in the next section when we compute the shear viscosity. In this section we will only consider the back reaction 
of the string on the geometry.}. 
More concretely, for the AdS-string part we will use
$L_{mm}^{[0]}$ as our metric perturbation computed using the zeroth order energy momentum tensor of the 
string \eqref{sourceF} whereas for the OKS-run part we will use $L_{mm}^{[1]}$ as the metric perturbation 
coming from the first order energy momentum tensor of the string \eqref{sourceF}. To avoid cluttering of 
formulae, we will henceforth label:
\bg\label{1and2}
S^{\rm eff({\rm AdS+string})}_{\rm total} \equiv {\cal S}^{(1)}, ~~~~ S^{\rm eff({\rm OKS+run})}_{\rm total}
\equiv {\cal S}^{(2)}
\nd 
Thus we can represent the wake of the quark by the following formal 
expressions\footnote{Although the energy momentum tensor in the following equation may seem {\it diagonal} 
in terms of $L_{mm}$ variables, they are in fact not diagonal in terms of the $l_{\mu\nu}$ variables. We 
simply find it convenient to express $T^{mn}$ in terms of the five independent metric fluctuations. Furthermore 
both terms of \eqref{actionAdS+KS} are restricted to four dimensional space-time.}:
\bg \label{actionAdS+KS}
 \langle T^{mn}\rangle_{\rm wake} &=&\lim_{u = 0}{\delta^{mn}\over \kappa} 
\cdot \frac{\delta_b S^{\rm eff}_{\rm total}}{\delta_b
{L}_{mm}}
-T^{mn}_{\rm quark}\nonumber\\
&=&\lim_{u = 0} {\delta^{mn}\over \kappa}\Bigg(\frac{\delta_b{\cal S}^{(1)}}{\delta_b
{L}^{[0]}_{mm}} +\frac{\delta_b {\cal S}^{(2)}}{\delta_b
{L}^{[1]}_{mm}}\Bigg) -T^{mn}_{\rm quark}
\nd
which would give us the result for the UV of the dual gauge theory because $\zeta = 0$ would take us to the 
boundary of our space\footnote{To be more precise $\zeta = 0$ takes us to $r = r_c$; and then once we add appropriate
UV degrees of freedom we can define our theory at the boundary $r \to \infty$.}, and the operation 
${\delta_b{\cal S}^{(i)}\over \delta_b L_{mm}^{[j]}}$ extracts the linear coefficient from the boundary action.
The energy momentum tensor of the quark 
can then be written as:
\bg\label{enrquark} 
T^{mn}_{\rm quark}(x,y,z,t)&=&m(T) U^{m} U^{n}\sqrt{1-v^2} 
~\delta^3(\overrightarrow{x}-\overrightarrow{v}\cdot t)
 \nd
where $U^m = (U^0, \overrightarrow{v})$ and $U^0$ is the energy of the quark and $\overrightarrow{v}$ is the 
three-velocity. Now
 at zeroth order, as mentioned before, the five dimensional effective
 action that we have is that of $AdS_5$ plus back reaction of the string. For an infinitely massive string 
i.e. $u_0 \equiv {1\over r_0} = 0$,\footnote{Recall that for our case the string ended at a finite 
distance $u_0 = \vert\mu\vert^{-2/3}$ (see \eqref{embedim} for details). Once the string is infinite, we are effectively 
putting the D7 brane at infinity i.e $u= u_0 = 0$.}
 just like in  \cite{Yaffe-1},using exactly the regularisation procedure of \cite{Kostas-1} \cite{Kostas-2}
 \cite{Kostas-3} and \cite{2}, the
 stress tensor  evaluated at $u=0$ is known to be:
\bg \label{StressTensor_gs=0}
 \langle T^{mn}\rangle_{\rm wake(AdS)}&=& \left(\lim_{u = 0}{\delta^{mn}\over\kappa}
\cdot \frac{\delta_b{\cal S}^{(1)}}{\delta_b
{L}^{[0]}_{mm}}\right)_{\rm FT}-{\widetilde T}^{mn}_{\rm quark} (\omega, \overrightarrow{q})\nonumber\\
&=&\int \frac{d^4q}{(2\pi)^4}c_o  \delta^{mn}s^{[0](4)}_{mm}(\omega,\overrightarrow{q})-
{\widetilde T}^{mn}_{\rm quark} (\omega, \overrightarrow{q})
\nd
where FT is the Fourier transform and 
${\widetilde T}^{mn}_{\rm quark} (\omega, \overrightarrow{q})$ are the quark energy-momentum Fourier modes
and the value of 
$c_o$ will be discussed later.
Note however that for a 
finite length string i.e. the string stretching from
 $u_0 \equiv \vert\mu\vert^{-2/3} \equiv 1$ to $u_h$ the metric perturbation due to
 the string goes to zero for $u < u_0$. This means that the previous analysis done in \cite{Yaffe-1} will not be 
valid for our case and we would require to regularise our effective action. This is where it may be more convenient 
to regularise the system to ${\cal O}(g_sN_f, g_sM^2/N)$ without splitting up the action.

In the following therefore we will give a brief discussion of our regularisation scheme using the warp factor 
\eqref{logr}.
We will start by analysing the background string configuration
by plugging in the $l_{\mu\nu}$ in the Einstein as well as the 
flux terms of the five 
dimensional action. Once the action is expressed in terms of $l_{\mu\nu}$ we expect that it will be equivalently 
rewritten in terms of $L_{mm}$. An example of this is given by \cite{Yaffe-1}, and we provide another example in
{\bf Appendix C}. Once the dust settles,  
the result in the most symmetric form is given by: 
\bg \label{KS7c} 
&&{\cal S}^{(1)}[\Phi]~=~\int \frac{d^4q}{(2\pi)^4 \sqrt{g(r_c)}}\int dr
~\Bigg\{{1\over 2} A^{mn}_1(r,q)\Big[\Phi^{[1]}_m(r,q)\Phi''^{[1]}_n(r,-q) + 
\Phi''^{[1]}_m(r,q)\Phi^{[1]}_n(r,-q)\Big]\nonumber\\
&& + B^{mn}_1(r,q) \Phi'^{[1]}_m(r,q)\Phi'^{[1]}_n(r,-q)
+{1\over 2} C^{mn}_1(r,q) \Big[\Phi'^{[1]}_m(r,q)\Phi^{[1]}_n(r,-q) + \Phi^{[1]}_m(r,q)\Phi'^{[1]}_n(r,-q)\Big]\nonumber\\
&& + D^{mn}_1(r,q)\Phi^{[1]}_m(r,q)\Phi^{[1]}_n(r,-q)+{\cal T}^m_1(r,q)\Phi^{[1]}_m(r,q)
+E^m_1\Phi'^{[1]}_m(r,q)+F^m_1\Phi''^{[1]}_m(r,q)\Bigg\}\nonumber\\
\nd
where $m,n=1,..., 5$, prime denotes differentiation with respect to $r$,\footnote{This is for simplicity. 
Prime could be more generic and denote derivatives wrt $r$ as well as $\overrightarrow{q}$. 
We will rectify this below. Notice also that
we have shifted to $r$ variable, but
the action could equivalently be expressed by $u$ variable.}; the script $[1]$ denote the {\it total} background to 
${\cal O}(g_sN_f, g_sM^2/N)$; 
and the explicit expressions for 
$A^{mn}_1,B^{mn}_1,C^{mn}_1,F^m_1$ for a specific case are given in {\bf Appendix D}. 
We have also defined $\Phi^{[1]}_m(r, q)$ 
in the following way (with $q_0 \equiv \omega \sqrt{g(r_c)}$ as before): 
\bg\label{phim}
\Phi^{[1]}_m(r,q)=\int \frac{d^4x}{(2\pi)^4 \sqrt{g(r_c)}}\;e^{i(q_0 t-q_1x-q_2y-q_3z)}L_{mm}(t,r,x,y,z)
\nd 
 We will see that the effective four dimensional boundary action is independent of $D^{mn}_1$ and ${\cal T}^m_1$
and hence we do not list their explicit expressions in appendix. Furthermore, note that the 
derivative terms in \eqref{KS7c} all come exclusively from $\sqrt{-G}R$, whereas fluxes contribute powers of 
$\Phi^{[1]}$ but no derivative interactions (this is also one of the reasons why F-theory seven branes will not 
change our conclusions provided we ignore the anomalous gravitational couplings \eqref{disclevel} or \eqref{effaction}). 
In the following we keep upto quadratic orders, and therefore the 
contributions from the fluxes will appear in $D^{mn}_1$ and $E^m_1$. Taking all these affects into account 
we 
compute the coefficients
 $A^{mn}_1,B^{mn}_1,C^{mn}_1,D^{mn}_1,E^m_1$, and $F^m_1$ for a specific example, 
whose values are given in {\bf Appendix D}.
However, as we mentioned before,  
we will be ignoring the back reaction of the string on the fluxes because we expect any 
additional components of fluxes coming from the string to be very small\footnote{This is easy to motivate from the 
fact that the size of our quark string is very small. One end of the string goes into the horizon of the black hole and 
the other end is on the D7 brane placed at $\vert\mu\vert^{2/3} \equiv 1$. Thus the $B_{NS}$ sources from the string 
will be negligible.
Furthermore fundamental string will not effect 
any of the background RR forms, at least in the limit \eqref{lcone} that we are considering. So it makes sense to ignore 
the effects of the string on fluxes. 
See section 3.2 for more details.}. The equation of
motion for $\Phi_n^{[1]}(r, -q)$ is given by:
\bg\label{eomphin}
&&{1\over 2} \Big[A_1^{mn}(r, q)\Phi_n^{[1]}(r, -q)\Big]''~ -~ \Big[B_1^{mn}(r, q)\Phi_n'^{[1]}(r, -q)\Big]' ~- ~
{1\over 2} \Big[C_1^{mn}(r, q)\Phi_n^{[1]}(r, -q)\Big]' \nonumber\\
&& ~~+~ D_1^{mn}(r, q)\Phi_n^{[1]}(r, -q) ~+~ {1\over 2} A^{mn}_1(r, q) \Phi_n''^{[1]}(r, -q)
~-~ {1\over 2} C_1^{mn}(r, q) \Phi_n'^{[1]}(r, -q)\nonumber\\ 
&&~~~~~~~~~~~~~~~~ + ~{\cal T}_1^m(r, q) ~-~ E_1'^m(r, q) ~+~ F''^m_1(r, q) = 0
\nd
The next few steps are rather standard and so we will quote the results. The variation of the action \eqref{KS7c} 
can be written in terms of the variations $\delta \Phi_m^{[1]}(r, q)$ and $\delta \Phi_n^{[1]}(r, -q)$ 
in the following way\footnote{Henceforth, unless mentioned otherwise, $\Phi_m^{[1]}, \Phi_n^{[1]}$ 
will always mean $\Phi_m^{[1]}(r, q)$ and 
$\Phi_n^{[1]}(r, -q)$ respectively. Similar definitions go for the variations $\delta\Phi_m^{[1]}$ 
and $\delta\Phi_n^{[1]}$.}:
\bg\label{varofac}
&&\delta{\cal S}^{(1)} = {1\over 2}\int \frac{d^4q}{(2\pi)^4 \sqrt{g(r_c)}}\int_{r_h}^{r_c} dr\Bigg\{
\Big[(A_1^{mn}\Phi_m^{[1]})'' - (2B_1^{mn}\Phi_m'^{[1]})' + C_1^{mn}\Phi_m'^{[1]} + 2D_1^{mn}\Phi_m^{[1]} \nonumber\\ 
&&+ A^{mn}_1 \Phi_m''^{[1]} - (C_1^{mn}\Phi_m^{[1]})'\Big]\delta\Phi_n^{[1]}
+ \Big[(A_1^{mn}\Phi_n^{[1]})'' - (2B_1^{mn}\Phi_n'^{[1]})' + C_1^{mn}\Phi_n'^{[1]} + 2D_1^{mn}\Phi_n^{[1]} \nonumber\\ 
&& + A^{mn}_1 \Phi_n''^{[1]} - (C_1^{mn}\Phi_n^{[1]})'\Big]\delta\Phi_m^{[1]} + 
2\left({\cal T}_1^m - E_1'^m + F''^m_1\right)\delta\Phi_m^{[1]}\nonumber\\
&&\partial_r\Big[A_1^{mn}\Phi_m^{[1]} \delta\Phi_n'^{[1]} - (A_1^{mn}\Phi_m^{[1]})'\delta\Phi_n^{[1]}
 + 2B_1^{mn}\Phi_m'^{[1]}\delta\Phi_n^{[1]} + 
C_1^{mn} \Phi_m \delta\Phi_n^{[1]} + 2B_1^{mn}\Phi_n'^{[1]} \delta\Phi_m^{[1]}\nonumber\\
&& +C_1^{mn}\Phi_n^{[1]}\delta\Phi_m^{[1]} + 2E_1^m \delta\Phi_m^{[1]} + 2F_1^m\delta\Phi_m'^{[1]} - 
2F_1'^m \delta\Phi_m^{[1]} + 
A_1^{mn} \Phi_n^{[1]} \delta\Phi_m'^{[1]} - (A_1^{mn}\Phi_n^{[1]})'\delta\Phi_m^{[1]}\Big]\Bigg\}\nonumber\\
\nd
which includes the equations of motion as well as the boundary term. We can then write the variation of the 
action $\delta{\cal S}^{(1)}$ in the following way: 
\bg \label{KS7c1}
&&\delta{\cal S}^{(1)}~ = ~ \int \frac{d^4q}{(2\pi)^4 \sqrt{g(r_c)}}
\Bigg\{ \int_{r_h}^{r_c} dr \Big[\left({\rm EOM \;for\; \Phi_n^{[1]}}\right)\delta \Phi^{[1]}_m 
+ \left({\rm EOM \;for\; \Phi_m^{[1]}}\right)\delta \Phi^{[1]}_n\Big] \nonumber\\
&&+ {1\over 2}\Big[(2B_1^{mn} - A_1^{mn}) (\Phi_m'^{[1]}\delta\Phi_n^{[1]} 
+ \Phi_n'^{[1]}\delta\Phi_m) + (C_1^{mn} - A_1'^{mn})
(\Phi_m^{[1]}\delta\Phi_n^{[1]} + \Phi_n^{[1]}\delta\Phi_m^{[1]}) \nonumber\\
&& + 2(E_1^{m} - F_1'^{m})\delta\Phi_m^{[1]} + A_1^{mn}\Phi_m^{[1]}\delta\Phi_n'^{[1]} 
+ A_1^{mn}\Phi_n^{[1]}\delta\Phi_m'^{[1]} + 2F_1^{m}\delta\Phi_m'^{[1]}\Big]_{\rm boundary}\Bigg\}\nonumber\\
\nd 
where by an abuse of notation by
the ``boundary'' here, and the next couple of pages (unless mentioned otherwise), 
we mean that the functions are all measured at $r_h$ and $r_c$ i.e the horizon and the 
cut-off respectively\footnote{Note that we have not carefully described the degrees of freedom at the boundary as yet.
For large enough $r_c$ we expect large degrees of freedom at the UV because of \eqref{doa}. This would mean that the 
contributions to various gauge theories from these degrees of freedom would go like $e^{-{\cal N}_{\rm eff}}$, which 
would be negligible. Thus unless we cut-off the geometry at $r = r_c$ and add UV caps with specified degrees of freedom
we are in principle only describing the parent cascading theory. For this theory of course ${\cal N}_{\rm eff}$ is 
infinite at the boundary, which amounts to saying that UV degrees of freedom don't contribute anything here.
We will, however, give a more precise description a little later.}. 
It is now easy to see why $D_1^{mn}$, ${\cal T}_1^{mn}$ and $E_0^{mn}$ etc do not 
appear in the boundary action. Finally, we need to add another boundary term to \eqref{KS7c1} to cancel of the 
term proportional to $\delta\Phi'_n$. This
is precisely the Gibbons-Hawking term \cite{Gibbons-Hawking}: 
\bg \label{KS7c2} 
{\cal K}_1 =- {1\over 2}\int \frac{d^4q}{(2\pi)^4 \sqrt{g(r_c)}}\Big(A^{mn}_1\Phi^{[1]}_m\Phi'^{[1]}_n + 
A^{mn}_1\Phi^{[1]}_n\Phi'^{[1]}_m 
+2 F^n_1\Phi'^{[1]}_n\Big)\Bigg{|}_{{\rm boundary}}
\nd
Taking the variation of \eqref{KS7c2} $\delta{\cal K}_1$ we get terms proportional to $\delta\Phi'^{[1]}$ as well as 
$\delta\Phi^{[1]}$. Adding $\delta{\cal K}_1$ to $\delta{\cal S}^{(1)}$ we can get rid of all the $\delta\Phi'^{[1]}$
terms from \eqref{KS7c1}. This means we can alternately state that the boundary theory should have the following 
constraints\footnote{One can impose similar constraints at the horizon also.}:
\bg\label{conts}
 \delta\Phi_m'^{[1]}(r_c, q) ~=~ \delta\Phi_n'^{[1]}(r_c, -q)~ =~ 0
\nd
Next, we can make a slight modification of the boundary integral \eqref{varofac} by specifying the coordinate $r$
to be from $r_h$ to certain $r_{\rm max} = r_c(1-\zeta)$ instead of just $r_c$ specified earlier. 
For different choices of $\zeta$ we can allow different UV completions at the boundary such that their degrees of 
freedom would match with the parent cascading theory only at $r_c(1-\zeta)$, while $r_h$ would
be associated with the characteristic temperature ${\cal T}$ of the parent cascading theory.

With all the above considerations we can present our final result for the boundary action. Putting the 
equations of motion constraints on \eqref{KS7c1}, as well as the derivative constraints \eqref{conts}, we can 
show that the variation \eqref{KS7c1} can come from the following boundary $3+1$ dimensional action:
\bg \label{action1}
&& {\cal S}^{(1)}~=~\int \frac{d^4q}{(2\pi)^4 \sqrt{g(r_{\rm max})}}
\Bigg\{\Big[ C^{mn}_1(r,q)-A'^{mn}_1(r,q)\Big] \Phi^{[1]}_m(r, q) \Phi^{[1]}_n(r, -q)\nonumber\\
&&~~~~~~+\Big[B^{mn}_1(r,q)-A^{mn}_1(r,q)\Big] \Big[\Phi'^{[1]}_m(r, q) \Phi^{[1]}_n(r, -q)+\Phi^{[1]}_m(r, q)
\Phi'^{[1]}_n(r, -q)\Big]\nonumber\\
&&~~~~~~~~~~~ +\Big(E^m_1 -F'^m_1\Big)\Phi^{[1]}_m(r, q)\Bigg\}\Bigg{|}_{r_h}^{r_c(1-\zeta)}
\nd
However the above action diverges, as one can easily check 
from the explicit expressions for 
$A^{mn}_1,B^{mn}_1,C^{mn}_1,E^m_1$ and $F^m_1$ for the specific case worked in {\bf Appendix D}. 
Indeed, comparing it to the known AdS results, we
observe that from the boundary there are terms proportional to $r_c^4$ in each of 
$C^{mn}_1,A'^{mn}_1,E^m_1$ and $F'^{m}_1$ and proportional to $r_c^5$ in $A^{mn}_1, B^{mn}_1$. 
As $r_c \to \infty$ the action diverges so as it stands $r_c$ cannot 
completely specify the UV degrees of freedom at the 
QCD scale $\Lambda_c$\footnote{We are a little sloppy here. Of course our theory is 
large $N$ QCD like only in the far IR. Nevertheless $r_c$ should specify the UV degrees of freedom 
as \eqref{doa} at scale $\Lambda_c$ for the resulting theory.}. 
Thus we need to
regularise/renormalise
it before taking functional derivative of it. This renormalisation procedure will give us a finite 
boundary theory from which one could get meaningful results of the dual gauge theory.

Once we express the warp factors in terms of power series in $r_{(\alpha)}$ \eqref{logr}
the renormalisability procedure becomes 
much simpler. 
Note however that this renormalisation is only 
in classical sense, as the procedure will involve removing the infinities in \eqref{action1} by adding counter-terms
to it. Comparing with the known AdS results, and the specific example presented in {\bf Appendix C}, one can argue that
the infinities in \eqref{action1} arise from the following three sources:
\bg\label{infinity}
&&1.~~~ C^{mn}_1(r_c,q)-A'^{mn}_1(r_c,q) ~ = ~ \sum_{\alpha} H_{\vert\alpha\vert}^{mn}(q)~ r_{c(\alpha)}^4 
~+~ {\rm finite~terms}\nonumber\\
&&2.~~~ B^{mn}_1(r_c,q)-A^{mn}_1(r_c,q)  ~ = ~ \sum_{\alpha} K_{\vert\alpha\vert}^{mn}(q)~ r_{c(\alpha)}^5 
~+~ {\rm finite~terms}\nonumber\\
&&3.~~~ E^m_1(r_c,q) -F'^m_1(r_c,q)  ~ = ~ \sum_{\alpha} I_{\vert\alpha\vert}^{m}(q) ~r_{c(\alpha)}^4 
~+ ~ {\rm finite~terms}
\nd
where in the above expressions we are keeping $\alpha$ arbitrary so that it can in general take both positive and 
negative values; and 
the finite terms above are of the form $r_{c(\alpha)}^{-n}$ with $n \ge 1$. Therefore
to regularise, first we write the metric perturbation also as a series in $1/r_{(\alpha)}$:
\bg \label{mpert1}
\Phi^{[1]}_n=\sum_{k=0}^{\infty} \sum_{\alpha}~{s_{nn}^{(k)[\alpha]}\over r_{(\alpha)}^k} ~\theta(r_0 - r)
\nd 
where 
the above relation could be easily derived using \eqref{phim}, \eqref{oand1} taking the background warp factor
correctly; and 
$r_0 = \vert\mu\vert^{2/3}$ is given by \eqref{embedim}.

Plugging in \eqref{mpert1} and \eqref{infinity} in \eqref{action1} we can easily extract the divergent parts of the 
action \eqref{action1}. Thus the counter-terms are given by:  
\bg\label{cterms}
&& {\cal S}^{(1)}_{\rm counter}=\int \frac{d^4q}{2(2\pi)^4 \sqrt{g(r_{\rm max})}}
\sum_{\alpha, \beta, \gamma} \Big\{{H}_{\vert\alpha\vert}^{mn}
\theta(r_0-r)\Big[s_{mm}^{(0)[\beta]}s_{nn}^{(0)[\gamma]}~r_{(\alpha)}^4 + 
\Big(s_{mm}^{(1)[\beta]}s_{nn}^{(0)[\gamma]}r_{(\alpha_1)}^3\nonumber\\ 
&& + s_{mm}^{(0)[\beta]}s_{nn}^{(1)[\gamma]}r_{(\alpha_2)}^3\Big) 
+ \Big(s_{mm}^{(2)[\beta]}s_{nn}^{(0)[\gamma]}r_{(\alpha_3)}^2 
+ s_{mm}^{(0)[\beta]}s_{nn}^{(2)[\gamma]}r_{(\alpha_4)}^2 + 
s_{mm}^{(1)[\beta]}s_{nn}^{(1)[\gamma]}r_{(\alpha_5)}^2\Big)\nonumber\\ 
&&+\Big(s_{mm}^{(3)[\beta]}s_{nn}^{(0)[\gamma]}r_{(\alpha_6)} + 
s_{mm}^{(0)[\beta]}s_{nn}^{(3)[\gamma]}r_{(\alpha_7)}
+ ~ s_{mm}^{(2)[\beta]}s_{nn}^{(1)[\gamma]}r_{(\alpha_8)} + 
s_{mm}^{(1)[\beta]}s_{nn}^{(2)[\gamma]}\Big)r_{(\alpha_9)} \Big]\nonumber\\
&& + {K}_{\vert\alpha\vert}^{mn}
\theta(r_0-r)\Big[-\big(s_{mm}^{(0)[\beta]}s_{nn}^{(1)[\gamma]}r_{(\alpha_{10})}^3+s_{mm}^{(1)[\beta]}s_{nn}^{(0)[\gamma]}r_{(\alpha_{11})}^3\big)
- \Big(2s_{mm}^{(1)[\beta]}s_{nn}^{(1)[\gamma]}r_{(\alpha_{12})}^2 \nonumber\\
&& +2 s_{mm}^{(0)[\beta]}s_{nn}^{(2)[\gamma]}r_{(\alpha_{13})}^2
+2 s_{nn}^{(0)[\beta]}s_{mm}^{(2)[\gamma]}r_{(\alpha_{14})}^2\Big) 
-~\Big(2s_{mm}^{(1)[\beta]}s_{nn}^{(2)[\gamma]}r_{(\alpha_{15})} 
+ s_{mm}^{(2)[\beta]}s_{nn}^{(1)[\gamma]}r_{(\alpha_{16})}\nonumber\\
&&+ 3s_{mm}^{(0)[\beta]}s_{nn}^{(3)[\gamma]}r_{(\alpha_{17})} 
+2s_{nn}^{(1)[\beta]}s_{mm}^{(2)[\gamma]}r_{(\alpha_{18})} 
+ s_{nn}^{(2)[\beta]}s_{mm}^{(1)[\gamma]}r_{(\alpha_{19})} + 
3s_{nn}^{(0)[\beta]}s_{mm}^{(3)[\gamma]}r_{(\alpha_{20})}\Big)\Big] \nonumber\\
&& ~~~~~~~~~ +I_{\vert\alpha\vert}^m \theta(r_0-r)\Big(s_{mm}^{(0)[\beta]} r_{(\alpha)}^4
+ s_{mm}^{(1)[\beta]} r_{(\alpha_1)}^3 + s_{mm}^{(2)[\beta]} r_{(\alpha_3)}^2 
+ s_{mm}^{(3)[\beta]} r_{(\alpha_6)}\Big)\Big\}
\nd
with an equal set of terms with $r_{(-\alpha_i)}$. In the above expression 
$r_{(\alpha_i)} \equiv r^{1- \epsilon_{(\alpha_i)}}$;
and as before, the integrand is defined at the horizon $r_h$ and the cutoff $r_c(1-\zeta)$; with the  
string stretching between $r_h$ and $r_0$. The other variables namely, 
$s_{mm}^{(k)[\beta]}, {H}_{\vert\alpha\vert}^{mn},K_{\vert\alpha\vert}^{mn}$ and ${I}_{\vert\alpha\vert}^m$ 
are independent of $r$ but functions of $q^i$. 
For one specific case their values 
are given in {\bf Appendix C}. Finally the $\epsilon_{(\alpha_i)}$ can be defined by the following procedure. Lets
start with the expression:
\bg\label{ster}
&&H^{mn}_{\vert\alpha\vert} s_{mm}^{(a)[\beta]}s_{nn}^{(b)[\gamma]}~r_{(\alpha_k)}^p ~\equiv~ 
H^{mn}_{\vert\alpha\vert} s_{mm}^{(a)[\beta]}s_{nn}^{(b)[\gamma]}~ 
{r^4_{(\alpha)}\over r^a_{(\beta)}r^b_{(\gamma)}}\nonumber\\
&&K^{mn}_{\vert\alpha\vert} s_{mm}^{(c)[\beta]}s_{nn}^{(d)[\gamma]}~r_{(\alpha_l)}^q ~\equiv~
K^{mn}_{\vert\alpha\vert} s_{mm}^{(c)[\beta]}s_{nn}^{(d)[\gamma]}~ 
{r^5_{(\alpha)}\over r^c_{(\beta)}r^d_{(\gamma)}}
\nd
{}from where one can easily infer:
\bg\label{finfer}
&&p ~=~ 4-a-b,~~~~ \epsilon_{(\alpha_k)} ~=~ {4\epsilon_{(\alpha)} - a \epsilon_{(\beta)} -b\epsilon_{(\gamma)}\over 
4-a-b}\nonumber\\
&&q ~=~ 5-c-d,~~~~ \epsilon_{(\alpha_l)} ~=~ {5\epsilon_{(\alpha)} - c \epsilon_{(\beta)} -d\epsilon_{(\gamma)}\over 
5-c-d}
\nd
Using this procedure we can determine all the $r_{(\alpha_i)}$ in the counterterm expression \eqref{cterms}. 

\noindent At this point the analysis of the theory falls into two possible classes. 

\noindent $\bullet$ The first class 
is to analyse the theory right at the 
usual boundary 
where $r_c \to \infty$. This is the standard picture where there are infinite degrees of freedom at the boundary, and 
the theory has a smooth RG flow from UV to IR till it confines (at least from the weakly coupled gravity dual). 

\noindent $\bullet$ The 
second class is to analyse the theory by specifying the 
degrees of freedom at generic $r_c$ and then defining the theories at the boundary. All these theories would meet the 
cascading theory at certain scales under RG flows. The gravity duals of these theories are the usual 
{\it deformed} conifold 
geometries cutoff at various $r_c$ with appropriate UV caps added (of course for $r < r_c$ the geometries change 
accordingly). 

The former is more relevant for the 
pure AdS/CFT case whereas the latter is more relevant for the present case\footnote{In both cases of course we need to 
add appropriate number of seven branes to get the finite F-theory picture. The holographic 
renormalisation procedure remains unchanged and the far IR physics remains unaltered. The UV caps affect mostly geometries
close to $r_c$, as expected.}.

For the pure AdS/CFT case without flavors $\epsilon_{(\alpha_i)} = 0$ 
(so that the subscript $\alpha_i$'s can be ignored from 
all variables), 
we can subtract the counter-terms \eqref{cterms} from the action \eqref{action1} to get 
the following renormalised action:
\bg \label{actionren}
 {\cal S}^{(1)}_{\rm ren}&=&{\cal S}^{(1)} - {\cal S}^{(1)}_{\rm counter}\nonumber\\
&=&\int \frac{d^4q}{(2\pi)^4}
\theta(r_0-r)\Big[H^{mn}\Big(s_{mm}^{(4)}s_{nn}^{(0)} + s_{mm}^{(3)}s_{nn}^{(1)} + 
s_{mm}^{(2)}s_{nn}^{(2)}+ s_{mm}^{(1)}s_{nn}^{(3)} \nonumber\\
&&+ s_{mm}^{(0)}s_{nn}^{(4)} \Big) - 
K^{mn}\Big(4s_{mm}^{(0)}s_{nn}^{(4)} + 3s_{mm}^{(1)}s_{nn}^{(3)} + 
4s_{mm}^{(2)}s_{nn}^{(2)}+s_{mm}^{(3)}s_{nn}^{(1)}\nonumber\\
&& 4s_{nn}^{(0)}s_{mm}^{(4)} + 3s_{nn}^{(1)}s_{mm}^{(3)} +s_{nn}^{(3)}s_{mm}^{(1)} \Big)
+I^m s_{mm}^{(4)}\Big]
\nd
where we have made all the ${\cal O}(1/r_c)$ terms vanishing, and in the limit $r_h$ small the small shifts to 
$s_{nn}^{(j)}$ given by $s_{nn}^{(3)} + {\cal O}(r_h^4)$ can also be ignored. Furthermore for more generic case where 
primes in \eqref{KS7c} denote derivatives wrt $r$ as well as $\overrightarrow{q}$, 
we can reinterpret the renormalised action 
\eqref{actionren} as the following new action:
\bg\label{newren}
{\cal S}^{(1)}_{\rm ren} &=&  \int \frac{d^4q}{(2\pi)^4} \Big[Z^{mn} \Phi_m(q) \Phi_n(-q)+
U^{mn} \big(\Phi_m(q) \Phi'_n(-q)+ \Phi'_m(q) \Phi_n(-q)\big)\nonumber\\
&+& Y^m \Phi_m(q) 
+ Y^n \Phi_n(-q) +V^m \Phi'_m(q) 
+ V^n \Phi'_n(-q) + X\Big] \nonumber\\
\nd
where $\Phi_m$ are now only functions of $\pm q$ with prime denoting derivatives wrt $\overrightarrow{q}$,
and we isolated all the $r$ dependences so that  
$X, Y, Z,U,V$ could be functions of $r$ and $\overrightarrow{q}$.
We can determine their functional form 
by comparing \eqref{newren} with \eqref{actionren}. For us however the most relevant part is the energy momentum
tensors which we could determine from \eqref{newren} by finding the coefficients $Y^m$ and $Y^n$. One can 
easily show that, upto a possible additive constant, $Y^m, Y^n$ are given by:
\bg\label{ymyn}
Y^m &=& H^{mn} s_{nn}^{(4)}-4K^{mn} s_{nn}^{(4)}, 
~~~~~ Y^n = H^{mn} s_{mm}^{(4)}-4K^{mn} s_{mm}^{(4)}\nonumber\\
V^m &=& K^{mn} s_{nn}^{(5)}, ~~~~~ V^n = K^{mn} s_{mm}^{(5)}
\nd
{}from where the wake of the quark can be shown to be exactly given by \eqref{StressTensor_gs=0} mentioned earlier. We 
also note that $c_o = H^{mn}\theta(r_0 - r)$ there. 

Now let us come to second class of theories 
 wherein we take any arbitrary $r = r_c$, with appropriate UV degrees of freedom such that they 
have good boundary descriptions satisfying all the necessary constraints. For these cases, once we 
subtract the counter-terms \eqref{cterms}, the renormalised action (specified by $r$) takes the following form:
\bg \label{actionren2}
 {\cal S}^{(1)}_{\rm ren}&=&{\cal S}^{(1)} - {\cal S}^{(1)}_{\rm counter}\nonumber\\
&=&\int \frac{d^4q}{2(2\pi)^4}\sum_{\alpha, \beta, \gamma}
\theta(r_0-r)\Big\{H_{\vert\alpha\vert}^{mn}\Big(s_{mm}^{(4)[\beta]}s_{nn}^{(0)[\gamma]} r^{4\epsilon_{(\beta)} - 
4\epsilon_{(\alpha)}}
+ s_{mm}^{(3)[\beta]}s_{nn}^{(1)[\gamma]} r^{3\epsilon_{(\beta)} + \epsilon_{(\gamma)} - 4\epsilon_{(\alpha)}}\nonumber\\
&&+s_{mm}^{(2)[\beta]}s_{nn}^{(2)[\gamma]} r^{2\epsilon_{(\beta)}+ 2 \epsilon_{(\gamma)} - 4\epsilon_{(\alpha)}}
+ s_{mm}^{(1)[\beta]}s_{nn}^{(3)[\gamma]} r^{\epsilon_{(\beta)}+ 3\epsilon_{(\gamma)} -4\epsilon_{(\alpha)}}
+ s_{mm}^{(0)[\beta]}s_{nn}^{(4)[\gamma]}r^{4\epsilon_{(\gamma)} - 4\epsilon_{(\alpha)}} \Big)\nonumber\\ 
&&-4K_{\vert\alpha\vert}^{mn}\Big(s_{mm}^{(0)[\beta]}s_{nn}^{(4)[\gamma]}r^{5\epsilon_{(\gamma)} - 
5\epsilon_{(\alpha)}}
+ s_{mm}^{(1)[\beta]}s_{nn}^{(3)[\gamma]}\big[r^{\epsilon_{(\beta)} + 4\epsilon_{(\gamma)} -5\epsilon_{(\alpha)}}  
+ r^{2\epsilon_{(\beta)} + 3\epsilon_{(\gamma)} -5\epsilon_{(\alpha)}}\big]\nonumber\\
&& + 
4s_{mm}^{(2)[\beta]}s_{nn}^{(2)[\gamma]}\big[r^{2\epsilon_{(\beta)} + 3\epsilon_{(\gamma)} -5\epsilon_{(\alpha)}}
+ r^{3\epsilon_{(\beta)} + 2\epsilon_{(\gamma)} -5\epsilon_{(\alpha)}}\big]
+ 4s_{nn}^{(0)[\beta]}s_{mm}^{(4)[\gamma]} r^{5\epsilon_{(\beta)} -5\epsilon_{(\alpha)}} \nonumber\\
&& +s_{mm}^{(3)[\beta]}s_{nn}^{(1)[\gamma]}\big[r^{4\epsilon_{(\beta)} + \epsilon_{(\gamma)} -5\epsilon_{(\alpha)}} +
r^{3\epsilon_{(\beta)} + 2\epsilon_{(\gamma)} -5\epsilon_{(\alpha)}}\big]\Big)
+I_{\vert\alpha\vert}^m s_{mm}^{(4)[\beta]} r^{5\epsilon_{(\beta)} - 5\epsilon_{(\alpha)}}\Big\}
\nd
defined at the cut-off and the horizon radii as usual. Notice now the appearance of 
$r^{m\epsilon_{(\alpha)} + n\epsilon_{(\beta)} + p\epsilon_{(\gamma)}}$ factors. One can easily show that:
\bg\label{shth}
{1\over 2}\big[r^{m\epsilon_{(\alpha)} + n\epsilon_{(\beta)} + p\epsilon_{(\gamma)}} + 
r^{-m\epsilon_{(\alpha)} -n\epsilon_{(\beta)} - p\epsilon_{(\gamma)}}\big]~ = ~ 1 ~ + ~ 
{\cal O}\big[\epsilon_{(\alpha, \beta, \gamma)}\big]^2
\nd
Since the warp factor $h$ is defined only for small values of 
$g_sN_f, g_sM^2/N, g^2_s N_fM^2/N$ we don't know the background (and 
hence the warp factor) for finite values of these quantities. 
Therefore for our case we can put ${\cal O}\big[\epsilon_{(\alpha, \beta, \gamma)}\big]^2$ 
to zero
so that the value in \eqref{shth} is identically 1. For finite values of these 
quantities both the warp factor and the background 
would change drastically and so new analysis need to be performed to holographically renormalise the theory. 
Our conjecture would be that once we know the background for finite 
values of $g_sN_f, g_sM^2/N$, the terms like 
\eqref{actionren2} would come out automatically renormalised by choice of our counterterms. We will discuss this in some 
details later.  
More elaborate exposition will be given in the sequel \cite{sequel}. 

Once this is settled, the renormalised action at the cut-off radius $r_c$ would only go as powers of 
$r^{-1}_{c(\alpha)}$. Thus we can express the total action as: 
\bg\label{renacto}
&&{\cal S}^{(1)}_{\rm ren} = \int \frac{d^4q}{(2\pi)^4} 
\sum_{\alpha, \beta}\Bigg\{\left(\sum_{j=0}^{\infty}{\widetilde{a}^{(\alpha)}_{mn(j)}\over r_{(\alpha)}^j}\right) 
\widetilde{G}^{mn} \Phi_m \Phi_n +
\left(\sum_{j=0}^{\infty}{\widetilde{e}^{(\alpha)}_{mn(j)}\over r_{(\alpha)}^j}\right) 
\widetilde{M}^{mn} (\Phi_m \Phi'_n\nonumber
\nd
\bg
&& +\Phi'_m \Phi_n) 
+{H}_{\vert\alpha\vert}^{mn}\Big[s_{nn}^{(4)[\beta]}\Phi_m + 
s_{mm}^{(4)[\beta]}\Phi_n\Big]+{K}_{\vert\alpha\vert}^{mn}\Big[-4s_{nn}^{(4)[\beta]}\Phi_m 
-4 s_{mm}^{(4)[\beta]}\Phi_n+s_{nn}^{(5)[\beta]}\Phi'_m\nonumber\\
&& +s_{mm}^{(5)[\beta]}\Phi'_n\Big]
+ \left(\sum_{j=0}^{\infty}\frac{\widetilde{b}^{(\alpha)}_{m(j)}}{r_{(\alpha)}^j}\right) 
\widetilde{J}^m\Phi_m + X[r_{(\alpha)}]\Bigg\}
\left[1-{r_h^4\over r_c^4(1-\zeta)^4}\right]^{-{1\over 2}}\theta(r_0-r)
\nd
where $\Phi_n$ are independent of $r$ with prime denoting derivatives wrt $\overrightarrow{q}$ henceforth; and 
the radial coordinate is measured at the horizon $r_h$ and the cutoff $r_c(1-\zeta)$ as before. The 
explicit expressions for the other coefficients
listed above, namely,
$\widetilde{G}^{mn},\widetilde{M}^{mn} ,\widetilde{a}^{(\alpha)}_{mn(j)}, \widetilde{J}^m,
\widetilde{e}^{(\alpha)}_{mn(j)}$ 
and $\widetilde{b}^{(\alpha)}_{m(j)}$ can be worked out easily from our earlier analysis (see
{\bf Appendix C} for one specific example). Note
that $X[r_{(\alpha)}]$ is a function independent of $\Phi^{[0]}_m$ and appears for generic renormalised action.

Now the generic form for the energy momentum tensor is evident from looking at the linear terms in the 
above action \eqref{renacto}. This is then given by:
\bg \label{wake}
&&T_0^{mm} \equiv 
\int \frac{d^4q}{(2\pi)^4}
\Bigg[({H}_{\vert\alpha\vert}^{mn}+ {H}_{\vert\alpha\vert}^{nm})s_{nn}^{(4)[\beta]} 
-4({K}_{\vert\alpha\vert}^{mn}+ {K}_{\vert\alpha\vert}^{nm})s_{nn}^{(4)[\beta]}+({K}_{\vert\alpha\vert}^{mn}+ 
{K}_{\vert\alpha\vert}^{nm})s_{nn}^{(5)[\beta]}\nonumber\\
&&~~~~~~~~~~~~~~~~~~~~~~~~~~~ + \left(\sum_{j=0}^{\infty}\frac{\widetilde{b}^{(\alpha)}_{n(j)}}
{r_{c(\alpha)}^j}\right)
\widetilde{J}^n \delta_{nm}\Bigg]\left(1- {r_h^4\over r_c^4}\right)^{-{1\over 2}}\theta(r_0-r_c)
\nd
at $r = r_c$ (we ignore the result at the horizon) and sum over ($\alpha, \beta$) is implied. 
This result should be compared to the ones derived in 
\cite{Yaffe-1} \cite{Kostas-1} \cite{Kostas-2}\cite{Kostas-3} \cite{2} which doesn't have any $r_c$ dependence. 
This is of course the first line of the above result. For the case studied in 
\cite{Yaffe-1} \cite{Kostas-1} \cite{Kostas-2}\cite{Kostas-3} \cite{2}
the boundary theory is defined with infinite degrees of 
freedom at UV (which we have been calling as the parent cascading theory). How do we then reproduce the results 
of those papers?   
Before we go about elucidating this, notice that the second line also has a $r_c$ independent
additive constant. This additive term is irrelevant for our purpose because the energy-momentum can always be 
shifted by a constant to absorb this factor. Thus once we specify the cutoff $r_c$ and the UV degrees of freedom, then
our result shows that the energy-momentum tensor not only
inherits the universal behavior of the parent cascading theory but there 
are additional corrections coming precisely from the added UV degrees of freedom at $r = r_c$. These corrections go as
$r_c^{-1}$ or as $e^{-{\cal N}_{\rm eff}}$ with ${\cal N}_{\rm eff}$ being the UV degrees of freedom 
at the cut-off (see \eqref{doa}). 
As long as these UV degrees of freedom are not infinite, they contribute to small corrections to 
the energy-momentum tensors of the various 
gauge theories. 

The above description is one of the key points of our paper, and tells us how we can distinguish our results from the 
standard AdS/QCD answers. Therefore
let us elaborate this in little more details. This will justify what we have been saying so far about UV caps, and 
put everything in a rigorous mathematical framework at least from the action point-of-view.  
In the process we will also be able to reconcile with the results 
of \cite{Yaffe-1} \cite{Kostas-1} \cite{Kostas-2}\cite{Kostas-3} \cite{2}.
The first important issue here is that we can study infinite number 
of UV completed theories in our full F-theory set-up. 
All of these theories have good boundary descriptions and have same degrees of 
freedom as the parent cascading theory at certain specified scales. The simplest UV complete theory (however with 
infinite degrees of freedom at UV) is of course the parent cascading theory that we will discuss in a moment.
The question now is to construct other possible theories by defining the degrees of freedom 
at the boundary. To do this observe that we defined the boundary theory using the identification:
\bg\label{bndiden}
\big[{\cal S}^{(1)}_{\rm ren}\big]_{r_h}^\infty ~ = ~ \big[{\cal S}^{(1)}_{\rm ren}\big]_{r_h}^{r_c} + 
\big[{\cal S}^{(1)}_{\rm ren}\big]_{r_c}^\infty
\nd 
where the boundary is at $r \to \infty$. For the boundary cascading theory the above expression simply means that 
\bg\label{casbnd}
 \big[{\cal S}^{(1)}_{\rm ren}\big]_{r_c}^\infty & = & - \int \frac{d^4q}{(2\pi)^4} \left(\sum_{j=0}^{\infty}
\frac{\widetilde{b}^{(\alpha)}_{n(j)}}{r_{c(\alpha)}^j}\right)
\widetilde{J}^n \Phi_n\theta(r_0-r_c) - \int \frac{d^4q}{(2\pi)^4} \left(\sum_{j=0}^{\infty} 
{B^{(\alpha)}_{n(j)}r_h^{4j} \over r_{c(\beta)}^j}\right)\Phi_n \nonumber\\
&& ~~~~~~~~~~~+~ {\{\widetilde{b}^{(\alpha)}_{n(j)}, {\cal O}(r_h^{4j})\}\over \infty}~{\rm factors} 
\nd
where the sign is crucial and sum over $\alpha$ is again implied (note (a) the cut-off dependence, and (b)
$r_{c(\beta)}$ is some function of $r_{c(\alpha)}$ that
one can determine easily). 
We now see that the contributions from the UV cap give the following values for $B_j$:
\bg\label{cfuv} 
&& B^{(\alpha)}_{n(0)} ~ = ~ B^{(\alpha)}_{n(1)} ~ = ~ B^{(\alpha)}_{n(2)} ~ = ~ B^{(\alpha)}_{n(3)} ~ = ~ 0 \nonumber\\
&& B^{(\alpha)}_{n(4)} ~ = ~ {1\over 2}\bigg[\widetilde{b}^{(\alpha)}_{n(0)}
\widetilde{J}^n \theta(r_0-r_c) + {\cal O}(H_{\vert\alpha\vert}^{nm}, 
K_{\vert\alpha\vert}^{nm}, s_{nn}^{[\alpha]})\bigg],~ .....
\nd
this means that in the limit of small $r_h$ we can ignore the contributions coming from $r_h^{4i}$. For this section
such an assumption will not change any of the results, so we will stick with this. From the next section onwards, 
we will restore back the ${\cal O}(r_h)$ dependences. 

Taking all of the above considerations, this 
implies that the contribution from $r > r_c $ {\it exactly} cancels the ${\cal O}(1/r_c)$ contributions 
coming in from the action measured from $r_h \le r \le r_c$, giving us the boundary energy-momentum tensor 
\bg\label{casenmo} 
\int \frac{d^4q}{(2\pi)^4} \sum_{\alpha, \beta}
\Big[({H}_{\vert\alpha\vert}^{mn}+ 
{H}_{\vert\alpha\vert}^{nm})s_{nn}^{(4)[\beta]} -4({K}_{\vert\alpha\vert}^{mn}+ 
{K}_{\vert\alpha\vert}^{nm})s_{nn}^{(4)[\beta]}+({K}_{\vert\alpha\vert}^{mn}+ 
{K}_{\vert\alpha\vert}^{nm})s_{nn}^{(5)[\beta]}\Big]\nonumber\\
\nd
which is the result derived in \cite{Yaffe-1} \cite{Kostas-1} \cite{Kostas-2}\cite{Kostas-3} \cite{2}.
The above way of reinterpreting the boundary contribution should tell us precisely how we could modify the boundary 
degrees of freedom to construct distinct UV completed theories. There are two possible ways we can achieve this:

%jwala

\noindent $\bullet$ From the geometrical perspective
 we can cutoff the deformed conifold background at $r = r_c$ and attach an 
appropriate UV ``cap'' from $r = r_c$ to $r \to \infty$ by carefully modifying the 
geometry at the neighborhood of the junction point. As an example, this UV cap could as well 
be another AdS background
from $r_c$ to $r \to \infty$. There are of course numerous other choices available from the F-theory limit. 
Each of these caps would give 
rise to distinct UV completed gauge theories.

\noindent $\bullet$ From the action perspective we could specify precisely the value of the action measured from 
$r_c$ to $r \to \infty$, i.e $\big[{\cal S}^{(1)}_{\rm ren}\big]_{r_c}^\infty$. The simplest case where this is 
zero gives rise to a boundary theory whose degrees of freedom at the UV is the one given in \eqref{edofk}. We will 
give an example of this towards the end of this section. To study more generic cases,
we need to see how much constraints we can put on our 
integral. One immediate constraint is the holographic renormalisability of our theory. This tells us that the 
value of the integral can only go as powers of $1/r_{c(\alpha)}$ otherwise we will not have finite actions. This in turn 
implies\footnote{Remember that there are additional ${\cal O}(r_h)$ contributions to the coefficients. For small 
$r_h$ they will only make the UV contributions more involved without changing any of the underlying physics as we 
saw above. We will therefore ignore them.}:
\bg\label{sifrom}
&&\Big[{\cal S}^{(1)}_{\rm ren}\Big]_{r_c}^\infty = \int \frac{d^4q}{(2\pi)^4}\sum_{\alpha}
\Bigg\{\left(\sum_{j=0}^{\infty}{\widetilde{A}^{(\alpha)}_{mn(j)}\over r_{c(\alpha)}^j}\right) 
\widetilde{G}^{mn} \Phi_m \Phi_n +
\left(\sum_{j=0}^{\infty}{\widetilde{E}^{(\alpha)}_{mn(j)}\over r_{c(\alpha)}^j}\right) \widetilde{M}^{mn}\\ 
&&\times (\Phi_m \Phi'_n 
+\Phi'_m \Phi_n) + \left(\sum_{j=0}^{\infty}\frac{\widetilde{B}^{(\alpha)}_{m(j)}}
{r_{c(\alpha)}^j}\right) 
\widetilde{J}^m\Phi_m + X[r_{(\alpha)}]\Bigg\}\theta(r_0-r_c) + {\rm finite~ terms}\nonumber
\nd
where by specifying the coefficients $\widetilde{A}^{(\alpha)}_{mn(j)}, 
\widetilde{E}^{(\alpha)}_{mn(j)}$ and $\widetilde{B}^{(\alpha)}_{m(j)}$ 
we can specify the precise 
UV degrees of freedom! The finite terms are $r_c$ independent and therefore would only provide finite shifts 
to our observables. They could therefore be scaled to zero. Notice also that the contributions from \eqref{sifrom} 
only renormalises the coefficients $\widetilde{a}^{(\alpha)}_{mn(j)}, 
\widetilde{e}^{(\alpha)}_{mn(j)}$ and $\widetilde{b}^{(\alpha)}_{m(j)}$
in \eqref{renacto}, 
and 
therefore the final expressions for all the physical variables for various UV completed theories could be written
directly from \eqref{renacto} simply by replacing the $1/r_c$ dependent coefficients by their renormalised values. 
This is thus our precise description of how to specify the UV degrees of freedom for various gauge theories 
in our setup (see {\bf figure 11} below). 
\begin{figure}[htb]
		\begin{center}
		\includegraphics[height=5.3cm]{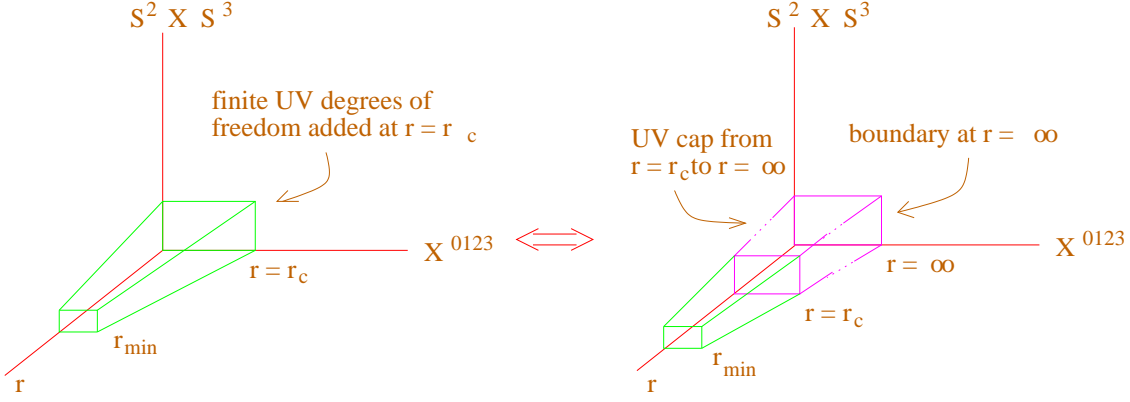}
		\caption{{The equivalence between two different ways of viewing the boundary theory at zero temperature. 
To the 
left we add finite UV degrees of freedom at $r = r_c$ of the deformed conifold geometry. Such a process is equivalent 
to the figure on the right where we cut-off the deformed conifold geometry at $r = r_c$ and
add a UV cap from $r = r_c$ to $r = \infty$. The boundary theory on the right 
has ${\cal N}_{\rm uv}$ degrees of freedom at $r = \infty$ and all physical quantities computed in either of these 
two pictures would only depend on ${\cal N}_{\rm uv}$ but not on $r = r_c$. At non-zero temperature the 
UV descriptions remain unchanged. Of course, for both cases one need to put in the corrections to IR from UV 
deformations.}\label{equivalence}} 
		\end{center}
		\end{figure}  

\noindent Once the UV descriptions are properly laid out, we can determine the form for 
$\widetilde{A}^{(\alpha)}, 
\widetilde{B}^{(\alpha)}$ and $\widetilde{E}^{(\alpha)}$ by writing Callan-Symanzik type equations for them. These are 
classical equations and do not capture any quantum behavior. Nevertheless they tell us how 
$\widetilde{A}^{(\alpha)}, 
\widetilde{B}^{(\alpha)}$ and $\widetilde{E}^{(\alpha)}$ would behave with the scale $r_c$ or equivalently $\mu_c$. For
$\widetilde{A}$ the equation is\footnote{The following equation is derived from the scale-invariance of 
$\big[{\cal S}^{(1)}_{\rm ren}\big]_{r_h}^\infty$.}:
\bg\label{calsym}
\mu_c {\partial \widetilde{A}^{(\alpha)}_{mn(j)} \over \partial \mu_c} ~ = ~ j\left[1-\epsilon_{(\alpha)}\right]
\left[\widetilde{A}^{(\alpha)}_{mn(j)} + 
\widetilde{a}^{(\alpha)}_{mn(j)}\right]
\nd
with similar equations for $\widetilde{B}^{(\alpha)}$ 
and $\widetilde{E}^{(\alpha)}$. These equations tell us that physical quantities are
independent of scales. The  
parent cascading theory is defined as the scale-invariant limits of \eqref{calsym}, i.e:  
\bg\label{abe}
\widetilde{A}^{(\alpha)}_{mn(j)} ~= ~ - \widetilde{a}^{(\alpha)}_{mn(j)}, ~~~ \widetilde{E}^{(\alpha)}_{mn(j)} 
~= ~ - \widetilde{e}^{(\alpha)}_{mn(j)}, ~~~ 
\widetilde{B}^{(\alpha)}_{m(j)} ~= ~ - \widetilde{b}^{(\alpha)}_{m(j)} 
\nd
%popo
The above relation gives us a hint how to express $\widetilde{A}^{(\alpha)}, 
\widetilde{B}^{(\alpha)}$ and $\widetilde{E}^{(\alpha)}$ in terms of 
${\cal N}_{\rm eff}$, the effective degrees of freedom at $r = r_c$ and ${\cal N}_{\rm uv}$, the effective
degrees of freedom at $r = \infty$ 
i.e the boundary:
\bg\label{result1}
 \widetilde{A}^{(\alpha)}_{mn(j)} &~= ~& - \widetilde{a}^{(\alpha)}_{mn(j)}~ +~ 
\hat{a}^{(\alpha)}_{mn(j)} e^{-j\left[{\cal N}_{\rm uv} - (1-\epsilon_{(\alpha)})
{\cal N}_{\rm eff}\right]}\nonumber\\
\widetilde{E}^{(\alpha)}_{mn(j)} &~= ~& - \widetilde{e}^{(\alpha)}_{mn(j)}~ 
+~ \hat{e}^{(\alpha)}_{mn(j)} e^{-j\left[{\cal N}_{\rm uv} - (1-\epsilon_{(\alpha)})
{\cal N}_{\rm eff}\right]}\nonumber\\
 \widetilde{B}^{(\alpha)}_{m(j)} &~= ~& - \widetilde{b}^{(\alpha)}_{m(j)}~ 
+~ \hat{b}^{(\alpha)}_{m(j)} e^{-j\left[{\cal N}_{\rm uv} - (1-\epsilon_{(\alpha)}){\cal N}_{\rm eff}\right]}
\nd
where the actual boundary degrees of freedom are specified by knowing $\hat{a}_{mn(j)}, \hat{e}_{mn(j)}$ and  
$\hat{b}_{m(j)}$ as well as
 ${\cal N}_{\rm uv}$. Since $j$ goes from 0 to $\infty$, there are infinite possible UV complete 
boundary theories 
possible\footnote{The connection of $j$ with UV completions come from the coefficients $\hat a_{mn(j)}^{(\alpha)}, 
\hat e_{mn(j)}^{(\alpha)}$ and $\hat b_{mn(j)}^{(\alpha)}$ etc. that depend on $j$. For different choices of 
these coefficients we can have different UV completions. In this sense $j$ and 
UV completions are related. See also the F-theory discussion presented towards the end of section 3.1.}. 
For very large ${\cal N}_{uv}$ (i.e ${\cal N}_{uv}\to \epsilon^{-n}, n >> 1$)
the boundary theories are similar to the original cascading 
theory. The various choices of 
($\hat{a}^{(\alpha)}_{mn(j)}(\overrightarrow{q}), \hat{e}^{(\alpha)}_{mn(j)}(\overrightarrow{q}), 
\hat{b}^{(\alpha)}_{m(j)}(\overrightarrow{q})$) 
tell us how the degrees of freedom
change from ${\cal N}_{\rm uv}$ to ${\cal N}_{\rm eff}$ under RG flow. The $\overrightarrow{q}$ dependence of all the 
quantities will tell us how the UV degrees of freedom affect IR physics. 
This is to be expected: addition of 
irrelevant operators do change IR physics, but not the far IR\footnote{The $\overrightarrow{q}$ dependences of the 
UV caps are also one-to-one correspondence to the changes in the local geometries near the cut-off radius 
$r_c$, as we discussed before. All in all this fits nicely with what one would have expected from UV degrees of 
freedom. Of course it still remains to verify the story from an actual supergravity calculation. We need to 
analyse the metric near the junction by studying the continuity and differentiability of the metric 
and see how far below $r = r_c$ we expect deformations from the UV caps. Various types of deformations will 
signal various set of irrelevant operators. 
Needless to say, the 
far IR physics remain completely unaltered.}.  
 
Therefore with this understanding of the boundary theories we can express the
energy-momentum tensor at the boundary with ${\cal N}_{uv}$ degrees of freedom at the boundary purely
in terms of gauge theory variables, as:
\bg \label{wakegt}
&&T_0^{mm} \equiv 
\int \frac{d^4q}{(2\pi)^4}
\Big[({H}_{\vert\alpha\vert}^{mn}+ {H}_{\vert\alpha\vert}^{nm})s_{nn}^{(4)[\beta]} 
-4({K}_{\vert\alpha\vert}^{mn}+ {K}_{\vert\alpha\vert}^{nm})s_{nn}^{(4)[\beta]}+({K}_{\vert\alpha\vert}^{mn}+ 
{K}_{\vert\alpha\vert}^{nm})s_{nn}^{(5)[\beta]}\nonumber\\
&&~~~~~~~~~~~~~~~~~~~~~~~~~~~~~~~~~~~ + \sum_{j=0}^{\infty}~\hat{b}^{(\alpha)}_{n(j)}(\overrightarrow{q}) 
\widetilde{J}^n  e^{-j{\cal N}_{\rm uv}}
\delta_{nm}\Big]
\nd
where sum over $\alpha$ is again implied, and
the first line is the universal property of the parent cascading theory inherited by our gauge theory. This 
part, common to all the theories, will be numerically different.
The second line specifies the precise 
degrees of freedom that we add at $r = r_c$ to describe the UV behavior of our theory at the boundary $r \to \infty$. 
Using this procedure, the final results of any physical quantities should be expressed only in terms of 
${\cal N}_{\rm uv}$ i.e the UV degrees of freedom\footnote{Restoring back the ${\cal O}(r_h)$ contributions would mean
that there should be an additional contribution to \eqref{wakegt} of the form 
$\sum_{j=0}^{\infty}~ G(\hat b^{(\alpha)}_n, H^{mn}_{\vert\alpha\vert}, K^{mn}_{\vert\alpha\vert}, s_{nn}^{[\alpha]})
{\cal T}^{4j} e^{-j{\cal N}_{uv}}$ where $G$ is a function
whose functional form could be inferred from the UV integral \eqref{sifrom}.}. 
 
%%%
The above analysis of holographic renormalisation tells us something very interesting. The final result of the 
renormalisation procedure is almost identical to the renormalisation procedure done by taking the highest 
positive integer power of $r$ and ignoring the log $r$ part! 
The only difference is the presence of coefficients like $H^{mn}_{\vert\alpha\vert}$
that depend explicitly on flavor degrees of freedom. This will be useful for us when we study shear viscosity in the 
next section.

Another question to ask regarding holographic renormalisation is the issue of corrections for finite
values of $g_sN_f, g_sM^2/N$. This is probably more difficult to tackle. However an easier question would be to ask
what happens if the UV behavior is governed by 
$r^{-p}$ instead of $r^{-p}_{(\alpha)}$ for a certain choice of an UV cap? 
 
To evaluate this observe that we can take the following ansatze for the background warp factor for large $r$
on the slice \eqref{sol}\footnote{The equation \eqref{lbaz} could be 
motivated by adding a suitable UV cap that allows an inverse $r$ dependences in 
\eqref{logr} which in turn stems from the finiteness of F-theory. 
At large $r$ the axio-dilaton contributions will
come from \eqref{jtau} that will back react on the geometry to give a finite asymptotic geometry. There are infinite 
such geometries possible. For a given geometry the coefficients of $r^{-p}$ terms would be fixed accordingly (by solving
the sugra equations with axio-dilaton and three and five-fluxes as sources) {\it provided}
these fluxes are also asymptotically
finite. 
For more details see the F-theory arguments presented earlier.}:
\bg\label{lbaz}
h(r) = {L^4\over r^4}\left[1 + \sum_{m, n, p} (g_sN_f)^m \left({g_sM^2\over N}\right)^n {\rm log}^p r\right] ~\to~ 
{L^4\over r^4}\left[1 + \sum_{p = 1}^{p_0} {c_{p}\over r^{p-4}}\right]
\nd
where $p_0$ is some given integer, $c_p$ are functions of $g_sN_f, g_sM^2/N$ and the arrow is motivated by the 
fact that, in the presence of a suitable UV cap the background is given by F-theory construction \cite{vafaF} that 
guarantees that the warp factor goes from ${\rm log}~r$ at small $r$ to inverse powers of $r$ at large $r$ to 
avoid the Landau poles.

In this background we expect the fluctuation to go like:
\bg\label{fluct1}
\Phi^{[1]}_m(r,q) = \int \frac{d^4x}{(2\pi)^4 \sqrt{g(r_c)}}\;e^{i(q_0 t-q_1x-q_2y-q_3z)}L_{mm}(t,r,x,y,z)
\nd
which is similar to the definition of $\Phi^{[1]}_m(r, q)$ given earlier in \eqref{phim} with 
$q_0 = \omega \sqrt{g(r_c)}$ as before. 
Our next step would be to write
the above action as an EOM part and a boundary part like as in \eqref{KS7c1} with appropriate changes. Once we do 
this we would require the Gibbons-Hawking boundary terms to cancel any unwanted  $\delta\Phi_m'^{[1]}$ parts from
the action. 

Taking all these into considerations the boundary action takes the usual form \eqref{action1}. Now 
the coefficients 
$C^{mn}_1, A^{mn}_1,B^{mn}_1, E^m_1$ and $F^m_1$ are again divergent, and from our previous consideration we know that the
divergences will be controlled by the highest {\it integer} power of $r$.  
Therefore we need to regularise and renormalise the action again. The divergences are:
\bg\label{infinity2}
&&1.~~~ C^{mn}_1(\zeta,q)-A'^{mn}_1(\zeta,q) ~ = ~\sum_{p=1}^{p_0}
\widetilde{H}_p^{mn}(q)~ r_{c}^{p}(1-\zeta)^{p}
+ {\rm finite~terms}\nonumber\\
&&2.~~~B^{mn}_1(\zeta,q)-A^{mn}_1(\zeta,q) ~ = ~ \sum_{p=1}^{p_0}
\widetilde{K}_p^{mn}(q)~ r_{c}^{p+1}(1-\zeta)^{p+1}
+ {\rm finite~terms}\nonumber\\
&&3.~~~ E^m_1(\zeta,q) -F'^m_1(\zeta,q) ~ = ~ \sum_{p=1}^{p_0} \widetilde{I}_p^{m}(q) 
~r_{c}^p(1-\zeta)^p + {\rm finite~terms}
\nd
To start off then, we write the metric perturbation
$\Phi_m^{[1]}$ as a power  
series in $\zeta$ where $\zeta$ is defined with the equation $r=r_c(1-\zeta)$:
\bg \label{metpert11}
\Phi_m^{[1]}(\zeta)&=&\theta(\zeta-\zeta_0)\sum_{k=0}^{\infty} {s_{mm}^{(k)} \over r_c^k (1-\zeta)^k}
\nd
which can again be easily derived by plugging \eqref{oand1} in \eqref{fluct1}. 

Once we have the mode expansions of the field, we can use this expansion directly in the action \eqref{KS7c} to 
determine the possible counter-terms. The infinities come from \eqref{infinity2}, much like the situation 
with \eqref{infinity}. Therefore its no surprise that we get the following counter-terms:
\bg\label{cterms2}
&& {\cal S}^{(2)}_{\rm counter}=\int \frac{d^4q}{2(2\pi)^4} \sum_{p} 
\Big\{\widetilde{H}_{p}^{mn}
\theta(\zeta-\zeta_0)\Big[s_{mm}^{(0)}s_{nn}^{(0)}~r_{c}^p (1-\zeta)^p\nonumber\\ 
&&~~~ + \Big(s_{mm}^{(1)}s_{nn}^{(0)} 
+ s_{mm}^{(0)}s_{nn}^{(1)}\Big)r_{c}^{p-1} (1-\zeta)^{p-1} + 
\Big(s_{mm}^{(2)}s_{nn}^{(0)} + s_{mm}^{(0)}s_{nn}^{(2)} \nonumber\\ 
&&~~~ + s_{mm}^{(1)}s_{nn}^{(1)}\Big)r_{c}^{p-2} (1-\zeta)^{p-2} + \Big(s_{mm}^{(3)}s_{nn}^{(0)} + 
s_{mm}^{(0)}s_{nn}^{(3)}
+ s_{mm}^{(2)}s_{nn}^{(1)}
+ s_{mm}^{(1)}s_{nn}^{(2)}\Big)r_{c}^{p-3}\nonumber\\
&& \times (1-\zeta)^{p-3}+ .... \Big] +
 {\widetilde{K}}_{p}^{mn}\theta(\zeta-\zeta_0)\Big[-\big(s_{mm}^{(0)}s_{nn}^{(1)}+s_{mm}^{(1)}
s_{nn}^{(0)}\big)~r_{c}^{p-1}(1-\zeta)^{p-1} \nonumber\\
&& - \Big(2s_{mm}^{(1)}s_{nn}^{(1)} +2 s_{mm}^{(0)}s_{nn}^{(2)}+2 s_{nn}^{(0)}s_{mm}^{(2)}
\Big)r_{c}^{p-2}(1-\zeta)^{p-2}
-~\Big(2s_{mm}^{(1)}s_{nn}^{(2)} + s_{mm}^{(2)}s_{nn}^{(1)}\nonumber\\
&& ~~~~~~~~ + 3s_{mm}^{(0)}s_{nn}^{(3)} +2s_{nn}^{(1)}s_{mm}^{(2)} + s_{nn}^{(2)}s_{mm}^{(1)} + 
3s_{nn}^{(0)}s_{mm}^{(3)}\Big)r_{c}^{p-3}(1-\zeta)^{p-3} + ....\Big] \nonumber\\
&&~~~~~~~~~~ +\widetilde{I}_{p}^m \theta(\zeta-\zeta_0)\Big[s_{mm}^{(0)} r_{c}^p (1-\zeta)^p 
 + s_{mm}^{(1)} r_{c}^{p-1} (1-\zeta)^{p-1}\nonumber\\
&&~~~~~~~~~ + s_{mm}^{(2)} r_{c}^{p-2} (1-\zeta)^{p-2} 
+ s_{mm}^{(3)} r_{c}^{p-3} (1-\zeta)^{p-3} + ....\Big]\Big\}
\nd
with of course another set of counter-terms defined at the horizon $r_h$ that we do not write here. Observe that we 
have written the counterterms wrt the highest power of $p$.  
Finally the renormalised action after some manipulations can be written, similar to \eqref{renacto} but with different
coefficients (whose explicit values will not be relevant for the present case), as:
\bg\label{renacto2} 
&& {\cal S}^{(2)}_{\rm ren}= \int \frac{d^4q}{(2\pi)^4} \sum_{p}
\Bigg\{\left[\sum_{j=0}^{\infty}{\widetilde{c}_{mn(j)}\over r_{c}^j (1-\zeta)^j}\right]
 \widetilde{L}^{mn} \Phi_m \Phi_n +
\left(\sum_{j=0}^{\infty}{\widetilde{f}_{mn(j)}\over r_{c}^j}\right) 
\widetilde{{\cal M}}^{mn} (\Phi_m \Phi'_n \nonumber\\
&&+\Phi'_m \Phi_n) 
+ \widetilde{H}_{p}^{mn}\Big[s_{nn}^{(p)}\Phi_m 
+ s_{mm}^{(p)}\Phi_n\Big]+{\widetilde{K}}_{p}^{mn}\Big[-4s_{nn}^{(p)}\Phi_m 
-4 s_{mm}^{(p)}\Phi_n+s_{nn}^{(p+1)}\Phi'_m \nonumber\\
&& ~~~~~~~+s_{mm}^{(p+1)}\Phi'_n\Big]
+ \left[\sum_{j=0}^{\infty}\frac{\widetilde{d}_{m(j)}}{r_{c}^j(1-\zeta)^j}\right] 
\widetilde{Q}^m\Phi_m + Y(r)\Bigg\}\theta(\zeta-\zeta_0)
\nd 
{}from here we can extract 
the energy-momentum tensor $T^{mm}$ once we add the right boundary degrees of freedom. This is exactly as we 
did before. The contribution from UV cap now will be:
\bg\label{uvcapco}
\left[{\cal S}^{(2)}_{\rm ren}\right]_{\zeta}^\infty &=& \int \frac{d^4q}{(2\pi)^4} 
\Bigg\{\left[\sum_{j=0}^{\infty}{\widetilde{C}_{mn(j)}\over r_{c}^j (1-\zeta)^j}\right]
 \widetilde{L}^{mn} \Phi_m \Phi_n +
\left(\sum_{j=0}^{\infty}{\widetilde{F}_{mn(j)}\over r_{c}^j}\right) 
\widetilde{{\cal M}}^{mn} (\Phi_m \Phi'_n \nonumber\\
&& ~~~ +\Phi'_m \Phi_n)
+ \left[\sum_{j=0}^{\infty}\frac{\widetilde{D}_{m(j)}}{r_{c}^j(1-\zeta)^j}\right] 
\widetilde{Q}^m\Phi_m \Bigg\}\theta(\zeta-\zeta_0) + {\rm finite~terms}
\nd
%patha
where by the range
$\zeta$ to $\infty$ we mean the coordinate range
$r_c(1-\zeta)$ to $r \to \infty$. As before, specifying the quantities $\widetilde{C}^{(\alpha)},
\widetilde{F}^{(\alpha)}$ 
and $\widetilde{D}^{(\alpha)}$ will in turn specify the boundary degrees of freedom. We also expect similar 
Callan-Symanzik type equations for them. Indeed:
\bg\label{casy}
\mu_c {\partial \widetilde{D}_{m(j)}\over \partial \mu_c} ~ 
= ~ j\left[\widetilde{D}_{m(j)} + \widetilde{d}_{m(j)}\right]
\nd
with similar relations for the others. This way we can compute the 
the total energy-momentum tensor of the system once we know the precise value of $p_0, c_p$ for the exact background. 
On the other hand, for the case that we know very well \eqref{prelim2}, the energy-momentum tensor 
that we should substitute in \eqref{goal}
to compute the wake will be:
\bg\label{wakeup}
&& T^{mm}_{{\rm medium} + {\rm quark}}    
 = \int \frac{d^4q}{(2\pi)^4}\sum_{\alpha, \beta}
\Bigg\{({H}_{\vert\alpha\vert}^{mn}+ {H}_{\vert\alpha\vert}^{nm})s_{nn}^{(4)[\beta]} 
-4({K}_{\vert\alpha\vert}^{mn}+ {K}_{\vert\alpha\vert}^{nm})s_{nn}^{(4)[\beta]}\nonumber\\
&& ~~~~~~~~~~~ +({K}_{\vert\alpha\vert}^{mn}+ {K}_{\vert\alpha\vert}^{nm})s_{nn}^{(5)[\beta]}
+\sum_{j=0}^{\infty}~\hat{b}^{(\alpha)}_{n(j)} \widetilde{J}^n  
\delta_{nm}  e^{-j{\cal N}_{\rm uv}} + {\cal O}({\cal T} e^{-{\cal N}_{uv}})\Bigg\}
\nd 
where ($\hat{b}^{(\alpha)}_{n(j)}, \hat{d}^{(\alpha)}_{n(j)}, 
{\cal N}_{\rm uv}$) together will specify the full boundary theory for a specific
UV complete theory. Observe that the result is completely independent of the cut-off that we imposed to do our analysis. 
In the limit ${\cal N}_{\rm uv} \to \epsilon^{-n}$ with $n >> 1$ i.e when the boundary degrees of freedom go to infinity  
as this way, we reproduce precisely the result of the parent cascading theory.

Before we end this section let us make the following observation. This is
a slightly more non-trivial example where ${\cal N}_{\rm uv} \approx 
{\cal N}_{\rm eff}$ with ${\cal N}_{\rm eff}$ being the degrees of freedom at $r = r_c$ given earlier 
in \eqref{edofk}. This means that the degrees of freedom don't change significantly as we go from far UV to the 
scale $\Lambda_c$. From the gravity dual perspective this is like adding a UV cap given by an AdS geometry. For 
such a theory \eqref{wakeup} can be exactly determined (both $\hat{b}^{(\alpha)}_{n(j)}, 
\hat{d}^{(\alpha)}_{n(j)}$ can be scaled to 
identity), and so precise prediction can be made\footnote{Interestingly 
for small 
enough $\Lambda_c$ we are almost in the IR where we expect the gauge theory to confine (at least at zero temperature). 
The UV of the gauge theory has very slow running, which can be constructed from our F-theory model by suitably 
choosing the coefficients $a_i, b_j$ in \eqref{jtau} such that axio-dilaton vanish. With vanishing axio-dilaton the 
geometry at UV approaches AdS. 
Therefore this set-up is almost like large $N$ flavored QCD with 
vanishing Beta function at UV and confinement at IR! Our method could then be used to evaluate the thermal quantities of 
this theory.}. 

In the next two sections where we deal with shear viscosity and viscosity-to-entropy ratio, we will find that all
the techniques that we discussed in this section will become very useful.

%\newpage 

\subsection{Shear Viscosity}

In this section we will compute the shear viscosity of the four dimensional theory
following some of the recent works \cite{Kats, buchel-1}. Our basic idea would be to use the Kubo formula \cite{GaleKapusta}:
\bg \label{SV-1}
\eta=\lim_{\omega\rightarrow 0}~\frac{1}{2\omega}\int dt d^3x\; e^{i\omega
t} \langle\left[ T_{23}(x), T_{23}(0)\right]\rangle =-\lim_{\omega\rightarrow
0}~\frac{{\rm Im}~G^R(\omega,0)}{\omega}
\nd
where $G^R(\omega,\overrightarrow{q})$ is the momentum space retarded propagator for
the operator $T_{23}$ at finite
temperature, defined by
\bg \label{SV-2}
{G}^R(\omega,\overrightarrow{q})= -i \int dt d^3x \; e^{i(\omega t-\overrightarrow{x}\cdot \overrightarrow{q})}
\theta(t) \langle\left[ T_{23}(x), T_{23}(0)\right]\rangle
\nd
In the following,
we will compute the Minkowski propagator $\langle T_{23}(t,\overrightarrow{x}) T_{23}(0,\overrightarrow{x}\rangle$ 
using gauge/gravity duality\footnote{Note that our prescription (\ref{KS16}) computes a path integral 
$\langle {\cal O}\phi \rangle_{\cal T} \sim \int \left[{\cal D O}\right]_{\cal T}
\;e^{\int_M^4 {\cal O}\phi +\cal{L}}$, 
(where ${\cal T}$ is the time ordering) 
 with a classical action $S_{\rm SUGRA}$, unaware of the time ordering. Therefore 
computing any commutator is rather subtle here. 
For this reason, we compute only the correlator  
$\langle T_{23}(\tau,\overrightarrow{x}) T_{23}(0,\overrightarrow{x})\rangle$
 and not it's commutator. Using this correlator, we will eventually obtain information about commutators and finally the
 viscosity $\eta$.}.   
However before we compute this explicitly, let us take a small detour to evaluate the higher order corrections 
to the effective action from the wrapped D7 brane in our theory. We have already given a brief discussion of this in 
\eqref{effaction}. It is now time to deal with this in some more details.

In the case of a D7 brane, the disc level action contains the term \cite{DJM}:
\bg\label{disclevel}
S_{\rm D7}^{\rm disc} = {1\over 192\pi g_s} \cdot {1\over (4\pi \alpha')^2} \int_{M^8} \left[ C_4 \wedge {\rm tr}~
(R \wedge R) - e^{-\phi}{\rm tr}~(R \wedge \ast R)\right]
\nd
where $C_4$ is the four-form, $R$ is the curvature two-form,
$\phi$ is the dilaton and $M^8$ is a non-trivial eight manifold forming the world-volume
of the D7 brane. 
The above action is $SL(2, {\bf Z})$ invariant which one can show by 
doing an explicit analysis \cite{DJM}. Since for our case the D7 wrap a non-trivial four-cycle, we can 
dimensionally reduce it over the four-cycle to get the following action:
\bg\label{dimred}
S_{\rm D7}^{\rm disc} = {1\over 16\pi^2} \int_{M^4} {\rm Re}\left[ {\rm log}~\eta(\tau) ~{\rm tr}\left(R \wedge 
\ast R - i R \wedge R\right)\right]
\nd
where $\eta(\tau)$ is the Dedekind function, and $\tau$ is the modular parameter defined by:
\bg\label{taudef}
\tau = {1\over g_s(4\pi\alpha')^2} \left(\int_{S^4} C_4 + i {\cal V}_4\right) \equiv {1\over g_s} (\tau_1 + i \tau_2)
\nd
with ${\cal V}_4$ being the volume of the four-cycle on which we have the wrapped D7 brane. An interesting
point here is that the above action can be {\it derived} from the following action that has two parts, one CP-even
and the other CP-odd \cite{BBG}:
\bg\label{cpevodd}
{1\over 32\pi^2} \int_{M^4} {\rm log} \vert \eta(\tau)\vert^2 {\rm tr}~(R \wedge \ast R) - {i \over 32\pi^2} 
\int_{M^4} {\rm log} ~{\eta(\tau) \over \eta(\bar\tau)}~{\rm tr}~(R \wedge R)
\nd
where the first part is CP-even and the second part is CP-odd. To compare \eqref{cpevodd} with \eqref{dimred} 
note that the Dedekind $\eta$ function has the following expansion in terms of $q \equiv e^{2\pi i \tau}$:
\bg\label{etaexpan}
&& {\rm log} \vert \eta(\tau)\vert^2 = - {\pi \over 6} \tau_2 - \left[ q + {3q^2 \over 2} + {4q^3\over 3} + ...
+ {\rm c.c}\right]\nonumber\\
&& {\rm log} ~{\eta(\tau) \over \eta(\bar\tau)} = + {i\pi\over 6}\tau_1 - \left[q + {3q^2 \over 2} + {4q^3\over 3} + ...
- {\rm c.c}\right]
\nd
Combining everything together we see that, upto powers of $q$ \eqref{cpevodd} and \eqref{dimred} are 
equivalent\footnote{Note that $q$ can be made small because both $C_4$ and ${\cal V}_4$ in \eqref{taudef} can be 
made small.}. 
However writing the action in terms of \eqref{cpevodd} instead of \eqref{dimred} has a distinct advantage: from D7 
point of view \eqref{cpevodd} captures the D3 instanton corrections in the system \cite{BBG, oog}. But there 
is a even deeper reason for writing the action as \eqref{cpevodd}. The CP-even and CP-odd terms can be $-$ in the 
case where the space is not a direct product of D7 world-volume times and normal space $-$ expanded further 
to incorporate Gauss-Bonet type interactions in the following way \cite{BBG}:
\bg\label{cpeven}
S_{\rm CP-even} = -\alpha_1 \int_{M^8} e^{-\phi} {\cal L}_{\rm GB} - T_7 \int_{M^8} e^{-\phi}\left[\sqrt{G} - 
{(4\pi^2\alpha')^2 \over 24} ~{\cal L}_R + {\cal O}(\alpha^{'4})\right]
\nd
where $\alpha_1$ is a constant, and ${\cal L}_{\rm GB}$ and ${\cal L}_R$ are respectively the Gauss-Bonnet and the 
curvature terms defined in the following 
way\footnote{The Gauss-Bonnet term is in general a topological invariant in four dimensions, but it 
is a total derivative at quadratic order in all dimensions. Therefore it wouldn't contribute to the equation of motion.}:
\bg\label{GBR}
&&{\cal L}_{\rm GB} = {\sqrt{G}\over 32\pi^2} \Big(R_{\alpha\beta\gamma\delta}R^{\alpha\beta\gamma\delta} - 
4 R_{\alpha\beta}R^{\alpha\beta} + R^2\Big)\nonumber\\
&& {\cal L}_R = {\sqrt{G}\over 32\pi^2}
\Big(R_{\alpha\beta\gamma\delta}R^{\alpha\beta\gamma\delta} - 2R_{\alpha\beta}R^{\alpha\beta} - 
R_{ab\gamma\delta}R^{ab\gamma\delta} + 2 R_{ab}R^{ab}\Big)
\nd
In the above note that the three curvature terms $R_{\alpha\beta}R^{\alpha\beta}$, $R_{ab}R^{ab}$ and $R^2$ are 
{\it not} the pull-backs of the bulk Ricci tensor. Furthermore we have used the notations ($\alpha, \beta$) to 
denote the world-volume coordinates, and ($a, b$) to denote the normal bundle. 

Thus from the CP-even terms, the coefficient of $R_{\alpha\beta\gamma\delta}R^{\alpha\beta\gamma\delta}$ is 
given by:
\bg\label{coeffeven}
c_3 \equiv {e^{-\phi}\sqrt{G} \over 32\pi^2}\left({4\pi^4 \alpha^{'2}\over 3} - \alpha_1\right)
\nd
which has an overall plus sign because $\alpha_1$ in many interesting cases tend to be zero (see \cite{BBG} for 
details on this).
However in general for certain exotic compactifications we expect 
$\alpha_1 << {4\pi^4 \alpha^{'2}\over 3}$. If we now compare this to \cite{Kats} we see that $c_3$, which
is the coefficient of $R_{\alpha\beta\gamma\delta}R^{\alpha\beta\gamma\delta}$ in \cite{Kats}, is positive. This would 
clearly mean that adding fundamental flavors lowers the viscosity to entropy bound!\footnote{We were motivated to 
carry out the above analysis from a comment by Aninda Sinha. His paper \cite{anindapaper} dealing with the 
violation of viscosity to entropy bound has appeared recently and has some overlap with this section.}     

The CP-odd term on the other hand has a standard expansion of the following form for the D7 brane \cite{GHM, DJM, BBG}:
\bg\label{cpodd}
S_{\rm CP-odd} = T_7 \int_{M^8} \left(C_8 + {\pi^2 \alpha^{'2}\over 24} C_4 \wedge {\rm tr}~R \wedge R\right)
\nd
where the first term gives the standard dual axionic charge of the D7 brane. Combining \eqref{cpeven} and 
\eqref{cpodd} we
get the full back reactions of the D7 brane upto ${\cal O}(\alpha^{'2})$. 

Having computed the back reactions of the embedded D7 brane, we can use this result to compute the shear viscosity. There
are three scenarios from which we can compute the viscosity using the gravity dual now:

\noindent $\bullet$ Allow a gravity dual that has no running (i.e no RG running in the gauge theory side), but 
incorporates the back reaction of the D7 brane \eqref{cpeven}.

\noindent $\bullet$ Allow a gravity dual that shows the RG running of the gauge theory, but does not incorporate 
the back reaction of the D7 brane (although may allow the CP-odd part \eqref{cpodd}).

\noindent $\bullet$ Allow a gravity dual that not only shows the RG running of the gauge theory, but also 
incorporates both the CP-even as well as CP-odd parts \eqref{cpeven} and \eqref{cpodd}. 

\noindent The first part has recently been done
 in \cite{Kats, violation, anindapaper}. The background remains AdS, but now there 
would be terms from the D7 brane \eqref{cpeven} that would lower the viscosity to entropy bound. However the second and 
the third part is not yet been addressed in the literature. As we show below, the second part is rather easy to 
do following the calculations of \cite{Kats} because we have already constructed the background in the previous 
section. 

The third part on the other hand is rather subtle because incorporating the higher order (curvature)${}^2$ corrections 
would change all the results that we derived earlier. In particular the mass and drag of the quark would need to 
be modified alongwith the wake of the quark. Plus the holographic renormalisability would also get modified because 
of the extra derivative terms from these corrections.
In this paper we will deal mostly with the second part of the 
above list and some calculations of the third part, but leave a more detailed analysis of the third part for the 
sequel \cite{sequel}.  

To start off the analysis, we need the correlation function of $T_{23}(x)$ and $T_{23}(0)$ to use it in the 
Kubo formula \eqref{SV-1}. From gauge/gravity duality we know that switching on $T_{23}$ in the gauge theory is 
equivalent to switching on graviton mode along $x^2 = x$ and $x^3 = y$ directions. Thus in the given OKS-BH 
background \eqref{leahl}, we switch on the following off-diagonal part:
\bg\label{metricmatrix}
\begin{pmatrix} g_{00}~~ & g_{0x}~~ & g_{0y}~~ & g_{0z} ~~ & g_{0r}\\
& & & & \\
g_{x0}~~ & g_{xx}~~ & g_{xy} ~~ & g_{xz} ~~ & g_{xr}\\& & & &\\
g_{y0} ~~& g_{yx}~~ & g_{yy}~~ & g_{yz} ~~ & g_{yr}\\& & & & \\
 g_{z0}~~ & g_{zx} ~~& g_{zy} ~~ & g_{zz} ~~ & g_{zr}\\ & & & &\\
 g_{r0}~~ & g_{rx} ~~& g_{ry} ~~ & g_{rz} ~~ & g_{rr}
\end{pmatrix} = {1\over \sqrt{h}}\begin{pmatrix} -g(r)~~ & 0~~ & 0~~ & 0 ~~ & 0\\
& & & & \\
0~~ & 1~~ & \phi(r,t) ~~ & 0 ~~ & 0\\& & & &\\
0 ~~& \phi(r,t)~~ & 1~~ & 0 ~~ & 0\\& & & & \\
 0~~ & 0 ~~& 0 ~~ & 1 ~~ & 0\\ & & & &\\
 0~~ & 0 ~~& 0 ~~ & 0 ~~ & {r^2 h \over g(r)}
\end{pmatrix}
\nd
where $h = h(r, \pi, \pi)$ and the internal space metric is independent of the radial coordinate $r$ 
as shown in \eqref{leahl}\footnote{Note that the above choice of the warp factor means that we have instinctively 
chosen the minima where $\theta_1 = \theta_2 = \pi$. Needless to say, since we are analysing the effect of the 
flavor on the viscosity, we took the point close to where the quark string originally was. Again a more detailed 
analysis could be performed directly from the ten-dimensional point of view, but such a analysis do not reveal 
any new physics.}. Note that the off-diagonal
gravitons are propagating in OKS-BH geometry with energy $\omega$. 

Since our goal is to compute the Fourier transform of 
$\langle T_{23}(t,\overrightarrow{x}) T_{23}(0,\overrightarrow{x})\rangle$, we can do this by first
writing the supergravity action in momentum space, treating it  as a functional of Fourier modes for
$\phi(r, t)$ where: 
\bg \label{SV-7a}
&& \phi(r,t)=\widetilde{\phi}(r, t)\bar{\phi}(t) 
\equiv \int d\omega\; e^{-i\omega \tau} \phi(r,\omega) = \int d\omega\; e^{-i\omega \sqrt{g}t} \phi(r,\omega) \nonumber\\
&& \phi(r,\omega)= \widetilde{\phi}(r, \vert\omega\vert)\bar{\phi}(\omega) 
 \nd 
where as before, we defined the Fourier transform using the curved space time $\tau \equiv \sqrt{g(r_c)} ~t$ and 
not simply $t$. Although this definition is precise for the theory at the cut-off $r = r_c$ only, we will use
it also for $r < r_c$ because in the end we will only provide description at the boundary (i.e $r \to \infty$)
where the results would 
be independent of the choice of the cut-off.

The way we proceed now is the following\footnote{This is almost similar to the procedure of \cite{Kats} whom we 
refer the readers for more details. Notice however that the theory considered by \cite{Kats} has no running 
but contains higher curvature-squared corrections, as mentioned above.}. 
We consider the metric fluctuation as in \eqref{metricmatrix} and plug this in the effective action \eqref{effaction}
but with $c_n = b_n = 0$ (we will modify this soon). We will also take a 
convention where $g(r_c) = 1$ in the subsequent analysis to avoid 
clutter. In the final result we will substitute the exact value of $g(r_c)$. Finally, 
we will call this resulting action as $S_{\rm SG}^{(2)}$ where 
the subscript (2) involves writing the action in terms of quadratic $\phi(r, \omega)$. The reason for doing this 
is because there exists a very useful relation for computing the shear viscosity (see for example \cite{S+Starinets}):
\bg \label{SV-7a2}
\lim_{\omega\rightarrow 0}
{\rm Im}~{G}_{11}^{\rm SK}(\omega,\overrightarrow{0})~ = ~
\lim_{\omega\rightarrow 0} 
\frac{2T}{\omega}~ {\rm Im}~G^{R}(\omega,\overrightarrow{0})  
\nd 
where ${G}_{ij}^{\rm SK}$ is the Schwinger-Keldysh propagator \cite{S+Starinets, H+Son}. Comparing this 
with our earlier Kubo formula \eqref{SV-1} we see that the shear viscosity is nothing but:
\bg\label{skbyt}
\eta ~ = ~-{1\over 2T}
\lim_{\omega\rightarrow 0}
~ {{\rm Im}~{G}_{11}^{\rm SK}(\omega,\overrightarrow{0}) }
\nd
Thus if we can write our effective supergravity action in the following way:
\bg \label{SV-7}
S_{\rm SG}^{(2)}[{\phi}(r_0, \omega)]~= ~ \frac{1}{2}\int \frac{d\omega d^3q}{(2\pi)^4}  ~ {\phi}_i(r_0, \omega)
{G}^{\rm SK}_{ij}(\vert\omega\vert,\overrightarrow{q}){\phi}_j(r_0, -\omega)
\nd 
where $r_0$ is a specified point, 
then taking the ${G}_{11}^{\rm SK}(\omega,\overrightarrow{0})$ part 
and using \eqref{skbyt} we can easily get our 
shear viscosity\footnote{Notice that there would be an overall volume factor of $T^{1,1}$ that would appear 
with the effective action. This factor is harmless and just modifies the Newton's constant in five 
dimensions.}.  Saying it a little differently, 
 we will be taking two functional derivatives of
$S_{\rm SG}^{(2)}[{\phi}(r_0, \omega)]$ with respect to ${\phi}(r_0, \omega)$ 
and thus are only interested in terms quadratic in
${\phi}(r_0, \omega)$ in the action. Of course as mentioned above, 
in real time formalism, we are concerned with the Schwinger-Keldysh propagator ${G}_{ij}^{\rm SK}$ of the
doublet fields $\phi_i(r, t),\phi_j(r, t)$. 
In the context of gauge/gravity duality, we follow the procedure outlined by \cite{H+Son} for
AdS/CFT correspondence and treat 
$\phi_1(r, t),\phi_2(r, t)$ as the perturbation $\phi(r, t)$
 and it's doublet in the four dimensional Minkowski space\footnote{We are a little sloppy here. Our background
is a deformation of the AdS space and therefore one might not expect the arguments of \cite{H+Son} to carry 
over exactly as above. However although the gauge/gravity duality argument for our case is more involved, we 
can still consider the $\phi_1$ perturbations to compute the Schwinger-Keldysh propagator because by definition
a propagator always appears sandwiched between the fields. See also \cite{sken2} for more generic 
approach.}.  
In ten dimensional gravity theory, $\phi_1(r)=\phi(r)$ is the 
 field in the R quadrant of the Penrose diagram while $\phi_2(r)$ is the field in the L quadrant. 
 For more details see \cite{H+Son} \cite{Israel} \cite{Maldacena-2} 
\cite{Balasubra} \cite{Horowitz-2} and \cite{sken2}.

Let us make this a bit more precise. Our aim is to get the effective action in the form \eqref{SV-7}. To this 
effect we take our metric \eqref{metricmatrix} and plug it in the five dimensional effective action \eqref{effaction}. 
The net result is the following action:
\bg \label{SV-8}
S_{\rm SG}^{(2)}&=&\frac{1}{8\pi G_N\sqrt{g(r_c)}}\int \frac{d\omega d^3q}{(2\pi)^4}\int_{r_h}^{r_c} dr
\Bigg[ A(r)\phi(r,-\omega) \phi''(r,\omega)+B(r)\phi'(r,-\omega) \phi'(r,\omega)\nonumber\\
&+&C(r)\phi(r,-\omega) \phi'(r,\omega)+D(r)\phi(r,-\omega) \phi(r,\omega)\Bigg] 
\nd
 where prime denotes derivative with respect to $r$ and the explicit expressions for 
 $A,B,C,D$ are given in {\bf Appendix E}. The five dimensional Newton's constant is given by:
\bg\label{nc}
G_N \equiv {\kappa_{10}^2 L^5 \over 4\pi V_{T^{1,1}}}
\nd 
 where volume of $T^{1,1}$ i.e $V_{T^{1,1}}$ is dimensionful  
and $\kappa_{10}$ is proportional to ten dimensional Newton's 
constant. 

The fluctuation $\phi(r, \omega)$ is not anything arbitrary of course. It satisfies the following Euler-Lagrange 
equation of motion:
\bg \label{SV-11a}
\phi''(r,\omega)+ \frac{A'(r)-B'(r)}{A(r)-B(r)}\phi'(r,\omega)+
\frac{2D(r)-C'(r)+A''(r)}{2\left[A(r)-B(r)\right]}\phi(r,\omega)=0
\nd
which we can derive from \eqref{SV-8} above. Once we plug in the values of $A, B$ etc., the above Euler-Lagrange
equation takes the following form:
\bg \label{SV-11b}
&& \phi''(r,\omega)+ \left[\frac{g'(r)}{g(r)}+\frac{5}{r} +{\cal M}(r)\right]\phi'(r,\omega)
+\left[\frac{\omega^2 g(r_c)\bar{h}(r)}{g(r)^2}+{\cal J}(r)\right]\phi(r,\omega)=0\nonumber\\
&&\bar{h}(r)\equiv \frac{L^4}{r^4}\Bigg\{1+\frac{3g_s N_f^2}{2\pi N}\left[1+\frac{3g_s N_f}{2\pi}
\left({\rm log}r+\frac{1}{2}\right)-\frac{g_s N_f}{4\pi}\right]
{\rm log}r\Bigg\}\nonumber\\
\nd
where ${\cal J}(r)$ and ${\cal M}$ are added to allow for most generic conditions on the scalar fields\footnote{In 
special cases we expect ${\cal J}(r)$ and ${\cal M}$ to 
vanish (see for example \cite{buch04}). 
However when this is not the case (as may arise for non-trivial UV completions of our model) we could 
expect a non-minimally coupled scalar field.}. 
As before, primes in \eqref{SV-11a} and \eqref{SV-11b}
denote derivatives wrt the five dimensional radial coordinate. 

Now as we mentioned above in \eqref{SV-7a}, $\phi(r, \omega)$ can be decomposed in terms of 
$\widetilde\phi(r, \vert\omega\vert)$ and 
$\bar\phi(\omega)$. Then 
as a trial solution, just like in \cite{Kats}, we first try $\widetilde{\phi}(r, \omega)= g(r)^{\gamma}$ and look at 
\eqref{SV-11b} for $r$ near $r_h$ where $g(r)\rightarrow 0$. Plugging this in \eqref{SV-11b} with $g(r) = 0$ we 
obtain the following expression  for $\gamma$:
\bg \label{SV-12} 
\gamma&=&\pm i\vert\omega\vert \sqrt{\frac{\bar{h}(r_h)g(r_c)}{16}}r_h\nonumber\\
&=&\pm i \frac{\vert\omega\vert}{4\pi T_c}
\nd
where 
in the last step we have used the definition of  
temperature $T_c$ as in (\ref{Temp4}). 

To get the solution with $g(r) \ne 0$  
we propose the following ansatze 
for the solution to \eqref{SV-11b}\footnote{Incidentally, this form for $\phi(r, \omega)$ can be shown to be 
exactly like \eqref{mpert1} discussed earlier. This will become clearer as we go along.}:
\bg\label{ansatsol}
\phi(r,\omega)~ = ~ g(r)^{\pm i \frac{\vert\omega\vert}{4 \pi T_c}}F(r, \vert\omega\vert)
\bar{\phi}(\omega)
\nd
Plugging this in \eqref{SV-11b} we see that the 
equation satisfied by $F(r, \vert\omega\vert)$ can be expressed in terms of $\gamma$ and $\gamma^2$ in the following way: 
\bg \label{SV-13}
&& F''(r, \vert\omega\vert)+\left({g'(r) \over g(r)}+ {5 \over r} + {\cal M} \right)F'(r, \vert\omega\vert)+
\left({\vert\omega\vert^2 g(r_c) {\bar h} \over
g^2(r)} + {\cal J}(r)\right) F(r, \vert\omega\vert)\\
&& + ~\gamma \Bigg\{{2g'(r)\over g(r)} F'(r, \vert\omega\vert) + 
\left[{g''(r) \over g(r)} + \left({5\over r} + {\cal M}\right){g'(r) \over 
g(r)}\right]F(r, \vert\omega\vert)\Bigg\} + \gamma^2 {g'^{2}(r) \over g^2(r)} F(r, \vert\omega\vert) = 0\nonumber
\nd
where the $\gamma^2$ terms come from both the last term in the above equation as well as the 
$\vert\omega\vert^2$ term above.
Furthermore,
note that the source ${\cal J}(r)\sim {\cal O}(g_s)+{\cal O}(g_s^2)$, so in the limit $g_s\rightarrow 0$ we find that 
(\ref{SV-13})
has a solution of the form $F(r, \vert\omega\vert) =c_1+c_2 g(r)^{-2\gamma}$ with $c_1, c_2$ constants. 
Then we expect the complete solution for $g_s\neq 0$ to be $F(r, \vert\omega\vert)=c_1+c_2
g(r)^{-2\gamma}+f(r, \vert\omega\vert)$. 
Demanding that $F(r, \vert\omega\vert)$ be regular at the horizon $r=r_h$ forces 
$c_2=0$ as $g(r_h)=0$. We choose $c_1=1$ and
$f ={\cal G}+\gamma{\cal H} + \gamma^2 {\cal K} + ...$ as a series solution in $\gamma$. Then our ansatze
for the solution to (\ref{SV-13}) becomes 
\bg\label{sersolo}
F(r, \vert\omega\vert) =1+{\cal G}(r) +\gamma{\cal H}(r) + \gamma^2 {\cal K}(r) + ...
\nd 
Once we 
plug in the ansatze \eqref{sersolo} in (\ref{SV-13}) we see that the resulting equation 
can be expressed as a series in $\gamma$:  
\bg\label{seringama}
&& ~~~~~~~ {\cal G}'' + \left({g'\over g} + {5\over r} + {\cal M} \right) G' + {\cal J}(1 + {\cal G}) \nonumber\\ 
&& +~\gamma \Bigg\{{\cal H}'' + \left({g'\over g} + {5\over r} + {\cal M} \right){\cal H}' + {\cal J} {\cal H} + 
{2g'\over g} {\cal G}' + \left[{g''\over g} + \left({5\over r} + {\cal M} \right){g'\over g}\right]
(1 + {\cal G})\Bigg\}\nonumber\\
&& +~\gamma^2 \Bigg\{{\cal K}'' +  \left({g'\over g} + {5\over r} + {\cal M} \right){\cal K}' + {\cal J} {\cal K} + 
{2g'\over g} {\cal H}' + \left[{g''\over g} + \left({5\over r} + {\cal M} \right){g'\over g}\right]{\cal H}\nonumber\\
&& ~~~~~~~~~~~~~~ + ~\left(\kappa_0 + {g'^{2}\over g^2}\right) (1 + {\cal G}) \Bigg\}\nonumber\\ 
&& + ~ \gamma^3 \Bigg\{{2g'\over g} {\cal K}' + 
\left[{g''\over g} + \left({5\over r} + {\cal M} \right){g'\over g}\right]{\cal K} + 
\left(\kappa_0 + {g'^{2}\over g^2}\right) {\cal H} + ...... \Bigg\} + {\cal O}(\gamma^4) ~ = ~ 0\nonumber\\
\nd
where we have avoided showing the explicit $r$ dependences of the various parameters to avoid clutter. We have also 
defined $\kappa_0$ in terms of the variables of \eqref{SV-12} in the following way:
\bg\label{kappa0}
\kappa_0 ~ \equiv~ - {16 \over {\cal T}^2 g^2}
\nd
Although the above equation \eqref{seringama} may look formidable there is one immediate simplification that could 
be imposed, namely, putting the coefficients of $\gamma^0, \gamma, \gamma^2, ...$ individually to zero. This is 
possible because one can view $\gamma$ to be an arbitrary parameter that can be tuned by choosing the graviton 
energy $\omega$ or the temperature $T_c$. This means that the zeroth order in $\gamma$ we will have the following 
equation: 
\bg \label{SV-13a}
{\cal G}''(r)+\left[{g'(r)\over g(r)}+ {5\over r} + {\cal M}(r) \right]{\cal G}'(r)
+{\cal J}(r)[1+{\cal G}(r)]~= ~ 0
\nd
In the above equation observe that 
the source ${\cal J}(r)$, for cases where it is non-zero, has a complicated structure with logarithms and powers of $r$. 
To simplify the 
subsequent expressions, let us choose to work near the cut-off $r = r_c$. This is similar to the spirit of the 
previous section where we eventually analysed the system from the boundary point of view. Then
to solve \eqref{SV-13a} near $r\sim r_c$ we can
switch to following coordinate system
\bg \label{SV-13a1}
r=r_c(1- {\zeta}) 
\nd 
Taylor expanding all the terms ${\cal J}(r),g(r),{1\over r_c^n(1-{\zeta})^n}$  in \eqref{SV-13a} 
about ${\zeta}=0$ and using similar arguments as the previous section, 
we obtain a power series solution for ${\cal G}$ as: 
\bg \label{SV-13a2}
{\cal G}(r) ~ = ~ \sum_{\alpha} \sum_{i=0}^{\infty}~ {{\tilde a}_i^{(\alpha)} \over r_{c(\alpha)}^{4i}(1-\zeta)^{4i}} ~ 
\equiv ~  \sum_{i=0}^{\infty}~a_i {\zeta}^i
\nd
Since (\ref{SV-13a}) is a second order differential equation, 
we can fix two coefficients and we choose $a_0=a_1=0$. Then the rest of $a_i's$ are
determined by equating coefficients of ${\zeta}^i$ on both sides of equation (\ref{SV-13a}). 
The exact solutions are listed in {\bf Appendix E}. Note that all $a_i$ are proportional to $g_s$ and in the limit
$g_s\rightarrow 0$, ${\cal G}\rightarrow 0$. 

To next order in $\gamma$ we have an equation for ${\cal H}$ that also depends on the solution that we got for 
${\cal G}$. 
The equation for ${\cal H}(r)$ can be taken from \eqref{seringama} as:
\bg \label{SV-13c}
&& {\cal H}''(r) + \left[{g'(r)\over g(r)} + {5\over r} + {\cal M}(r) \right]{\cal H}'(r) + {\cal J}(r) {\cal H}(r) =  
-{2g'(r)\over g(r)} {\cal G}'(r)\nonumber\\
&& ~~~~ - \left\{{g''(r)\over g(r)} + \left[{5\over r} + {\cal M}(r) \right]
{g'(r)\over g(r)}\right\}
\left[1 + {\cal G}(r)\right]
\nd 
To solve this we make 
the coordinate transformation (\ref{SV-13a1}) and plug in the series solution for ${\cal G}(r)$ given above. 
The final 
result for ${\cal H}$ can again be expressed as a series solution in ${\zeta}$ in the following way:
\bg \label{SV-13c1}
{\cal H}(r)~ = ~ \sum_{\alpha}\sum_{i=0}^{\infty}~ {{\tilde b}_i^{(\alpha)} \over r_{c(\alpha)}^{4i}(1-\zeta)^{4i}} ~
\equiv ~ \sum_{i=0}^{\infty}b_i {\zeta}^i  
\nd 
We again set $b_0=b_1=0$ and following similar ideas used to solve for ${\cal G}$, 
we determine all $b_i's$ by equating coefficients in(\ref{SV-13c}). The exact
solution is given in {\bf Appendix E}. Again note that all $b_i$ are of 
at least ${ \cal O}(g_s)$ and thus with $g_s\rightarrow
0$, ${\cal H}\rightarrow 0$.

Finally the second order in $\gamma$ is a much more involved equation that uses results of the previous two equations 
to determine ${\cal K}$. This is given by:
\bg\label{SV-13c2} 
&& {\cal K}''(r) +  \left({g'(r)\over g(r)} + {5\over r} + {\cal M}(r) \right){\cal K}'(r) + {\cal J}(r) {\cal K}(r)  =  
- {2g'(r)\over g(r)} {\cal H}'(r)\\
&& ~~~ - \left[{g''(r)\over g(r)} + \left({5\over r} + {\cal M}(r) \right){g'(r)\over g(r)}\right]{\cal H}(r)
 - \left(\kappa_0 + {g'^{2}(r)\over g^2(r)}\right) \left[1 + {\cal G}(r)\right]\nonumber
\nd
which could also be solved using another series expansion in ${\zeta}^i$ (we haven't attempted it here). 
Therefore combining 
(\ref{SV-13a2}) and (\ref{SV-13c1}) we finally have the
solution for the metric perturbation: 
\bg \label{SV-13e}
\widetilde{\phi}(r, \vert\omega\vert)_\pm= g(r)^{\pm i\frac{\vert\omega\vert }{4\pi T_c}}
\left[1+{\cal G}(r)\pm i{\vert\omega\vert  \over 4{\pi T_c}} {\cal H}(r) 
- {\vert\omega\vert^2 \over 16 \pi^2 T_c^2} 
{\cal K}(r) + ....\right]
\nd     
We can analyse this in the regime where the gravitons have very small energy, i.e $\omega \to 0$ or equivalently 
$\gamma \to 0$. In this limit 
we can Taylor expand  $\widetilde{\phi}(r, \vert\omega\vert)$ about $\gamma = 0$ to give us the two possible solutions: 
\bg \label{SV-13f} 
\widetilde{\phi}(r, \vert\omega\vert)_\pm &=& 1+{\cal G}(r)\pm i\frac{\vert\omega\vert }{4\pi T_c}
\Big\{{\cal H} (r) + [1+{\cal G}(r)] {\rm log}~g(r)\Big\}\\
&& - {\vert\omega\vert^2 \over 16 \pi^2 T_c^2}\Big\{{\cal K}(r) + {\cal H}(r) ~{\rm log}~g(r) + [1 + {\cal G}(r)]
{\rm log}^2 g(r)\Big\} + {\cal O}(\vert\omega\vert^3)\nonumber
\nd
%therimal
which 
consequently means that to the first order in $\omega$ the off diagonal gravitational perturbation at low energy is 
given by two possible solutions corresponding to positive and negative frequencies as: 
\bg \label{SV-13g}
\phi(r,\omega)_\pm = \left[1+{\cal G}(r)\right] {\bar \phi}(\omega) \pm i\frac{\vert\omega\vert } {4 \pi T_c} 
\Big\{{\cal H} (r) + [1+{\cal G}(r)] {\rm log}~g(r)\Big\} \bar{\phi}(\omega) 
\nd
As is well known following, say, \cite{Unruh} \cite{H+Son}, we can define field on the right 
${\bf R}$ and left ${\bf L}$ quadrant of the 
Kruskal plane in terms of $\phi_+(r, \omega)$ and $\phi_-(r, \omega)$ in the following way:
\bg \label{SV-13g1}
\phi_{{\bf R},\pm}(\omega,r)&=& \phi_{\pm}(\omega,r) ~~~ {\rm in ~~ {\bf R}} \nonumber\\ 
 &=&  0 ~~~ {\rm in ~~ {\bf L}} \nonumber\\
 \phi_{{\bf L},\pm}(\omega,r)&=& \phi_{\pm}(\omega,r) ~~~ {\rm in ~~ {\bf L}}\nonumber\\
&=& 0 ~~~ {\rm in ~~ {\bf R}} 
\nd  
Now  $\phi_{{\bf R},\pm},\phi_{{\bf L},\pm}$ contain positive and negative frequency modes but a 
certain linear combination of  
$\phi_{{\bf R},\pm},\phi_{{\bf L},\pm}$
 gives purely positive or purely negative frequency modes in the entire Kruskal plane \cite{Unruh} \cite{H+Son}. 
Furthermore imposing that positive frequency modes are
 infalling at the horizon in ${\bf R}$ quadrant and negative frequency modes are outgoing at the horizon in ${\bf R}$ 
fixes two combinations :
\bg \label{SV-13g2}
\phi_{\rm pos}&=&e^{\omega/T_c}\phi_{{\bf R},-}(\omega,r) +e^{\omega/2T_c}\phi_{{\bf L},-}(\omega,r)\nonumber\\
\phi_{\rm neg}&=&\phi_{{\bf R},+}(\omega,r) +e^{\omega/2T_c}\phi_{{\bf L},+}(\omega,r)
\nd  
With (\ref{SV-13g2}) we see that we can define fields in ${\bf R}({\bf L})$ 
quadrant as linear combination of positive and negative frequency modes 
\bg \label{SV-13g3}
\phi_{\bf R}(\omega,r)&\equiv& {{\tilde a}_0}\big[\phi_{{\bf R},+}(\omega,r) 
- e^{\omega/T_c}\phi_{{\bf R},-}(\omega,r)\big]\equiv\phi_1\nonumber\\
\phi_{\bf L}(\omega,r) &\equiv&{{\tilde a}_0}e^{\omega/2T_c}\big[\phi_{{\bf L},+}(\omega,r) 
- \phi_{{\bf L},-}(\omega,r)\big]\equiv\phi_2
\nd
where we have identified $\phi_{\bf R}(\phi_{\bf L})$ with the thermal field $\phi_1(\phi_2)$ 
defined on the complex time contour which familiarly
appears in the Schwinger-Keldysh propagators of real time thermal field theory. Here ${{\tilde a}_0}$ is a constant.  
The final physical quantity that we
will extract from here will only depend on ${\cal T}$, as we will show soon. 

%therimal
Having got the graviton fluctuations $\phi(r, \omega) \equiv \phi_{\bf R}(\omega,r)$, 
we are almost there to compute the viscosity $\eta$ using 
\eqref{skbyt}. Our next step would be to compute the Schwinger-Keldysh propagator 
$G^{\rm SK}_{11}(0, \overrightarrow{q})$. All we now need is to write the action \eqref{SV-8} as \eqref{SV-7} and 
from there extract the Schwinger-Keldysh propagator. This analysis is almost similar to the one that we 
did in the previous section, so we could be brief (see also \cite{buchel-1}). The action \eqref{SV-8} can be used to 
get the boundary action once we shift $\phi(r, \omega)$ to $\phi(r, \omega) + \delta \phi(r, \omega)$ in the 
following way:
\bg\label{boundary}
S_{\rm SG}^{(2)}(\phi &+&\delta\phi) =\frac{g(r_c)^{-1/2}}{8\pi G_N}\int \frac{d\omega d^3q}{(2\pi)^4}\int_{r_h}^{r_c} dr
\Big\{A(r)\phi(r,-\omega) \phi''(r,\omega)+B(r)\phi'(r,-\omega) \phi'(r,\omega)\nonumber\\
&&+C(r)\phi(r,-\omega) \phi'(r,\omega)+D(r)\phi(r,-\omega) \phi(r,\omega) + \Big[2A(r) \phi''(r, \omega)\\
&&- 2B(r) \phi''(r, \omega) - 2 B'(r) \phi'(r, \omega) - C'(r) \phi(r, \omega) + 2D(r) \phi(r, \omega) \nonumber\\
&&+ A''(r) \phi(r, \omega) + 2 A'(r) \phi'(r, \omega)\Big]\delta\phi(r, -\omega) + 
\partial_r \Big[2B(r) \phi'(r, \omega)\delta\phi(r, -\omega) \nonumber\\
&&+ C(r) \phi(r, \omega) \delta\phi(r, -\omega) + A(r) \phi(r, \omega) \delta\phi'(r, -\omega) 
- \partial_r\left(A(r) \phi(r, \omega)\right)\delta\phi(r, -\omega)\Big]\Big\}\nonumber
\nd
Plugging in the background value of $\phi(r, \omega)$ will tell us that only the boundary term survives. And as before,
 to cancel the $A(r) \phi(r, \omega) \delta\phi'(r, -\omega)$ we will have to add the Gibbons-Hawking term to the 
action \cite{Gibbons-Hawking}. The net result is the following boundary action:
 \bg \label{SV-9}
S_{\rm SG}^{(2)}&=&{g(r_c)^{-1/2}\over 8\pi G_N}\int \frac{d\omega d^3q}{(2\pi)^4}
\phi(r,-\omega) \Bigg\{\frac{1}{2}\Big[C(r)-A'(r)\Big]
+\Big[ B(r)-A(r)\Big] {\phi'(r,-\omega)\over \phi(r,-\omega)}\Bigg\} \phi(r,\omega)\Bigg\vert_{r_h}^{r_c}\nonumber\\
&\equiv& {1\over 8\pi G_N\sqrt{g(r_c)}}\int \frac{d\omega d^3q}{(2\pi)^4}{\cal F}(\omega,r)\Bigg{|}_{r_h}^{r_c} 
\nd 
Now
comparing (\ref{SV-7}) with \eqref{SV-9} we see that the terms between the braces combine to give us the required 
Schwinger-Keldysh propagator:
\bg \label{SV-10}
{G}^{\rm SK}_{11}(0, \overrightarrow{q})&=& \lim_{\omega \to 0}~{1\over 4\pi G_N\sqrt{g(r_c)}}
\frac{{\cal F}(\omega,r)}{\phi_1(r,\omega)\phi_1(r,-\omega)}
\Bigg{|}_{r_h}^{r_c}\\
& = & \lim_{\omega \to 0} ~{1\over 4\pi G_N\sqrt{g(r_c)}}\Bigg\{
\frac{1}{2}\Big[C(r)-A'(r)\Big] +\Big[ B(r)-A(r)\Big] {\phi_1'(r,-\omega)\over \phi_1(r,-\omega)}\Bigg\}
\Bigg{|}_{r_h}^{r_c}\nonumber
\nd 
where we assume\footnote{At this point one might worry that the solution for $\phi_1$ is only known around 
$r_c$. That this is not the case can be seen in the following way:
Integration by parts gives \eqref{SV-10} which says one only needs to
know the value of the field $\phi_1$ at $r_c $ and $r_h$. The solution for
$\tilde{\phi_1} ={\phi_1\over {\bar{\phi_1}}}$ is given in \eqref{SV-13e} from which it is
clear that $\phi_1(r_h)=0$ as $g(r_h)=0$.  Furthermore to know $\eta$ we only
need to know the imaginary part of \eqref{SV-10}, which is evaluated using \eqref{vislim}
in \eqref{SV-10} and using boundary values of $\phi_1(r_c)$ and $\phi_1(r_h)$.}
that $\phi_1(r_h, \omega) = 
{\tilde a}_0\big[\phi_{{\bf R},+}(r_h, \omega) - e^{\omega/T_c}\phi_{{\bf R},-}(r_h, \omega)\big]$. Now
to evaluate the shear viscosity from the above result we need to perform two more steps:

\noindent $\bullet$ Evaluate the contributions from the UV cap that we attach from $r = r_c$ to 
$r = \infty$.  

\noindent $\bullet$ Take the imaginary part of the resulting {\it total} Schwinger-Keldysh propagator. This should 
give us result independent of the cut-off. 

\noindent To evaluate the first step i.e
contributions from the UV cap, we need to see precisely the singularity structure of 
$S_{\rm SG}^{(2)}$. The second step would then be to extract the imaginary part of SK propagator from there. Since the 
imaginary part can only come from the second term of \eqref{SV-9}, we only need to evaluate:
\bg\label{visuv}
\lim_{\omega \to 0} ~{1\over 4\pi G_N \sqrt{g(r_c)}}
~\Big[ B(r)-A(r)\Big] {\phi_1'(r,-\omega)\over \phi_1(r,-\omega)}
\Bigg{|}_{r_c}^{\infty}
\nd 
with $\phi(r, -\omega)$ being the graviton fluctuation in the regime $r > r_c$. To analyse this let us first 
consider a case where $g_s \to 0$ and (${\cal G}(r), {\cal H}(r), {\cal K}(r), ...$) $\to 0$. In this limit we 
expect for $r_h \le r \le r_c$:
\bg\label{vislim}
&& B(r) - A(r) ~ =  -{1\over 2 g_s^2} ~g(r) r^5 ~ + ~ {\cal O}(g_s N_f) \\
&& \phi_1(r, \omega) ~ = ~ {\tilde a}_0\Bigg[-{\omega\over T_c}\left(1+{\cal G} - i{\vert\omega\vert \over 4\pi T_c} 
{\cal H}\right) + i{\vert\omega\vert \over 2\pi T_c}{\cal H} + i{\vert \omega \vert \over 2\pi T_c} {\rm log}~g (1 + 
{\cal G})\Bigg]\nonumber\\
&& {\phi_1'(r, -\omega)\over \phi_1(r, -\omega)} = {g'(r)\over g(r)} \left({2\pi\over 4\pi^2 + {\rm log}^2 g(r)}\right)
\nonumber
\nd
The above considerations would mean that the contribution to the viscosity, $\eta_1$, 
for this simple case without incorporating
the UV cap will be:
\bg\label{visfirst}
\eta_1 & = & {r_h^4 \over 2\pi T_c g_s^2 G_N \sqrt{g(r_c)}} \Bigg({1\over 4\pi + {1\over \pi} ~{\rm log}^2 
g(r_c)}\Bigg) = {{\cal T}^3 L^2 \over 2 g_s^2 G_N}\left({1\over 4\pi + {1\over \pi} ~{\rm log}^2 g(r_c)}\right)\nonumber\\
\nd
where we have used the relations $\pi T_c \sqrt{g(r_c)} = \left[r_h \sqrt{{\bar h}(r_h)}\right]^{-1}$ and 
${\bar h}(r_h) \approx {L^4 \over r_h^4}$ in this limit. This helps us to write everything in terms of ${\cal T}$ and 
not the scale dependent temperature $T_c$. In fact as we show below, once we incorporate the contributions from the 
UV cap, the $r_c$ dependence of the above formula will also go away and the final result will be completely independent 
of the cut-off. Note that in the limit $r_c \to \infty$ we recover the result for the cascading theory.    

Combining all the ingredients together, the contribution to the viscosity in the limit where 
(${\cal G}(r), {\cal H}(r), {\cal K}(r), ...$) etc are non-zero can now be presented succinctly as (although 
$\eta_1$ below doesn't have any real meaning on the gauge theory side as this is an intermediate quantity):
\bg\label{vishu} 
\eta_1 = {r_h^5 \sqrt{{\bar h}(r_h)}\over 2 g_s^2 G_N} \left\{ {1 + {r_c^5 g(r_c) \over 4 r_h^4} 
\left[{{\cal H}' \over 1 + {\cal G}} - {{\cal H}{\cal G}' \over (1+{\cal G})^2}\right]\over 
4\pi + {1\over \pi} \left[{\rm log}~g(r_c) + {{\cal H} \over 1 + {\cal G}}\right]^2}\right\}
\nd
Note that the above expression is exact for our background at least in the limit 
where we take the leading order $r^5$ singularity 
of the background. This is motivated from our detailed discussion that we gave in the previous section. Note that 
the second term in the action \eqref{SV-9} is exactly the second equation of the set \eqref{infinity} whose 
singularity structure has been shown to be renormalisable. Thus taking the leading order singularity $r^5$ instead of 
the actual $r_{(\alpha)}^5$ will not change anything if we carefully compensate the coefficients with appropriate
$g_sN_f, g_sM^2/N$ factors!  
 
But this is still not the complete expression as we haven't 
added the contributions from the UV cap. Before we do that, we want to re-address
the singularity structure of the
above expression. The worrisome aspect is the existence of $r_c^5$ factor in \eqref{vishu}. Does that create a problem for
our case? 

The answer turns out to be miraculously no, because of the form of ${\cal H}$ and ${\cal G}$ given in \eqref{SV-13c1} 
and \eqref{SV-13a2}. This, taking only the leading powers of $r_c$, yields:
\bg\label{sigcal}
{\cal H}' ~ = ~ -{4{\tilde b}_1 \over r_c^5} -{8{\tilde b}_2 \over r_c^9} + ...., 
~~~~{\cal G}' ~ = ~ -{4{\tilde a}_1 \over r_c^5} -{8{\tilde a}_2 \over r_c^9} + .... 
\nd
killing the $r_c^5$ dependence in \eqref{vishu}\footnote{It is now easy to see why ${\tilde b}_1 = {\tilde a}_1 = 0$ is
consistent. For non-zero ${\tilde b}_1, {\tilde a}_1$ there would have been additional ${\rm log}~r$ terms from 
$r_{(\alpha)}^5$. These would have made the theory non-renormalisable. Thus holographic renormalisability would 
demand ${\tilde b}_1 = {\tilde a}_1 = 0$ from the very beginning $-$ consistent with what we choose earlier.}. 
This would make $\eta_1$ completely finite and all the $r_c$ 
dependences would go as ${\cal O}(1/r_c)$. Therefore we expect the contribution to the viscosity from the UV cap to 
go like:
\bg\label{eta1n}
\eta_2 ~ \equiv ~ \eta\big\vert_{r_c}^\infty ~ = ~ \sum_{i = 0}^\infty ~{G_i \over r_c^{4i}}
\nd
where the total viscosity will be defined as $\eta \equiv \eta_1 + \eta_2$. As this is a physical quantity  
we expect it to be independent of the scale. Therefore 
\bg\label{pq}
 {\partial \eta\over \partial r_c} ~ = ~ 0
\nd
which will give us similar Callan-Symanzik type equations, as discussed in the previous section,
from where we could derive the precise forms for 
$G_i$ in \eqref{eta1n}. Finally when the dust settles, the result for shear viscosity can be expressed as:
\bg\label{shearvo} 
\eta ~ = ~ {{\cal T}^5 \sqrt{{\bar h}({\cal T})}\over 2 g_s^2 G_N} \Bigg[{1 + \sum_{k = 1}^\infty
\alpha_k e^{-4k {\cal N}_{\rm uv}} \over 4\pi + {1\over \pi}~{\rm log}^2 \left(1 - 
{\cal T}^4 e^{-4{\cal N}_{uv}}\right)}\Bigg]
\nd
where $\alpha_k$ are functions of ${\cal T}$ that can be easily determined from the coefficients 
(${\bar a}_i, {\bar b}_i$) in \eqref{SV-13a2} and \eqref{SV-13c1} or ($a_i, b_i$) worked out in {\bf Appendix E}; and
${\bar h}({\cal T}) \equiv {L^4\over {\cal T}^4} + {\cal O}(g_s, N_f, M)$.  
Observe that the final result for shear viscosity is completely independent of $r_c$ and $T_c$; and only depend 
on ${\cal T}$ and the degrees of freedom at the UV i.e through $e^{-{\cal N}_{uv}}$. Needless to say, for large enough 
${\cal N}_{uv}$ (which is always the case for our case because ${\cal N}_{uv} \to \epsilon^{-n}, n \ge 1$), 
the shear viscosity is only sensitive to the characteristic 
temperature ${\cal T}$ of the cascading theory. The interesting thing however is that the shear viscosity with finite
but large enough ${\cal N}_{uv}$ can be {\it smaller} than or {\it equal to}
the shear viscosity with ${\cal N}_{uv} \to \epsilon^{-n}, n >> 1$ i.e for the 
parent cascading theory provided:
\bg\label{smlim}
\alpha_k ~\le ~ {1\over 4\pi^2} \sum_{n \in {\bf Z}} ~{{\cal T}^{4k}\over n(k-n)}, ~~~~~ n \le k, ~~ k \in {\bf Z}
\nd
in the limit of small characteristic temperature ${\cal T}$. 
This will have effect on the viscosity to entropy ratio, to which we turn next.

%\newpage

\subsection{Viscosity to entropy ratio}

Our final set of analysis will be to calculate the viscosity to the entropy ratio for the above two cases i.e one with 
only RG flow, and the other with both RG flow and curvature squared corrections. As usual the former is easier to handle
so we discuss this first. 

Starting with the type IIB supergravity action in ten dimension i.e the $S_{\rm OKS-BH}$ in \eqref{KS77} 
the entropy is given by the Wald's formula \cite{wald1},\cite{wald2},\cite{wald3},\cite{wald4}
\bg \label{e2}
&& {\cal S}=-2\pi \oint dxdydz d^5{\cal M} \sqrt{{\cal P}}\frac{\partial{{\cal L}_{10}}}{\partial {R}_{abcd}}\epsilon^{ab}
\epsilon^{cd}
\nd
where the integral is over the eight dimensional surface of the horizon at $r=r_h,{\cal L}_{10}$ is the lagrangian density of the action in
\eqref{KS77},
${\cal P}_{ab},a,b=1..8$ 
 is the induced $8\times 8$ metric at horizon, $\epsilon_{ab}$ is the binormal normalized to  
$\epsilon_{ab}\epsilon^{ab}=-2$. Finally using
 explicit expression for the metric \eqref{bhmet} and \eqref{bhmet2}, we have
 \bg \label{e2a}
 s &= &
{{\cal S}\over V_3} = -\frac{\pi r_h^5}{108 V_3\kappa_{10}^2} \oint dxdydz ~d^5{\cal M}~
{\rm sin}~\theta_1~{\rm sin}~\theta_2 ~\sqrt{h(r_h,\theta_1,\theta_2)}~\frac{\partial{{\cal L}_{10}}}{\partial {R}_{abcd}}\epsilon^{ab}
\epsilon^{cd}\nonumber\\
&= & \frac{r_h^3 L^2}{2 g_s^2 G_N} \Bigg\{1+\frac{3g_s M^2}{2\pi N}\left[1+\frac{3g_s N_f}{2\pi}
\left({\rm log}~r_h+\frac{1}{2}\right)-\frac{g_s N_f}{4\pi}\right]
{\rm log}~r_h\Bigg\}^{1/2}
\nd
where $V_3$ is the infinte three dimensional volume 
%and in the last step we approximated 
%\bg \label{e2c}
 %\int d^5{\cal M}\frac{\rm{sin}(\theta_1)\rm{sin}(\theta_2)}{108}\sqrt{h(r_h,\theta_1,\theta_2)}\frac{\partial{{\cal L}_{10}}}{\partial {R}_{abcd}}\epsilon^{ab}
%\epsilon^{cd}=-2\sqrt{h(r_h,\pi,\pi)}V_{T^{1,1}}/L^5
%\nd
 and we have used the definition of five dimensional Newton's constant $G_N$ introduced in \eqref{nc} as well as 
$\bar h(r_h)$ introduced in \eqref{SV-11b}. 
The relation above is consistent
 with the notion that the effective five dimensional warp factor is well approximated with taking a slice
 $\theta_i=\pi,\phi_i=\psi=0$. The results would only differ by a ${\cal O}(g_sN_f)$ term which, in our approximation, is
very small. 
Note that the definiton of temperature depends on the effective five dimensional warp factor
 and as the approximation in \eqref{e2a} is consistent with it,  
our computation of entropy and temperature are consistent.

Once we replace $r_h$ by the characteristic temperature ${\cal T}$, we see that the entropy is only sensitive to the 
temperature and is independent of any other scale of the theory. Since the above result is also independent of 
${\cal N}_{uv}$ it would seem that the Wald formula only gives the entropy for the theory 
with ${\cal N}_{uv} = \infty$ i.e for the parent 
cascading theory\footnote{This can be argued by observing that fact that in 
a renormalisable theory, like ours, the 
dependences on degrees of freedom go like ${\cal O}\left(e^{-{\cal N}_{uv}}\right)$ corrections as we saw in 
the previous sections.}. The interesting question now would be to ask what is the entropy for the theory whose UV 
description is different from the parent cascading theory? In other words, what is the effect of the UV cap attached 
at $r = r_c$ on the entropy? 

To evaluate this, observe first that in finite temperature 
gauge theory, entropy density of a thermalized medium having stress tensor 
$\langle T^{\mu\nu}\rangle ={\rm diagonal} (\epsilon,P,P,P)$ is given by
\bg \label{entropyC1}
s=\frac{\epsilon +P}{T}
\nd 
where $\epsilon$ is the energy density, $P \equiv P_x=P_y=P_z$ is the pressure of the medium and 
T being the temperature. With our gravity dual
we can compute the stress tensor $\langle T^{pq}_{\rm med}\rangle$
(and thus the energy $\epsilon= \langle T_{\rm med}^{00}\rangle$ and 
the pressure $P= \langle T^{11}_{\rm med}\rangle$) of the medium through equation of the form (\ref{KS17}), i.e
\bg \label{entropyC2}
\langle T^{pq}_{\rm med}\rangle = \frac{\delta_b {\bf S}_{\rm total}}{\delta_b {\bf g}_{pq}}
\nd 
where again $p,q=0,1,2,3$ and ${\bf g}_{pq}$ is the four dimensional metric obtained from  
the ten dimensional OKS-BH metric
$g_{ij},i,j=0,1..,9$; and $\delta_b$ operation has been defined earlier. 
There are two ways by which we could get a four-dimensional metric from the corresponding 
ten-dimensional one. The first way is to integrate out the $\theta_i, \phi_i$ directions to get the four-dimensional 
effective theory. This is because the warp factor for our case is dependent on the $\theta_i$ directions. The second way 
is to work on a slice in the internal space. The slice is coordinated by choosing some specific values for the internal
angular coordinates. Such a choice is of course ambiguous, and we can only rely on it if the physical quantities that we 
want to extract from our theory is not very sensitive to the choice of the slice. Clearly the first way is much more 
robust but unfortunately not very easy to implement. We will therefore follow the second way by choosing the  
the five dimensional slice 
as $\theta_1=\theta_2=\pi, \psi=\phi_1=\phi_2=0$
and thus obtaining 
\bg \label{entropyC2}
{\bf g}_{\mu\nu}~\equiv~ g_{\mu\nu}(\theta_i=\pi,\psi=\phi_i=0)
\nd 
with $\mu,\nu=0,1,2,3,4$. The next step would be to evaluate all the fluxes and the axio-dilaton on the slice. To do this 
we define:
\bg \label{entropyC3}
&&|{\bf H}_3|^2~=~ |H_3|^2(\theta_i=\pi,\psi=\phi_i=0);~~~~~~|{\bf F}_3|^2~ = ~|\widetilde{F}_3|^2(\theta_i=\pi,\psi=\phi_i=0)\nonumber\\
&& |{{\bf F}}_5|^2~ =~ |\widetilde{F}_5|^2(\theta_i=\pi,\psi=\phi_i=0);~~~~~~~|{\bf F}_1|^2~=~|F_1|^2(\theta_i=\pi,\psi=\phi_i=0)\nonumber\\
&&{\bf \Phi}~=~\Phi(\theta_i=\pi,\psi=\phi_i=0)
\nd
Once the fluxes have been defined, we need the description for ${\bf S}_{\rm total}$ in \eqref{entropyC2}. This is easily
obtained from (\ref{KS77}) as:
\bg \label{entropyC3}
{\bf S}_{\rm total}&=&\frac{1}{2\kappa_{5}^2}\int d^{5}x~e^{-2{\bf \Phi}} \sqrt{-{\bf g}}
\Bigg({\bf R}-4\partial_i {\bf \Phi} \partial^j {\bf \Phi}-\frac{1}{2}|{\bf H}_3|^2\Bigg)\nonumber\\
&-&\frac{1}{2\kappa^2_5} \int d^{5}x
\sqrt{-{\bf g}}\Bigg(|{\bf F}_1|^2+|{{\bf F}}_3|^2+\frac{1}{2}|{\bf F}_5|^2\Bigg)
\nd
 with ${\bf R}$ being the Ricci-scalar for ${\bf g}_{\mu\nu}$ and 
${\bf g}={\rm det}~{\bf g}_{\mu\nu}$.
%\bg \label{entropyC4}
%\bar{R}=R(g_{ij}),~~~~~~~\theta_i=\pi,\psi=\phi_i=0
%\nd
Note that in the definition for the {\it slice} sources ${\bf H}_3, {\bf F}_1, {\bf F}_3,
{\bf F}_5$ and ${\bf R}$, 
we still have $g_{ij}, i,j\geq 5$ which we evaluate at 
$\theta_i=\pi, \psi=\phi_i=0$, treating them simply as functions and not metric degrees of freedom. 

To complete the background we need the line element. Here we will encounter some subtleties regarding the choice of the 
black-hole factors and the corresponding $g_s N_f$ type corrections to them. 
With the definition of ${\bf g}_{\mu\nu}$ the line element is:
\bg \label{entropyC5}
&& ds^2~=~-\frac{\bar{g}_1(r)}{\sqrt{h(r,\pi,\pi)}}dt^2+\frac{\sqrt{h(r,\pi,\pi)}}{\bar{g}_2(r)}dr^2+\frac{1}{\sqrt{h(r,\pi,\pi)}}d\overrightarrow{x}^2\\
&&\bar{g}_1(r)~=~g_1(r,\theta_1=\pi,\theta_2=\pi)=1-\frac{r_h^4}{r^4}+\sum_{i,j=0}^{\infty}\alpha_{ij}\frac{{\rm log}^i(r)}{r^j} = 1 + \sum_{j, \alpha} {\sigma_j^{(\alpha)}\over r^j_{(\alpha)}}\nonumber\\
&&\bar{g}_2(r)~=~g_2(r,\theta_1=\pi,\theta_2=\pi)=1-\frac{r_h^4}{r^4}+\sum_{i,j=0}^{\infty}\beta_{ij}\frac{{\rm log}^i(r)}{r^j}= 1 + \sum_{j, \alpha} {\kappa_j^{(\alpha)}\over r^j_{(\alpha)}}\nonumber
\nd
where $\alpha_{ij},\beta_{ij}$ are all of ${\cal O}(g_sN_f,g_sM)$ and only involve the parameters of the theory 
namely, $r_h, L$ and $\mu$ from the
embedding equation (\ref{embedim}) and guarantees that $\frac{\alpha_{ij}}{r^j},\frac{\beta_{ij}}{r^j}$ are 
dimensionless. On the other hand $\sigma_j^{(\alpha)}, \kappa_j^{(\alpha)}$ can incorporate zeroth orders in $g_sN_f$. 
However note that so far we have been assuming $g_1(r) \approx g_2(r) = g(r)$, ignoring their 
inherent $\theta_i$ dependences, and also the inequality stemming from the choices of $\alpha_{ij}$ and $\beta_{ij}$. 
This will be crucial in what follows, so we will try to keep the black hole factors unequal. These considerations do 
not change any of our previous results of course. 

%jaslea

Now looking at the form of the metric, knowing the warp factor $h(r,\pi,\pi)$ and $\bar{g}_i(r)$, just like before we 
can expand the line
element as $AdS_5$ line element plus ${\cal O}(g_sN_f,g_sM)$ corrections. We can then rewrite the line element 
\eqref{entropyC5} as: 
\bg \label{entropyC6}
&&ds^2~=~ -\frac{r^2}{L^2}\left[g(r)+l_1\right]dt^2+\frac{\sqrt{h(r,\pi,\pi)}}
{\bar{g}_2(r)}dr^2+\frac{r^2}{L^2}\left(1+l_2\right) d\overrightarrow{x}^2\nonumber\\
&&l_1(r)~=~\sum_{i,j=0}^{\infty}\gamma_{ij}\frac{{\rm log}^i(r)}{r^j}\nonumber\\
&&l_2(r)~=~\sum_{i,j=0}^{\infty}\zeta_{ij}\frac{{\rm log}^i(r)}{r^j}
\nd 
where again $\gamma_{ij},\zeta_{ij}$ are of ${\cal O}(g_sN_f,g_sM)$ and we are taking $h(r, \pi, \pi) =  
{L^4\over r^4} + {\cal O}(g_sN_f,g_sM)$. Such a way of writing the local line element tells us that there are two 
induced four-dimensional metrics at any point $r$ along the radial direction:
\bg\label{indme}
{\bf g}^{(0)}_{pq} ~\equiv~ {\rm diagonal}~(-g(r),1, 1, 1), ~~~~~~~ 
{\bf g}^{(1)}_{pq} ~\equiv~ {\rm diagonal}~(-l_1, l_2, l_2, l_2)
\nd
where we haven't shown the $r^2/L^2$ dependences. The reason for specifically isolating the four-dimensional part is to 
show that we can
study the system from boundary point of view where the dynamics will be governed by our choice of the boundary degrees 
of freedom. It should also be clear, from four-dimensional point of view, the metric choice ${\bf g}^{(0)}_{pq}$ 
is directly related to the AdS geometry whereas the other choice ${\bf g}^{(1)}_{pq}$ is the deformation due to 
extra fluxes and seven branes. This decomposition is similar to the decomposition that we studied earlier. 

The above decomposition also has the effect of simplifying our calculations of the energy momentum tensor 
$\langle T^{pq}_{\rm med}\rangle$. We can rewrite the total energy momentum tensor as the sum of two parts, one coming
from the AdS space and the other coming from the deformations, in the following way: 
\bg \label{entropyC7}
\langle T^{pq}_{\rm med} \rangle &=&\frac{\delta_b {\bf S^{[0]}}_{\rm total}}{\delta_b {\bf g}_{pq}^0}
~+ ~ \frac{\delta_b {\bf S}^{[1]}_{\rm total}}{\delta_b{\bf g}_{pq}^1}\nonumber\\
&\equiv& \langle T^{pq}_{\rm med}\rangle_{\rm AdS}~+ ~ \langle T^{pq}_{\rm med}\rangle_{\rm def} \nonumber\\
{\bf S} _{\rm total}&=&{\bf S}^{[0]}_{\rm total}~ + ~ {\bf S}^{[1]}_{\rm total}
\nd 
%popo
where ${\bf S}^{[0]}_{\rm total}$ is zeroth order in $g_sN_f,g_sM$ and  ${\bf S}^{[1]}_{\rm total}$ is higher order in 
$g_sN_f,g_sM$.
Note that $\langle T^{pq}_{\rm med}\rangle_{\rm AdS} =\frac{\delta_b {\bf S}^{[0]}_{\rm total}}
{\delta_b {\bf g}_{pq}^0}$ 
is the well known AdS/CFT result obtained from the analysis of \cite{Kostas-1} \cite{Kostas-2}\cite{Kostas-3}\cite{2}
in the limit $r_c \to \infty$. 
With the
${\cal O}(1/r)$ series expansion of our metric ${\bf g}_{00}^0=1-{r_h^4}/{r^4}, {\bf g}_{11}^0= {\bf g}_{22}^0
= {\bf g}_{33}^0=1$,
the result at the boundary is  
\bg \label{entropyC8}
\langle T^{00}_{\rm med}\rangle_{\rm AdS} &=& \frac{r_h^4}{2g_s^2 G_N}=\frac{{\cal T}^4}{2 g_s^2 G_N}\nonumber\\
\langle T^{mn}_{\rm med}\rangle_{\rm AdS} &=&0~~~~~~~~~~~~~m,n=1,2,3
\nd
This only gives the CFT stress tensor as we evaluate the tensor on the AdS boundary at infinity, reproducing the expected
first term of \eqref{e2a}. 
How do we then 
evaluate the ${\cal O}(g_s N_f, g_s M)$
contributions from the deformed AdS part i.e the energy momentum tensor 
$\langle T^{pq}_{\rm med}\rangle_{\rm def}$ at any $r = r_c$ cut-off in the geometry? 

In fact the procedure to evaluate exactly such a result has already been discussed in the last two sections: namely  
the wake analysis in section 3.3 and shear viscosity analysis in section
3.4. Therefore without going into any details, the final answer
after integrating by parts, adding appropriate Gibbons-Hawkings terms and then using the equation of motion for
 ${\bf g}_{pq}^{[1]}$, we have
\bg \label{action1a} 
&& {\bf S}^{[1]}_{\rm total}~=~{1\over 8\pi G_N}\int \frac{d^4q}{(2\pi)^4 \sqrt{g(r_{\rm max})}}
\Bigg\{\Big[ \bar{C}^{mn}_1(r,q)-\bar{A}^{'mn}_1(r,q)\Big] \Phi^{[1]}_m(r, q) \Phi^{[1]}_n(r, -q)\nonumber\\
&&~~~~~~+\Big[\bar{B}^{mn}_1(r,q)-\bar{A}^{mn}_1(r,q)\Big] \Big[\Phi'^{[1]}_m(r, q) \Phi^{[1]}_n(r, -q)+\Phi^{[1]}_m(r, q)
\Phi'^{[1]}_n(r, -q)\Big]\nonumber\\
&&~~~~~~~~~~~~~~~~ +\Big(\bar{E}^m_1 -\bar{F}'^m_1\Big)\Phi^{[1]}_m(r, q)\Bigg\}\Bigg{|}_{r_h}^{r_{\rm max}}
\nd
where $r_{\rm max} \equiv r_c(1-\zeta)$. The values of the coefficients are given in {\bf Appendix F}. The above form is
exactly as we had before, and so all we now need is to get the mode expansion for $\Phi^{[1]}_m$. Note however that 
the subscript $m$ can take only two values, namely 
$m =0,1$ as there are only two distinct fields ${\bf g}_{00}^{[1]}$ and 
${\bf g}_{11}^{[1]}= {\bf g}_{22}^{[1]}={\bf g}_{33}^{[1]}$. Therefore our proposed mode expansion is:
\bg\label{mexp}
\Phi^{[1]}_m~ =~ {\bf g}_{mm}^{[1]}~= ~\sum_{\alpha}\sum_{i=0}^{\infty}
\frac{{\bf s}_{mm}^{(i)[\alpha]}}{r_{c(\alpha)}^i(1-\zeta)^i}
\nd
Just like the wake and shear viscosity analysis, the action in (\ref{action1a}) is
divergent due to terms of ${\cal O}(r_c^4), {\cal O}(r_c^3)$ and hence we need to renormalise the action. 
The equations for renormalisation are identical to the set of equations
\eqref{cterms}$--$\eqref{ymyn}, and therefore we analogously subtract 
the counter terms to obtain the following renormalised action:
\bg \label{actionren1a}
&& {\bf S}^{[1]}_{\rm ren}= {1\over 8\pi G_N} 
\int \frac{d^4q}{(2\pi)^4}\left[1-{r_h^4\over r_c^4(1-\zeta)^4}\right]^{-{1\over 2}} \sum_{\alpha, \beta}
\Bigg\{\left(\sum_{i=0}^{\infty}{\widetilde{{\cal A}}^{(\alpha)}_{mn(i)[1]}\over r_{(\alpha)}^i}\right) 
\widetilde{{\cal G}}^{mn[1]} \Phi_m
\Phi_n \nonumber\\ && + X[r_{(\alpha)}]
+ \left(\sum_{i=0}^{\infty}{\widetilde{{\cal E}}^{(\alpha)}_{mn(i)[1]}\over r_{(\alpha)}^i}\right) 
\widetilde{{\cal M}}^{mn[1]} (\Phi_m
\Phi'_n+\Phi'_m \Phi_n) 
+{H}_{\vert\alpha\vert}^{mn[1]}\Big[s_{nn}^{(4)[\beta]}\Phi_m +
{s}_{mm}^{(4)[\beta]}\Phi_n\Big] \nonumber\\
&& +{K}_{\vert\alpha\vert}^{mn[1]}\Big[-4\tilde{s}_{nn}^{(4)[\beta]}\Phi_m 
-4 {s}_{mm}^{(4)[\beta]}\Phi_n+{s}_{nn}^{(5)[\beta]}\Phi'_m+{s}_{mm}^{(5)[\beta]}\Phi'_n\Big]
+ \left(\sum_{i=0}^{\infty}\frac{\widetilde{b}^{(\alpha)}_{m(i)[1]}}{r_{(\alpha)}^i}\right) 
\Phi_m\Bigg\}\nonumber\\
\nd
where the radial coordinate is measured at the two boundaries $r_h$ and $r_c(1-\zeta)$ and $\Phi_m$ are independent 
of $r$ as before. Note
that $X[r_{(\alpha)}]$ is a function independent of $\Phi_m$ and appears for generic renormalised action.

Now the generic form for the energy momentum tensor is evident from looking at the linear terms in the 
above action \eqref{actionren1a}. This is again the same as before. However now we also need the entropy from the 
energy-momentum tensor as in \eqref{entropyC1}. The result for the energy-momentum tensor   
at $r = r_c$ is given by:
\bg\label{energymeda}
&&{\langle T^{mm}_{\rm med}\rangle_{\rm def}}\equiv 
{1\over 8\pi G_N} \int \frac{d^4q}{(2\pi)^4}{1\over \sqrt{g(r_c)}}\sum_{\alpha, \beta} 
\Bigg[({H}_{\vert\alpha\vert}^{mn[1]}+ {H}_{\vert\alpha\vert}^{nm[1]}){s}_{nn}^{(4)[\beta]} 
-4({K}_{\vert\alpha\vert}^{mn[1]} \nonumber\\
&&+{K}_{\vert\alpha\vert}^{nm[1]}){s}_{nn}^{(4)[\beta]}+({K}_{\vert\alpha\vert}^{mn[1]}+
{K}_{\vert\alpha\vert}^{nm[1]}){s}_{nn}^{(5)[\beta]}
+\left(\sum_{i=0}^{\infty}\frac{\widetilde{b}^{(\alpha)}_{n(i)[1]}}{r_{c(\alpha)}^i}\right)
 \delta_{nm}\Bigg]
\nd
The  explicit expressions for the coefficients
listed above, namely,
${H}_{\vert\alpha\vert}^{mn[1]}, 
{K}_{\vert\alpha\vert}^{mn[1]},\widetilde{b}^{(\alpha)}_{n(i)[1]}$ and ${s}_{nn}^{(i)[1]}$ are given
in {\bf Appendix F}.

To complete the story we need the contribution from the UV cap. This is similar to our earlier results. The final 
expression for the ratio of the energy-momentum tensor to the temperature takes the simple form:
\bg\label{finent} 
&&{\langle T^{mm}_{\rm med}\rangle_{\rm def}\over T_b} \equiv 
{\pi {\cal T}\sqrt{h({\cal T})}\over 8\pi G_N}\int \frac{d^4q}{(2\pi)^4}\sum_{\alpha, \beta} 
\Big[({H}_{\vert\alpha\vert}^{mn[1]}+ 
{H}_{\vert\alpha\vert}^{nm[1]}){s}_{nn}^{(4)[\beta]} -4({K}_{\vert\alpha\vert}^{mn[1]} \nonumber\\
&&+{K}_{\vert\alpha\vert}^{nm[1]}){s}_{nn}^{(4)[\beta]}+({K}_{\vert\alpha\vert}^{mn[1]}+
{K}_{\vert\alpha\vert}^{nm[1]}){s}_{nn}^{(5)[\beta]}
+ \sum_{j=0}^{\infty}~{\widetilde{b}^{(\alpha)}_{n(j)[1]}}\delta_{nm} {e^{-j{\cal N}_{uv}}} 
\Big]
\nd
We would like to make a few comments here: First, observe that the final result is independent of our choice of 
cut-off. Secondly, in the string frame there should be a $1/g_s^2$ dependence. Finally, we can pull out a 
${\cal T}^4$ term because the coefficients have an explicit $r_h^4$ dependences (see {\bf Appendix F}). This means that 
both from the AdS and the deformed calculations performed above we can show that the entropy is of the form:
\bg\label{entad}
s ~ = ~ {{\cal T}^5 \sqrt{h({\cal T})}\over 2 g_s^2 G_N}
\left[1 + {\cal O}\left(g_sN_f, g_s M, e^{-{\cal N}_{uv}}\right)\right]
\nd 
where the first part is from \eqref{entropyC8} and the second part is from \eqref{finent}. The result for the parent 
cascading theory is \eqref{e2a}, and so we should regard \eqref{entad} as the entropy for the theory with 
${\cal N}_{uv}$ degrees of freedom at the boundary. Of course in the limit ${\cal N}_{uv} \to \epsilon^{-n}, n >> 1$ 
we should recover
the entropy formula \eqref{e2a} for the parent theory. All in all we see that the correction due to 
${\cal N}_{uv}$ degrees of freedom only goes as $e^{-{\cal N}_{uv}}$, so in practice this is always small for the 
type of ${\cal N}_{uv}$ that we consider here.  
This means that we can use the entropy for the parent cascading theory to estimate the viscosity by entropy ratio 
for a system with ${\cal N}_{uv}$ degrees of freedom at the UV as:
\bg \label{final} 
\frac{\eta}{s}~=~ {\left[{1 + \sum_{k = 1}^\infty
\alpha_k e^{-4k {\cal N}_{\rm uv}} \over 4\pi + {1\over \pi}~{\rm log}^2 \left(1 - 
{\cal T}^4 e^{-4{\cal N}_{uv}}\right)}\right]}
\nd
\noindent where we see that the boundary entropy term \eqref{e2a} neatly cancels the ${\cal T}^3$ coefficient in the 
viscosity \eqref{shearvo} to give us the precise bound of ${1\over 4\pi}$ when ${\cal N}_{\rm uv} \to \infty$. Of 
course from our other analysis \eqref{entad} we might expect a
${\cal O}\left(g_sN_f, g_s M, e^{-{\cal N}_{uv}}\right)$ contribution that 
would make 
\eqref{final} saturate the celebrated bound ${1\over 4\pi}$ if the total entropy density factors 
compensate the factors coming from the viscosity. This would seem consistent with, for example,  
\cite{jhul}\footnote{Provided of course if we assume that $\alpha_k$'s are more general now, being functions of 
${\cal T}, g_sM, g_sN_f$. This way even for non-zero $M, N_f$, whenever we have ${\cal N}_{uv} \to \infty$ the 
bound is exactly ${1\over 4\pi}$.}. 
In fact our conjecture would be for non-zero $M, N_f$ and 
${\cal N}_{uv} \to \epsilon^{-n}, n >> 1$, the bound is exactly saturated\footnote{Note that any possible deviations
from ${1\over 4\pi}$ due to \eqref{sigcal} in \eqref{vishu} {\it cannot} happen because the underlying holographic 
renormalisability will make ${\tilde a}_1 = {\tilde b}_1 =0$, as discussed earlier. Thus the bound in itself is a rather 
strong result.}  
i.e ${\eta\over s} = {1\over 4\pi}$.

Our second and final step would be to 
incorporate both the RG flow as well as curvature square corrections. As we 
discussed before the curvature squared corrections are typically of the form
$c_3 R_{\mu\nu\rho\sigma}R^{\mu\nu\rho\sigma}$ with $c_3$ being the coefficient \eqref{coeffeven} 
that we computed before. 

The crucial point here is that (see \cite{Kats} where this has also been recently emphasised) in the presence of 
curvature squared corrections the five dimensional metric itself changes to:  
\bg \label{metricR^2}
ds^2=\frac{-g_1(r)}{\sqrt{h(r, \pi, \pi)}}dt^2+\frac{\sqrt{h(r, \pi, \pi)}}{g_2(r)}dr^2+\frac{d\overrightarrow{x}^2}{\sqrt{h(r, \pi, \pi)}}
\nd 
where the black hole factors $g_i$ are no longer given by \eqref{grdef} or its simplified version \eqref{simlificazion}. 
They take the following forms:
\bg\label{grdefnow}
g_1(r)&=&1-\frac{r_h^4}{r^4}+\alpha + \gamma \frac{r_h^8}{r^4}+\widetilde{\alpha}_{mn}\frac{{\rm log}^m r}{r^n}\nonumber\\
g_2(r)&=&1-\frac{r_h^4}{r^4}+\alpha+\gamma\frac{r_h^8}{r^4}+ \widetilde{\beta}_{mn}\frac{{\rm log}^m r}{r^n}
\nd
where $\widetilde{\alpha}_{mn},\widetilde{\beta}_{mn}$ are all of ${\cal O}(g_sM,g_sN_f)$ and 
can be worked out with some effort (we will not derive their explicit forms here). Similarly we could also 
express \eqref{grdefnow} in terms of inverse powers of $r$ to have good asymptotic behavior.    
Observe that we can still impose $g_1 \approx g_2$ because the corrections are to ${\cal O}(g_sN_f, g_s M)$, although 
all our previous analysis have to be changed in the 
presence of curvature corrections because the explicit values of $g_i(r)$ have changed.  
We will address these issues in the sequel \cite{sequel}. Finally ($\alpha, \gamma$) 
are given by 
\bg\label{alga}
\alpha&=&\frac{4c_3\kappa}{3L^2};~~~~~~~~~\gamma=\frac{4c_3\kappa}{L^2} 
\nd
At this point one might get worried that the metric perturbation on this background would become very complicated. 
On the contrary our analysis becomes rather simple once we 
ignore terms of ${\cal O}(c_3g_sM, c_3g_sN_f)$ (which is a valid approximation with $c_3<<1$). In this limit 
the metric perturbation can be written simply as a {\it linear} combination of  the
terms proportional to $c_3$ in $\Phi$ which appears in \cite{Kats} and our solution
 (\ref{SV-13e}, \ref{SV-13f}) derived for RG flow. The final result is:
 \bg \label{SV-13f1} 
\tilde{\phi}(r, \vert\omega\vert)_{\pm,R^2} &=& 1\pm i\frac{\vert\omega\vert}{4\pi T_c}
\Bigg\{{\cal H} (r) + [1+{\cal G}(r)] {\rm log}~g(r) + \frac{\alpha r^8+\gamma r_h^8}{r^8 g(r)}-\alpha+4\gamma
\frac{r_h^4}{r^4}\Bigg\}\nonumber\\
&+ & {\cal G}(r) - {\vert\omega\vert^2 \over 16 \pi^2 T^2_c}\Big\{{\cal K}(r) 
+ {\cal H}(r) ~{\rm log}~g(r) + [1 + {\cal G}(r)]
{\rm log}^2 g(r)\Big\}\nonumber\\
&&~~~~~~~~~~ + {\cal O}(\vert\omega\vert^3) +  {\cal O}(c_3 g_s N_f) + {\cal O}(c_3 g_s M)\nd
where we have written \eqref{grdef} as;
\bg\label{mother}
&& g_1(r) ~=~ g(r) ~+~ {\cal O}(c_3 g_s N_f)~ +~ {\cal O}(c_3 g_s M)\nonumber\\ 
&& g_2(r) ~=~ g(r) ~+~ {\cal O}(c_3 g_s N_f) ~+~ {\cal O}(c_3 g_s M)
\nd
with $g(r)$ being the usual black hole factor defined in \eqref{simlificazion}. Of course as emphasised above, this is 
valid only in the limit $c_3 << 1$, which at least for our background seems to be the case (see \eqref{coeffeven}).
 
The above corrections are not the only changes. The  
entropy computed earlier
also gets corrected and therefore the horizon can no longer be at $r=r_h$. To evaluate the 
correction to entropy we again ignore the terms of 
${\cal O}(c_3g_sM,c_3g_sN_f)$. In this limit the correction terms are precisely given by the analysis of \cite{Kats} and
are proportional to the $c_3$ factor \eqref{coeffeven} as expected. 
This means that the final result for 
$\eta/s$ including all the ingredients i.e RG flows,
Riemann square corrections as well as the contributions from the UV caps; is given by:   
\bg \label{finala} 
\frac{\eta}{s} &=&~ {\left[{1 + \sum_{k = 1}^\infty
\alpha_k e^{-4k {\cal N}_{\rm uv}} \over 4\pi + {1\over \pi}~{\rm log}^2 \left(1 - 
{\cal T}^4 e^{-4{\cal N}_{uv}}\right)}\right]}\nonumber\\
&-&\frac{c_3\kappa}{3 L^2 \left(1- {\cal T}^4 e^{-4{\cal N}_{uv}}\right)^{3/2}}
 \left[\frac{{B_o}(4\pi^2-{\rm log}^2 ~C_o)+4\pi{A_o}~{\rm log}~C_o}{\Big(4\pi^2-{\rm
log}^2~C_o\Big)^2+16\pi^2~{\rm
log}^2~C_o}\right]
\nd
where we see two things: one, the bound is completely independent of the cut-off $r = r_c$ in the geometry, and two, 
the bound {\it decreases} in the presence of curvature square corrections {even} when 
${\cal N}_{uv} \to \epsilon^{-n}$ with $n = {\cal O}(1)$.\footnote{In \cite{Cherman:2007fj}, 
non-relativistic systems that appear to have no lower bound
were constructed.
However, these are systems which necessarily require large chemical
potentials and low temperature.
In highly relativistic system created at high energy colliders such as the
RHIC or the LHC,
chemical potentials are small and temperature is high. Our discussion here
assumes that the system under
discussion has such properties so that the use of thermodynamic identity
$\varepsilon + P = Ts$ is valid.
Hence, our discussion here is in no direct conflict with the models
constructed in \cite{Cherman:2007fj}.}
The constants appearing in \eqref{finala} are defined as:
\bg\label{constants}  
{C_o}&=&1-{{\cal T}^4}{e^{-4{\cal N}_{uv}}}\nonumber\\
{A_o}&=&-18 {\cal T}^8 e^{-8{\cal N}_{uv}}+\left(3{\cal T}^8 e^{-8{\cal N}_{uv}}
- 47 {\cal T}^4 e^{-4{\cal N}_{uv}}\right){\rm log}~{C_o}+ 26 {\cal T}^4 e^{-4{\cal N}_{uv}} \nonumber\\
&+&24 \left(1 + {\cal T}^2 e^{-2{\cal N}_{uv}}\right){\rm log}~{C_o} \nonumber\\
{B_o}&=&- 88 \pi {\cal T}^8 e^{-8{\cal N}_{uv}}+ 48 \pi {\cal T}^4 e^{-4{\cal N}_{uv}} +48  
\nd
This 
is consistent with \cite{Kats}, and the only violation of $\eta/s$ may be entirely from the $c_3$ factor provided the 
increase in bound 
from the first term of \eqref{finala} is negligible, as we discussed 
earlier for \eqref{final}. This means in particular:
\bg\label{jolba}
{\eta\over s} ~ = ~ {1\over 4\pi} ~-~ n_b c_3 ~+~ {\cal O}\left({\cal T}e^{-{\cal N}_{uv}}\right)
\nd
where $n_b$ can be extracted from \eqref{finala}.  
In this paper however we will not 
study the subsequent implication of this result, for example
whether there exists a causality violation in our theory due to the curvature corrections as in \cite{violation}. 
We hope to 
address this in the sequel \cite{sequel}. 

To see the explicit behavior of $\eta/s$ we can plot the function \eqref{finala} (see {\bf figure 12}). 
Of course our result is 
valid for infinitely large UV degrees of freedom i.e ${\cal N}_{uv} \to \epsilon^{-n}, n \ge 1$, 
but we can extrapolate the result to see what properties we get for small enough UV degrees of freedom. 
Incidentally with the geometry cut-off at small $r_c$ and an AdS cap attached from $r= r_c$ to $r = \infty$ would 
start resembling standard QCD as we briefly mentioned earlier. For such a scenario we expect the 
UV degrees of freedom to be smaller than the UV degrees of freedom 
for the parent cascading theory i.e ${\cal N}_{uv}$ would approach infinity at a smaller rate.   
 
\section{Conclusions}
In this work we have studied the dynamic response of a strongly-coupled, strongly-interacting medium to a fast quark. The drag coefficient and the wake resulting from the parton-medium interaction was evaluated. The calculations were performed by constructing a gravity dual inspired by the Klebanov-Strassler model with D7 branes to take into account fundamental quarks. The gauge dual then has a running coupling constant, unlike models stemming from a pure AdS geometry, with features close to that of QCD. This procedure in fact allows the consideration of a family of gauge theories which have a well-defined completion in the UV, beyond a cut-off scale. We could also show that physical results were independent of the choice of this cut-off. We have applied this model to the calculation of  the drag force and the wake left by a moving quark in the strongly interacting medium were computed. In addition, we have evaluated the ratio of shear viscosity to entropy density  - $\eta/s$ - and shown the violation of the bound conjectured in Ref. \cite{pss}, for a range of parameter values. It is important to test and verify the robustness of this limit, and we view the current work as contributing to this effort.  More work is needed in order to identify the size and extent of this violation, at the moment this violation is only parametric and our parameters need to have better defined physical origins. In the end, one may need to rely on the empirical identification of key quantities, like transport coefficients for example \cite{Luzum}. In this regard, the role of heavy ion experiments at RHIC and at the LHC can't be overestimated. 

\begin{figure}[htb]\label{ETAoS}
		\begin{center}
                \includegraphics[height=13cm,angle=-90]{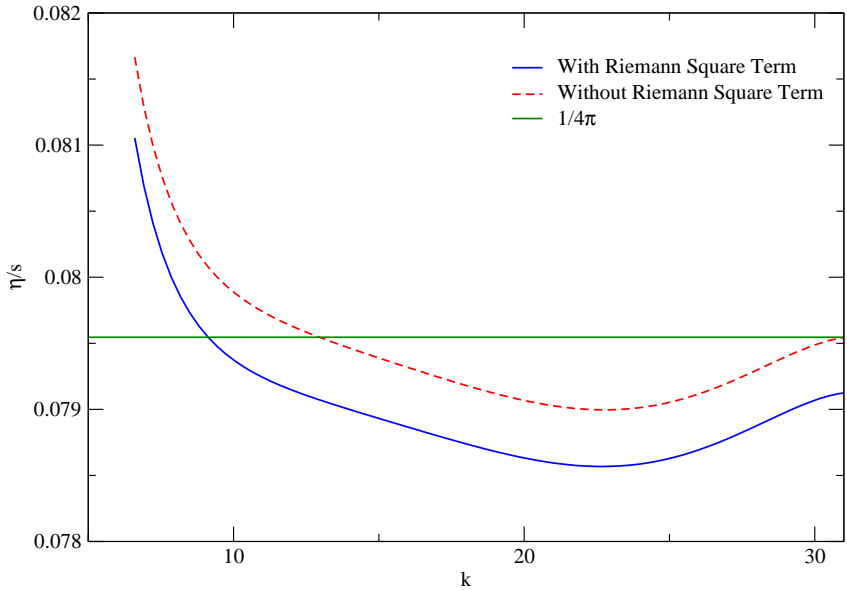}
                		\caption{{Plot for $\eta/s$ with and without Riemann Square term. 
The x-axis is defined as 
$k = {e^{{\cal N}_{\rm uv}}\over {\cal T}}$ where ${\cal N}_{\rm uv}$ is the UV degrees of freedom and 
${\cal T}$ is the characteristic temperature of the cascading theories. For the parent cascading theory $k \to \infty$
and we see a violation of the bound (the solid blue line). As $k$ decreases (assuming this is possible) the red dashed 
line dips slightly below the $1/4\pi$ axis, but the solid blue line remains considerable below the $1/4\pi$ axis. For 
$k$ sufficiently small the bound is not violated. However all the models that we studied in this 
paper can {\it only} realise
the large $k$ limit.}}  
		\end{center}
		\end{figure}

\noindent {\bf Note added:} In our recent paper \cite{jpsi}, we have given a concrete example of a UV complete 
theory that consistently proves all the statements that we made here. Our new geometry has a UV cap given by 
an asymptotic AdS-Schwarzchild geometry. The IR dynamics of our theory is captured by the OKS-BH geometry. The 
deformation of the OKS-BH geometry at the boundary is indeed captured by an 
{\it interpolating} geometry that ties smoothly,
in the presence of sources and fluxes, to the asymptotic AdS-Schwarzchild geometry. All the relevant details like 
changes in stress-tensor, entropy etc due to the UV cap can now be calculated. In addition to that we have been 
able to study zero temperature linear confinement as well as high temperature quarkonium suppression and melting from
our UV complete theory. Our results are also consistent with many of the earlier predictions made in the literature
using completely different techniques than ours \cite{reyyee, brambilla, polstra, boschi}.

\section*{Acknowledgements}

We are especially grateful to Matt Strassler for patiently explaining the subtleties of cascading theories, and to Ofer
Aharony for many clarifications about boundary degrees of freedom, and other things. 
We would also like to thank Alex Buchel, Simon Caron-Huot, Paul 
Chessler, 
Andrew Frey, Jaume Gomis, Evgeny Kats, Ingo Kirsch,  
Rob Myers, Peter Ouyang,
Omid Saremi, Aninda Sinha, Diana Vaman and Larry Yaffe for many helpful discussions and 
correspondence. M.M would like to thank specifically the 
organizers and participants 
of {\it Strong and Electroweak Matter 2008 (SEWM)} as well the {\it McGill Workshop on AdS/QCD Duality (2008)} - where  
preliminary versions of this work were presented -  for comments and criticisms. K.D would like to thank the 
organizers and the participants of Banff workshop on  {\it Gauge Fields, Cosmology and Mathematical 
String Theory (2009)} for 
many useful comments. 
This work was supported in 
part by the Natural Sciences and Engineering Research Council of Canada, and in part by McGill University.

\newpage

%\section{Appendix}
\appendix
\section{Back reaction effects in the AdS Black-Hole background: A toy example}

%\subsection{Back reaction effects in the AdS Black-Hole background: A toy example}

The following analysis, although independent of the main calculations in the paper, serves as an interesting 
warm-up example where we can study metric perturbations due to fluxes and D7 brane in a controlled AdS background. 
This will prepare us for the analysis of the next section where we study metric perturbations in the non-trivial 
OKS-BH geometry due to fluxes, D7 brane and strings. 
   
To start-off consider an $AdS_5$ {Black Hole} (AdS-BH) metric given in the following way:
\bg \label{1}
ds^2&=&\frac{1}{u^2}\Big[-f(u)dt^2+dx_1^2+dx_2^2+dx_3^2+\frac{du^2}{f(u)} \Big]\nonumber\\
&=&g^{\rm f}_{\mu\nu}(u)dx^\mu dx^\nu
\nd
with $f(u)=1-{u^4\over u_h^4}$, $u_h$ is black hole horizon, $u=0$ is the boundary, and 
$\mu\; {\rm or} \; \nu=0,1,2,3,4$. 
The above metric is asymptotically AdS \cite{2} as
\bg\label{asadsa}
u^2 g^{\rm f}_{\mu\nu}\arrowvert_{u=0}~=~\eta_{ij}
\nd
where $\eta_{ij}$ is the Minkowski metric, with $i\;{\rm or}\; j=0,1,2,3$. Note that we have expressed the metric 
in terms of $u \equiv {1\over r}$ coordinate. We could as well express everything in terms of $r$ coordinates (as we 
did in the main text). 

To this background, first let us now add a D7 brane by switching off the black-hole factor. The
D brane Born Infeld (DBI) action for D7 brane in 5 dimensional AdS-BH background is given by \cite{3}
\bg \label{2}
S_{\rm DBI}~=~ \int d^5x {\cal L}(x) ~=~
g_s\int d^5x  \sqrt{g(x)} ~{\rm cos}^3\Phi(x) ~\sqrt{1+g^{\mu\nu}(x)\partial_\mu \Phi(x)\partial_\nu \Phi(x)}\nonumber\\
\nd
where the metric $g_{\mu\nu}$ is the metric of the resulting background, i.e
$g_{\mu\nu}(x)~=~g^{{\rm f}=1}_{\mu\nu}(x)$.
In the above, $g_s$ is coupling constant, and 
$\Phi(x)$ is a scalar field describing the D7 brane. The Euler-Lagrangian equation for the $\Phi(x)$
field leads to:
\bg \label{4}
\square \Phi +3 {\rm tan}~\Phi -\frac{1}{2} \frac{g^{\mu\nu}\partial_{\mu}
\partial_{\nu} \Big(g^{\alpha\beta}\partial_{\alpha}\Phi\partial_{\beta}\Phi\Big)}{1+g^{\alpha\beta}\partial_{\alpha}\Phi\partial_{\beta}\Phi}=0
\nd
where $\square =\frac{\partial_\mu\Big(\sqrt{g} g^{\mu\nu}\partial_{\nu}\Big)}{\sqrt{g}}$. As (\ref{4}) is a second order
non linear differential equation, the solution can be written in the form \cite{3}: 
\bg \label{5}
\Phi(u,\overrightarrow{x},t)~ = ~ u \sum_{i=0}^{\infty}\Big[\phi_i(\overrightarrow{x},t) u^i~+~ 
\psi_i(\overrightarrow{x},t) u^i ~{\rm log}~ u\Big] 
\nd 
Now considering first order ${\cal O}(g_s)$ in the perturbation and terms only upto ${\cal O}(u^3)$, we find (see 
\cite{3} for details),
\bg\label{6}
\Phi(u,\overrightarrow{x},t)~= ~ c u ~+~\frac{c^3 u^3}{6}
\nd
where $c$ is a constant. Using the lagrangian (\ref{2}) and the solution (\ref{6}), 
we can find the stress energy tensor for 
the D7 brane:
\bg \label{7}
T^{\mu}_\nu(x)&=&\frac{\partial {\cal L}}{\partial (\partial_\mu \Phi(x))}\partial_\nu \Phi(x)-{\cal L}\delta^\mu_\nu\nonumber\\
&=&\sum_{i=1}^{\infty}g_s^i~T^{\mu(i)}_\nu(x)
\nd
where in the last line we have expressed this in powers of $g_s$ to make it more generic. In fact, as we will see below, 
the energy momentum tensor can be assumed to come from various sources that include the D7 brane as well as background 
fluxes. 

Once we introduce D7 brane as well as the black-hole factor $f$ we need to modify the scenario. In particular
there will be non-trivial axio-dilaton background that we have been ignoring. 
This will generally back-react
on the geometry and make the system non-AdS. To preserve the AdS like configuration we will need more D7 branes 
(and also orientifold planes so that 
we could lift the system to a F-theory configuration \cite{vafaF}) 
as well as other background fluxes. Let us denote the action of
all the fluxes etc. in five dimensional space as $S_{\rm fluxes}$, the action of the O-planes as $S_{\rm planes}$ and 
the action of the string from the D7 branes to the black-hole as $S_{\rm strings}$. 
This means that our total action is:  
\bg \label{8}
S_{total}=\int d^5x \sqrt{G}~R~+~S_{\rm DBI} ~ + ~ S_{\rm fluxes} ~ + ~ S_{\rm planes} ~ + ~ S_{\rm strings}
\nd
where $R$ is the Ricci scalar for the metric $G_{\mu\nu} = g^{\rm f}_{\mu\nu} + g_s h_{\mu\nu}$, 
and $S_{\rm DBI}$ now include the total action of all the D7 
branes. We then expect the 
Euler-Lagrange equation for $G_{\mu\nu}$ using the action 
(\ref{8}) determines the Einstein tensor: 
\bg \label{9}
{\cal G}_{\mu\nu}(x)~=~ \sum_{i=1}^{\infty}~g_s^i~{\cal G}_{\mu\nu}^{(i)}(x)~
=~ \sum_{i=1}^{\infty}~g_s^i~ {\cal T}^{(i)}_{\mu\nu}(x)
\nd
where the energy-momentum tensor is determined by varying the actions $S_{\rm DBI} + S_{\rm fluxes} + S_{\rm planes}$
wrt the background metric $G_{\mu\nu}$. Now
equating coefficients of $g_s$ in the above expansion, we have the following equation for $h_{\mu\nu}$: 
\bg \label{10}
{\cal G}_{\mu\nu}^{(1)}[h_{\mu\nu}(x)]~= ~ {\cal T}^{(1)}_{\mu\nu}(x)
\nd
Notice that as $h_{\mu\nu}$ is symmetric tensor, it has fifteen degrees of freedom. 
We use coordinate transformations to fix ten degrees of
freedom and treat $h_{\mu\nu}$ as a diagonal matrix. 
Consider the coordinate transformation $x^a\rightarrow x'^a=x^a+g_s e^a$. Under this
transformation 
\bg\label{jomad}
h_{\mu\nu}~\rightarrow ~h_{\mu\nu}~-~ D_\mu e_\nu~ - ~ D_\nu e_\mu
\nd
which is a  
gauge transformation. Thus we can fix
five components (say $h_{01}=h_{02}=h_{03}=h_{04}=h_{12}=0$) with the above transformations. 
Just like in electromagnetism, this does not completely fix the gauge. We can add to $e^a$
another set of functions $e'^a$ which obey $\square e'^a=0$ and 
leaves Einstein tensor unchanged. Demanding that $e'^a$ also leaves the fixed
components invariant, we can further fix another five components 
(say $h_{13}=h_{14}=h_{23}=h_{24}=h_{34}=0$). Thus we have
diagonal $h_{\mu\nu}$. Observe that this is not the most generic choice, but will suffice for this toy example 
provided we can have the energy-momentum tensors to be diagonal.
We will also assume that the fluxes and branes/planes are independent of 
$\overrightarrow{x},t$ and thus the metric perturbations $h_{\mu\nu}$ they induce
should only depend on $u$ and we should have $h_{11}=h_{22}=h_{33}$. 
With all these, we have 
\bg\label{haluaK} h_{\mu\nu} ~ = ~ 
\begin{pmatrix} k(u)~~ & 0~~ & 0~~ & 0 ~~ & 0\\
& & & & \\
0~~ & h(u)~~ & 0 ~~ & 0 ~~ & 0\\& & & &\\
0 ~~& 0~~ & h(u)~~ & 0 ~~ & 0\\& & & & \\
0~~ & 0 ~~& 0 ~~ & h(u) ~~ & 0\\ & & & &\\
0~~ & 0 ~~& 0 ~~ & 0 ~~ & l(u)
\end{pmatrix}
\nd
We observe that (\ref{10}) is a second order non linear differential equation for $h_{\mu\nu}$. 
This will have a general solution:
\bg \label{11}
m(u)= \frac{1}{u}\sum_{j}^{\infty}\Big[m_j u^j~+~ \tilde{m}_j u^j ~{\rm log~u}\Big]      
\nd
where $m(u)=k(u), h(u), l(u)$ and $m_j=k_j, h_j, l_j$, $\tilde{m}_j=\tilde{k}_j, \tilde{h}_j, \tilde{l}_j$ are constants.

As we mentioned before the above form of the metric \eqref{haluaK} will make sense if the sources can also be 
made diagonal. In the presence of fluxes, branes, planes and non-trivial profile of the strings this doesn't 
look very difficult to achieve. We will therefore assume that the energy-momentum tensors are also diagonal, and 
the right hand side of (\ref{10}) can be written as a Lorentz series as:
\bg \label{12}
{\cal T}^{(1)}_{\mu\nu}(u)~ = ~ \frac{t^{(1)}_{\mu\nu(-7)}}{u^7}
~+ ~ \frac{t^{(1)}_{\mu\nu(-6)}}{u^6}~+ ~\cdots ~ + ~ t^{(1)}_{\mu\nu(3)}u^3+\cdots
\nd
Once \eqref{12} is made diagonal, we can satisfy \eqref{10} easily. 
The explicit expressions for the Einstein tensor are:
\bg \label{13}
{\cal G}_{00}^{(1)}(u)&=&\frac{3}{4u^2u_h^{12}}\Big[\Big(-4u^8u_h^8+2u^{12}u_h^4+2u^4u_h^{12}\Big)\frac{d^2h(u)}{du^2}
+\Big(6u^{11}u_h^4-
8u^7u_h^8+2u^3u_h^{12}\Big)\frac{dh(u)}{du}\nonumber\\
&-&8\Big((u_h^4-u^4)u^2u_h^8\Big)h(u)
+\Big(2u^3u_h^{12}-6u^7u_h^8+6u^{11}u_h^4-2u^{15}\Big)\frac{dl(u)}{du}
+\Big(12u^2u_h^{12}\nonumber\\
&-&36u^6u_h^8
+36u^{10}u_h^4-12u^{14}-8(u_h^4-u^4)(2u^2u_h^8-2u^6)\Big)l(u)
-8u^2u_h^{12}k(u)\Big]\\ 
{\cal G}_{ii}^{(1)}(u)&=&\frac{1}{B(u)}\Big[\Big(-4u^8u_h^{16}+4u^{16}u_h^4-12u^{12}u_h^{8}+12u^{8}u_h^{12}\Big)\frac{d^2h(u)}{du^2}
\nonumber\\
&+&\Big(-4u^{3}u_h^{16}+28u^7u_h^{12}-44u^{11}u_h^{8}+20u^{15}u_h^{4}\Big)\frac{dh(u)}{dr}\nonumber\\
&+&\Big(80u^{6}u_h^{12}
-32u^2u_h^{16}-64u^{10}u_h^{8}+16u^{14}u_h^{4}-(96u^4u_h^{16}(u_h^4-u^4)^4)B^{-1}(u)\Big)h(u)\nonumber\\
&+&\Big(20u^7u_h^{12}-6u^3u_h^{16}
-24u^{11}u_h^8+12u^{15}u_h^4-2u^{19}\Big)\frac{dl(u)}{du}
+\Big(-36u^2u_h^{16}-176u^{10}u_h^8\nonumber\\
&+&88u^{14}u_h^4
+136u^{6}u_h^{12}-12u^{12}u_h^{18}-(192u^4u_h^{12}(u_h^4-u^4)^5)B^{-1}(u)\Big)l(u)\nonumber\\
&+&\Big(2u^4u_h^{16}+2u^{12}u_h^8-4u^{8}u_h^{12}\Big)\frac{d^2k(u)}{du^2}
+\Big(-2u^{11}u_h^{8}+2u^{3}u_h^{16}\Big)\frac{dk(u)}{du}\nonumber\\
&+&\Big(40u^2u_h^{16}-24u^6u_h^{12}-(96u^4u_h^{20}(u_h^4-u^4)^3)/B(u)\Big)k(u)\Big]\\ 
{\cal G}_{44}^{(1)}(u)&=&\frac{3}{C(u)}\Big[\Big(-2u^7+6u^{3}u_h^4\Big)\frac{dh(u)}{du} +\Big(32u^4u_h^8C^{-1}(u)\Big)k(u)
\nonumber\\
&& ~~~~~~~~~~~+~\Big(-4u^6-4u^2u_h^4+64u^4u_h^4(-u_h^4-u^4)C^{-1}(u)\Big)h(u)\Big]
\nd
where we have defined $B(u)$ and $C(u)$ in the following way:
\bg\label{bcdefin}
B(u)~= ~ 4u^2u_h^8(u_h^4-u^4)^2,~~~~~~~~~C(u)~=~ -4u^2(u_h^4-u^4)
\nd 
with $ii~= ~11, 22, 33$ from here on due to our assumed diagonal nature of the perturbation. Once the perturbations 
become non-diagonal the analysis will become more involved as we saw in section 3.3. 
 
By matching coefficients of various powers of $u$ in (\ref{10}) we can solve for $h_{\mu\nu}$ order by order. 
We obtain equations
for the constants $m_j, \tilde{m}_j$ by plugging in (\ref{11}) in (\ref{10}) and using the explicit expressions
(\ref{13}) onwards
and the expansion (\ref{12}). Doing so, we observe that it is sufficient to have $m_j, \tilde{m}_j = 0$ for $j<-6$. 

The equations obtained by matching coefficients of $\frac{1}{u^j}$ for $-6\leq j \leq -3$
are:
\bg \label{15}
 \frac{4t^{(1)}_{00(j-1)}}{3} ~ = ~ 
2(-3-2j+j^2)h_{j}+(-6+2j)l_{j}-8k_{j}+2(j^2+j-2)\tilde{h}_j+2\tilde{l}_j \nonumber
\nd
\bg\label{jokla}
&&4t^{(1)}_{ii(j-1)} ~ = ~ -4(3-2j+j^2)h_{j}-6(-3+j)l_{j}+2(-23-2j+j^2)k_{j}\nonumber\\
&& ~~~~~~~~~~~~~~~~~~~~~~~~ -4(-2+j+j^2)\tilde{h}_j
-6\tilde{l}_j+2(-2+j+j^2)\tilde{k}_j\nonumber\\
&& \frac{4t^{(1)}_{44(j-1)}}{3} ~ = ~ -(-2+6j)h_j+(6+2j)k_{j}-6\tilde{h}_j+2\tilde{k}_j
\nd
where because of the form of ${\cal G}_{\mu\nu}$ all the above equations are independent of $u_h$. This will not be
the case for $j > -3$. Finally,  
matching coefficients of $\frac{{\rm log}~u}{u^j}$ again for $-6\leq j \leq -3$ gives us: 
\bg \label{16}
&&2(-3-j)\tilde{h}_j+2(-3+j)\tilde{l}_j-8\tilde{k}_j~=~0\nonumber\\
&& -4(2-j)\tilde{h}_j-6(-3+j)\tilde{l}_j+2(-23-j)\tilde{k}_j~=~0\nonumber\\
&& (-2+6j)\tilde{h}_j+(-6-2j)\tilde{k}~=~0
\nd 
For a given $-6\leq j \leq -3$, we have six equations and six unknowns in (\ref{15}) and (\ref{16}) 
and thus we can solve exactly. For $j>-3$, as we mentioned earlier,
equations governing $m_j, \tilde{m}_j$ will in general depend\footnote{Also on $u_h$. This can be easily checked by 
plugging in the mode expansion for $m_j, {\tilde m}_j$ in ${\cal G}_{\mu\nu}$ given above.} on
$m_{j-p}, \tilde{m}_{j-p}$ for 
$0<p\leq j+6$. Once we solve for $m_{j-p},\tilde{m}_{j-p}$, we can obtain exact solutions for
all $m_j,\tilde{m}_j$ with $j > -3$. This way we have exact solutions for all $j$. 
For $-6\leq j \leq -3$ the solutions are: 
\bg \label{17}
h_j&=&\frac{(3+j)}{(2j^3-3j^2-62j-5)}4t^{(1)}_{00(j-1)}+
\frac{j^3-9j^2-61j+5}{(2j^3-3j^2-62j-5)(-3+j)}t^{(1)}_{ii(j-1)}\nonumber\\
&+&\frac{(-1+3j)}{((2j^3-3j^2-62j-5)}t^{(1)}_{44(j-1)}\\
l_j&=&\frac{(3+j)}{3(2j^3-3j^2-62j-5)}t^{(1)}_{00(j-1)}-
\frac{(-5-21j+j^2+j^3)}{((2j^3-3j^2-62j-5)(-3+j))}t^{(1)}_{ii(j-1)}\nonumber\\
&+&\frac{(-1+3j)}{3(2j^3-3j^2-62j-5)}t^{(1)}_{44(j-1)}\\
k_j&=&-\frac{(j^2-2j-35)}{3(2j^3-3j^2-62j-5)}t^{(1)}_{00(j-1)}
+\frac{(45+68j-30j^2-4j^3+j^4)}{((-3+j)(2j^3-3j^2-62j-5))}t^{(1)}_{ii(j-1)}\nonumber\\
&+&\frac{(-2j-15+j^2)}{3(2j^3-3j^2-62j-5)}t^{(1)}_{44(j-1)}\\
\tilde{h}_j&=&\frac{(2j^4+9j^3+11j^2-21j-181)}{2((2j^3-3j^2-62j-5)(j^2+30j+1))}t^{(1)}_{00(j-1)}\nonumber\\
&-&\frac{(107j^5+324j^4-1932j^3-3614j^2+697j-1790)}{2((j^2+30j+1)(-3+j)^2(2j^3-3j^2-62j-5))}t^{(1)}_{ii(j-1)}\nonumber\\
&+&\frac{(6j^4+7j^3-17j^2+7j+77)}{2((2j^3-3j^2-62j-5)(j^2+30j+1))}t^{(1)}_{44(j-1)}
\nd
\bg \label{18}
\tilde{l}_j&=&\frac{(2j^4+9j^3+11j^2-21j-181)}{6((2j^3-3j^2-62j-5)(j^2+30j+1))}t^{(1)}_{00(j-1)}\nonumber\\
&-&\frac{(33j^5+32j^4-444j^3+1286j^2+515j-590)}{2((j^2+30j+1)(-3+j)^2(2j^3-3j^2-62j-5))}t^{(1)}_{ii(j-1)}\nonumber\\
&+&\frac{(6j^4+7j^3-17j^2+7j+77)}{6((2j^3-3j^2-62j-5)(j^2+30j+1))}t^{(1)}_{44(j-1)}\\
\tilde{k}_j&=&\frac{(-13j^3-99j^2+177j+2131)}{6((2j^3-3j^2-62j-5)(j^2+30j+1))}t^{(1)}_{00(j-1)}\nonumber\\
&-&\frac{(2j^7+50j^6-343j^5-1342j^4+5682j^3-2384j^2-4869j+6980)}{2((j^2+30j+1)(-3+j)^2(2j^3-3j^2-62j-5))}t^{(1)}_{ii(j-1)}\nonumber\\
&-&\frac{(31j^3-33j^2+71j+907)}{6((2j^3-3j^2-62j-5)(j^2+30j+1))}t^{(1)}_{44(j-1)}
\nd
This way we get the background in the limit where the back reactions from fluxes, branes, planes and strings are small. In 
the presence of off-diagonal energy-momentum tensor, the analysis will have to change but the underlying physics 
will remain unchanged.

\newpage

\section{Operator equations for metric fluctuations}

As we discussed in section 3.3, 
in the limit where the background fluxes, including the 
effects of the D7 brane, 
are very small
the equation \eqref{lalbe}
can be presented as an operator equation of the following form:
\bg \label{KS7b}
\kappa\; \triangle_{\mu\nu}^{\alpha\beta} l_{\alpha\beta}(x) \approx 
T_{\mu\nu}^{\rm string}(x)
\nd
where $x$ is a generic five-dimensional coordinate and $T_{\mu\nu}^{\rm string}(x)$ is the energy momentum of the 
string. 
In this final form,
the operator $\triangle_{\mu\nu}^{\alpha\beta}$ is a second order differential operator derived from
(\ref{KS7a}), with $\mu,\nu,\alpha, \beta = 0,1,2,3,4$. 

Recall now that
where we have used coordinate transformations to fix the 
five components of $l_{\mu\nu}$, namely
$l_{4\mu}=0,\mu=0,1,2,3,4$. An additional residual gauge transformation allow us to 
eliminate another five degrees of
freedom and we end up with five physical degrees of freedom i.e. five independent metric perturbation using
which all other components could be expressed. Alternatively one can use certain combinations of the fifteen 
components to write five independent degrees of freedom for the metric fluctuations.

The above result is easy to demonstrate for the AdS space, as has already been discussed in \cite{Yaffe-1}. For non-AdS
spaces this is not so easy to construct. Therefore
in the following we will take the complete set of ten components and using them we will determine the 
triangle operator $\triangle_{\mu\nu}^{\alpha\beta}$ in \eqref{lalbe}. If the ten components that can be 
labelled as a set: 
\bg\label{tencomp2}
l_n = \big\{l_{00},~~l_{01},~~l_{02},~~l_{03},~~l_{11},~~l_{12},~~l_{13},~~l_{22},~~l_{23},~~l_{33}\big\}
\nd
then the operator $\triangle_{\mu\nu}^{\alpha\beta}$ in \eqref{lalbe} gives rise to 77 equations that we present 
below. The warp factor $h$ appearing in these equations can be  
taken to be $h=h(r,\pi,\pi)$ because we will analyse fluctuations close to the string and therefore our
choice of background will be:
\bg
ds^2&=&L_{\mu\nu} dx^{\mu} dx^{\nu}\nonumber\\
L_{00}(t,r,x,y,z)&=&\frac{-g(r)+\kappa l_{00}(t,r,x,y,z)}{h(r)^{1/2}}\nonumber\\
L_{01}(t,r,x,y,z)&=&\frac{\kappa l_{01}(t,r,x,y,z)}{h(r)^{1/2}} \nonumber\\
 L_{02}(t,r,x,y,z)  &=& \frac{\kappa l_{02}(t,r,x,y,z))}{h(r)^{1/2}} \nonumber\\
L_{03}(t,r,x,y,z)&=&\frac{\kappa l_{03}(t,r,x,y,z)}{h(r)^{1/2}} \nonumber\\
L_{11}(t,r,x,y,z)&=&\frac{h(r)^{1/2}+\kappa l_{11}(t,r,x,y,z)}{g(r)}\nonumber\\
L_{12}(t,r,x,y,z)  &=& \frac{\kappa l_{12}(t,r,x,y,z)}{h(r)^{1/2}} \nonumber
\nd
\bg
L_{13}(t,r,x,y,z)  &=& \frac{\kappa l_{13}(t,r,x,y,z)}{h(r)^{1/2}} \nonumber\\
L_{22}(t,r,x,y,z)&=&\frac{1+\kappa l_{22}(t,r,x,y,z)}{h(r)^{1/2}}\nonumber\\
L_{23}(t,r,x,y,z)   &=&   \frac{\kappa l_{23}(t,r,x,y,z)}{h(r)^{1/2}}\nonumber\\
L_{33}(t,r,x,y,z)&=&\frac{1+\kappa l_{33}(t,r,x,y,z)}{h(r)^{1/2}}
\nd
Therefore using all the considerations, the explicit forms for $\triangle_{\mu\nu}^{\alpha\beta}$ can be presented as:
\bg \label{Lin_einstein_00}
&&\triangle^{00}_{00}=-\frac{1}{16h^{\frac{7}{2}}}\left(12gh^{\frac{3}{2}}\frac{\partial^2 h}{\partial r^2}-21\left(\frac{\partial h}{\partial
r}\right)^2g\sqrt{h}+6\frac{\partial h}{\partial r}\frac{\partial g}{\partial r}h^{\frac{3}{2}}\right)\nonumber\\
&&\triangle^{11}_{00}=-\frac{1}{16h^{\frac{7}{2}}}\left(12g^2h\frac{\partial^2 h}{\partial r^2}+8gh^3\left(\frac{\partial^2 }{\partial
x^2}+\frac{\partial^2 }{\partial y^2}\right)-24g^2\frac{\partial h}{\partial r}+6g^2h\frac{\partial h}{\partial r}\frac{\partial }{\partial r}
+6gh\frac{\partial h}{\partial r}\frac{\partial g}{\partial r}\right)\nonumber\\
&&\triangle^{22}_{00}=-\frac{1}{16h^{\frac{7}{2}}}\left(8gh^{\frac{7}{2}}\left(\frac{\partial^2 }{\partial
y^2}+\frac{\partial^2 }{\partial z^2}\right)+8g^2h^{\frac{5}{2}}\frac{\partial^2 }{\partial
r^2}-10g^2h^{\frac{3}{2}}\frac{\partial h}{\partial r}\frac{\partial }{\partial
r}+4gh^{\frac{5}{2}}\frac{\partial g}{\partial r}\frac{\partial }{\partial r}\right)\nonumber\\
&&\triangle^{33}_{00}=-\frac{1}{16h^{\frac{7}{2}}}\left(8g^2h^{\frac{5}{2}}\frac{\partial^2 }{\partial r^2}+8gh^{\frac{7}{2}}
\left(\frac{\partial^2 }{\partial x^2}+\frac{\partial^2 }{\partial z^2}\right)-10g^2h^{\frac{3}{4}}\frac{\partial h}{\partial
r}\frac{\partial }{\partial r}+4gh^{\frac{5}{2}}\frac{\partial }{\partial r}\right)\nonumber\\
&&\triangle^{12}_{00}=-\frac{1}{16h^{\frac{7}{2}}}\left(16g^2h^{\frac{5}{2}}\frac{\partial^2 }{\partial r\partial x}+20g^2h^{\frac{3}{2}}\frac{\partial h}{\partial
r}\frac{\partial }{\partial x}-8gh^{\frac{5}{2}}\frac{\partial g}{\partial
r}\frac{\partial }{\partial x}\right)\nonumber\\
&&\triangle^{13}_{00}=-\frac{1}{16h^{\frac{7}{2}}}\left(16g^2h^{\frac{5}{2}}\frac{\partial^2 }{\partial r\partial y}+20g^2h^{\frac{3}{2}}\frac{\partial h}{\partial
r}\frac{\partial }{\partial y}-8gh^{\frac{5}{2}}\frac{\partial g}{\partial
r}\frac{\partial }{\partial y}\right)\nonumber\\
&&\triangle^{13}_{00}=-\frac{1}{16h^{\frac{7}{2}}}\left(-16gh^{\frac{7}{2}}\frac{\partial^2 }{\partial x\partial y}
\right)\nonumber\\
%\nd 
%\bg \label{Lin_einstein_11}
&&\triangle^{02}_{11}=-\frac{1}{8hg^2}\left(8h^2\frac{\partial^2 }{\partial t \partial x}\right)\nonumber\\
&&\triangle^{03}_{11}=-\frac{1}{8hg^2}\left(8h^2\frac{\partial^2 }{\partial t \partial y}\right)\nonumber\\
&&\triangle^{00}_{11}=-\frac{1}{8hg^2}\left(-4h^2\left(\frac{\partial^2 }{\partial z^2}+\frac{\partial^2 }{\partial y^2}+\frac{\partial^2
}{\partial x^2}\right)-3\frac{\partial h}{\partial
r}\frac{\partial g}{\partial r}+3g\frac{\partial h}{\partial
r}\frac{\partial }{\partial r}\right)\nonumber\\
&&\triangle^{33}_{11}=-\frac{1}{8hg^2}\left(-4h^2\frac{\partial^2 }{\partial t^2}+4h^2g\left(\frac{\partial^2 }{\partial x^2}+\frac{\partial^2
}{\partial z^2}\right)-3g^2\frac{\partial h}{\partial r}\frac{\partial }{\partial r}+2gh{\partial g}{\partial r}\frac{\partial }{\partial r}\right)\nonumber\\
&&\triangle^{22}_{11}=-\frac{1}{8hg^2}\left(-4h^2\frac{\partial^2 }{\partial t^2}+4h^2g\left(\frac{\partial^2 }{\partial y^2}+\frac{\partial^2
}{\partial z^2}\right)-3g^2\frac{\partial h}{\partial r}\frac{\partial }{\partial r}+2gh\frac{\partial g}{\partial r}\frac{\partial }{\partial r}\right)\nonumber\\
&&\triangle^{23}_{11}=-\frac{1}{8hg^2}\left(-8h^2g\frac{\partial^2 }{\partial x \partial y}\right)\nonumber\\
&&\triangle^{13}_{11}=-\frac{1}{8hg^2}\left(6g^2\frac{\partial h }{\partial r}\frac{\partial  }{\partial y}-4gh\frac{\partial g}{\partial r}\frac{\partial  }{\partial y}\right)\nonumber
\nd
\bg
&&\triangle^{12}_{11}=-\frac{1}{8hg^2}\left(6g^2\frac{\partial h }{\partial r}\frac{\partial  }{\partial x}-4gh\frac{\partial g}{\partial
r}\frac{\partial  }{\partial x}\right)\nonumber\\
&&\triangle^{01}_{11}=-\frac{1}{8hg^2}\left(6g\frac{\partial h }{\partial r}\frac{\partial  }{\partial t}\right)
\nonumber\\
&&\triangle^{11}_{22}=\frac{1}{16h^{\frac{7}{2}}g^2}\Bigg(-4g^2h^2\frac{\partial g }{\partial r}\frac{\partial  }{\partial
r}+6g^3h\frac{\partial h}{\partial r}\frac{\partial  }{\partial r}-24g^3\left(\frac{\partial g }{\partial r}\right)^2+18g^2h\frac{\partial g
}{\partial r}\frac{\partial g }{\partial r}\nonumber\\
&&~~~~~~~~~~+8g^2h^3\left(\frac{\partial^2  }{\partial z^2}+\frac{\partial^2  }{\partial y^2}-\frac{\partial^2 
}{\partial t^2}\right)+12g^3h\frac{\partial^2 h }{\partial r^2}-8g^2h^2\frac{\partial^2 g }{\partial r^2}\Bigg)\nonumber\\
&&\triangle^{22}_{22}=\frac{1}{16h^{\frac{7}{2}}g^2}\left(21g^3\sqrt{h}\left(\frac{\partial h}{\partial
r}\right)^2-16h^{\frac{3}{2}}g^2\frac{\partial h}{\partial r}\frac{\partial g}{\partial r}-12g^3h^{\frac{3}{2}}\frac{\partial^2 h}{\partial
r^2}+8g^2h^{\frac{5}{2}}\frac{\partial^2 g}{\partial r^2}\right)\nonumber\\
&&\triangle^{13}_{22}=\frac{1}{16h^{\frac{7}{2}}g^2}\left(20g^3h^{\frac{3}{2}}\frac{\partial h}{\partial r}-16g^2h^{\frac{5}{2}}\frac{\partial
g}{\partial r}\frac{\partial }{\partial y}-16g^3h^{\frac{5}{2}}\frac{\partial^2 }{\partial y \partial r}\right)\nonumber\\
&&\triangle^{01}_{22}=\frac{1}{16h^{\frac{7}{2}}g^2}\left(-20h^{\frac{3}{2}}g^2\frac{\partial }{\partial t}+8gh^{\frac{5}{2}}\frac{\partial g
}{\partial r}\frac{\partial }{\partial t}+16h^{\frac{5}{2}}g^2\frac{\partial^2 }{\partial r \partial t}\right)\nonumber\\
&&\triangle^{00}_{22}=\frac{1}{16h^{\frac{7}{2}}g^2}\Bigg(10h^{\frac{3}{2}}g^2\frac{\partial h}{\partial r}\frac{\partial }{\partial
r}+4gh^{\frac{5}{2}}\frac{\partial g}{\partial r}\frac{\partial }{\partial r}-4h^{\frac{5}{2}}\left(\frac{\partial g}{\partial
r}\right)^2-10gh^{\frac{3}{2}}\frac{\partial g}{\partial r}\frac{\partial h}{\partial r}\nonumber\\
&&~~~~~~~~~~ -8h^{\frac{5}{2}}g^2\frac{\partial^2 }{\partial r^2}-8h^{\frac{7}{2}}\left(\frac{\partial^2 }{\partial y^2}+\frac{\partial^2 }{\partial
z^2}\right)+8gh^{\frac{5}{2}}\frac{\partial^2 g}{\partial r^2}\Bigg)\nonumber\\
&&\triangle^{33}_{22}=\frac{1}{16h^{\frac{7}{2}}g^2}\left(8g^2h^{\frac{5}{2}}\frac{\partial g}{\partial r}\frac{\partial }{\partial
r}+8g^3h^{\frac{5}{2}}\frac{\partial^2}{\partial r^2}-8h^{\frac{7}{2}}g\frac{\partial^2}{\partial
t^2}+8g^2h^{\frac{7}{2}}\frac{\partial^2}{\partial z^2}-10g^3h^{\frac{3}{2}}\frac{\partial h}{\partial r}\frac{\partial }{\partial r}\right)\nonumber\\
&&\triangle^{13}_{22}=\frac{1}{16h^{\frac{7}{2}}g^2}\left(-16h^{\frac{5}{2}}g^2\frac{\partial g}{\partial r}\frac{\partial }{\partial
y}-16g^3h^{\frac{5}{2}}\frac{\partial^2 }{\partial r \partial y}\right)\nonumber\\
&&\triangle^{03}_{22}=\frac{1}{16h^{\frac{7}{2}}g^2}\left(16h^{\frac{7}{3}}g\frac{\partial^2 }{\partial t \partial y} \right)\nonumber\\ 
&&\triangle^{12}_{33}=\frac{1}{16h^{\frac{7}{2}}g^2}\left(20g^3h^{\frac{3}{2}}\frac{\partial h }{\partial r}\frac{\partial
}{\partial  x}-16g^3h^{\frac{5}{2}}\frac{\partial g }{\partial r}\frac{\partial }{\partial 
x}-16g^3h^{\frac{5}{2}}\frac{\partial^2 }{\partial r \partial x}\right)\nonumber\\
&&\triangle^{22}_{33}=\frac{1}{16h^{\frac{7}{2}}g^2}\left(-10g^3h^{\frac{3}{2}}\frac{\partial h }{\partial r}\frac{\partial
}{\partial r}+8g^2h^{\frac{5}{2}}\frac{\partial g }{\partial r}\frac{\partial
}+8g^3h^{\frac{5}{2}}\frac{\partial^2}{\partial r^2}+8h^{\frac{7}{2}}g^2\left(\frac{\partial^2 }{\partial
z^2}-\frac{\partial^2 }{\partial t^2}\right)\right)\nonumber\\
&&\triangle^{33}_{33}=\frac{1}{16h^{\frac{7}{2}}g^2}\left(21g^3\sqrt{g}\left(\frac{\partial h }{\partial
r}\right)^2-16g^2h^{\frac{3}{2}}\frac{\partial h }{\partial r}\frac{\partial g }{\partial
r}-12g^3h^{\frac{3}{2}}\frac{\partial^2 h }{\partial r^2}+8h^{\frac{5}{2}}g^2\frac{\partial^2 g }{\partial r^2}\right)\nonumber\\
&&\triangle^{11}_{33}=\frac{1}{16h^{\frac{7}{2}}g^2}\Bigg(-4g^2h^2\frac{\partial g }{\partial r}\frac{\partial }{\partial
r}+6g^3h\frac{\partial h }{\partial r}\frac{\partial }{\partial r}-24hg^3\left(\frac{\partial h }{\partial
r}\right)^2+18g^2h\frac{\partial h }{\partial r}\frac{\partial g}{\partial r}\nonumber\\
&&~~~~~~~~~~-8g^2h^2\frac{\partial^2 g }{\partial r^2}
+12g^3\frac{\partial^2 h }{\partial r^2}+8h^3g^2\left(\frac{\partial^2 }{\partial x^2}+\frac{\partial^2 }{\partial
z^2}-\frac{\partial^2 }{\partial t^2}\right)\Bigg)\nonumber\\
&&\triangle^{01}_{33}=\frac{1}{16h^{\frac{7}{2}}g^2}\left(-20h^{\frac{3}{2}}g^2\frac{\partial h }{\partial
r}\frac{\partial}{\partial t}+8gh^{\frac{5}{3}}\frac{\partial g }{\partial r}\frac{\partial }{\partial
t}+16h^{\frac{5}{2}}g^2\frac{\partial^2 }{\partial r \partial t}\right)\nonumber
\nd
\bg
&&\triangle^{00}_{33}=\frac{1}{16h^{\frac{7}{2}}g^2}\Bigg(10h^{\frac{3}{2}}g^2\frac{\partial h }{\partial r}\frac{\partial}{\partial r}
+4gh^{\frac{5}{2}}\frac{\partial g}{\partial r}\frac{\partial}{\partial r}-4h^{\frac{5}{3}}\left(\frac{\partial g}{\partial
r}\right)^2-16h^{\frac{3}{2}}g\frac{\partial g}{\partial r}\frac{\partial h}{\partial r}\nonumber\\
&& ~~~~~~~~~~-8h^{\frac{7}{2}}g\left(\frac{\partial^2}{\partial x^2}+\frac{\partial^2}{\partial
z^2}\right)-8h^{\frac{5}{2}}g^2\frac{\partial^2}{\partial r^2}+8gh^{\frac{5}{2}}\frac{\partial^2 g}{\partial r^2}\Bigg)\nonumber\\
&&\triangle^{02}_{33}=\frac{1}{16h^{\frac{7}{2}}g^2}\left(16h^{\frac{7}{2}}g\frac{\partial^2 }{\partial t \partial x}\right)\nonumber\\
&&\triangle^{12}_{33}=\frac{1}{16h^{\frac{7}{2}}g^2}\left(-16h^{\frac{5}{2}}g^3\frac{\partial^2 }{\partial r \partial x}\right)\nonumber\\
&&\triangle^{01}_{01}=\frac{1}{16h^{3}g}\left(8h^3g\left(\frac{\partial^2}{\partial x^2}+\frac{\partial^2}{\partial
y^2}+\frac{\partial^2}{\partial z^2}\right)-21g^2\left(\frac{\partial h}{\partial r}\right)^2+12g^2h\frac{\partial^2
h}{\partial r^2}+6gh\frac{\partial h}{\partial r}\frac{\partial g}{\partial r}\right)\nonumber\\
&&\triangle^{12}_{01}=\frac{1}{16h^{3}g}\left(-8h^3g\frac{\partial^2}{\partial t \partial x}\right)\nonumber\\
&&\triangle^{13}_{01}=\frac{1}{16h^{3}g}\left(-8h^3g\frac{\partial^2}{\partial t \partial y}\right)\nonumber\\
&&\triangle^{02}_{01}=\frac{1}{16h^{3}g}\left(8h^3\frac{\partial g}{\partial r} \frac{\partial}{\partial
x}-8h^3g\frac{\partial^2}{\partial r \partial x}\right)\nonumber\\
&&\triangle^{03}_{01}=\frac{1}{16h^{3}g}\left(8h^3\frac{\partial g}{\partial r} \frac{\partial}{\partial
y}-8h^3g\frac{\partial^2}{\partial r \partial y}\right)\nonumber\\
&&\triangle^{22}_{01}=\frac{1}{16h^{3}g}\left(8h^3g\frac{\partial^2}{\partial t \partial r} -4h^3\frac{\partial g}{\partial
r}\frac{\partial }{\partial t}\right)\nonumber\\
&&\triangle^{33}_{01}=\frac{1}{16h^{3}g}\left(-8h^3\frac{\partial g}{\partial r}+8h^3g\frac{\partial^2 }{\partial r\partial
t }\right)\nonumber\\
&&\triangle^{11}_{01}=\frac{1}{16h^{3}g}\left(6h^{\frac{3}{2}}g\frac{\partial h}{\partial r}\frac{\partial}{\partial
t}\right)\nonumber\\
&&\triangle^{23}_{02}=\frac{1}{16h^{3}}\left(8h^3\frac{\partial^2 }{\partial t \partial y}\right)\nonumber\\
&&\triangle^{33}_{02}=\frac{1}{16h^{3}}\left(-8h^3\frac{\partial^2 }{\partial t \partial y}\right)\nonumber\\
&&\triangle^{02}_{02}=\frac{1}{16h^{3}}\Bigg(-8h^3\left(\frac{\partial^2 }{\partial y^2}+\frac{\partial^2 }{\partial
z^2}\right)+21g\left(\frac{\partial h}{\partial r}\right)^2-8gh^2\frac{\partial^2 }{\partial r^2}\nonumber\\
&&~~~~~~~~~~-12gh\frac{\partial^2 h }{\partial r^2}-16h\frac{\partial g }{\partial r}\frac{\partial h }{\partial r}
+10gh\frac{\partial h }{\partial r}\frac{\partial}{\partial r}+8h^2\frac{\partial^2 g }{\partial r^2}\Bigg)\nonumber\\
&&\triangle^{03}_{02}=\frac{1}{16h^{3}}\left(8h^3\frac{\partial^2  }{\partial x \partial y}\right)\nonumber\\
&&\triangle^{11}_{02}=\frac{1}{16h^{3}}\left(-8h^{\frac{5}{2}}\frac{\partial^2  }{\partial t \partial x}\right)\nonumber\\
&&\triangle^{01}_{02}=\frac{1}{16h^{3}}\left(8gh^2\frac{\partial^2  }{\partial r \partial x}-10gh\frac{\partial h }{\partial
r}\frac{\partial }{\partial x}+8h^2\frac{\partial g }{\partial r}\frac{\partial }{\partial x}\right)\nonumber
\nd
\bg
&&\triangle^{12}_{02}=\frac{1}{16h^{3}}\left(8gh^2\frac{\partial^2  }{\partial r \partial t}-10gh\frac{\partial h }{\partial
r}\frac{\partial  }{\partial t}\right)\nonumber\\
&&\triangle^{12}_{03}=\frac{1}{16h^{3}}\left(8h^3\frac{\partial^2}{\partial x \partial y}\right)\nonumber\\
&&\triangle^{22}_{03}=\frac{1}{16h^{3}}\left(-8h^3\frac{\partial^2}{\partial t \partial y}\right)\nonumber\\
&&\triangle^{03}_{03}=\frac{1}{16h^{3}}\Bigg(-8h^3\left(\frac{\partial^2}{\partial z^2}+\frac{\partial^2}{\partial
x^2}\right)+21g\left(\frac{\partial h}{\partial r}\right)^2-8gh^2\frac{\partial^2}{\partial r^2}-12gh\frac{\partial^2
h}{\partial r}\nonumber\\
&&~~~~~~~~~~+10gh\frac{\partial h}{\partial r}\frac{\partial }{\partial r}-16h\frac{\partial h}{\partial r}\frac{\partial g}{\partial
r}+8h^2\frac{\partial^2 g}{\partial r^2}\Bigg)\nonumber\\
&&\triangle^{23}_{03}=\frac{1}{16h^{3}}\left(-8h^3\frac{\partial^2}{\partial t \partial x}\right)\nonumber\\
&&\triangle^{11}_{03}=\frac{1}{16h^{3}}\left(-8h^{\frac{5}{2}}\frac{\partial^2}{\partial t \partial y}\right)\nonumber\\
&&\triangle^{01}_{03}=\frac{1}{16h^{3}}\left(8h^2g\frac{\partial^2}{\partial r \partial y}-10gh\frac{\partial h}{\partial
r}\frac{\partial }{\partial y}+8h^2\frac{\partial g}{\partial r}\frac{\partial }{\partial y}\right)\nonumber\\
&&\triangle^{13}_{03}=\frac{1}{16h^{3}}\left(8h^2g\frac{\partial^2}{\partial r \partial t}-10gh\frac{\partial h}{\partial
r}\frac{\partial }{\partial t}\right)\nonumber\\
&&\triangle^{12}_{12}=\frac{1}{16h^{3}g^2}\Bigg(-8h^3g^2\left(\frac{\partial^2}{\partial y^2}
+\frac{\partial^2}{\partial z^2}\right)+21g^3\left(\frac{\partial h}{\partial r}\right)^2+8h^3g\frac{\partial^2}{\partial
t^2}\nonumber\\
&&~~~~~~~~~~-12g^3h\frac{\partial^2 h}{\partial r^2}+8h^2g^2\frac{\partial^2 g}{\partial r^2}-16hg^2\frac{\partial h}{\partial
r}\frac{\partial g}{\partial r}\Bigg)\nonumber\\
&&\triangle^{13}_{12}=\frac{1}{16h^{3}g^2}\left(8h^3g^2\frac{\partial^2}{\partial x \partial y}\right)\nonumber\\
&&\triangle^{33}_{12}=\frac{1}{16h^{3}g^2}\left(-8h^3g^2\frac{\partial^2}{\partial r \partial x}\right)\nonumber\\
&&\triangle^{23}_{12}=\frac{1}{16h^{3}g^2}\left(8h^3g^2\frac{\partial^2}{\partial r \partial y}\right)\nonumber\\
&&\triangle^{02}_{12}=\frac{1}{16h^{3}g^2}\left(-8h^3g\frac{\partial^2}{\partial r \partial t}\right)\nonumber\\
&&\triangle^{00}_{12}=\frac{1}{16h^{3}g^2}\left(-4h^3\frac{\partial g}{\partial r}\frac{\partial }{\partial
x}+8h^3g\frac{\partial^2}{\partial r \partial x}\right)\nonumber\\
&&\triangle^{01}_{12}=\frac{1}{16h^{3}g^2}\left(-8h^3g\frac{\partial^2}{\partial t \partial x}\right)\nonumber\\
&&\triangle^{11}_{12}=\frac{1}{16h^{3}g^2}\left(-6h^{\frac{3}{2}}g^2\frac{\partial h}{\partial r}\frac{\partial }{\partial
x}+4h^{\frac{5}{2}}g\frac{\partial g}{\partial r}\frac{\partial }{\partial x}\right)\nonumber\\
&&\triangle^{13}_{13}=-\frac{1}{16h^{3}g^2}\Bigg(8h^3g^2\left(\frac{\partial^2}{\partial x^2}
+\frac{\partial^2}{\partial z^2}\right)-21g^3\left(\frac{\partial h}{\partial r}\right)^2-8h^3g\frac{\partial^2}{\partial
t^2}\nonumber\\
&&~~~~~~~~~~+12g^3h\frac{\partial^2 h}{\partial r^2}-8h^2g^2\frac{\partial^2 g}{\partial r^2}+16hg^2\frac{\partial h}{\partial
r}\frac{\partial g}{\partial r}\Bigg)\nonumber
\nd
%jhuta
\bg
&&\triangle^{12}_{13}=-\frac{1}{16h^{3}g^2}\left(-8h^3g^2\frac{\partial^2}{\partial x \partial y}\right)\nonumber\\
&&\triangle^{22}_{13}=-\frac{1}{16h^{3}g^2}\left(8h^3g^2\frac{\partial^2}{\partial r \partial y}\right)\nonumber\\
&&\triangle^{03}_{13}=-\frac{1}{16h^{3}g^2}\left(8h^3g\frac{\partial^2}{\partial r \partial t}\right)\nonumber\\
&&\triangle^{00}_{13}=-\frac{1}{16h^{3}g^2}\left(4h^3\frac{\partial g}{\partial r}\frac{\partial }{\partial
y}-8h^3g\frac{\partial^2}{\partial r \partial y}\right)\nonumber\\
&&\triangle^{23}_{13}=-\frac{1}{16h^{3}g^2}\left(-8h^3g^2\frac{\partial^2}{\partial r \partial x}\right)\nonumber\\
&&\triangle^{13}_{13}=-\frac{1}{16h^{3}g^2}\left(8h^3g\frac{\partial^2}{\partial t \partial y}\right)\nonumber\\
&&\triangle^{13}_{13}=-\frac{1}{16h^{3}g^2}\left(6h^{\frac{3}{2}}g^2\frac{\partial h}{\partial r}\frac{\partial }{\partial
y}-4h^{\frac{5}{2}}g\frac{\partial g}{\partial r}\frac{\partial }{\partial y}\right)\nonumber\\
&&\triangle^{23}_{23}=-\frac{1}{16h^{3}g}\Bigg(-8h^3\frac{\partial^2}{\partial t^2}+8h^3g\frac{\partial^2}{\partial
z^2}-21g^28g^3\left(\frac{\partial h}{\partial r}\right)^2+8g^2h^2\frac{\partial^2}{\partial r^2}+12g^2h\frac{\partial^2
h}{\partial r^2}\nonumber\\
&&~~~~~~~~~~-10g^2h\frac{\partial h}{\partial r}\frac{\partial }{\partial r}+8gh^2\frac{\partial g}{\partial r}\frac{\partial
}{\partial r}+16gh\frac{\partial h}{\partial r}\frac{\partial g}{\partial r}-8h^2g\frac{\partial^2 g}{\partial r^2}\Bigg)\nonumber\\
&&\triangle^{03}_{23}=-\frac{1}{16h^{3}g}\left(8h^3\frac{\partial^2}{\partial t \partial x}\right)\nonumber\\
&&\triangle^{00}_{23}=-\frac{1}{16h^{3}g}\left(-8h^3\frac{\partial^2}{\partial x \partial y}\right)\nonumber\\
&&\triangle^{02}_{23}=-\frac{1}{16h^{3}g}\left(8h^3\frac{\partial^2}{\partial t \partial y}\right)\nonumber\\
&&\triangle^{11}_{23}=-\frac{1}{16h^{3}g}\left(8h^{\frac{5}{2}}g\frac{\partial^2}{\partial y \partial x}\right)\nonumber\\
&&\triangle^{13}_{23}=-\frac{1}{16h^{3}g}\left(-8g^2h^2\frac{\partial^2}{\partial r \partial x}-8gh^2\frac{\partial
g}{\partial r}\frac{\partial }{\partial x}+10g^2h\frac{\partial h}{\partial r}\frac{\partial }{\partial x}\right)\nonumber\\
&&\triangle^{12}_{23}=-\frac{1}{16h^{3}g}\left(-8g^2h^2\frac{\partial^2}{\partial r \partial y}-8gh^2\frac{\partial
g}{\partial r}\frac{\partial }{\partial y}+10g^2h\frac{\partial h}{\partial r}\frac{\partial }{\partial y}\right)\nonumber
\nd
The above therefore summarises all the fluctuation operators for the OKS-BH background. As one can see, the situation here
is much more involved than the AdS-BH case. One might however try to simplify the 77 equations by imposing some 
symmetry in the background, much like the one that we discussed in the previous appendix. In the next appendix we will 
consider such a simplification for the OKS-BH background by taking diagonal perturbations.

%popo

%\newpage

\section{An example with diagonal perturbations in OKS-BH background}

As we discussed in the previous section, the full analysis with all ten components of the metric fluctuations is rather 
difficult. In section 3.3 we did present a partial analysis taking all the components into account. However in that 
section we couldn't provide precise numerical answers to the metric fluctuations because of the 
underlying complexity of the problem. In this section we will take a middle path where we will only consider  
{\it diagonal} perturbations for the metric fluctuations much like what we did in 
{\bf Appendix A} for the AdS case. This 
means that with 
the choice of coordinates and the string profile given by (\ref{KS10}) and (\ref{KS12}) 
we can now formally obtain the source in
(\ref{KS7b}) provided we take into account not only the energy momentum tensor of the string but also other sources (see
below). However 
since the string moves in the $x$ direction, the perturbation created
in $y$ and $z$ directions are equal and therefore we can demand: 
\bg\label{demand}
l_{22}~ = ~ l_{33}
\nd
This would mean that we have only four independent components of
$l_{\mu\nu}$ and it is enough to consider four independent equations (\ref{KS7b}) sourced by the
energy momentum tensors that can come from various sources like the string, fluxes, D7 branes as well as O7 planes:
\bg\label{source}
T_{\mu\nu}^{\rm total} \equiv T_{\mu\nu}^{\rm string} ~+~ T_{\mu\nu}^{\rm fluxes} ~+~ T_{\mu\nu}^{\rm planes} ~+~
T_{\mu\nu}^{\rm branes}
\nd
At this stage we will assume that the total energy momentum tensor from all the above sources {\it guarantee} a diagonal
perturbation in the system. We will also assume that the sources are all expressed in terms of the variable $u$
where in terms of $u$, the UV is at $u = 0$ whereas IR is at 
$u = \infty$. Therefore in the regime close to $u = \infty$ we expect certain aspects of QCD to be revealed from the 
gravity dual \eqref{metric1}. 

We will now solve (\ref{KS7b}) order by order in
$g_sN_f, g_sM^2/N$ and $g^2_sN_f M$.
At zeroth order in $g_sN_f, g^2_s N_f M, g_sM^2/N$, the warp factor becomes $h(u)=L^4u^4$ and the metric (\ref{metric1}) 
reduces to that of $AdS_5$. Hence at zeroth order in
$g_sN_f, g_sM^2/N, g_s^2M N_f$ our analysis will be similar to the AdS/CFT calculations \cite{Yaffe-1}.    
As (\ref{KS7b}) is a second order non linear partial differential equation, 
we can solve it by Fourier decomposing $x,y,z,t$ dependence of $l_{\mu\nu}$ and 
writing it as a Taylor series in $u$ in the following way:
\bg \label{metricSolu}
l_{\mu\nu} ~ = ~ l_{\mu\nu}^{[0]}~+ ~ g_s N_f (a~+b~g_sM)l_{\mu\nu}^{[1]}
\nd
where the subscript $[0], [1]$ refer to the zeroth and the first order in ($g_sN_f$).\footnote{Henceforth it also means
we are keeping terms upto ${\cal O}(g_sM^2/N, g_s^2 N_f M)$.} 
The zeroth order can 
then be succinctly presented as a Fourier series in the following way\footnote{Note that in all the subsequent mode 
expansions we will be ignoring the $\sqrt{g(u_c)}$ dependences. Therefore for us $u_c$ is close to the actual boundary
so that $\tau \approx t$. A more careful analysis has been presented in section 3.3 wherein we took all the subtleties
into account.}:
\bg\label{fseries}
l_{\mu\nu}^{[0]}(t,u,x,y,z)~= ~ \sum_{k=0}^{\infty}\int \frac{dq_1 dq_2dq_3 d\omega}{(2\pi)^4} 
\Bigg[e^{-i(\omega t-q_1x-q_2y-q_3z)} s^{(k)[0]}_{\mu\nu}(\omega,q_1,q_2,q_3)u^k\Bigg]\nonumber\\
\nd
where $s^{(k)[0]}_{\mu\nu}$ are expansion coefficients of the solution $l_{\mu\nu}^{[0]}$. The constant coefficients 
($a, b$) in  
 $l_{\mu\nu}^{[1]}$ can be worked out easily. 

Similarly, we can also write the source in Fourier space as:
\bg \label{sourceF2}
T_{\mu\nu}^{\rm total}~= ~T_{\mu\nu}^{[0]{\rm total}}+T_{\mu\nu}^{[1]{\rm total}}
\nd
where as before, [0, 1] refer to the zeroth and first orders in ($g_sN_f$) respectively. The zeroth order 
can then be written as:
\bg\label{zert}
T_{\mu\nu}^{[0]{\rm total}}(t,u,x,y,z) =\int \frac{dq_1 dq_2dq_3 d\omega}{(2\pi)^4} \;\;e^{-i(\omega t-q_1x-q_2y-q_3z)} 
t_{\mu\nu}^{[0]}(\omega,u,q_1,q_2,q_3)\nonumber\\
\nd
where $t_{\mu\nu}^{[0]}$ are expansion coefficients of source  
$T_{\mu\nu}^{[0]{\rm total}}$ at zeroth order in $g_sN_f$. 
These coefficients are obtained by using explicit expressions 
for $T_{\mu\nu}^{\rm total}(x^\mu)$ given above in (\ref{source}). 
In terms of matrices we then expect the metric
fluctuations $l_{ii}$, $i = 0, 1, 2, 3$ to take the following form:
\begin{equation} \label{metfla}
\begin{pmatrix}~~ 0~~ & \Delta^{11}_{01}~~ & \Delta^{22}_{01}~~ & \Delta^{33}_{01} ~~\\
& & & \\
~~0~~ & \Delta^{11}_{02}~~ & 0 ~~ & \Delta^{33}_{02} ~~ \\& & & \\
~~0 ~~& \Delta^{11}_{03}~~ & \Delta^{22}_{03}~~ & \Delta^{33}_{03} ~~ \\& & & \\
~~ \Delta^{00}_{12}~~ & \Delta^{11}_{12} ~~& 0~~ & \Delta^{33}_{12} ~~
\end{pmatrix} \begin{pmatrix}~~ l_{00}\\& & & \\~~ l_{11} \\& & & \\~~ l_{22} 
\\& & & \\ ~~l_{33}\end{pmatrix}
= {1\over \kappa}
\begin{pmatrix} ~~T_{01}\\& & & \\ ~~ T_{02} 
\\& & & \\ ~~ 0 \\& & & \\ ~~ T_{12}\end{pmatrix}
\end{equation}
where we have only switched on the 01, 02 and 12 components of the sources; and 
$\Delta^{\alpha\beta}_{\mu\nu}$ are now defined wrt the variable $u$ instead of $r$, the radial coordinate.  

Manipulating the above matrix equation and using  
the explicit expressions for  $t_{\mu\nu}^{[0]}$, we can extract a relation for all $s^{(k)[0]}_{22}$ 
at zeroth order in $g_sN_f,g_sM$ (\ref{KS7b}) as:
\bg \label{perteq1}  
\sum_{k}\Bigg[\frac{3k u}{2}
+ \Bigg(-\frac{u^2}{4g} \frac{dg}{du}
-\frac{3u}{g}
+u^2\frac{dg}{du} +\frac{g }{2}\Bigg)\Bigg] u^k ~s^{(k)[0]}_{22}
~ = ~ -{i{\cal A}^{0}(\omega,u,\overrightarrow{q})\over \omega ~u^4 L^4}
\nd
where ${\cal A}^0$ can be given in terms of a series in $u^j$ in the 
following way:
\bg \label{sperteq1}
{\cal
A}^0~=~\delta(\omega-vq_1)\theta(u-u_0)\sum_{j=4}^{\infty}\tilde{\zeta}_{j}u^j
\nd
where $\tilde{\zeta}_j$ are $u$-independent constants that could be determined from \eqref{source} once we have the 
explicit expressions for all the terms in \eqref{source}. 
We can now use the above equations \eqref{perteq1} and 
\eqref{metfla} to write a relation between $s_{11}^{(k)[0]}$ and $s_{22}^{(k)[0]}$ in the following way:
\bg \label{perteq2}
\sum_{k}\left[s^{(k)[0]}_{11}+\frac{s^{(k)[0]}_{22}}{g}\right]u^{k}={\cal B}^{0}(\omega,u,\overrightarrow{q})
\nd
where again ${\cal B}^0$ can also be expressed in series like 
\eqref{sperteq1}. The above two set of equations have infinite number 
of variables. They can be solved if we know a generating function. Such a function could be determined for 
$s_{33}^{(k)[0]}$ in terms of $s_{11}^{(k)[0]}$ and $s_{22}^{(k)[0]}$ as:
\bg \label{perteq3}
s^{(k)[0]}_{33}={\cal C}^{0}(\omega,u,\overrightarrow{q})-g s^{(k)[0]}_{11}- s^{(k)[0]}_{22}
\nd
Thus if we know $s_{11}^{(k)[0]}$ and $s_{22}^{(k)[0]}$ we can determine $s_{33}^{(k)[0]}$. In the following we 
will determine these coefficients using Green's function. However before we go into it, let us write the last 
equation relating $s_{00}^{(k)[0]}$ to the other coefficients: 
\bg \label{perteq4}
&& \sum_k\Bigg[\Bigg(\frac{-2q_1u}{g} \frac{dg}{du}+4 q_1 k\Bigg)s^{(k)[0]}_{00} + 
\Bigg(2 gq_1 \frac{dg}{du}  -12 g^2q_1\Bigg) s^{(k)[0]}_{11}u^k-8g q_1k
s^{(k)[0]}_{33}\Bigg] u^{k-1} \nonumber\\
&& ~~~~~~~~~ ={{\cal D}^{0}(\omega,u,\overrightarrow{q}) \over L^4 u^6}
\nd
where ${\cal C}^{0},{\cal D}^{0}$ are determined from the source (\ref{source}) like \eqref{sperteq1} above. 
Note that we could also write \eqref{perteq4} in terms 
of $s_{11}^{(k)[0]}$ and $s_{22}^{(k)[0]}$ using the generating function \eqref{perteq3}. 

%jhuta
To solve the set of equations
\eqref{perteq1} to \eqref{perteq4} we will be 
using Green's functions. Since all the equations are given in terms of series in $u^k$, 
we first write the delta function as\footnote{Such a way of expressing the delta function 
can be motivated from the standard
completeness relation in quantum mechanics. Here of course the coefficient $b_i$ are some specified integers.} 
\bg\label{delfn}
\delta(u)=\sum_i b_i u^i 
\nd
so that we can equate coefficients on 
both sides of the equation.
Using this it is straightforward to show that the 
the Green's function for 
(\ref{perteq1}) is given by:
\bg \label{perteq5}
{\cal G}_{22}^{0}(u,\omega,\overrightarrow{q})~ = ~ \sum_{i=-4}^{\infty}c^0_i(\omega,\overrightarrow{q})u^i
\nd 
where $c_i^0$ are given, in terms of the $b_i$ coefficients appearing in the delta function \eqref{delfn}, 
in the following way:
\bg \label{sperteq2}
&& c_{-4}^0~=~-\frac{2ib_0}{\omega L^4};~~~~c_{-3}^0~=~-\frac{2ib_1+36ib_0}{\omega L^4};~~~~
c_{-2}^0~=~-\frac{2ib_2+30ib_1+540ib_0}{\omega L^4}\nonumber\\
&& c_{-1}^0~=~-\frac{2ib_3+24ib_2+360ib_1+6480ib_0}{\omega L^4} \nonumber\\
&& c_{0}^0~=~\frac{2ib_4+18ib_3+216ib_2+3240ib_1+58320ib_0}{\omega L^4}
\nd
Observe that the lower limit of the sum is from $i = -4$ because of the 
$u^{-4}$ suppression in the LHS of \eqref{perteq1}.

It is now easy to write down the  
solution for the metric perturbation $l_{22}^{[0]}$ using the Green's function as:
\bg \label{perteq6}
l_{22}^{[0]}(t,u,x,y,z)&=&\int \frac{d\omega d^3q}{(2\pi)^4}\;e^{-i(\omega t-q_1x-q_2y-q_3z)} \int_{0}^{u}du'{\cal A}^{0}(u'){\cal
G}_{22}^0(u',\omega,\overrightarrow{q})\nonumber\\
&\equiv& \sum_{k = 0}^{\infty} \int \frac{d\omega d^3q}{(2\pi)^4}\;e^{-i(\omega t-q_1x-q_2y-q_3z)}
s_{22}^{(k)[0]}(\omega,\overrightarrow{q})\theta(u-u_0)u^k
\nd
where explicit expressions for $s_{22}^{(k)[0]}$ are given below. Observe that 
with $s_{22}^{(k)[0]}$ known, we obtain  $s_{11}^{(k)[0]}$ using (\ref{perteq2}). 
Then knowing $s_{22}^{(k)[0]},s_{11}^{(k)[0]}$
we obtain $s_{33}^{(k)[0]}$ using (\ref{perteq3}) and finally $s_{00}^{(k)[0]}$ using (\ref{perteq4}). The explicit 
expressions for $s_{22}^{(k)[0]}$ are:
\bg \label{sperteq6}
s_{22}^{(0)[0]}&=&\delta(\omega-vq_1)\bigg[-\tilde{\zeta}_4c_{-4}^0u_0+(\tilde{\zeta}_4c_{-3}^0+\tilde{\zeta}_5c_{-4}^0)\frac{u_0^2}{2}
-(\tilde{\zeta}_4c_{-2}^0+\tilde{\zeta}_5c_{-3}^0)\frac{u_0^3}{3} \nonumber\\
&& ~~~~~~~~+(\tilde{\zeta}_4c_{-1}^0
+\tilde{\zeta}_5c_{-2}^0)\frac{u_0^4}{4}
-(\tilde{\zeta}_4c_{0}^0+\tilde{\zeta}_5c_{-1}^0+\tilde{\zeta}_8c_{-4}^0)\frac{u_0^5}{5}\bigg]\nonumber\\
s_{22}^{(1)[0]}&=&\delta(\omega-vq_1)\tilde{\zeta}_4c_{-4}^0\nonumber\\
s_{22}^{(2)[0]}&=&\frac{1}{2}\delta(\omega-vq_1)(\tilde{\zeta}_4c_{-3}^0+\tilde{\zeta}_5c_{-4}^0)\nonumber\\
s_{22}^{(3)[0]}&=&\frac{1}{3}\delta(\omega-vq_1)(\tilde{\zeta}_4c_{-2}^0+\tilde{\zeta}_5c_{-3}^0)\nonumber\\
s_{22}^{(4)[0]}&=&\frac{1}{4}\delta(\omega-vq_1)(\tilde{\zeta}_4c_{-1}^0+\tilde{\zeta}_5c_{-2}^0)\nonumber\\
s_{22}^{(5)[0]}&=&\frac{1}{5}\delta(\omega-vq_1)(\tilde{\zeta}_4c_{0}^0+\tilde{\zeta}_5c_{-1}^0+\tilde{\zeta}_8c_{-4}^0)
\nd
Observe also
that all $s_{22}^{(k)[0]}$ are proportional to $\delta(\omega-q_1v)$ which will eventually produce the Mach
cone. Finally, 
using (\ref{perteq6}) in \eqref{perteq2}, \eqref{perteq3}, and \eqref{perteq4}, we
can obtain rest of the metric perturbations at zeroth order in $g_s$ (although we do not present the explicit expressions
for the rest of the zeroth order perturbations here). 

%mrkoremi
Now we solve for the metric perturbation at linear order in $g_sN_f,g_sM$ where it is easiest to switch to 
coordinate $u=\frac{1}{r_c(1-\zeta)}$, 
so that the entire manifold is now described by $0 \leq \zeta \leq 1$ with $r_c$ arbitrarily large and 
we can get a meaningful 
Taylor series expansion of the logarithms and other functions appearing in the equation. 
As we did for the zeroth order cases, we can decompose the first order 
in $g_sN_f,g_sM$ fluctuations via the following Fourier series:
\bg \label{perteq11a}
l_{\mu\nu}^{[1]}(t,\zeta,x,y,z)&=&\sum_{k=0}^{\infty}\int \frac{dq_1 dq_2dq_3 d\omega}{(2\pi)^4} 
\Bigg[e^{-i(\omega t-q_1x-q_2y-q_3z)} s^{(k)[1]}_{\mu\nu}(\omega,q_1,q_2,q_3)\zeta^k\Bigg]\nonumber\\
\nd
where $s_{\mu\nu}^{(k)[1]}$ are the corresponding Fourier modes. These Fourier modes will 
eventually appear in the final equations for $l_{\mu\nu}^{[1]}$. In fact we would also need the zeroth 
order perturbations $l_{\mu\nu}^{[0]}$ in the equations. Therefore we decompose $l_{\mu\nu}^{[0]}$ as:
\bg\label{zerord}
l_{\mu\nu}^{[0]}(t,\zeta,x,y,z)&=& \int \frac{d\omega d^3q}{(2\pi)^4}\;e^{-i(\omega t-q_1x-q_2y-q_3z)}
l_{\mu\nu}^{(k)[0]}(\omega,\zeta,\overrightarrow{q})\zeta^k
\nd
where $l_{\mu\nu}^{(k)[0]}$ are now the Fourier modes for $l_{\mu\nu}^{[0]}$. These modes have one-to-one 
correspondence with $s_{\mu\nu}^{(k)[0]}$ given earlier and can be related by coordinate transformations.
Using these equation \eqref{metfla} reads:
\bg \label{perteq11}  
&&~~~~~~~~~~~~~~~\sum_{k}\Bigg[\frac{3i\omega \widetilde{h}k}{2r_c \zeta}l_{22}^{(k)[0]} +
\frac{3i\omega kL^4}{2r_c^5 \zeta(1-\zeta)^4}s^{(k)[1]}_{22}\\
&&~~~~~~~~~~~~~~~~~~~~~ +\frac{i\omega l_{22}^{(k)[0]}}{r_c}\Bigg(-\frac{\widetilde{h}}{4g}\frac{dg}{d\zeta}
-\frac{3}{4g}\frac{d\widetilde{h}}{d\zeta}
+\widetilde{h}  \frac{dg}{d\zeta}
+\frac{g\widetilde{h}r_c}{2}\Bigg)
\nonumber\\
&&+{i\omega s^{(k)[1]}_{22}L^4 \over r_c^4 (1-\zeta)^4}
\Bigg(-\frac{1}{4r_c g}\frac{dg}{d\zeta}
-\frac{3}{r_c(1-\zeta)g}
+{1\over r_c}\frac{dg}{d\zeta} + {g\over 2}\Bigg)\Bigg]\zeta^k ={\cal A}^{1}(\omega,\overrightarrow{q})\nonumber
\nd
where as before as in \eqref{perteq1}, we could separate the equations relating $s_{22}^{(k)[1]}$ and $l_{22}^{(k)[0]}$ 
from the rest of the other Fourier modes. We have also defined $\widetilde h$ as:
\bg \label{perteq11b}
\widetilde{h}=\frac{L^4}{r_c^4(1-\zeta)^4}\Big[A~{\rm log}~r_c(1-\zeta)- B~
{\rm log}^2~r_c(1-\zeta)\Big]
\nd
with $A$ and $B$ are defined in \eqref{metric1}. The other variables appearing above have already been 
defined earlier. Using these, and using the appropriate Green's function we can easily determine these Fourier modes. 
For example the explicit expressions 
for $s_{22}^{(k)[1]}$ can be determined by first writing the delta 
function as:
\bg \label{delta1}
\delta(1/r)~=~\sum_{i=0}^{\infty}\frac{\tilde{b}_i}{r^i} ~\equiv~ \sum_{j=0}^{\infty}\bar{b}_j\zeta^i
\nd
with $\bar{b}_j$ defined in the following way:
\bg\label{barbi} 
\bar{b}_j~=~ \sum_{i=0}^{\infty}\frac{i(i+1)...(i+1-j)\tilde{b}_i}{r_c^i}
\nd
Then solve for the Greens function for (\ref{perteq11}) to obtain 
\bg \label{perteq5}
{\cal
G}_{22}^{1}(\zeta,\omega,\overrightarrow{q})~= ~\sum_{i=1}^{\infty}c^1_i(\omega,\overrightarrow{q})\zeta^i
\nd
where the coefficients appearing above are defined as:
\bg\label{codelo}
c^1_1&=&-l_{22}^{(1)[0]}\left[A{\rm log}(r_c)-B{\rm log}^2(r_c)\right]-\frac{2i\bar{b}_0r_c^5}{3\omega L^4}\nonumber\\
c^1_2&=&-4c^1_1-l_{22}^{(1)[0]}\left[4A{\rm log}(r_c)-4B{\rm log}^2(r_c)-A+2B{\rm log}(r_c)\right]-2l_{22}^{(2)[0]}
\left[A{\rm log}(r_c)-B{\rm
log}^2(r_c)\right]\nonumber\\
&&~~~+\frac{2c^1_1r_c}{3}\left[\frac{r_h^4}{r_c^5g(r_c)}-\frac{3}{4r_cg(r_c)}-\frac{4r_h^4}{r_c^5}+\frac{g(r_c)}{2}\right]
+\frac{2l_{22}^{(1)[0]}r_c}{3}\Bigg\{\frac{\left[A{\rm log}(r_c)-B{\rm log}^2(r_c)\right]r_h^4}{r_c^5g(r_c)}\nonumber\\
&&~~~-\frac{12\left[A{\rm log}(r_c)-B{\rm
log}^2(r_c)\right]-3A+6B{\rm log}(r_c)}{4r_cg(r_c)}+\frac{4\left[A{\rm log}(r_c)-B{\rm
log}^2(r_c)\right]r_h^4}{r_c^5}\nonumber\\
&&~~~+\frac{L^4g(r_c)\left[A{\rm log}(r_c)-B{\rm log}^2(r_c)\right]}{2}\Bigg\}
-\frac{2ir_c^5\bar{b}_1}{3\omega L^4}
\nd
Once everything is laid up, we can write
the source in (\ref{perteq11}) as a power series in $\zeta$ i.e. ${\cal A}^1=\sum_j \tilde{a}_j\zeta^j$ 
with $\tilde{a}_j$ derivable from \eqref{source}. This would finally give us the required Fourier coefficients 
to first order in $g_sN_f$ as\footnote{We are only solving upto ${\cal O}(\zeta^3)$.}:
\bg 
s_{22}^{(0)[1]}&=&\delta(\omega-vq_1)\left[-\frac{c_1^1\tilde{a}_0 \zeta_0^2}{2}-\frac{(c_2^1\tilde{a}_0+c_1^1\tilde{a}_1)
\zeta_0^3}{3}\right]\nonumber\\
s_{22}^{(2)[1]}&=&\delta(\omega-vq_1)\left(\frac{c_1^1\tilde{a}_0}{2}\right)\nonumber\\
s_{22}^{(3)[1]}&=&\delta(\omega-vq_1)\left[\frac{(c_2^1\tilde{a}_0+c_1^1\tilde{a}_1)}{3}\right]
\nd 
with the following Fourier decomposition:
\bg \label{perteq6a}
l_{22}^{[1]}(t,\zeta,x,y,z)&=&\int \frac{d\omega d^3q}{(2\pi)^4}\;e^{-i(\omega t-q_1x-q_2y-q_3z)} \int_{0}^{\zeta}d\zeta'{\cal A}^{1}(\zeta'){\cal
G}_{22}^1(\zeta',\omega,\overrightarrow{q})\nonumber\\
&\equiv& \sum_{k = 0}^{\infty} \int \frac{d\omega d^3q}{(2\pi)^4}\;e^{-i(\omega t-q_1x-q_2y-q_3z)}
s_{22}^{(k)[1]}(\omega,\overrightarrow{q})\theta(\zeta-\zeta_0)\zeta^k
\nd
Once we know these modes, we can use them to write the 
relation for $s^{(k)[1]}_{11}$ in the following way:
\bg \label{perteq21}
\sum_{k}
\left[s^{(k)[1]}_{11}+\frac{s^{(k)[1]}_{22}}{g}\right]\zeta^{k}={\cal B}^{1}(\omega,\overrightarrow{q})
\nd
with ${\cal B}^{1}$ given in the appendix. Observe that the above equation has exactly the same form as 
\eqref{perteq2} except that these modes are written for linear order perturbations. It is then no surprise that the 
generating function for $s_{33}^{(k)[1]}$ takes exactly the same form as in \eqref{perteq3}:
\bg \label{perteq31}
s^{(k)[1]}_{33}={\cal C}^{1}(\omega,\overrightarrow{q})-g s^{(k)[1]}_{11}- s^{(k)[1]}_{22}
\nd
Finally
once we know the Fourier modes $s_{ii}^{(k)[1]}$ with $i = 1, 2, 3$ we can use them to write the equation for 
$s_{00}^{(k)[1]}$. The equation turns out to be rather involved with all zeroth and first order coefficients appearing
together. Nevertheless one can present the following form for the equation:
\bg \label{perteq41}
&&~~~~~~~~ \sum_k\Bigg[-2\frac{dg}{d\zeta}\frac{\widetilde{h}q_1}{r_cg}
l_{00}^{(k)[0]}\zeta^{k}-2\frac{dg}{d\zeta}\frac{L^4q_1}{r_c^5(1-\zeta)^4g}s^{(k)[1]}_{00}\zeta^{k}
\nonumber\\
&&+4q_1\widetilde{h}l_{00}^{(k)[0]}k\zeta^{k-1}\frac{1}{r_c}+4q_1L^4\frac{1}{r_c^5(1-\zeta)^4}s^{(k)[1]}_{00}k\zeta^{k-1}
+2g\frac{dg}{d\zeta}\frac{\widetilde{h}}{r_c}q_1 l_{11}^{(k)[0]}\zeta^k\nonumber\\
&&~~~~~~~~~+2g\frac{dg}{d\zeta}\frac{L^4}{r_c^5(1-\zeta)^4}q_1
s^{(k)[1]}_{11}\zeta^k
-3g^2\frac{d\widetilde{h}}{d\zeta}\frac{1}{r_c}q_1 
l_{11}^{(k)[0]}\zeta^k\nonumber\\
&&~~~~~~~~~~ -12g^2\frac{L^4}{r_c^5(1-\zeta)^5}q_1s^{(k)[1]}_{11}\zeta^k
-8\frac{g\widetilde{h}q_1}{r_c}l_{33}^{(k)[0]}k\zeta^{k-1}\nonumber\\
&&~~~~~~~~~~~~~-8\frac{gL^4q_1}{r_c^5(1-\zeta)^4}s^{(k)[1]}_{33}k\zeta^{k-1}\Bigg]
={\cal D}^{1}(\omega,\zeta,\overrightarrow{q})
\nd
where ${\cal D}^1,{\cal C}^1$ is determined from the source (\ref{source}). Plugging in the other Fourier modes, we have been able to solve for 
all the $s_{00}^{(k)[1]}$ modes using the corresponding 
Green's function (we don't present the results here). 

To conclude therefore,
using the equations (\ref{perteq21}), (\ref{perteq31}) and (\ref{perteq41}) 
we find $s_{00}^{(k)[1]}, s_{11}^{(k)[1]}$ and $s_{33}^{(k)[1]}$ etc. which in turn give us
$l_{00}^{[1]}(t,\zeta,x,y,z), l_{11}^{[1]}(t,\zeta,x,y,z)$ and $l_{33}^{[1]}(t,\zeta,x,y,z)$ etc.
using equations like (\ref{perteq6a}).

\newpage
%chamar

\section{Coefficients in (\protect\ref{KS7c}) and  (\protect\ref{KS7c1})}

\noindent A sample of the coefficients appearing in \eqref{KS7c1} are given below 
for the simplified OKS-BH geometry with 
approximate diagonal perturbations:
\bg \label{sperteq2} 
A^{00}_1&=&\frac{e^{-2\phi}}{32g^3h^{13/4}}\left(16h^2g^2\right)\nonumber\\
A^{10}_1&=&\frac{e^{-2\phi}}{32g^3h^{13/4}}\left(-16h^2g^4\right)\nonumber\\
A^{02}_1&=&A^{20}_1=\frac{e^{-2\phi}}{32g^3h^{13/4}}\left(16h^2g^3\right)\nonumber\\
A^{03}_1&=&A^{30}_1=\frac{e^{-2\phi}}{32g^3h^{13/4}}\left(32h^2g^3\right)\nonumber\\
A^{12}_1&=&\frac{e^{-2\phi}}{32gh^{13/4}}\left(16g^3h^2\right)\nonumber\\
A^{13}_1&=&\frac{e^{-2\phi}}{32gh^{13/4}}\left(32g^3h^2\right)\nonumber\\
A^{22}_1&=&\frac{e^{-2\phi}}{32g^2h^{13/4}}\left(16g^3h^2\right)\nonumber\\
A^{32}_1&=&A^{23}_1=\frac{e^{-2\phi}}{32g^2h^{13/4}}\left(-32g^3h^2\right)\nonumber\\
C^{00}_1&=&\frac{e^{-2\phi}}{32g^3h^{13/4}}\left(-40h^2g\frac{dg}{dr}-24h\frac{dh}{dr}g^2\right)\nonumber\\
C^{10}_1&=&\frac{e^{-2\phi}}{32g^3h^{13/4}}\left(8h^2g^3\frac{dg}{dr}+16h\frac{dh}{dr}g^4\right)\nonumber\\
C^{01}_1&=&\frac{e^{-2\phi}}{32g^3h^{13/4}}\left(-8h^2g^3\frac{dg}{dr}+24h\frac{dh}{dr}g^4\right)\nonumber\\
C^{20}_1&=&\frac{e^{-2\phi}}{32g^3h^{13/4}}\left(-24h\frac{dh}{dr}g^3\right)\nonumber\\
C^{02}_1&=&\frac{e^{-2\phi}}{32g^3h^{13/4}}\left(-24h\frac{dh}{dr}g^3-8h^2g^2\frac{dg}{dr}\right)\nonumber\\
C^{03}_1&=&\frac{e^{-2\phi}}{32g^3h^{13/4}}\left(-48h\frac{dh}{dr}g^3-16h^2g^2\frac{dg}{dr}\right)\nonumber\\
C^{30}_1&=&\frac{e^{-2\phi}}{32g^3h^{13/4}}\left(-48h\frac{dh}{dr}g^3\right)\nonumber\\
C^{11}_1&=&\frac{e^{-2\phi}}{32gh^{13/4}}\left(48h\frac{dh}{dr}g^4-24g^3h^2\frac{dg}{dr}\right)\nonumber\\
C^{12}_1&=&\frac{e^{-2\phi}}{32gh^{13/4}}\left(-16h\frac{dh}{dr}g^3+8h^2g^2\frac{dg}{dr}\right)\nonumber\\
C^{21}_1&=&\frac{e^{-2\phi}}{32gh^{13/4}}\left(32h^2g^2\frac{dg}{dr}-24h\frac{dh}{dr}g^3\right)\nonumber
\nd
\bg
C^{13}_1&=&\frac{e^{-2\phi}}{32gh^{13/4}}\left(-32h\frac{dh}{dr}g^3+16h^2g^2\frac{dg}{dr}\right)\nonumber\\
C^{31}_1&=&\frac{e^{-2\phi}}{32gh^{13/4}}\left(64h^2g^2\frac{dg}{dr}-48g^3h\frac{dh}{dr}\right)\nonumber\\
C^{32}_1&=&\frac{e^{-2\phi}}{32g^2h^{13/4}}\left(48g^3h\frac{dh}{dr}-32g^2h^2\frac{dg}{dr}\right)\nonumber\\
C^{23}_1&=&\frac{e^{-2\phi}}{32g^2h^{13/4}}\left(-32g^2h^2\frac{dg}{dr}+48g^3h\frac{dh}{dr}\right)\nonumber\\
C^{22}_1&=&\frac{e^{-2\phi}}{32g^2h^{13/4}}\left(-24g^3h\frac{dh}{dr}+16h^2g^2\frac{dg}{dr}\right)\nonumber
\nd
where 
$h$ is the warp factor measured on the slice \eqref{sol}, 
and $\phi$ is the dilaton. To get the explicit expressions for $A^{ij}_0, C^{ij}_0$ in 
\eqref{KS7c} we need to  
replace $h$ with $L^4/r^4$ and $e^{-2\phi}$ with $1/g_s^2$ in the above expressions. 
We have written the nonzero $A^{ij}_{1},C^{ij}_{1}$ and therefore the terms not appearing above are all zeroes.
Note also that $A^{ij}_{k}=B^{ij}_{k}, k = 0, 1$ and 
since we don't explicitly know the sources and the full background, we don't know the explicit 
expressions for the terms $D^{ij}_k, E^i_k,F^i_k$ etc. 

\newpage
\section{Detailed Viscosity Analysis}
As in the previous section, here we work out the 
coefficients in the quadratic action in (\ref{SV-8}) upto ${\cal O}(g_sN_f, g_sM^2/N)$: 
\bg \label{C-1}
&&A=\frac{1}{g_s^2}2g(r)r^5\left(1-\frac{3N_f g_s {\rm log}r}{2\pi}+\frac{N_f g_s}{\pi}\right)\nonumber\\
&&B=\frac{1}{g_s^2}\frac{3}{2}g(r)r^5\left(1-\frac{3N_f g_s {\rm log}r}{2\pi}+\frac{N_f g_s}{\pi}\right)\nonumber\\
&&C=\frac{1}{2g_s^2}\Bigg[r^4\left(\frac{3g_s M^2}{2\pi N}\right)+24r^4-8r_h^4+r_h^4\left(\frac{3g_s M^2}{2\pi
N}\right)
+\frac{N_fg_s}{4\pi}\Bigg\{150r^4\left(\frac{3g_s M^2}{2\pi N}\right){\rm log}r\nonumber\\
&&~~~~-96r^4+6r^4g\left(\frac{3g_s M^2}{2\pi N}\right)
+32r_h^4-48r_h^4{\rm log}r-6r_h^4\left(\frac{3g_s M^2}{2\pi N}\right){\rm log}r  \Bigg\}\Bigg]\nonumber\\
&&D=\frac{\tilde{V}}{g_s^2}\Bigg[-g(r)r^3\left(\frac{3g_s M^2}{2\pi N}\right)
\left\{1-\left(\frac{3g_s M^2}{2\pi N}\right){\rm log}r\right\}+\frac{N_fg(r)r^3}{16\pi}\left(\frac{3g_s^2 M^2}{2\pi N}\right)
\left(-28{\rm log}r-5\right)\nonumber\\
&&+\frac{81g_s^2M^2g(r)r^3\alpha'^2}{8L^4}+\frac{1}{4r^3}\Bigg\{16+48\left(\frac{3g_s M^2}{2\pi N}\right)^2({\rm log}r)^2
-8\left(\frac{3g_s M^2}{2\pi N}\right)+8\left(\frac{3g_s M^2}{2\pi N}\right)^2{\rm log}r\nonumber\\
&&+\left(\frac{3g_s M^2}{2\pi N}\right)^2\Bigg\}-\frac{r^3}{8\pi}\left\{-30\left(\frac{3g_s M^2}{2\pi N}\right)
({\rm log}r)^2-3\left(\frac{3g_s M^2}{2\pi N}\right){\rm log}r+12{\rm log}r+2\right\}\Bigg]\nonumber\\
&&{\cal M}(r)=-\frac{3g_sN_f}{2\pi
r}-g_s^2N_f^2\left(\frac{3{\rm log}(r)-2}{2\pi}\right)\left[\frac{3{\rm log}(r)-2}{2\pi}\left(\frac{g'(r)+\frac{5g(r)}{r}}{g}\right)+\frac{3}{2\pi r}\right]\nonumber\\
&&{\cal J}(r)=\frac{1}{r^2}\left[d_0+d_1{\rm log}(r)+d_2 ({\rm log}(r))^2\right]+\frac{rh^4}{r^6}\left[e_0+e_1{\rm
log}(r)+d_2 ({\rm log}(r))^2\right] +\frac{rh^8f_0}{r^{10}}\nonumber
\nd
where ${\cal J}(r)$ and ${\cal M}(r)$ appear in \eqref{SV-11b}. The 
coefficients appearing in the equations above are defined as:
\bg
&&d_0=-4g_s N_f+\frac{729g_s^2 M^4\alpha'^2}{16L^4\pi^2 N^2}+\frac{75g_s^2M^2 N_f}{16\pi N}-\frac{122g_s N_f}{4\pi}\nonumber\nonumber\\
&&d_1=4 \left(\frac{3g_s M^2}{2\pi N}\right)^2-28\frac{3g_s^2 M^2 N_f}{16\pi^2 N}-\frac{159g_s^2 M^2 N_f}{4\pi^2 N}-\frac{12g_s
N_f}{4\pi}\nonumber\\
&&d_2=-120\left(\frac{3g_s M^2}{2\pi N}\right)^2+30\left(\frac{3g_s^2 M^2 N_f}{8\pi^2 N}\right)\nonumber\\
&&e_0=\frac{-6g_s M^2}{2\pi N}+15\frac{3g_s^2 N_f M^2}{8\pi^2 N}-\frac{122g_s N_f}{4\pi}+\left(\frac{3g_s M^2}{2\pi N}\right)^2\frac{1}{2}
\nonumber\\
&&~~~~~~~~~~~~-\frac{24N_f g_s+\frac{9g_s^2 M^2 N_f}{2\pi N}}{4\pi}\nonumber
\nd 
\bg 
&&e_1=2\left(\frac{3g_s M^2}{2\pi N}\right)^2-\frac{159g_s^2 M^2 N_f}{4\pi^2 N}-\frac{12g_s
N_f}{4\pi}\nonumber\\
&&f_0=\left(\frac{3g_s M^2}{2\pi N}\right)^2\frac{1}{2}-\frac{24N_f g_s+\frac{9g_s^2 M^2 N_f}{2\pi N}}{4\pi}\nonumber 
\nd
Finally the 
perturbation coefficients $a_i, b_j$ appearing in the mode expansions for ${\cal G}$ and ${\cal H}$ in 
\eqref{SV-13a2} and \eqref{SV-13c1} respectively are given to ${\cal O}(g_sN_f, g_sM^2/N)$ as:
\bg 
a_0&=&a_1=0\nonumber\\
a_2&=&\frac{1}{2}\left[\frac{6g_s M^2}{\pi N}+\frac{122N_fg_s+12N_fg_s{\rm log}r_c}{4\pi}
+\frac{r_h^4}{r_c^4}\left(\frac{3g_s M^2}{\pi N}+\frac{122N_fg_s+12N_fg_s{\rm log}r_c}{4\pi}\right)\right]\nonumber\\
b_0&=&b_1=b_2=0\nonumber\\
b_3&=&\frac{8r_h^4}{3r_c^4}\left(1+\frac{r_h^4}{r_c^4}\right)a_2\nonumber
\nd

\newpage

\section{Detailed Entropy Analysis}
The coefficients in (\ref{action1a}), imposing an UV cut-off at $r = r_c$, are: 
\bg 
\bar{C}_1^{11}&=&\frac {r^6e^{-2\phi}}{32\sqrt{gg^{-1}_2}hg^3L^4r^4h^{1/4}}
\left(80rhg^2+8r^2hg^2g^{-1}_2\frac{dg_2}{dr}-4r^2g^2\frac{dh}{dr}-48r^2h\frac{dg}{dr}\right)\nonumber\\
\bar{C}^{12}_1&=&\frac{r^6e^{-2\phi}}{32\sqrt{gg^{-1}_2}hg^3L^4r^4h^{1/4}}
\left(-240g^3hr+12\frac{dh}{dr}g^3r^2-24r^2hg^3g^{-1}_2\frac{dg_2}{dr}+24r^2hg^2\frac{dg}{dr}\right)\nonumber\\
\bar{C}^{21}_1&=&\frac{r^6e^{-2\phi}}{32\sqrt{gg^{-1}_2}hg^3L^4r^4h^(1/4)}
\left(-240g^3hr+12\frac{dh}{dr}g^3r^2-24r^2hg^3g^{-1}_2\frac{dg_2}{dr}+48r^2hg^2\frac{dg}{dr}\right)\nonumber\\
\bar{C}^{22}_1&=&\frac{r^6e^{-2\phi}}{32\sqrt{gg^{-1}_2}hg^3L^4r^4h^{1/4}}
\left(-240g^4hr+12\frac{dh}{dr}g^4r^2-24r^2hg^3\frac{dg}{dr}-24r^2hg^4g^{-1}_2\frac{dg_2}{dr}\right)\nonumber\\
\bar{A}^{11}_1&=&\frac{r^6e^{-2\phi}}{32\sqrt{gg^{-1}_2}hg^3L^4r^4h^{1/4}}\left(16g^2r^2h\right)\nonumber\\
\bar{A}^{22}_1&=&\frac{r^6e^{-2\phi}}{32\sqrt{gg^{-1}_2}hg^3L^4r^4h^{1/4}}\left(-48g^4r^2h\right)\nonumber\\
\bar{A}^{21}_1&=&\frac{r^6e^{-2\phi}}{32\sqrt{gg^{-1}_2}hg^3L^4r^4h^{1/4}}\left(-48g^3r^2h\right)\nonumber\\
\bar{A}^{12}_1&=&\bar{A}^{21}_1\nonumber\\
\bar{B}^{11}_1&=&\frac{r^6e^{-2\phi}}{32\sqrt{gg^{-1}_2}hg^3L^4r^4h^{1/4}}\left(16g^2r^2h\right)\nonumber\\
\bar{B}^{12}_1&=&\frac{r^6e^{-2\phi}}{32\sqrt{gg^{-1}_2}hg^3L^4r^4h^{1/4}}\left(-48g^3r^2h\right)\nonumber\\
\bar{E}^1_1&=&\frac{r^6e^{-2\phi}}{8hg^2r^4L^4h^{1/4}}
\left(-40g^2hr+8r^2hg\frac{dg}{dr}+2g^2r^2\frac{dh}{dr}-4r^2g^2h{g^{-1}_2}\frac{dg_2}{dr}\right)\nonumber\\
\bar{E}^2_1&=&\frac{r^6e^{-2\phi}}{8hg^2r^4L^4h^{1/4}}
\left(6g^3r^2\frac{dh}{dr}-120g^3hr-12g^3g^{-1}_2hr^2\frac{dg_2}{dr}-12r^2g^2h\frac{dg}{dr}\right)\nonumber\\
\bar{F}^1_1&=&\frac{r^6e^{-2\phi}}{8hg^2r^4L^4h^{1/4}}\left(-8g^2r^2h\right)\nonumber\\
\bar{F}^2_1&=&\frac{r^6e^{-2\phi}}{8hg^2r^4L^4h^{1/4}}\left(-24g^3r^2h\right)\nonumber
\nd 
The nonzero coefficients in (\ref{energymeda}) are (taking the approximation
$g_1=g_2= 1-\frac{r_h^4}{r^4}+\alpha+\gamma\frac{r_h^8}{r^8}$)
\bg
&&H^{11[1]}=\frac{1}{32g_s^2\bar{\alpha}^3L^5(1+A{\rm log}~r+B{\rm log}^2 r)^{1/4}}
\Big[80\bar{\alpha}^2
-\frac{4\bar{\alpha}^2}{1+A{\rm log}~r+B{\rm log}^2 r}\{-4(1+A{\rm log}~r+B{\rm log}^2 r)\nonumber\\
&&+A+2B{\rm log}~r\}\Big]
-\frac{2}{L^5g_s^2\bar{\alpha}(1+A{\rm log}~r+B{\rm
log}^2 r)^{1/4}}
-\frac{4(1+A{\rm log}~r+B{\rm log}^2 r)-A-2B{\rm log}~r}{8\bar{\alpha}g_s^2L^5(1+A{\rm log}~r+B{\rm
log}^2r)^{5/4}}\nonumber
\nd
\bg 
&&+\frac{3g_sN_f}{4L^5g_s^2\bar{\alpha}\pi(1+A{\rm log}~r+B{\rm log}^2 r)^{1/4}}\nonumber\\
&&H^{12[1]}=\frac{1}
{32g_s^2L^5(1+A{\rm log}~r+B{\rm log}^2r)^{5/4}}\Big[-240(1+A{\rm log}~r+B{\rm log}^2r)
\nonumber\\
&&+12\{-4(1+A{\rm log}~r+B{\rm log}^2r)+A+2B{\rm log}~r\}\Big]
+\frac{6}{L^5g_s^2(1+A{\rm log}~r+B{\rm log}^2r)^{1/4}}
\nonumber\\
&&-\frac{3[-4(1+A{\rm log}~r+B{\rm log}^2r)+A+2B{\rm log}~r]}{8L^5g_s^2(1+A{\rm log}~r+B{\rm log}^2r)^{5/4}}
-\frac{9g_sN_f}{4L^5\bar{\alpha}\pi(1+A{\rm log}~r+B{\rm log}^2 r)^{1/4}}\nonumber\\
&&K^{21[1]}=-\frac{3}{2L^5g_s^2(1+A{\rm log}~r+B{\rm log}^2r)}\nonumber
\nd
The other coefficients $\widetilde{b}_{n(i)[1]}$ in \eqref{energymeda} are defined as:
\bg
&&\widetilde{b}_{0(0)[1]}=\frac{4gr_h^4}{2L^5g_s^2(1+A{\rm log}~r+B{\rm log}^2r)^{1/4}}\nonumber\\
&&\widetilde{b}_{0(4)[1]}=-\frac{32g \kappa c_3/L^2r_h^8}{2L^5g_s^2(1+A{\rm log}~r+B{\rm log}^2r)^{1/4}}\nonumber\\
&&\widetilde{b}_{1(0)[1]}=\frac{1}
{8L^5g_s^2(1+A{\rm log}~r+B{\rm log}^2r)^{5/4}}\Big[-6r_h^4\{-4(1+A{\rm log}~r+B{\rm log}^2r)+A+2B{\rm log}~r\}
\nonumber\\
&&+120r_h^4(1+A{\rm log}~r+B{\rm log}^2r)-96r_h^4(1+A{\rm log}~r+B{\rm log}^2r)+6r_h^4\{-4(1+A{\rm log}~r+B{\rm
log}^2r)\nonumber\\
&&+A+2B{\rm log}~r\}+\frac{72g_sN_f}{2\pi} r_h^4(1+A{\rm log}~r+B{\rm log}^2r)\Big]\nonumber\\
&&\widetilde{b}_{1(4)[1]}=\frac{1}
{8L^5g_s^2(1+A{\rm log}~r+B{\rm log}^2r)^{5/4}}\Big[24\kappa c_3L^{-2}r_h^8\{-4(1+A{\rm log}~r+B{\rm log}^2r)+A+2B{\rm log}
~r\}\nonumber\\
&&-480r_h^8\kappa c_3L^{-2}(1+A{\rm log}~r+B{\rm log}^2r)+384r_h^8\kappa c_3L^{-2}(1+A{\rm log}~r+B{\rm log}^2r)
\nonumber\\
&&-24\kappa c_3L^{-2}r_h^8\{-4(1+A{\rm log}~r+B{\rm log}^2r)+
A+2B{\rm log}~r\}\nonumber\\
&&-144(1+A{\rm log}~r+B{\rm log}^2r)g_sN_f L^{-2}r_h^8\kappa c_3/\pi\Big]\nonumber
\nd
where $A=\frac{3g_sM^2}{N}\left(1+\frac{3g_sN_f}{4\pi}\right)$, $B=\frac{9g_s^2M^2N_f}{8\pi^2N}$,
$\bar{\alpha}=1+\alpha=1+\frac{4c_3\kappa}{3L^2}$. We also have 
\bg
&&{s}_{00}^{(0)[1]}=-\frac{1}{2}\left(A{\rm log}~r+B{\rm log}^2r\right)\nonumber\\
&&{s}_{00}^{(4)[1]}=\frac{r_h^4}{2}\left(A{\rm log}~r+B{\rm log}^2r\right)\nonumber\\
&&{s}_{11}^{(0)[1]}=-A{\rm log}~r-B{\rm log}^2r\nonumber
\nd  
with every other ${s}_{nn}^{(i)[1]}=0$. Note that (a) all coefficients are suppressed as 
${\cal O}(g_sN_f, g_sM^2/N)$ as expected, and (b) an appropriate UV cap will make all ${s}_{mn}^{(k)[1]}$
independent of $r$, as we discussed in the main text.

\newpage
%\section*{References}

\end{document}